\documentclass[times,review,preprint]{elsarticle}

\usepackage{url}
\usepackage{xcolor}

\usepackage{jcomp}
\usepackage{framed,multirow}

\usepackage{amssymb}
\usepackage{latexsym}

\usepackage{natbib}
\usepackage{amsmath,amssymb}
\usepackage{psfrag,pstool,graphicx,epsfig}
\usepackage{color,psfrag,colordvi,graphics}
\usepackage{multirow}
  \usepackage{framed} % Framing content
    \usepackage{multicol} % Multiple columns environment
\usepackage{tikz}

\usepackage{nomencl} % Nomenclature package
    \makenomenclature
    \renewcommand*\nompreamble{\begin{multicols}{2}}
    \renewcommand*\nompostamble{\end{multicols}}

\def\Rey{\mbox{\text{Re}}}

\newdefinition{rmk}{Remark}

\usepackage{amsthm}

\newcommand{\be}{{\bf e}}
\newcommand{\mbD}{{\mathbb{D}}}

\newcommand{\bu}{{\bf u}}
\newcommand{\bv}{{\bf v}}
\newcommand{\bw}{{\bf w}}

\newcommand{\btau}{\boldsymbol\tau}

\newcommand{\bX}{{\bf X}}
\newcommand{\bY}{{\bf Y}}
\newcommand{\dive}{\nabla\cdot}

\newcommand{\bpsi}{\boldsymbol\psi}
\newcommand{\bPsi}{\boldsymbol\Psi}

\newcommand{\bnu}{\boldsymbol\nu}

\def\bx{{\bf x}}

\definecolor{newcolor}{rgb}{.8,.349,.1}

\journal{Journal of Computational Physics}

\begin{document}

\verso{J. Venkatesan and S. Ganesan}

\begin{frontmatter}

\title{Local projection stabilized finite element modeling of viscoelastic two-phase flows}

\author{Jagannath Venkatesan}
\ead{jagannathv@iisc.ac.in}

\author{Sashikumaar Ganesan\corref{cor1}}
\ead{sashi@iisc.ac.in}
 
\cortext[cor1]{Corresponding author}

\address{Computational Mathematics Group, Department of Computational and Data Sciences,
 Indian Institute of Science, Bangalore-560012, India}

% \received{1 May 2013}
% \finalform{10 May 2013}
% \accepted{13 May 2013}
% \availableonline{15 May 2013}
\communicated{S. Ganesan}

\begin{keyword}
\KWD \\
 Rising bubble \\
 Viscoelastic fluids \\
 Giesekus model \\
 Finite elements \\
 ALE approach \\
 Local Projection Stabilization 
\end{keyword}

\begin{abstract}
A three-field local projection stabilized finite element method is developed for computations of a 3D-axisymmetric buoyancy driven bubble rising in a liquid column in which either the bubble or the liquid column can be viscoelastic.
The two-phase flow is described by the time-dependent incompressible Navier--Stokes equations, whereas the viscoelasticity is modeled by the Giesekus constitutive equation in a time-dependent domain.
The arbitrary Lagrangian Eulerian~(ALE) formulation with finite elements is used to solve the governing equations in the time-dependent domain.
The interface-resolved moving meshes in ALE allows to incorporate the interfacial tension force and jumps in the material parameters accurately.
An one-level Local Projection Stabilization~(LPS), which is based on an enriched approximation space and a discontinuous projection space, where both spaces are defined on a same mesh is used to stabilize the model equations. 
The stabilized numerical scheme allows us to use equal order interpolation spaces for the velocity and the viscoelastic stress, whereas inf-sup stable finite elements are used for the velocity and the pressure.
A comprehensive numerical investigation is performed for a Newtonian bubble rising in a viscoelastic fluid and a viscoelastic bubble rising in a Newtonian fluid.
The influence of the viscosity ratio, Newtonian solvent ratio, Giesekus mobility factor and the E\"{o}tv\"{o}s number on the bubble dynamics are analyzed.
The numerical study shows that a Newtonian bubble rising in a viscoelastic fluid experiences an extended trailing edge with a cusp-like shape and  also exhibits the negative wake phenomena.
However, a viscoelastic bubble rising in a Newtonian fluid develops an indentation around the rear stagnation point with a dimpled shape.
\end{abstract}

\end{frontmatter}

\begin{table*}[ht!]
      \begin{framed}
        \printnomenclature
      \end{framed}
    \end{table*}

\section{Introduction}
Multiphase flows of two immiscible fluids are encountered in many industrial processes such as enhanced oil recovery, emulsions in colloid and interface science, polymer blends, droplet based microfluidics, plastic profile extrusion and medical applications in the case of blood pumps.
Viscoelasticity plays a prominent role in the aforementioned applications. 
The fundamental understanding of the effects of viscoelasticity in multiphase flows is crucial as these effects directly impact the design and optimization of engineering processes  subjected to complex interfacial flow dynamics.
Therefore, scientific studies on a single bubble rising in a fluid column due to buoyancy with viscoelastic effects are highly demanded.

Due to the inherent complexity of viscoelastic fluids and the resulting analytic intractability of the mathematical models, theoretical predictions of rising viscoelastic bubble behaviour are very challenging or nearly impossible to obtain.
The effects of viscoelasticity on the bubble behavior have been investigated experimentally by a few researchers~\cite{Liu95, Sostarecz03, Pilz07, Amirnia13, Xu17}. 
With recent advancement in numerical techniques and computational capabilities using high performance computing, the use of high-fidelity numerical simulations is an useful and viable tool to understand the complex flow dynamics.

In spite of significant progress made in the development of numerical methods for simulation of viscoelastic single-phase flows, computational methods for viscoelastic two-phase flows is gaining rapid attention only very recently~\cite{Daulet15, Figueiredo16, Walters16, Habla11, Zainali13}.  
Numerical computations of incompressible viscoelastic flows involve simultaneous solution of the Navier--Stokes equations and an equation for the evolution of viscoelastic stresses. 
Mathematical models for the evolution of viscoelastic stresses can be classified into two categories: kinetic theory models and continuum mechanics models. 
The kinetic theory approach attempts to model the polymer dynamics by using a coarse-grained description of polymer chains by representing them as chains of springs or rods which eventually lead to the Fokker--Planck equation.  
Continuum approach attempts to provide constitutive differential equations, where the micro properties are obtained empirically.  
Oldroyd-B~\cite{Oldroyd}, Giesekus~\cite{Giesekus}, finitely extensible non-linear elastic~(FENE-P~\cite{FENEP}, FENE-CR~\cite{FENECR}), Phan-Thien-Tanner (PTT) \cite{PTT} and eXtended Pom-Pom~(XPP)~\cite{XPP} are the commonly used  continuum models in the literature.
In this study, we use the continuum models as they are computationally less expensive compared to the kinetic theory models.
In particular, we consider the Giesekus constitutive model as it models shear-thinning and elasticity together.

In addition to the challenges associated with the viscoelastic flows, the main challenge in the numerical simulation of interface flows is the tracking/capturing of the moving interface.
Further, precise inclusion of the interfacial tension force and the local curvature on the interface is very challenging. 
Moreover, care needs to be taken to handle the jumps in the material properties~(viscosity, density, relaxation time of polymers) across the interface.
Most importantly the numerical scheme should not induce spurious velocities and should conserve the mass.
Further, the advective nature of the viscoelastic constitutive equation becomes dominant when the Weissenberg number (measure of the elasticity of fluid) is high.
This necessitates the use of an accurate and robust stabilized numerical scheme to avoid global oscillations in the numerical solution. 

We now briefly review some of the numerical schemes used to simulate viscoelastic two-phase flows and the list is not exhaustive.
Pillapakkam et. al.~\cite{Pillapakkam01, Pillapakkam07}  developed a finite element code based on level-set method to examine the transient motion of bubbles rising in a viscoelastic liquid modeled by the Oldroyd--B equation. 
Further, Chinyoka et. al.~\cite{Chinyoka05} investigated an Oldroyd--B droplet deforming under simple shear using volume-of-fluid and finite difference method.
In addition, Habla et. al.~\cite{Habla11} developed a volume-of-fluid methodology using the OpenFOAM CFD toolbox to simulate transient and steady-state viscoelastic droplet flow in shear and elongational flows. 
Further, Harvie et. al.~\cite{Harvie08} studied the dynamics of an Oldroyd--B droplet passing through a microfluidic contraction using volume-of-fluid and finite volume method. 
Moreover, Yue et. al.~\cite{Yue05, Yue06} introduced a phase field method for computing interfacial dynamics in viscoelastic fluids using finite elements. 
In addition, Zhang et. al.~\cite{Zhang10} proposed a moving finite element method based on phase-field method to simulate interfacial dynamics of two-phase viscoelastic flows.
You et. al.~\cite{You08, You09} proposed a finite volume based boundary-fitted grid method for computations of an axisymmetric bubble rising in viscoelastic fluids using FENE-CR model.
Further, Chung et. al.~\cite{Chung08, Chung09} implemented a finite element-front tracking method to understand the effects of viscoelasticity using Oldroyd--B model on drop deformation in simple shear and 5:1:5 planar contraction/expansion micro-channels.
In addition, Mukherjee et. al.~\cite{Mukherjee10, Mukherjee11} numerically investigated the deformation of an Oldroyd--B drop in a Newtonian fluid using a front-tracking finite difference method. 
Moreover, Zainali et. al.~\cite{Zainali13} presented an improved smoothed particle hydrodynamics method for simulation of a buoyancy driven Newtonian bubble rising in an Oldroyd--B fluid.
Further, Vahabi and Sadeghy~\cite{Vahabi14} developed a weakly compressible smoothed particle hydrodynamics method for simulating bubble rising in Oldroyd--B fluids.
In addition, Lind and Phillips~\cite{Lind10} used a boundary element method to study the dynamics of rising gas bubbles. 
Moreover, Walters and Phillips~\cite{Walters16} developed a non-singular boundary element method for modeling bubble dynamics in viscoelastic fluids.
Recently, Izbassarov and Muradoglu~\cite{Daulet15, Daulet16} proposed a front tracking method for the simulation of viscoelastic two-phase flow systems in a buoyancy and pressure driven flow through a capillary tube with/without sudden contraction and expansion using Oldroyd--B, FENE-CR and FENE-MCR models.

In this paper, we present an arbitrary Lagrangian Eulerian~(ALE) based finite element scheme for computations of a buoyancy driven 3D-axisymmetric bubble rise in a fluid column with viscoelastic effects using Giesekus model.
The choice of ALE approach avoids fast distortion of meshes, which is the case in Lagrangian method. 
Since, the interface is resolved by the computational mesh, the interfacial force and the different material properties in different phases can be incorporated very accurately in the ALE approach.
The spurious velocities, which might arise due to the approximation errors of the pressure and the interfacial force, can be suppressed by using this approach~\cite{GMT}. 
We use the tangential gradient operator technique to treat the local curvature in a semi-implicit manner~\cite{GANJCP15} and it avoids explicit computation of the curvature.
Further, in contrast to the standard approach of using the differential equations in the cylindrical coordinates and seeking a suitable variational form, we derive the 3D-axisymmetric weak form directly from the weak form in 3D-Cartesian coordinates, refer~\cite{GAN07, ViscoelasticDropLPS18}.
Since the advective nature of the viscoelastic constitutive equation becomes dominant when the Weissenberg number is high, an appropriate stabilized numerical scheme needs to be used.
In the context of stabilization schemes for viscoelastic flows, several schemes such as the Streamline Upwind Petrov Galerkin (SUPG) method~\cite{BH82}, Discrete Elastic Viscous Stress Splitting (DEVSS)~\cite{Guenette95, Fortin00}, Discontinuous Galerkin (DG) method \cite{Fortin89b}, Galerkin Least Squares~\cite{BEHR06} and Variational Multiscale method~\cite{MASUD10, Codina14} have been proposed in the literature.  
Further, Log-Conformation reformulation method~\cite{Fattal05} has also been used in several computations of viscoelastic two-phase flows. 
Recently, a three-field Local Projection Stabilized~(LPS) finite element scheme for simulation of viscoelastic fluid flows in fixed domains has been presented by Venkatesan and Ganesan~\cite{VJSGLPS17}.
In this work, we extend the LPS scheme proposed in~\cite{VJSGLPS17} for finite element computations of 3D-axisymmetric  viscoelastic two-phase flows.
Local Projection Stabilization is used in the numerical scheme to handle the convective nature of the viscoelastic constitutive equation and to use equal order interpolation spaces for the velocity and the viscoelastic stress.

The novelty of the present work can be summarized as follows~:
\begin{itemize}
 \item Arbitrary Lagrangian--Eulerian approach with finite elements for 3D-axisymmetric viscoelastic two-phase flows.
 \item Local Projection Stabilization method to handle the advective nature of viscoelastic flows with moving interface.
 \item The Giesekus constitutive model is used for understanding the rising bubble phenomena with shear thinning and elastic  effects.
 \item Comprehensive numerical investigation of the rising bubble dynamics is performed with viscoelastic effects using the following metrics~: bubble shape, sphericity of bubble, diameter of the bubble at the axis of symmetry, kinetic energy, elastic energy, rise velocity, center of mass of the bubble and viscoelastic stress contours.  
 \end{itemize}

The paper is organized as follows.
The governing equations for buoyancy driven viscoelastic two-phase flows and its dimensionless form are presented in Section~\ref{mathmodel}.
Section~\ref{NumericalScheme} describes the proposed numerical scheme.
We first introduce the ALE formulation for time-dependent domains and the governing equations are rewritten in the ALE frame. 
Further, we derive the  variational form of the model equations and its axisymmetric form using cylindrical coordinates.
The spatial and temporal discretization used in the numerical scheme are then outlined.
The linearization strategy and the linear elastic mesh update technique for handling the inner mesh points in the computational domain is then explained. 
Section~\ref{Results} is concerned with the computational results. 
The numerical scheme is first validated for a Newtonian bubble rising in a Newtonian fluid column using a benchmark configuration. 
Then, we perform a grid independence test for the same benchmark configuration.
Further, a comprehensive numerical investigation on the Newtonian bubble rising in a viscoelastic fluid and a viscoelastic bubble rising in a Newtonian fluid is presented.
We study the influence of the viscosity ratio, Newtonian solvent ratio, Giesekus mobility factor and the E\"{o}tv\"{o}s number on the rising bubble dynamics.
Finally, a brief summary of the proposed numerical scheme and the key observations are presented in Section~\ref{Summary}.

\section{Mathematical Model}\label{mathmodel}
\subsection{Governing Equations}
\begin{figure}[ht!]
\begin{center}
\unitlength5mm
\footnotesize
\begin{picture}(6,12.0)
\put(3,5.7){\makebox(0,0){\includegraphics[width=8.5cm]{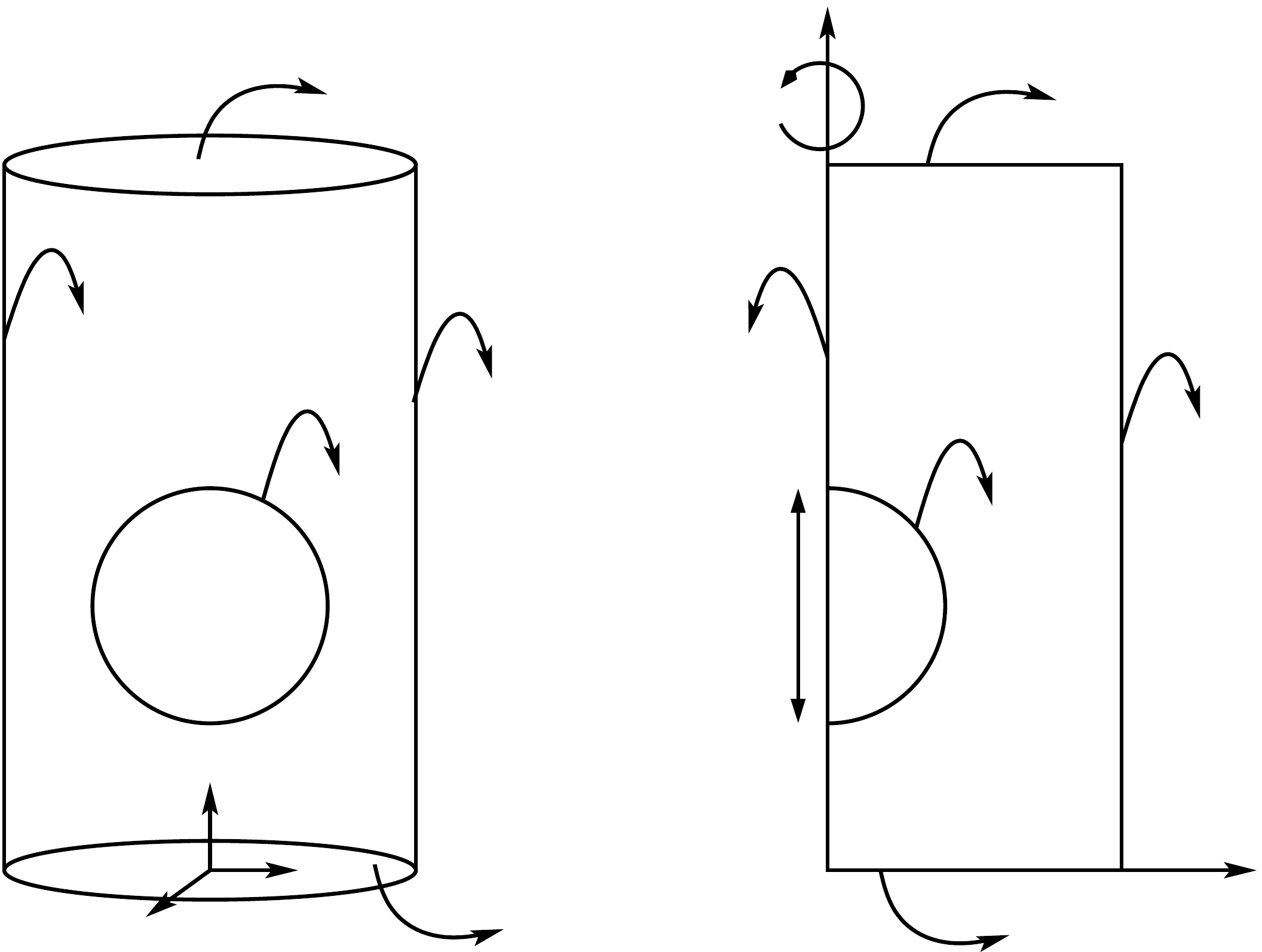}}}
\put(8.8 ,10.5){$\Gamma_D$}
\put(3.9 ,7.3){$\Gamma_{\text{Axial}}$}
\put(8.2 ,-0.7){$\Gamma_D$}
\put(10.5 ,6.0){$\Gamma_N$}
\put(7.5,4.8){$\Gamma_F(t)$}
\put(5.85 ,3.9){$\Phi_1(t)$}
\put(7.0 ,7.5){$\Phi_2(t)$}
\put(11.5 ,0.25){$\bf r$}
\put(5.5 ,12.1){$\bf z$}
\put(-3.9 ,-0.5){$\bf x$}
\put(-1.4 ,0.3){$\bf y$}
\put(-2.8 ,1.6){$\bf z$}
\put(4.65 ,3.7){$D$}
\put(5.0, -0.2){$(0,0)$}
\put(8.85, -0.2){$(0.5,0)$}
\put(3.7, 1.95){$(0,0.25)$}
\put(3.95, 5.65){$(0,0.5)$}
\put(4.1, 9.6){$(0,h_c)$}
\put(9.75, 9.7){$(0.5,h_c)$}
\put(-3.0 ,3.9){$\Omega_1(t)$}
\put(-3.0 ,7.5){$\Omega_2(t)$}
\put(-1.5,5.3){$\Gamma_F(t)$}
\put(1.4 ,-0.6){$\Gamma_D$}
\put(-1.0 ,10.6){$\Gamma_D$}
\put(0.9 ,6.5){$\Gamma_N$}
\put(-4.5 ,7.4){$\Gamma_N$}
\end{picture}
\end{center}
\caption{Computational model of viscoelastic two-phase flow.\label{domain}}
\end{figure}
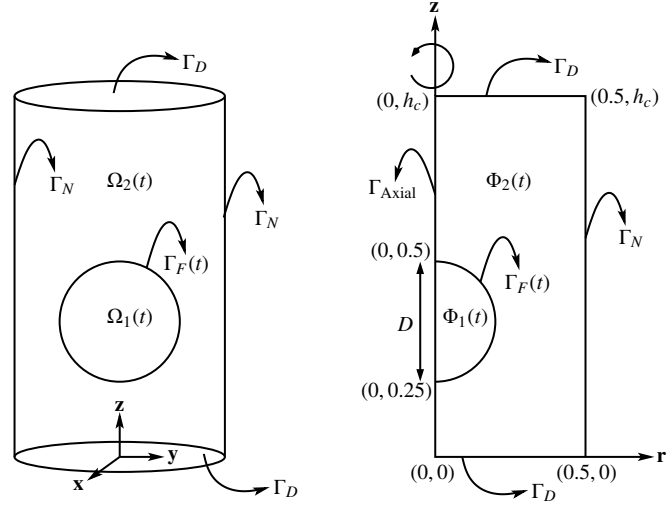

We consider a two-phase viscoelastic flow (either phase can be viscoelastic) in a bounded domain $\Omega \subset \mathbb{R}^3$ with a Lipschitz continuous boundary $\partial \Omega$.
We assume that the fluid is incompressible, immiscible and the material properties such as density, viscosity and relaxation time of polymers are constant. 
The schematic representation of the computational model is shown in Fig.~\ref{domain}.
The computational domain is denoted by $\Omega(t) := \Omega_1(t) \cup \Gamma_F(t) \cup \Omega_2(t)$, where a liquid droplet filling $\Omega_1(t)$ is completely surrounded by another liquid filling the domain $\Omega_2(t)$.
Further, the interface between the two liquids is denoted by $\Gamma_F(t)$, whereas $\Gamma_{\text{Axial}}$, $\Gamma_D$ and $\Gamma_N$  denote the symmetry of axis, Dirichlet and Neumann boundaries, respectively.
Note that the boundary of the computational domain $\Omega(t)$ is fixed over time. 
Here, $t$ is the time in a given time interval $[0,\text{I}]$ with an end time I.

The fluid flow in $\Omega(t)$ is described by the time-dependent incompressible Navier--Stokes equations~:
\begin{equation}
\begin{aligned} \label{NSE1}
 \rho_k \left( \frac{\partial \bu}{\partial t} +  \left( \bu \cdot \nabla\right) \bu \right) - \nabla \cdot \mathbb{T}_k(\bu,p,\btau_p) &=  \rho_k \, g \, \be \quad &\text{in} \quad \Omega_k(t) \times (0,\text{I}], \\  
 \nabla \cdot \bu &= 0 \quad &\text{in} \quad \Omega_k(t) \times (0,\text{I}],
\end{aligned}
\end{equation}
for $k=1,2$. Here, $\bu$ is the fluid velocity, $p$ is the pressure in the fluid, $\btau_p$ is the viscoelastic conformation stress, $g$ is the gravitational constant, $\be$ is an unit vector in the direction opposite to the gravitational force and $\rho_k$ is the density of fluid in $\Omega_k(t)$, $k=1,2$, respectively.
For an incompressible viscoelastic fluid, the stress tensor $\mathbb{T}_k(\bu, p, \btau_p)$ and the velocity deformation tensor $\mathbb{D}(\bu)$ are given by 
\begin{align*}
 \mathbb{T}_k(\bu, p, \btau_p) = 2 \mu_{s,k} \mathbb{D}(\bu) - p \mathbb{I} + \frac{\mu_{p,k}}{\lambda_k} \left( \btau_p - \mathbb{I} \right), \quad \mathbb{D}(\bu) = \frac{1}{2}\left( \nabla \bu + \nabla \bu^{\text{T}} \right),
\end{align*}
where  $\mu_{s,k}$ is the Newtonian solvent viscosity, $\mu_{p,k}$ is the polymeric viscosity, $\mu_{0,k} = \mu_{s,k} + \mu_{p,k}$ is the total viscosity,  $\mathbb{I}$ is the identity tensor and $\lambda_k$ is the relaxation time of the polymers  in $\Omega_k(t)$, $k=1,2$, respectively.

The Giesekus model~\cite{Giesekus}  is adopted as a constitutive equation for the viscoelastic stresses and it is given by 
\begin{align}
 \frac{\partial \btau_p}{\partial t} + \left( \bu \cdot \nabla \right) \btau_p - \nabla \bu^{\text{T}}\cdot \btau_p - \btau_p \cdot \nabla \bu + \frac{1}{\lambda_k} \left[ \left(\btau_p - \mathbb{I} \right) + \alpha_k \left( \btau_p - \mathbb{I} \right)^2 \right] &= 0 \quad &\text{in} \quad \Omega_k(t) \times (0,\text{I}], \label{Giesekus1}
\end{align}
for $k=1,2$ , where  $\alpha_k$ is the Giesekus mobility factor.
Note that, one can obtain the Oldroyd--B constitutive equation~\cite{Oldroyd} by setting the Giesekus mobility parameter to zero, i.e. $\alpha_k = 0$ in \eqref{Giesekus1}.
The coupled Navier--Stokes~\eqref{NSE1} and Giesekus constitutive~\eqref{Giesekus1} equations are closed with initial and boundary conditions. 
At time t = 0, we specify the conformation stress tensor $\btau_{p,0}$  and the divergence-free velocity field $\bu_0$ over the entire computational domain $\Omega_0$, i.e.,
\begin{align*}
 \Omega(0) = \Omega_0, \quad \bu(\cdot , 0) = \bu_0 \; \text{in} \; \Omega_0, \quad \btau_p(\cdot, 0) = \btau_{p,0} \; \text{in} \; \Omega_0.
\end{align*}
On the interface $\Gamma_F(t)$, we impose the kinematic condition
\begin{align}\label{kinematiccondition}
 \bu \cdot \bnu_F &= \bw \cdot \bnu_F  \quad  \text{on} \quad \Gamma_F (t)\times (0,\text{I}],
\end{align}
and force balancing conditions
\begin{align*}
 [|\bu|] = 0, \quad [ | \mathbb{T}(\bu, p, \btau_p) | ] \cdot \bnu_F = \nabla_{\Gamma_F} \cdot \mathbb{S}_{\Gamma_F}  \quad  \text{on} \quad \Gamma_F (t)\times (0,\text{I}].
\end{align*}
Here, $\bw$ is the domain velocity, $\bnu_F$  is an unit normal vector  on $\Gamma_F(t)$ and $[|\cdot|]$ denotes the jump of a function at the interface.
Further, we define the surface gradient of a scalar function $\psi$ and the surface divergence of a vector function $\bv$ on the interface $\Gamma_F(t)$ by
\begin{align*}
 \nabla_{\Gamma_{\bnu_F}} \psi = \mathbb{P}_{\bnu_F} \nabla \psi, \quad \nabla_{\Gamma_{\bnu_F}} \cdot \bv = \text{tr} \left( \mathbb{P}_{\bnu_F} \nabla \bv \right),
\end{align*}
where $\mathbb{P}_{\bnu_F} = \mathbb{I} - \bnu_F \otimes \bnu_F$ is the projection onto the tangential plane of $\Gamma_F(t)$. 
The interface stress tensor $\mathbb{S}_{\Gamma_F}$ is modeled by $\mathbb{S}_{\Gamma_F} = \sigma \mathbb{P}_{\bnu_F}$, where $\sigma$ is the interfacial tension.
Next, we assume that the boundary $\partial \Omega$ := $\Gamma_D \cup \Gamma_N$ of the computational domain $\Omega(t)$ is fixed in time and we impose the no-slip condition
\begin{align*}
  \bu = 0 \quad  \text{on} \quad \Gamma_D \times (0,\text{I}],
\end{align*}
and the free slip condition
\begin{align}\label{freeslipbc}
   \btau_N \cdot \mathbb{T}_2 (\bu, p, \btau_p) \cdot \bnu_N = 0, \quad \bu \cdot \bnu_N = 0 \quad  \text{on} \quad \Gamma_N \times (0,\text{I}], 
\end{align}
where  $\btau_N$ and $\bnu_N$ are unit tangential and normal vectors respectively on $\Gamma_N$.

\subsection{Non-dimensional form of the governing equations}
Let L and $\text{U}_\infty$ be the characteristic length and velocity, respectively.
We now define the following dimensionless variables 
\begin{align*}
 \tilde{x} = \frac{x}{\text{L}}, \quad \tilde{\bu} = \frac{\bu}{\text{U}_{\infty}}, \quad \tilde{\bw} = \frac{\bw}{\text{U}_{\infty}}, \quad \tilde{t} = \frac{t \text{U}_{\infty}}{\text{L}}, \quad \tilde{p} = \frac{p}{\rho_2 \text{U}_\infty^2}, \quad \tilde{I} = \frac{I \text{U}_{\infty}}{\text{L}}, \quad \tilde{\btau} = \btau, \quad \varepsilon = \frac{\mu_{0,2}}{\mu_{0,1}}.
\end{align*}
Here, $\varepsilon$ is the ratio between the total viscosity of outer and inner phases. 
In addition, we define the non-dimensional density $\rho$, Newtonian solvent ratio $\beta$, Giesekus mobility factor $\alpha$, Reynolds number $\text{Re}$ and Weissenberg number $\text{Wi}$ as
\[
 \rho = \begin{cases}
             \rho_1 / \rho_2   &\forall ~\bx \in \Omega_1(t), \\
             1    &\forall ~\bx \in \Omega_2(t),
            \end{cases}
            \quad 
    \beta = \begin{cases}
             \beta_1 = \mu_{s,1} / \mu_{0,1} &\forall ~\bx \in \Omega_1(t), \\
             \beta_2 = \mu_{s,2} / \mu_{0,2} &\forall ~\bx \in \Omega_2(t),
            \end{cases}
             \quad 
              \alpha = \begin{cases}
             \alpha_1    &\forall ~\bx \in \Omega_1(t), \\
             \alpha_2   &\forall ~\bx \in \Omega_2(t),
            \end{cases}
\]
\[
 \text{Re} = \begin{cases}
             \varepsilon \text{Re}_2 &\forall ~\bx \in \Omega_1(t), \\
             \text{Re}_2  &\forall ~\bx \in \Omega_2(t),
            \end{cases}
            \quad 
             \text{Re}_2 = \frac{\rho_2\text{U}_\infty \text{L}}{\mu_{0,2}},
            \quad 
            \text{Wi} = \begin{cases}
              \text{Wi}_1 = \lambda_1\text{U}_\infty / \text{L}   &\forall ~\bx \in \Omega_1(t), \\
             \text{Wi}_2 = \lambda_2\text{U}_\infty / \text{L}  &\forall ~\bx \in \Omega_2(t),
            \end{cases}  .
\]
Using these non-dimensional parameters in the governing equations and omitting the tilde afterwards, we obtain the dimensionless form of the governing equations for the two-phase viscoelastic flow as 
\begin{equation}
\begin{aligned}\label{nondimeqns}
 \rho  \left( \frac{\partial \bu}{\partial t} +  \left( \bu \cdot \nabla \right) \bu \right) -  \nabla \cdot \mathbb{T}(\bu, p, \btau_p) &=   \frac{\rho \, \be}{\text{Fr}} \enskip &\text{in} \enskip &\Omega(t) \times (0,\text{I}], \\
 \nabla \cdot \bu &= 0 \enskip &\text{in} \enskip &\Omega(t) \times (0,\text{I}], \\
\frac{\partial \btau_p}{\partial t} + \left( \bu \cdot \nabla \right) \btau_p - \nabla \bu^{\text{T}}\cdot \btau_p - \btau_p \cdot \nabla \bu + \frac{1}{\text{Wi}} \left[ \left(\btau_p - \mathbb{I} \right) + \alpha \left( \btau_p - \mathbb{I} \right)^2 \right] &= 0 \enskip &\text{in} \enskip &\Omega(t) \times (0,\text{I}], \\
\bu \cdot \bnu_F = \bw \cdot \bnu_F, \quad [ | \mathbb{T}(\bu, p, \btau_p)|] \cdot \bnu_F = \frac{1}{\text{We}} \nabla_{\Gamma_{\bnu_F}} \cdot \mathbb{P}_{\bnu_F}, \quad [|\bu|] &= 0   \enskip  &\text{on} \enskip &\Gamma_F (t)\times (0,\text{I}], \\
 \bu &= 0 \enskip  &\text{on} \enskip &\Gamma_D \times (0,\text{I}], \\
 \btau_N \cdot \mathbb{T}_2 (\bu, p, \btau_p) \cdot \bnu_N = 0, \quad \bu \cdot \bnu_N &= 0 \enskip  &\text{on} \enskip &\Gamma_N \times (0,\text{I}],
\end{aligned}
\end{equation}
with the dimensionless numbers~(Froude and Weber numbers, respectively)
\begin{align*}
 \text{Fr} = \frac{\text{U}_\infty^2}{\text{L}g}, \quad \text{We} = \frac{\rho_2 \text{U}_\infty^2 \text{L}}{\sigma},
\end{align*}
and the dimensionless stress tensor
\begin{align*}
 \mathbb{T}(\bu, p, \btau_p) = \frac{2 \beta}{\text{Re}} \mathbb{D}(\bu) - p \mathbb{I} + \frac{\left( 1 - \beta\right)}{\text{Re}\text{Wi}} \left( \btau_p - \mathbb{I}  \right).
\end{align*}
Often the characteristic velocity in interface flows is chosen as $\text{U}_\infty = \sqrt{\text{L}g}$ and
in this case, the Weber number will become E\"{o}tv\"{o}s number,
\begin{align*}
 \text{Eo} = \frac{\rho_2 g \text{L}^2}{\sigma}
\end{align*}
and the Froude number will reduce to one.

\section{Numerical Scheme}\label{NumericalScheme}
\subsection{Arbitrary Lagrangian--Eulerian (ALE) formulation for time-dependent domain}
The time-dependent sub-domains and the interface are tracked using the arbitrary Lagrangian--Eulerian~(ALE) approach with moving meshes.
Let $\hat{\Omega} := \hat{\Omega}_1 \cup \hat{\Gamma}_F \cup \hat{\Omega}_2$ be a reference domain of $\Omega(t)$ and then, we define a family of ALE mappings
\begin{align*}
 \mathcal{A}_t:\hat{\Omega} \rightarrow \Omega(t),  \qquad  \mathcal{A}_t(\bY)= \bX(\bY, t), \qquad t\in(0,\text{I}),
\end{align*}
where $\bX \in \Omega(t)$ and $\bY \in \hat{\Omega}$ are the Eulerian and ALE coordinates, respectively.
In computations, we take the previous time-step domain as the reference domain. 
To rewrite the model equations into a non-conservative ALE form, the time derivative has to be replaced with the time derivative on the reference frame and it results in an addition of convective domain velocity term in the equations, for more details we refer to~\cite{GANJCP15, GAN07, ViscoelasticDropLPS18}. 
Incorporating it, the ALE form of the time-dependent Navier--Stokes equations can be written as~: 
\begin{align}\label{NSEALE}
 \left.\nabla \cdot \bu = 0, \quad \rho \left( \frac{\partial \bu}{\partial t}\right|_{\hat{\Omega}} +  \left( (\bu - \bw) \cdot \nabla \right) \bu \right) -  \nabla \cdot \mathbb{T}(\bu, p, \btau_p) &=  \frac{ \rho \, \be}{\text{Fr}} \quad &\text{in} \quad \Omega(t) \times (0,\text{I}],
\end{align}
whereas, the ALE form of the Giesekus constitutive equation is given by
\begin{align}\label{GiesekusALE}
\left.\frac{\partial \btau_p}{\partial t}\right|_{\hat{\Omega}} + \left( (\bu - \bw) \cdot \nabla \right) \btau_p - \nabla \bu^{\text{T}}\cdot \btau_p - \btau_p \cdot \nabla \bu + \frac{1}{\text{Wi}} \left[ \left(\btau_p - \mathbb{I} \right) + \alpha \left( \btau_p - \mathbb{I} \right)^2 \right] = 0 \quad \text{in} \quad \Omega(t) \times (0,\text{I}].
\end{align}
Further, we assume that the topology of the computational domain does not change during the computations.

\subsection{Variational formulation}
Let $\text{L}^2(\Omega(t))$ and $\text{H}^1(\Omega(t))$ be the standard Sobolev spaces and $(\cdot,\cdot)$ be the inner product in $\text{L}^2(\Omega(t))$ and its vector/tensor-valued versions, respectively. 
We define the velocity, pressure and viscoelastic stress spaces as 
\begin{align*}
V(\Omega(t))&:=\left\{ \, \bv\in \text{H}^1(\Omega(t))^3 \, : \, \bv \cdot \bnu_N = 0 \; \text{on} \; \Gamma_N, \quad \bv = 0 \; \text{on} \; \Gamma_D \, \right\}, \\
 Q(\Omega(t))&:=\left\{\, q \in \text{L}^2(\Omega(t)) \, : \, \int_{\Omega} q \, dx = 0 \, \right\}, \\
 S(\Omega(t))&:= \left\{ \, \bpsi = [\psi_{ij}], \, 1 \leq i,j \leq 3 \, : \quad  \psi_{ij} \in H^1(\Omega(t)),  \quad {\psi_{ij} = \psi_{ji}} \, \right\}. 
\end{align*}
We now multiply the ALE form of the mass and momentum balance equations~\eqref{NSEALE} by test functions $q \in Q$ and $\bv \in V$, respectively and integrate over the computational domain $\Omega(t)$.
Then, applying integration by parts to the stress tensor term over the sub-domain $\Omega_1(t)$, we get
\begin{equation}
\begin{aligned}\label{intpartomega1}
 - \int_{\Omega_1(t)} \nabla \cdot \mathbb{T}_1 (\bu, p, \btau_p) \cdot \bv~ dx = &\int_{\Omega_1(t)} \frac{2\beta}{\text{Re}} \, \mathbb{D}(\bu) : \mathbb{D}(\bv)~ dx - \int_{\Omega_1(t)} p~ (\nabla \cdot \bv)~ dx \\
 &+ \int_{\Omega_1(t)} \frac{\left( 1 - \beta\right)}{\text{Re}\text{Wi}} \, \btau_p : \mathbb{D}(\bv)~ dx + \int_{\Gamma_F(t)} \bv \cdot \mathbb{T}_1 (\bu, p, \btau_p) \cdot \bnu_F~ d\gamma_F,
\end{aligned}
\end{equation}
and over the sub-domain $\Omega_2(t)$, we obtain
\begin{equation}
\begin{aligned}\label{intpartomega2}
 - \int_{\Omega_2(t)} \nabla \cdot \mathbb{T}_2 (\bu, p, \btau_p) \cdot \bv~ dx = &\int_{\Omega_2(t)} \frac{2\beta}{\text{Re}} \mathbb{D}(\bu) : \mathbb{D}(\bv)~ dx - \int_{\Omega_2(t)} p~ (\nabla \cdot \bv)~ dx \\
 &+ \int_{\Omega_2(t)} \frac{\left( 1 - \beta\right)}{\text{Re}\text{Wi}} \btau_p : \mathbb{D}(\bv)~ dx - \int_{\partial \Omega_2(t)} \bv \cdot \mathbb{T}_2 (\bu, p, \btau_p) \cdot \bnu~ d\gamma.
\end{aligned}
\end{equation}
Rewriting the boundary integral in~\eqref{intpartomega2} into integral over $\Gamma_D$, $\Gamma_N$ and $\Gamma_F(t)$, we get
\begin{equation}
\begin{aligned}\label{splitboundaryint}
 - \int_{\partial \Omega_2(t)} \bv \cdot \mathbb{T}_2 (\bu, p, \btau_p) \cdot \bnu~ d\gamma = - \int_{\Gamma_D} &\bv \cdot \mathbb{T}_2 (\bu, p, \btau_p) \cdot \bnu_D~ d\gamma_D - \int_{\Gamma_N} \bv \cdot \mathbb{T}_2 (\bu, p, \btau_p) \cdot \bnu_N~ d\gamma_N \\
 &- \int_{\Gamma_F(t)} \bv \cdot \mathbb{T}_2 (\bu, p, \btau_p) \cdot \bnu_F~ d\gamma_F.
\end{aligned}
\end{equation}
Since the velocity space is chosen such that $\bv=0$ on $\Gamma_D$, the integral over $\Gamma_D$ in~\eqref{splitboundaryint} vanishes.
Further, using the orthonormal decomposition, we split the test function $\bv$ as
\begin{align*}
 \bv = \left( \bv \cdot \bnu_N \right) \bnu_N + (\bv \cdot \btau_N) \btau_N,
\end{align*}
in the integral over $\Gamma_N$ in~\eqref{splitboundaryint} and the integral becomes,
\begin{equation}
 \begin{aligned}\label{splitgammaN}
  - \int_{\Gamma_N} \bv \cdot \mathbb{T}_2 (\bu, p, \btau_p) \cdot \bnu_N~ d\gamma_N = - \int_{\Gamma_N} &\left( \bv \cdot \bnu_N \right) \left( \bnu_N \cdot \mathbb{T}_2 (\bu, p, \btau_p) \cdot \bnu_N\right)~ d\gamma_N \\
  &- \int_{\Gamma_N} \left( \bv \cdot \btau_N \right) \left( \btau_N \cdot \mathbb{T}_2 (\bu, p, \btau_p) \cdot \bnu_N\right)~ d\gamma_N.
 \end{aligned}
\end{equation}
Since the velocity space is chosen such that $\bv \cdot \bnu_N = 0$ on $\Gamma_N$, the first integral in~\eqref{splitgammaN} vanishes and further, incorporating the free slip condition~\eqref{freeslipbc}, the second integral in~\eqref{splitgammaN} also vanishes.
After summing up the interface $\Gamma_F(t)$ integrals in equations~\eqref{intpartomega1} and~\eqref{intpartomega2}, and further incorporating the force balancing condition~($4^{th}$ equation in \eqref{nondimeqns}) and applying integration by parts, we obtain
 \begin{align}
 \int_{\Gamma_F(t)} &\bv \cdot \mathbb{T}_1 (\bu, p, \btau_p) \cdot \bnu_N~ d\gamma_F - \int_{\Gamma_F(t)} \bv \cdot \mathbb{T}_2 (\bu, p, \btau_p) \cdot \bnu_N~ d\gamma_F \nonumber \\
 &= - \int_{\Gamma_F(t)} \bv \cdot [|\mathbb{T} (\bu, p, \btau_p)|] \cdot \bnu_N~ d\gamma_F = - \frac{1}{\text{We}} \int_{\Gamma_F(t)} \bv \cdot \left(\nabla_{\Gamma_{\bnu_F}} \cdot \mathbb{P}_{\bnu_F}\right)~ d\gamma_F = \frac{1}{\text{We}} \int_{\Gamma_F(t)}  \mathbb{P}_{\bnu_F} : \left(\nabla_{\Gamma_{\bnu_F}} \bv \right)~ d\gamma_F. \label{curvatureterm}
 \end{align}
Thus, the variational form of the Navier--Stokes equations read~: \\

\noindent For given $\Omega_0$, $\bu_0$, $\bw$, $\btau_{p,0}$, find $(\mathbf u, p)  \in V\times Q$ such that
\begin{equation}
\begin{aligned} \label{weakNSE}
  \left(\rho \frac{\partial \mathbf u}{\partial t},\bv \right)_{\hat{\Omega}} + a(\hat{\mathbf u} - \bw;\mathbf u,\bv) - b(p,\bv) + c(\btau_p, \bv) &= f_1(\bv)   \\
b(q,\mathbf u) &= 0 
  \end{aligned}
\end{equation}
for all  $(\bv,q) \in V\times Q$ , where
\begin{align*}
a(\hat{\mathbf u}- \bw;\mathbf u,\bv)&= \int_{\Omega(t)}\rho \left( \left( (\hat{\mathbf u} - \bw)\cdot\nabla \right) \mathbf u \right) \cdot \bv~ dx + \int_{\Omega(t)} \frac{2\beta}{\text{Re}}  \,  \mbD(\mathbf u):\mbD(\bv)~ dx  \\ 
b(q,\bv)&=\int_{\Omega(t)} q \,(\nabla\cdot \bv)~ dx \\ 
c(\btau_p, \bv)&=\int_{\Omega(t)} \frac{\left( 1 - \beta\right)}{\text{Re}\text{Wi}}  \, \btau_p : \mathbb{D} (\bv) \,dx \\
f_1(\bv) &= \frac{1}{\text{Fr}} \int_{\Omega(t)} \rho \, (\be \cdot \bv) \,dx - \frac{1}{\text{We}} \int_{\Gamma_F(t)}  \mathbb{P}_{\bnu_F} : \left(\nabla_{\Gamma_{\bnu_F}} \bv \right)~ d\gamma_F.
\end{align*}

Next, to derive a variational form of the Giesekus equation, we multiply the ALE form of Giesekus equation~\eqref{GiesekusALE} by a test function $\bpsi \in S$ and integrate over the computational domain $\Omega(t)$.
The variational form of the Giesekus equation read~: \\

\noindent For given $\Omega_0$, $\bu_0$, $\bw$, $\btau_{p,0}$, find $\btau_p \in S$ such that
\begin{align} \label{weakGiesekus}
  \left( \frac{\partial \btau_p}{\partial t},\bpsi \right)_{\hat{\Omega}} + d(\hat{\mathbf u} - \bw;\btau_p,\bpsi) + e(\hat{\btau}_p; \btau_p, \bpsi) &= f_2(\bpsi)   
  \end{align}
for all  $\bpsi \in S$ , where
\begin{align*}
 d(\hat{\bu} -\bw; \btau_p, \bpsi)&= \int_{\Omega(t)}  ( \left( (\hat{\bu} - \bw) \cdot \nabla \right) \btau_p ) : \bpsi \,dx  -\int_{\Omega(t)} (\nabla \hat{\bu}^{\text{T}} \cdot \btau_p + \btau_p \cdot \nabla \hat{\bu}) : \bpsi \,dx \\
e(\hat{\btau}_p; \btau_p, \bpsi)&= \int_{\Omega(t)} \frac{\alpha}{\text{Wi}} \, \left( \hat{\btau}_p \cdot \btau_p \right) : \bpsi \,dx + \int_{\Omega(t)} \frac{(1-2\alpha)}{\text{Wi}} \, \btau_p : \bpsi \,dx \\
f_2(\bpsi) &= \int_{\Omega(t)} \frac{(1-\alpha)}{\text{Wi}} \, \mathbb{I} : \bpsi \,dx.
\end{align*}
Since the coupled two-phase viscoelastic flow system is solved in a monolithic approach, we rewrite the variational formulation as follows~: \\
\noindent For given $\Omega_0$, $\bu_0$, $\bw$ and $\btau_{p,0}$, find $(\mathbf u, p, \btau_p)  \in V\times Q\times S $ such that
\begin{align}
\left(\rho\frac{\partial \bu}{\partial t},\bv \right)_{\hat{\Omega}} + \left(\frac{\partial \btau_p}{\partial t},\bpsi \right)_{\hat{\Omega}} + A(((\hat{\bu}- \bw), \hat{\btau}_p); (\bu,p,\btau_p),(\bv, q, \bpsi))  = f_1(\bv) + f_2(\bpsi) \label{WeakForm}
\end{align}
for all  $(\bv,q, \bpsi) \in V\times Q\times S$ , where
\begin{align*}
A(((\hat{\bu}-\bw), \hat{\btau}_p); (\bu,p,\btau_p),(\bv, q, \bpsi)) = a(\hat{\bu} - \bw;\mathbf u,\bv) - b(p,\bv) + c(\btau_p, \bv)  
                                                   + b(q,\bu) + d(\hat{\bu} - \bw; \btau_p, \bpsi) + e(\hat{\btau}_p; \btau_p, \bpsi).     
\end{align*}

\subsection{3D-axisymmetric formulation}
The considered domain is rotational symmetric and thus we consider a 2D meridian domain $\Phi(t)$ of $\Omega(t)$ with a 3D-axisymmetric configuration.
The axisymmetric formulation allows us to reduce the space-dimension of the problem by one and hence, we use two-dimensional finite elements for approximating the velocity, pressure and viscoelastic stress.
Further, the computational cost and complexity of mesh movement will drastically be reduced by using the 3D-axisymmetric formulation.
In the meridian domain $\Phi$(t), the unknown components of the velocity and the symmetric viscoelastic conformation stress tensor are given by
\[
\bu = (u_r,~u_z)^{\text{T}} \quad \text{and}  \quad \btau_p=
\begin{bmatrix}
 \tau_{rr} & \tau_{rz}\\
  \tau_{zr} & \tau_{zz} 
\end{bmatrix}
\text{ with } \tau_{zr} = \tau_{rz}.
\]
The boundary of the meridian domain $\Phi(t)$ is given by $\partial \Phi_1(t) := \Gamma_F(t) \cup \Gamma_{\text{Axial}}$ and $\partial \Phi_2(t) := \Gamma_F(t) \cup \Gamma_{\text{Axial}} \cup \Gamma_D \cup \Gamma_N$.
In contrast to the standard approach of starting with the differential equations in cylindrical coordinate form and deriving a suitable variational formulation, 
we derive the 3D-axisymmetric weak form in the meridian domain $\Phi(t)$ directly from the weak form~\eqref{WeakForm} defined in 3D-Cartesian coordinates. 
To achieve this, we transform the volume and surface integrals in~\eqref{WeakForm} into area and line integrals by introducing cylindrical coordinates and imposing irrotational, axisymmetric conditions as described in~\cite{GAN07, ViscoelasticDropLPS18}. 
This approach leads naturally to boundary conditions along the  rotational axis 
\begin{equation} \label{AxialBC}
 u_r = 0, \quad \frac{\partial u_z}{\partial r} = 0  \quad \text{on} \quad \Gamma_{\text{Axial}}(t),
\end{equation}
which are already partly included in the weak form.
Further, we define the velocity, pressure and viscoelastic conformation stress spaces in the 2D meridian domain $\Phi(t)$ as 
\begin{align*}
\widetilde{V}(\Phi(t))&:=\left\{ \, \bv\in \text{H}^1(\Phi(t))^2 \, : \; \bv \cdot \bnu_N = 0 \; \text{on} \; \Gamma_N, \quad \bv = 0 \; \text{on} \; \Gamma_D, \quad v_r = 0 \; \text{on} \; \Gamma_{\text{Axial}} \, \right\}, \\
\widetilde{Q}(\Phi(t))&:=\left\{ \, q \in \text{L}^2(\Phi(t)) \, : \, \int_{\Omega} q \, dx = 0 \, \right\}, \\
\widetilde{S}(\Phi(t))&:= \left\{ \, \bpsi = [\psi_{ij}], \quad 1 \leq i,j \leq 2 \, : \quad  \psi_{ij} \in \text{H}^1(\Phi(t)),  \quad {\psi_{ij} = \psi_{ji}} \, \right\}. 
\end{align*}

\subsection{Spatial and temporal discretization}
Let $\{\mathcal{T}_{h}\}$ be a partition of the meridian domain $\Phi(t)$ into an interface resolved triangular mesh using the mesh generator Triangle~\cite{TRI96, TRI02}.
The diameter of a cell $K \in \mathcal{T}_{h}$ is denoted by $h_K$. 
The mesh parameter $h$ is defined by $h = \max \{ h_K \,| \, K \in \mathcal{T}_{h}\}$.
The discrete form of the meridian domain $\Phi$ is given by $\Phi_h := \bigcup_{K \in \mathcal{T}_{h}} K$, whereas $\hat{\Phi}_h$ denotes the reference domain of $\Phi_h$.  
Further, let $V_h\subset \widetilde{V}$, $Q_h\subset \widetilde{Q}$ and $S_h\subset \widetilde{S}$  be the conforming finite element spaces on $\mathcal{T}_{h}$.
The standard Galerkin finite element approximation of the variational problem~\eqref{WeakForm} reads~: \\ 

\noindent For given $\Phi_0$, $\bu_0$, $\bw_h$ and $\btau_{p,0}$, find $(\bu_h, p_h, \btau_{p,h})  \in V_h\times Q_h\times S_h $ such that
\begin{align}
\left(\rho \frac{\partial \mathbf u_h}{\partial t},\bv_h \right)_{\hat{\Phi}_h} + \left(\frac{\partial \btau_{p,h}}{\partial t},\bpsi_h \right)_{\hat{\Phi}_h} + A(((\hat{\mathbf u}_h - \bw_h), \hat{\btau}_{p,h}); (\bu_h,p_h,\btau_{p,h})&,(\bv_h, q_h, \bpsi_h))  = f_1(\bv_h) + f_2(\bpsi_h) \label{WeakFormSG}
\end{align}
for all  $(\bv_h, q_h, \bpsi_h) \in V_h\times Q_h\times S_h$. 
Here, $(\cdot,\cdot)$ denotes the inner product in $\text{L}^2\left(\Phi(t)\right)$ and its vector/tensor valued versions respectively.
The choice of finite element spaces for the velocity, pressure and viscoelastic stress is subject to the following two inf-sup conditions, 
  \begin{align}
  \inf_{q_h \in Q_h} \sup_{\bv_h \in V_h} \frac{\big( q_h, \dive \bv_h \big)}{{ \Arrowvert q_h \Arrowvert}_{Q_h} {\Arrowvert\bv_h\Arrowvert}_{V_h}} \geq \zeta_1 > 0, \quad \inf_{\bv_h \in V_h} \sup_{\btau_{p,h} \in S_h} \frac{\big( \btau_{p,h}, \mathbb{D}(\bv_h) \big)}{{ \Arrowvert \btau_{p,h} \Arrowvert}_{S_h} {\Arrowvert\bv_h\Arrowvert}_{V_h}} \geq \zeta_2 > 0. \label{infsup}
 \end{align}
The standard Galerkin approach for solving the coupled Navier--Stokes and Giesekus constitutive problem may suffer in general from two shortcomings. 
First, the constitutive equation is highly advection dominated at high Weissenberg numbers.
Second, the finite element spaces should satisfy these two discrete inf-sup conditions~\eqref{infsup} simultaneously to have a control over $p_h$ and $\mathbb{D}(\bu_h)$.
One way to overcome these difficulties is to use a stabilized formulation. 
In this work, we add symmetric stabilization terms to the standard Galerkin formulation~\eqref{WeakFormSG} by using one-level Local Projection Stabilization~(LPS) method.
LPS was initially proposed for the Stokes problem by Becker and Braack \cite{BB01},  and later it has been extended for transport~\cite{BB04} and Oseen~\cite{BB06} problems.
Recently, LPS technique has been used by Venkatesan and Ganesan~\cite{ViscoelasticDropLPS18, VJSGLPS17}  for the simulation of viscoelastic fluid flows. 
The one-level LPS scheme~\cite{VJSGLPS17, MAT07, GAN08, GAN10} is based on enrichment of approximation spaces and it allows us to perform the computations on a single mesh as the approximation and the projection spaces are defined on the same mesh. 
We use mapped finite element spaces in the computations, where the enriched approximation spaces on the reference cell $\hat{K}$ are given by
\begin{align*}
 P_r^{bubble}\left(\hat{K}\right) := P_r\left(\hat{K}\right) \oplus \left( \hat{b}_{\triangle} \cdot P_{r-1}\left(\hat{K}\right) \right),
\end{align*}
with $r\geq2$.
Here, $\hat{b}_{\triangle}$ is a cubic polynomial bubble function on the reference triangle.

Let $Y_h$   denote the approximation space and $D_h$ be the discontinuous projection space defined on $\mathcal{T}_{h}$. 
Let $D_h(K):=\{d_h|_K : d_h \in D_h\}$  and $\pi_K : Y_h(K) \rightarrow D_h(K)$ be the local $\text{L}^2$-projection into $D_h(K)$. 
Further, we define the global projection $\pi_h : Y_h \rightarrow D_h$ by $(\pi_h y)|_K := \pi_K(y|_K)$. 
The fluctuation operator $\kappa_h: Y_h \rightarrow Y_h$ is given by $\kappa_h := id - \pi_h$, where $id$ is the identity mapping.
We apply these operators to vector/tensor valued functions in a component-wise manner.
Adding symmetric stabilization terms  to the variational problem~\eqref{WeakFormSG}, leads to the following variational form~: \\

\noindent For given $\Phi_0$, $\bu_0$, $\bw_h$ and $\btau_{p,0}$, find $(\mathbf u_h, p_h, \btau_{p,h})  \in V_h\times Q_h\times S_h $ such that
\begin{equation}
 \begin{aligned}\label{LPSweakform}
\left(\rho \frac{\partial \mathbf u_h}{\partial t},\bv_h \right)_{\hat{\Phi}_h} + \left(\frac{\partial \btau_{p,h}}{\partial t},\bpsi_h \right)_{\hat{\Phi}_h} + A(((\hat{\bu}_h - \bw_h),& \hat{\btau}_{p,h}); (\bu_h,p_h,\btau_{p,h}),(\bv_h, q_h, \bpsi_h))  \\
  &+ S_1(\bu_h,\bv_h) +  S_2(\btau_{p,h},\bpsi_h)  = f_1(\bv_h) + f_2(\bpsi_h) 
\end{aligned}
\end{equation}
for all  $(\bv_h, q_h, \bpsi_h) \in V_h\times Q_h\times S_h$, where 
\begin{align*}
 S_1(\bu_h,\bv_h) &=  \sum_{K \in \mathcal{T}_{h}}  \varsigma_1 \left< \kappa_h \mathbb{D}(\bu_h), \kappa_h \mathbb{D}(\bv_h) \right>_K \\
 S_2(\btau_h,\bpsi_h) &=  \sum_{K \in \mathcal{T}_{h}} \varsigma_2 \left< \kappa_h \left( \dive \btau_h \right), \kappa_h \left( \dive \bpsi_h \right) \right>_K  \quad + \sum_{K \in \mathcal{T}_{h}} \varsigma_3 \left< \kappa_h  \nabla \btau_h, \kappa_h  \nabla \bpsi_h \right>_K.
\end{align*}
Here, $\varsigma_1 = (1-\beta) c_1 h_K, \varsigma_2 = c_2 h_K, \varsigma_3 = c_3 h_K$, with $c_1$, $c_2$ and $c_3$ being user-chosen constants.
This scheme allows us to use inf-sup stable finite elements for the velocity and pressure spaces, and equal order interpolation spaces for the velocity and viscoelastic stress.
For more details on LPS for viscoelastic fluid flows we refer to~\cite{ViscoelasticDropLPS18, VJSGLPS17}.

The finite elements should be chosen in such a way that the mass should be conserved well and spurious velocities, if there are any should be suppressed~\cite{GMT}.
Hence, we use the following triplet $\left(V_h, Q_h, S_h \right) = \left( P_2^{bubble}, P_1^{disc}, P_2^{bubble} \right)$.
By using discontinuous pressure approximation on interface resolved meshes, spurious velocities can be avoided during the computations~\cite{GMT}.
Moreover, the first integral moments of the divergence of velocity field vanishes element-wise with discontinuous pressure approximation and it leads to a better mass conservation.   
Further, in order to suppress the spurious velocities generated by the curvature approximation error, we use the tangential gradient operator technique with isoparametric finite elements for velocity approximation. 

Let $0=t^0<t^1<\dots <t^N=\rm{I}$ be a decomposition of the time interval $[0, \text{I}]$, and $\delta t= t^{n+1} - t^{n}$, $n=0,\ldots,N-1$, be a uniform time step.
We use the first-order implicit Euler method for the time discretization of the coupled system~\eqref{LPSweakform} in the time interval $\left( t^n , t^{n+1} \right)$.
An implicit handling of the curvature term~\eqref{curvatureterm} is needed to obtain unconditional stability and however, it is too complicated as well.
Thus, as in~\cite{EB1}, we use a semi-implicit approximation of the curvature  
\begin{align*}
 - \frac{1}{\text{We}} \int_{\Gamma_F^{n+1}}  \mathbb{P}_{\bnu_F^{n+1}} : \left( \nabla_{\Gamma_{\bnu_F}} \bv_h \right) ~ d\gamma_F &= - \frac{1}{\text{We}} \int_{\Gamma_F^n} \left[ \mathbb{P}_{\bnu_F^n} + \delta t \nabla_{\Gamma_{\bnu_F}} \bu_h^{n+1} \right] : \left( \nabla_{\Gamma_{\bnu_F}} \bv_h \right) ~ d\gamma_F \\
 &= - \frac{1}{\text{We}} \int_{\Gamma_F^n} \mathbb{P}_{\bnu_F^n}  : \left( \nabla_{\Gamma_{\bnu_F}} \bv_h \right) ~ d\gamma_F - \frac{ \delta t}{\text{We}} \int_{\Gamma_F^n} \left( \nabla_{\Gamma_{\bnu_F}} \bu_h^{n+1} \right) : \left( \nabla_{\Gamma_{\bnu_F}} \bv_h \right) ~ d\gamma_F.
 \end{align*}
The first term in the above equation is an explicit term and it stays on the right hand side of the weak formulation, whereas the second term is an implicit term and it goes to the left hand side.
Note that the implicit term is symmetric and positive semi-definite and thus it improves the stability of the discrete system compared to a fully explicit approach.

\subsection{Linearization and mesh movement}
In each time step $(t^n,t^{n+1})$, the non-linear terms in~\eqref{LPSweakform} are handled by an iteration of fixed point type.
Let $\bu_{h,0}^{n+1} = \bu_h^{n}$, $\btau_{p,h,0}^{n+1} = \btau_{p,h}^{n}$ and $\bw_{h,0}^{n+1} = \bw_h^{n}$. 
In computations, we adopt the following linearization strategy~: 
\begin{align*}
 a\left( \bu_h^{n+1} - \bw_h^{n+1}; \bu_h^{n+1}, \bv_h  \right) \approx &a\left( \bu^{n+1}_{h,m-1} - \bw^{n+1}_{h,m-1}; \bu^{n+1}_{h,m}, \bv_h  \right)  \\
 d\left( \bu_h^{n+1} - \bw_h^{n+1}; \btau_{p,h}^{n+1}, \bpsi_h \right) \approx  &d\left( \bu_{h,m-1}^{n+1} - \bw_{h,m-1}^{n+1}; \btau_{p,h,m}^{n+1}, \bpsi_h \right) + d\left( \bu_{h,m}^{n+1} - \bw_{h,m}^{n+1}; \btau_{p,h,m-1}^{n+1}, \bpsi_h \right) \\
 &- d\left( \bu_{h,m-1}^{n+1} - \bw_{h,m-1}^{n+1}; \btau_{p,h,m-1}^{n+1}, \bpsi_h \right) \\
 e\left(\btau_{p,h}^{n+1}; \btau_{p,h}^{n+1}, \bpsi_h \right) \approx  &e\left(\btau_{p,h,m-1}^{n+1}; \btau_{p,h,m}^{n+1}, \bpsi_h \right),
\end{align*}
where, $m = 1,2,...,\text{M}$, with M being the maximum allowed number of nonlinear iterations. 
The linearized  system of algebraic equations are solved using the Multifrontal Massively Parallel Sparse  (MUMPS)  direct solver \cite{MUMPS1,MUMPS2}.
In computations, the non-linear iterations are continued until the residual of the monolithic system~\eqref{LPSweakform} becomes less than the threshold value of $10^{-7}$.

For the mesh movement, we use the linear elastic mesh update technique.
Let ${\bf Z}_k^{n}$ be the vertices on the boundary $\partial \Phi_k^n$.
We first advect the boundary vertices  using the computed flow velocity as follows~:
\begin{align*}
 {\bf Z}_k^{n+1} = {\bf Z}_k^{n} + \delta t \, \bu_k^{n+1}.
\end{align*}
Then, based on the displacement of the boundary vertices ${\bf d}_k^{n+1} = {\bf Z}_k^{n+1} - {\bf Z}_k^n$, the inner points are displaced in a prescribed way to preserve the mesh quality in each domain separately.
The displacement~$\bPsi_k^{n+1}$ of the inner mesh points in both the phases are obtained by solving the following linear elasticity problem with the displacement of boundary vertices as a Dirichlet boundary condition, i.e., \\
Find $\bPsi_k^{n+1} \in  \text{H}^1 \left(\Phi_k^{n} \right)$, such that
\begin{equation}\label{Elast}
\begin{array}{rcll}
\vspace{2mm}
\nabla\cdot\mathbb{S}(\bPsi_k^{n+1}) & = & 0 & \mbox{in }{\Phi_k^n}, \\
\bPsi_k^{n+1}  & = &  {\bf d}_k^{n+1}   & \mbox{on } \partial \Phi_k^n, \\
\end{array}
\end{equation}
for k=1,2, 
where 
$
 \mathbb{S}(\bPsi) = \lambda_{\text{L}1}(\dive \bPsi)\mathbb{I} + 2\lambda_{\text{L}2} \mathbb{D}(\bPsi).
$
Here, $\lambda_{\text{L}1}$ and $\lambda_{\text{L}2}$ are Lame constants, and in  computations we use $\lambda_{\text{L}1}=\lambda_{\text{L}2} =1$. 
Continuous piecewise linear $P_1$ elements on the same triangular mesh as for solving the flow equations are used for the solution of~\eqref{Elast}.
Once the displacement vector $\bPsi_k^{n+1}$ is known for each phase, the mesh velocity is then computed as
$
\bw_k^{n+1} =  \bPsi_k^{n+1}/ \delta t.
$

Even though the elastic mesh update technique is used to preserve the mesh quality, the quality of the mesh becomes poor after several time steps due to large deformation in each subdomain. 
In such an instant, we need to remesh the domain.
We have implemented an automatic remeshing algorithm to remesh the domain when the minimum angle of any triangular cell in the mesh is less than $15^\circ$.
During remeshing the points on the interface are equally re-distributed using interpolated cubic splines and the new mesh is constructed using the mesh generator Triangle~\cite{TRI96, TRI02}.
The solutions are then interpolated from the old to the newly generated mesh.
Further, to minimize the interpolation error, we solve the monolithic system~\eqref{LPSweakform} with the interpolated values as initial guess and $\bw = 0$ before moving to the next time step.
The proposed numerical scheme for the simulation of viscoelastic two-phase flows is implemented in our in-house finite element code ParMooN~\cite{ParMooN1}.

\section{Numerical Results}\label{Results}
In this section we present the numerical results of 3D-axisymmetric buoyancy driven viscoelastic two-phase flows using the proposed numerical scheme.
In order to validate the numerical scheme, computations are performed with 2D planar configuration for buoyancy driven Newtonian bubble rising in a Newtonian fluid column and compared with the benchmark results~\cite{HYS09}.
We simultaneously perform a grid independence test for the benchmark configuration.
Next, we present a detailed numerical investigation for a buoyancy driven Newtonian bubble rising in a viscoelastic fluid column.
We examine the effects of viscosity ratio~($\varepsilon$), Newtonian solvent ratio~($\beta$), Giesekus mobility factor~($\alpha$) and E\"{o}tv\"{o}s number~(Eo) on the flow dynamics of the rising bubble.
Further, we also investigate the flow dynamics of a viscoelastic bubble rising in a Newtonian fluid column. 
Key flow features are explained using the visualization of viscoelastic stress profiles.
Further, to assist in describing the temporal evolution of the rising bubble quantitatively, we use the following metrics~: bubble shape, diameter of bubble at the axis of symmetry~($D|_{r=0}$), sphericity, kinetic energy, elastic energy, center of mass~($z$ coordinate) and rise velocity.    
Let $|\Omega_1(t)| := 2 \pi \int_{\Phi_1(t)} r~dr~dz$ be the volume of the bubble.
The sphericity of the bubble is given by 
\begin{align*}
 \text{Sphericity} = \frac{\text{surface area of the volume-equivalent sphere}}{\text{surface area of the bubble}} = \frac{A_e}{A}.
\end{align*}
The surface area of volume-equivalent sphere and surface area of the bubble are calculated as follows~: 
\begin{align*}
 A_e = 4\pi \left( \frac{3}{4\pi} |\Omega_1(t)|  \right)^{2/3}, \quad A = 2 \pi \int_{\partial \Phi_1(t)} r~dl.
\end{align*}
For a perfectly spherical bubble, the sphericity will be one and for any other deformed bubble it will be less than one. 
It is a good quantitative measure of the bubble deformation. 
The kinetic and elastic energies of the bubble are computed as follows~:
\begin{align*}
 E_{kinetic} = \frac{2 \pi}{|\Omega_1(t)|}\int_{\Phi_1(t)} \left(\bu \cdot \bu\right)~r~dr~dz,   \quad
E_{elastic} = \frac{2 \pi}{|\Omega_1(t)|} \int_{\Phi_1(t)} tr (\btau_p)~r~dr~dz.
\end{align*}
Further, the rise velocity and center of mass~($z$ coordinate) of the bubble are given by~:
\begin{align*}
 \text{Rise velocity} = \frac{2 \pi}{|\Omega_1(t)|}\int_{\Phi_1(t)} u_z~r~dr~dz  , \quad
 \text{Center of mass} = \frac{2 \pi}{|\Omega_1(t)|}\int_{\Phi_1(t)} z~r~dr~dz   .
\end{align*}

\subsection{Grid independence test and validation}
In this section, we first perform a grid independence test for the proposed numerical scheme and then validate the numerical results using benchmark solutions~\cite{HYS09} of a 2D planar rising bubble.
We consider a Newtonian bubble rising in a Newtonian fluid column with the following benchmark parameters~(refer test case-1 in Table~1 of~\cite{HYS09})~: $\rho_1$~=~100, $\rho_2$~=~1000, $\mu_{0,1}$~=~1, $\mu_{0,2}$~=~10, $g$~=~0.98, $\sigma$~=~24.5, $D$~=~0.5 and $h_c$~=~2.0. 
Using the characteristic length $\text{L}$~=~$1$ and characteristic velocity $\text{U}_{\infty}$~=~$\sqrt{\text{L}g}$, we get the following dimensionless quantities $\Rey_2$~=~$99$, Eo~=~$40$, $\rho_1/\rho_2$~=~0.1, $\varepsilon$~=~10, $\beta_1$~=~1 and $\beta_2$~=~1.
In order to identify a grid that provides a grid independent solution, we consider five different meshes of varying mesh sizes. 
In particular, we vary the number of degrees of freedom~(DOFs) on the interface. 
The characteristics of these meshes are tabulated in Table~\ref{DOFsGrid}.   
The time step length is set as $\delta t$~=~0.0005 and the computations are performed till I~=~3.0.

 \begin{table}[h!]
 \begin{center}
 \begin{tabular}{cccccc} 
 \hline
 Mesh& DOFs on $\Gamma_F$ & $h_{0}$ &Cells  & Total DOFs  \\
 \hline
 L1   & 100  & 0.015705380 &1,837 &16,793 \\
 L2   & 200   & 0.007853659 &2,576 &23,454 \\
 L3   & 400   & 0.003926950 &3,767 & 34,183 \\
 L4   & 600   & 0.002617982 &4,980  & 45,094 \\
 L5   & 800   & 0.001963490 &6,237  & 56,425 \\
 \hline 
 \end{tabular}
 \end{center}
 \caption{Grid independence test~: characteristics of triangular meshes.} \label{DOFsGrid}
 \end{table}

\begin{figure*}[ht!]
\begin{center}
\unitlength1cm
\begin{picture}(14.5,8.3)
\put(1.9,6.2){\makebox(0,0){\includegraphics[width=8.5cm,height=4.2cm,keepaspectratio]{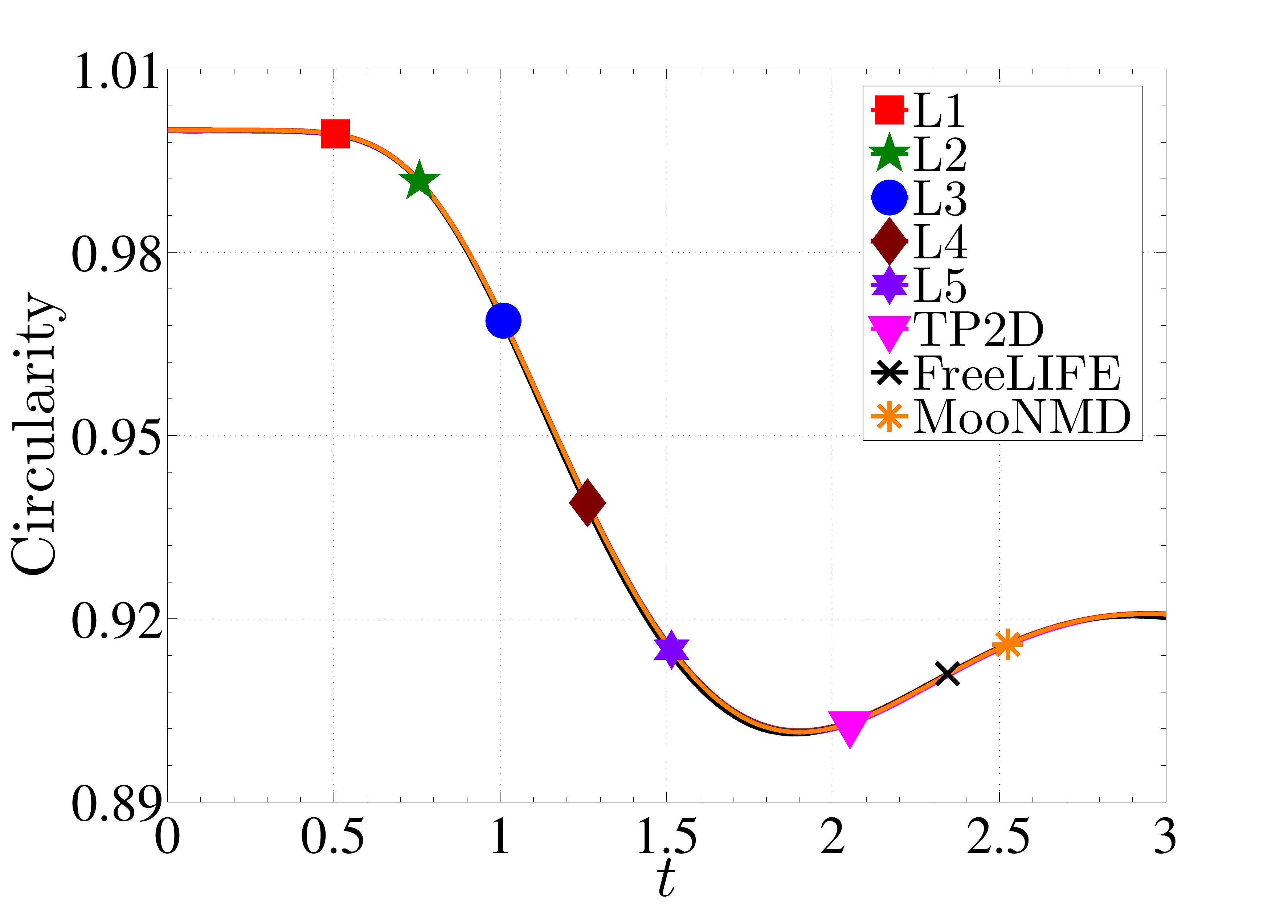}}}
\put(7.4,6.2){\makebox(0,0){\includegraphics[width=8.5cm,height=4.2cm,keepaspectratio]{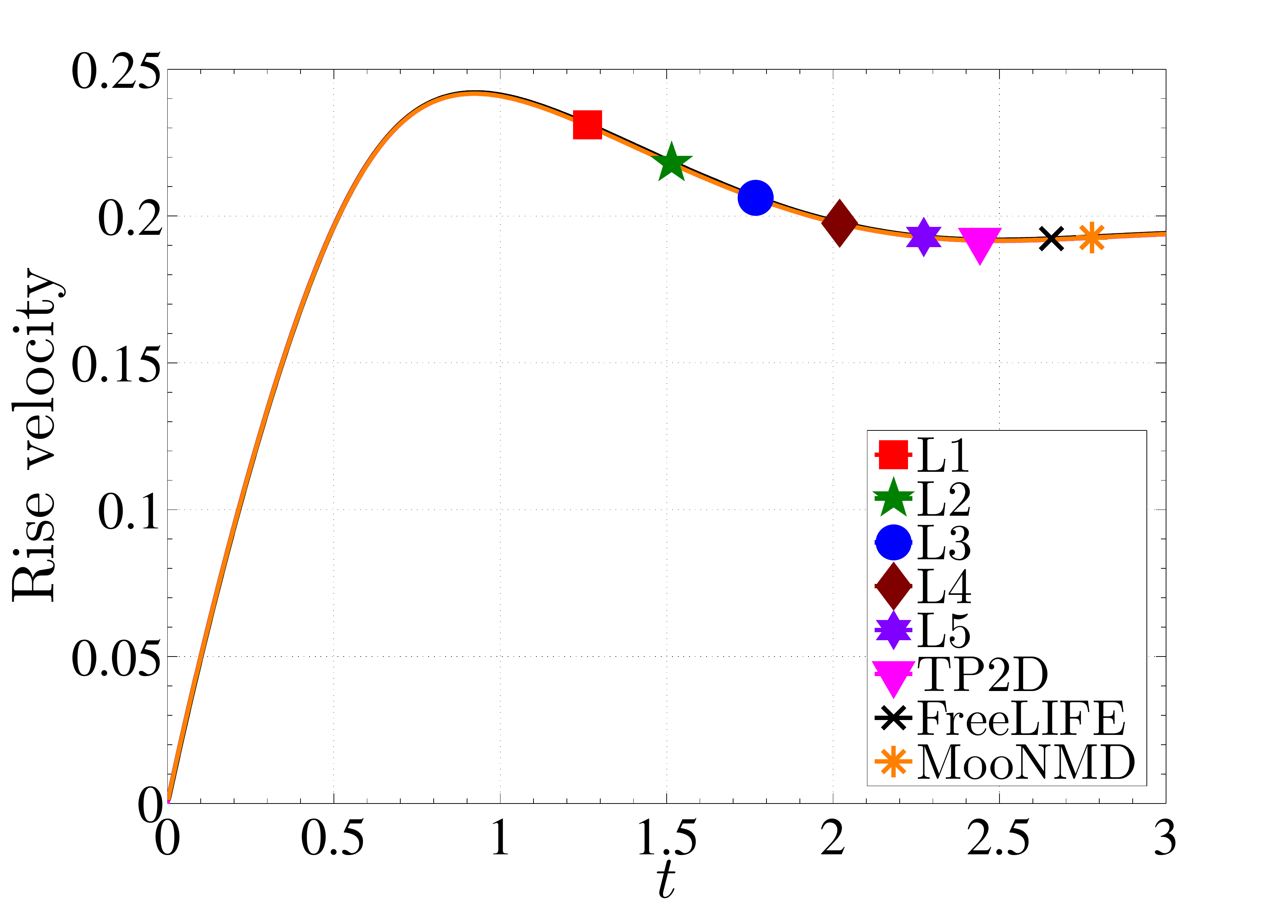}}}
\put(12.9,6.2){\makebox(0,0){\includegraphics[width=8.5cm,height=4.2cm,keepaspectratio]{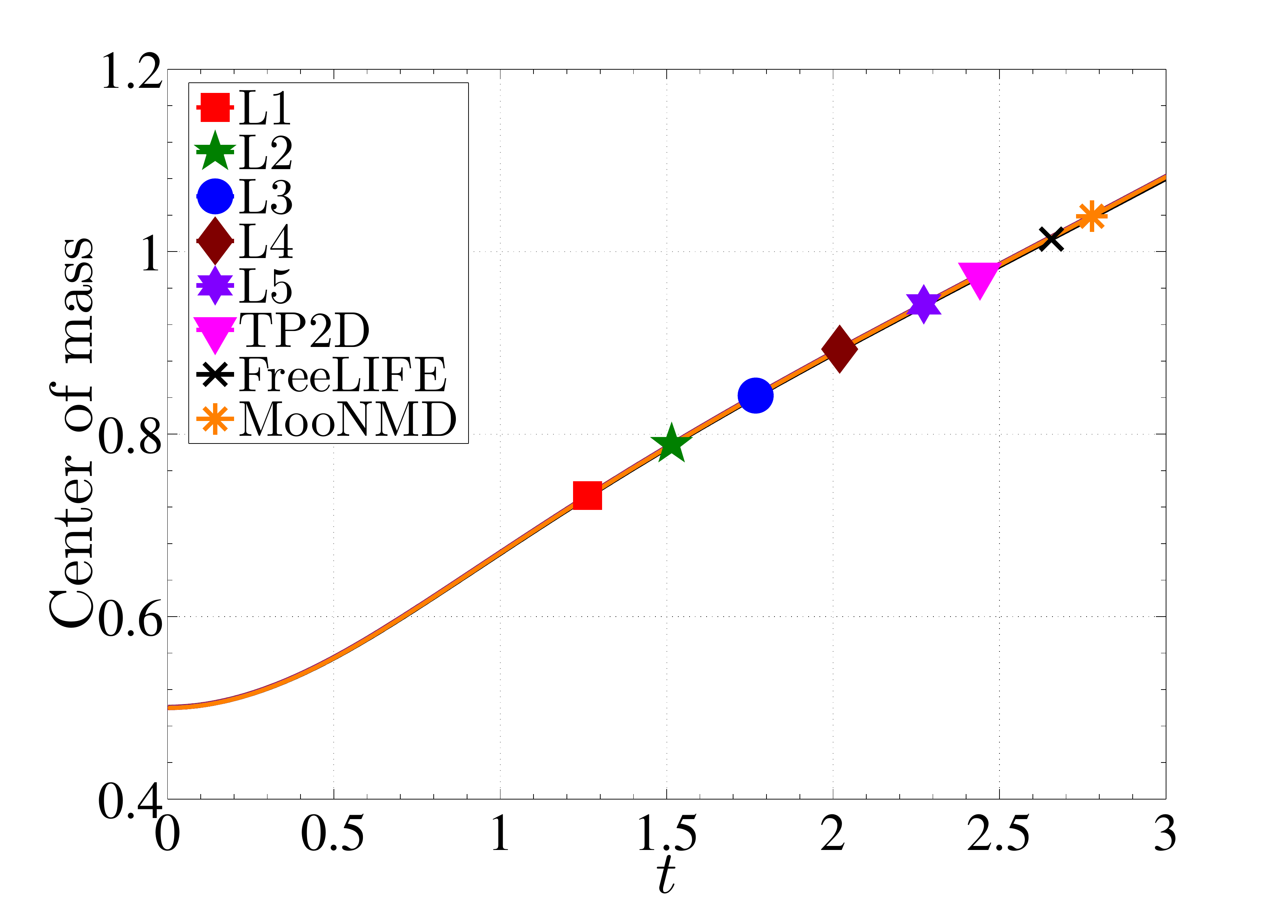}}}
\put(1.9,1.6){\makebox(0,0){\includegraphics[width=8.5cm,height=4.2cm,keepaspectratio]{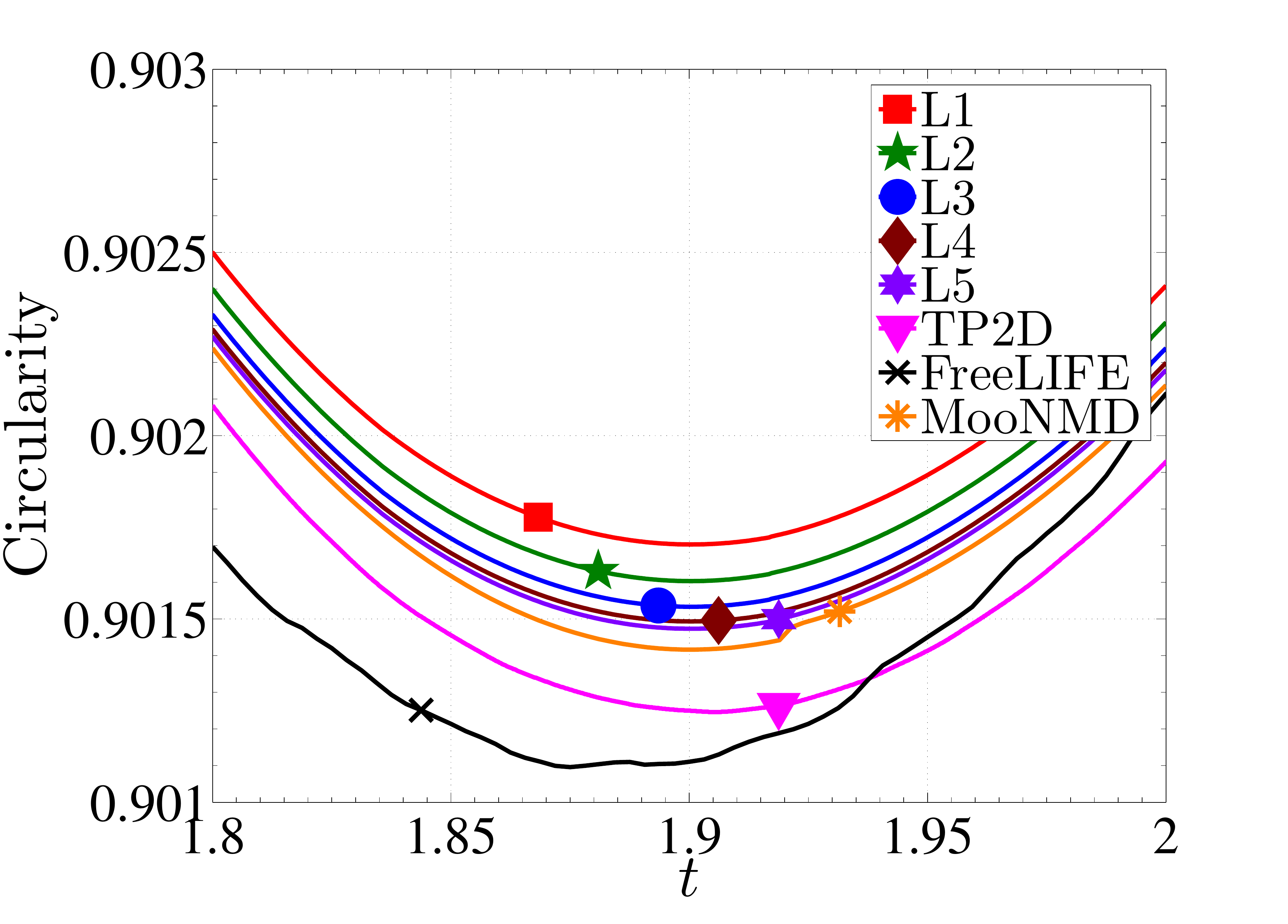}}}
\put(7.4,1.6){\makebox(0,0){\includegraphics[width=8.5cm,height=4.2cm,keepaspectratio]{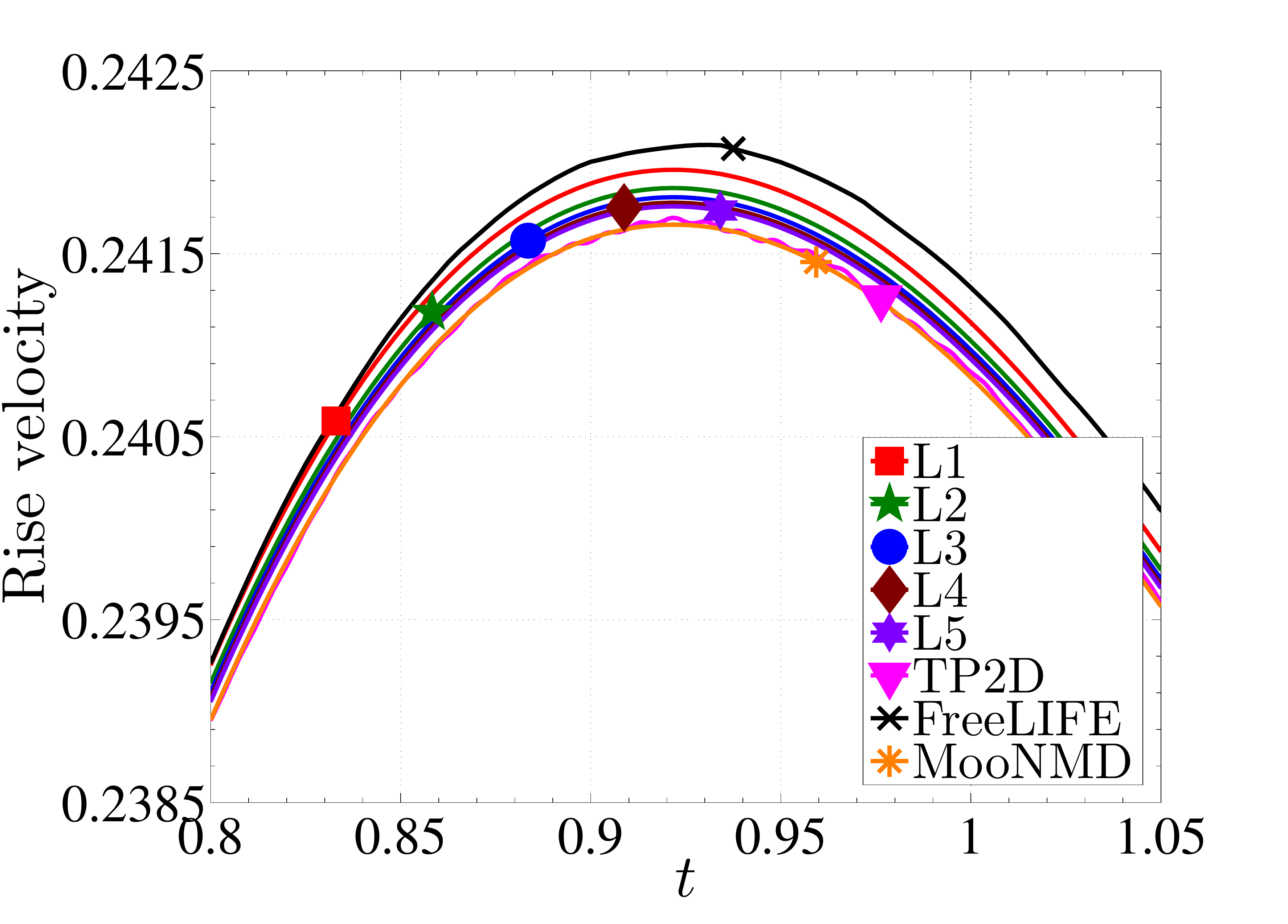}}}
\put(12.9,1.6){\makebox(0,0){\includegraphics[width=8.5cm,height=4.2cm,keepaspectratio]{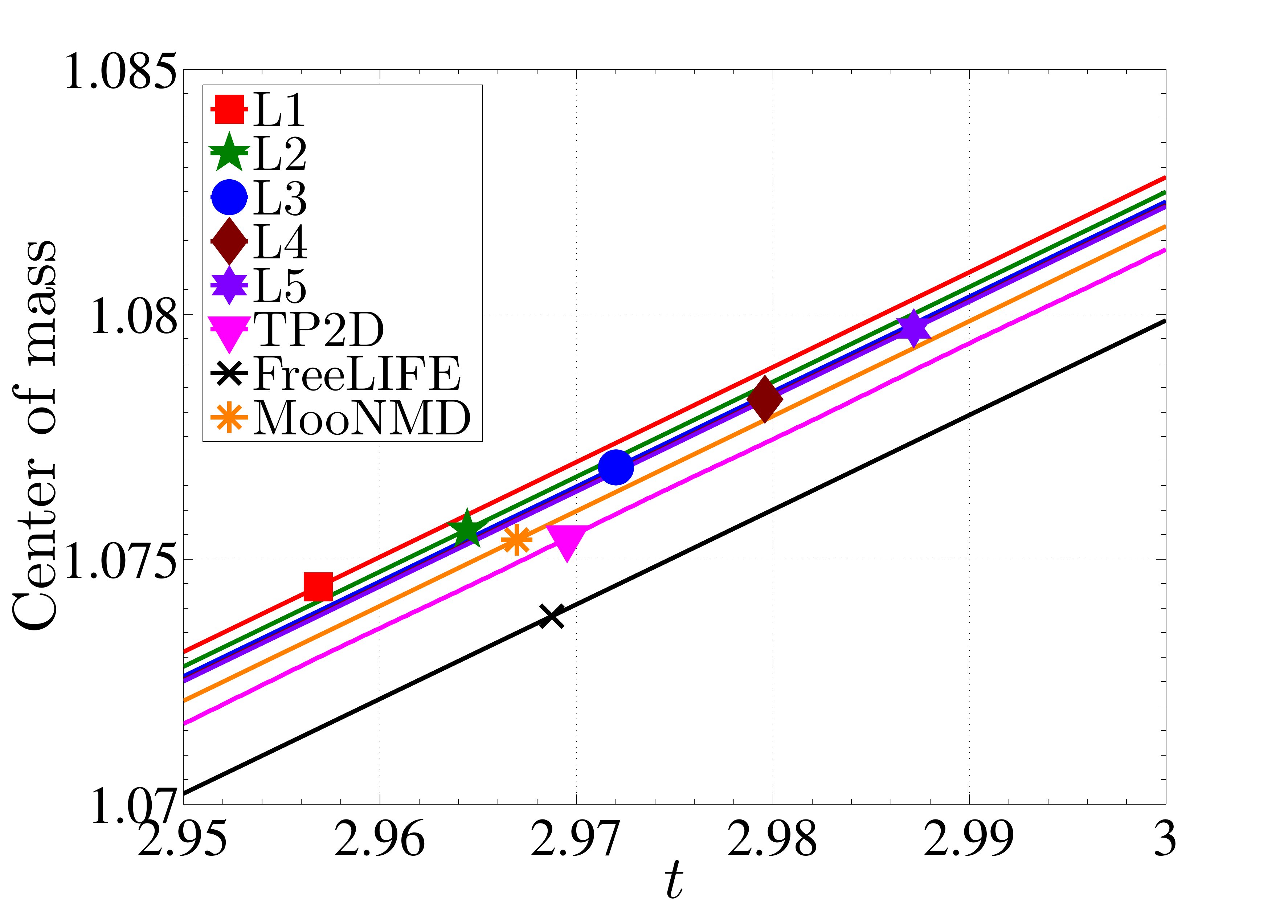}}}
\put(1.85,8.1){$(a)$}
\put(7.3, 8.1){$(b)$}
\put(12.8, 8.1){$(c)$}
\put(1.85,3.5){$(d)$}
\put(7.3, 3.5){$(e)$}
\put(12.8, 3.5){$(f)$}
\end{picture}
\end{center}
\caption{Grid independence test and validation~: temporal evolution of circularity~(a), (d), rise velocity~(b), (e) and center of mass~(c), (f) of a Newtonian bubble rising in a Newtonian fluid column using five different meshes compared with benchmark solutions~\cite{HYS09}.} \label{Plots_Benchmark}
\end{figure*}
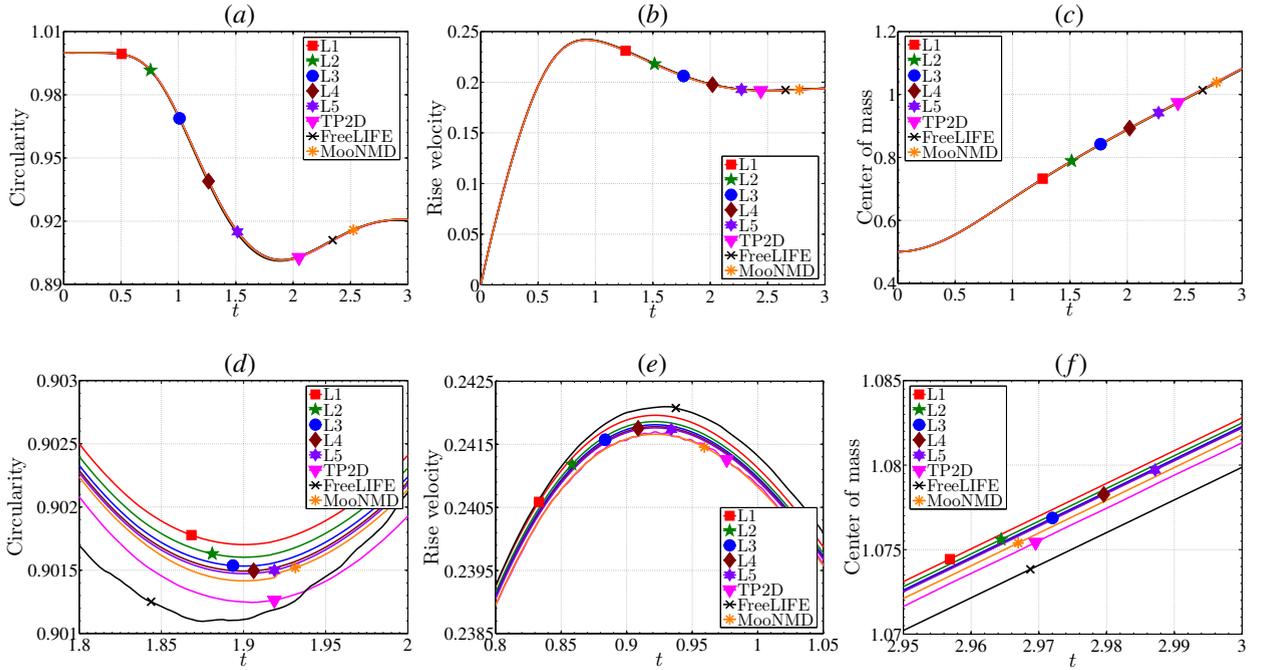

Fig.~\ref{Plots_Benchmark} depicts the convergence behaviour of the temporal evolution of circularity, rise velocity and center of mass of the rising bubble  with different meshes. 
From the zoomed plots (refer Fig.~\ref{Plots_Benchmark}~(d), (e) and (f)), we can observe that the considered flow variables gradually tend to a grid independent value when the mesh becomes finer. 
In particular, the numerical results obtained with the mesh L4 is quite close to those obtained with the mesh L5, which shows the grid independence of the numerical solution. 
In order to have a fine balance between the computational cost and the accuracy, all numerical results in the following sections are obtained with the mesh L4.
Note that we have presented the grid independence test for a 2D Planar configuration. 
However, the same convergence behavior is also observed with L4 and L5 meshes in 3D-axisymmetric configuration.
Further, the benchmark solutions are also plotted in Fig.~\ref{Plots_Benchmark} and our results agree well with the benchmark results.
In order to quantitatively compare our numerical solutions with the benchmark results, the minimum circularity, time at minimum circularity, maximum rise velocity, time at maximum rise velocity and center of mass at $t$~=~3.0 are tabulated in Table~\ref{Bench_Compare}.
We can observe that our results agree well with those in the literature~\cite{HYS09}.

  \begin{table}[h!]
 \begin{center}
 \begin{tabular}{cccccc} 
 \hline
 Reference& L5 (Current work) & TP2D & FreeLIFE  & MooNMD  \\
 \hline
 $\min$(Circularity)   & 0.9015  & 0.9013 & 0.9011 & 0.9013 \\
 $t|_{\min(\text{Circularity})}$  & 1.9005  & 1.9041 &1.8750 &1.9000 \\
 $\max$(Rise velocity)   & 0.2418   & 0.2417 & 0.2421 & 0.2417 \\
  $t|_{\max(\text{Rise velocity})}$  & 0.9214   & 0.9213 & 0.9313 & 0.9239 \\
 Center of mass at $t=3.0$   & 1.0822   & 1.0813 & 1.0799  & 1.0817 \\
 \hline 
 \end{tabular}
 \end{center}
 \caption{Newtonian bubble rising in a Newtonian fluid column~: comparison of results with the benchmark solutions in the literature~\cite{HYS09}.} \label{Bench_Compare}
 \end{table}

\subsection{Newtonian bubble rising in a viscoelastic fluid column}
In this section, we consider a 3D-axisymmetric Newtonian bubble rising in a viscoelastic fluid column due to buoyancy. 
We designate a base case to systematically examine the effects of various flow parameters.
The base case is defined as~: $\Rey_2$~=~$10$, Eo~=~$400$, $\text{Wi}_2$~=~25, $\rho_1/\rho_2$~=~0.1, $\varepsilon$~=~10, $\beta_1$~=~1.0, $\beta_2$~=~0.75, $\alpha_2$~=~0.1, $D$~=~0.5 and $h_c$~=~2.0.
The computational domain is triangulated into an interface resolved mesh using the mesh generator Triangle~\cite{TRI96, TRI02} based on constrained Delaunay triangulation.
We limit the maximum area of each cell in the mesh to 0.001 during the triangulation (initially and as well as during the remeshing).
This results in 1835 and 3111 cells in the initial inner and outer domains respectively. 
The finite element spaces used in computations for the velocity / pressure / viscoelastic stress are  $P_2^{bubble}$ / $P_1^{disc}$ / $P_2^{bubble}$.
This choice of initial mesh and finite element spaces results in 49742 velocity, 14838 pressure and 74613 viscoelastic degrees of freedom. 
Further, we use a constant time step $\delta t = 0.0005$ and 600 degrees of freedom on the interface with $h_0 = 0.002617982$, where $h_0$ is the mesh size at $t$~=~0.
In computations, the number of cells and the number of degrees of freedom might change during the remeshing.
Further, the stabilization constants used in computations are $c_1$~=~0.005, $c_2$~=~0.005 and $c_3$~=~0.005.
In order to avoid the effect of the presence of the wall at the top of the domain, simulations were stopped when the bubble reaches a constant velocity or when its velocity begins to decrease due to the proximity of the top surface.

\begin{figure*}
\begin{center}
\unitlength1cm
\begin{picture}(20,20.5)

\put(1.5,-1.1){\makebox(3,6){\includegraphics[trim=1.6cm 1.5cm 7.6cm 1.0cm, clip=true,width=4.3cm]{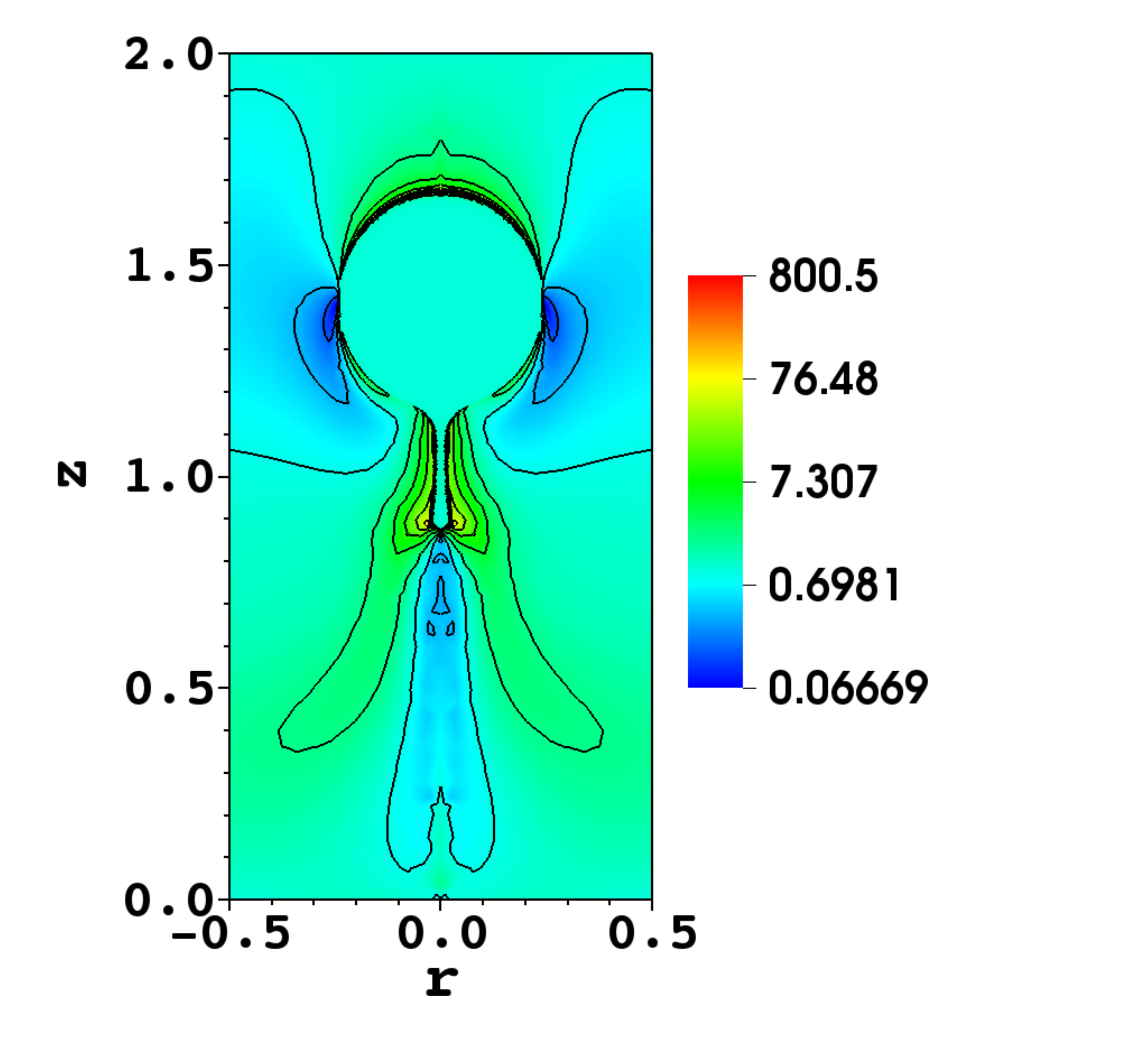}}}
\put(6.5,-1.1){\makebox(3,6){\includegraphics[trim=1.6cm 1.5cm 7.6cm 1.0cm, clip=true,width=4.3cm]{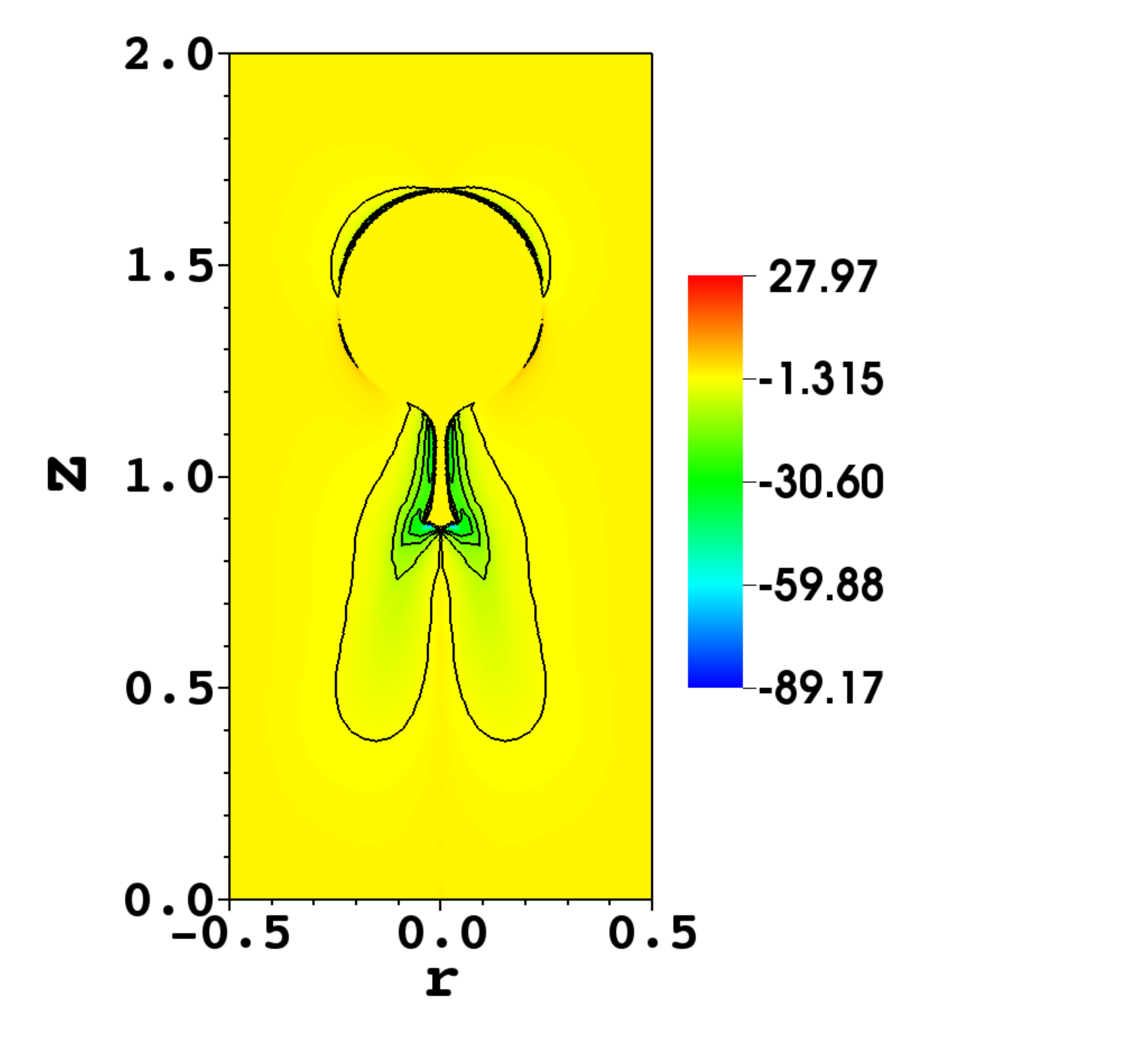}}}
\put(11.5,-1.1){\makebox(3,6){\includegraphics[trim=1.6cm 1.5cm 7.6cm 1.0cm, clip=true,width=4.3cm]{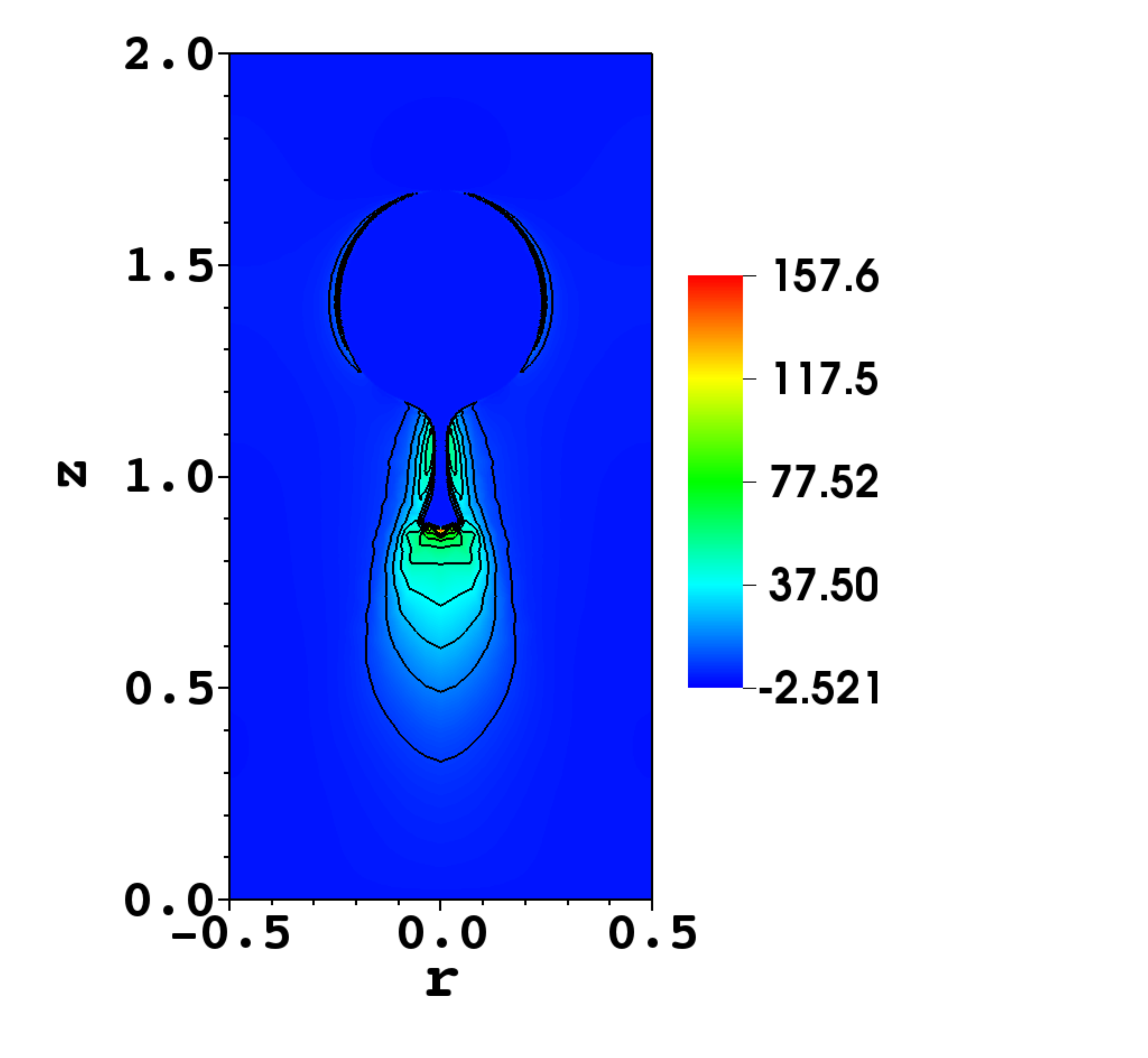}}}

\put(1.5,4.2){\makebox(3,6){\includegraphics[trim=1.6cm 1.5cm 7.6cm 1.0cm, clip=true,width=4.3cm]{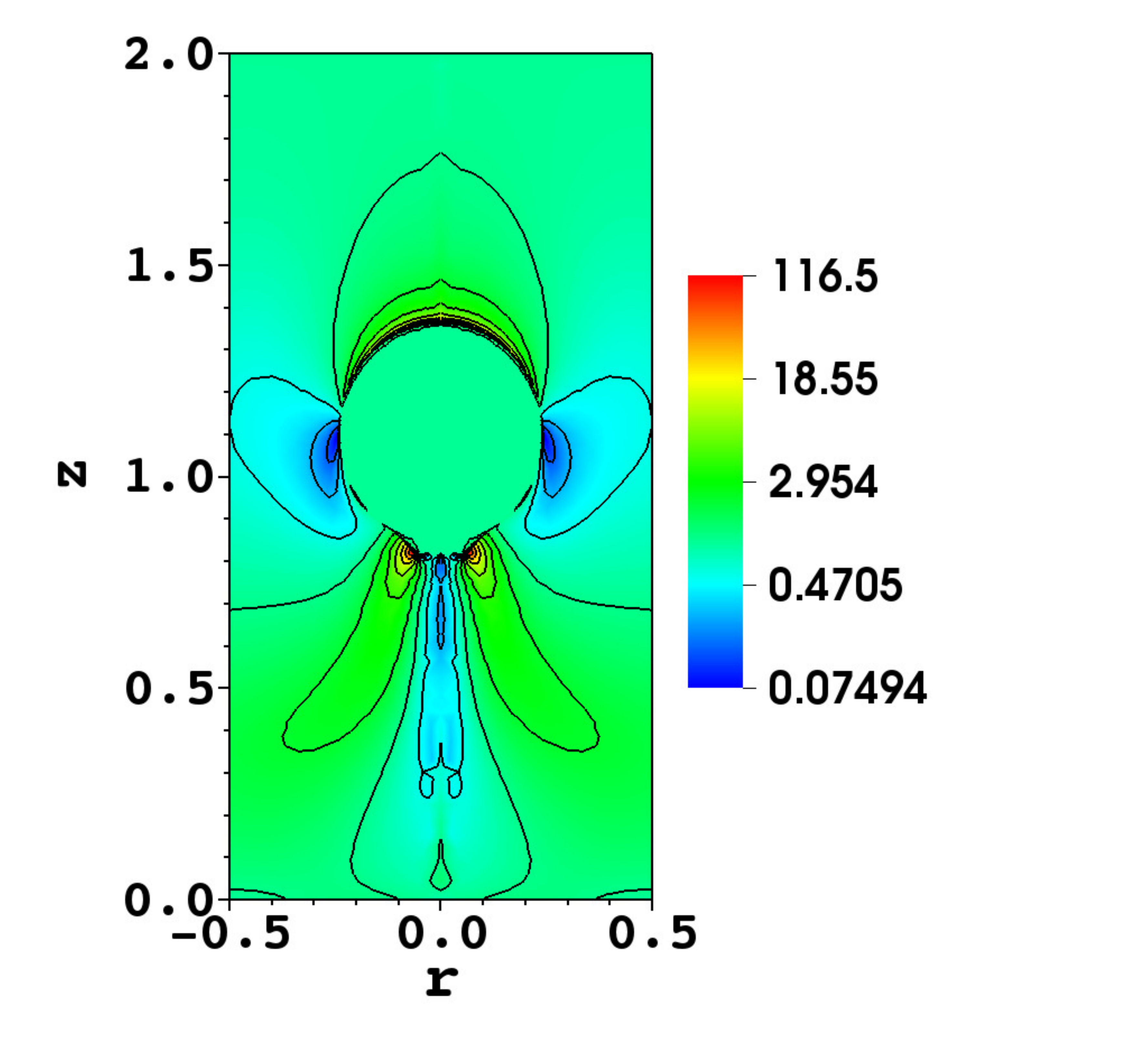}}}
\put(6.5,4.2){\makebox(3,6){\includegraphics[trim=1.6cm 1.5cm 7.6cm 1.0cm, clip=true,width=4.3cm]{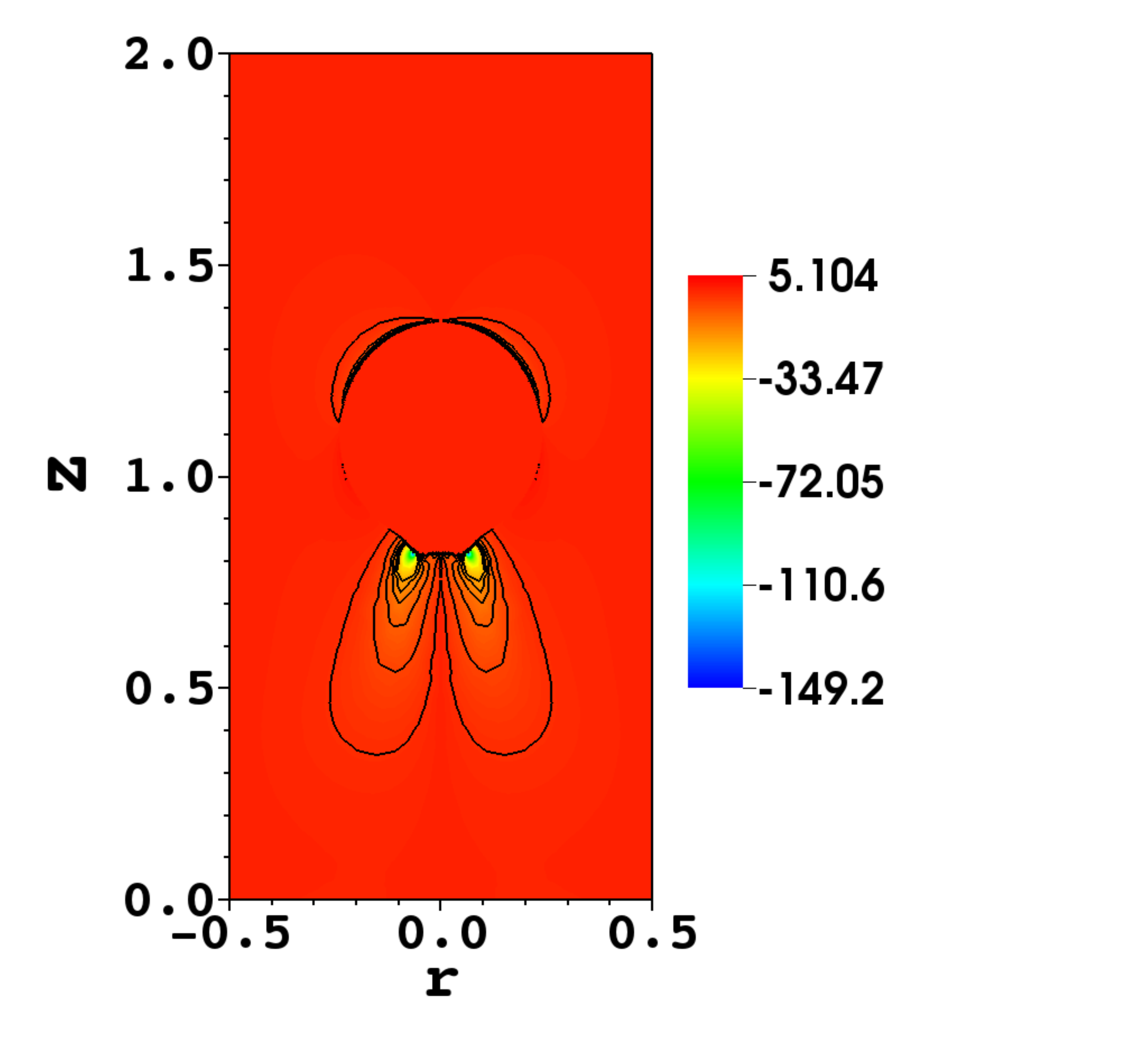}}}
\put(11.5,4.2){\makebox(3,6){\includegraphics[trim=1.6cm 1.5cm 7.6cm 1.0cm, clip=true,width=4.3cm]{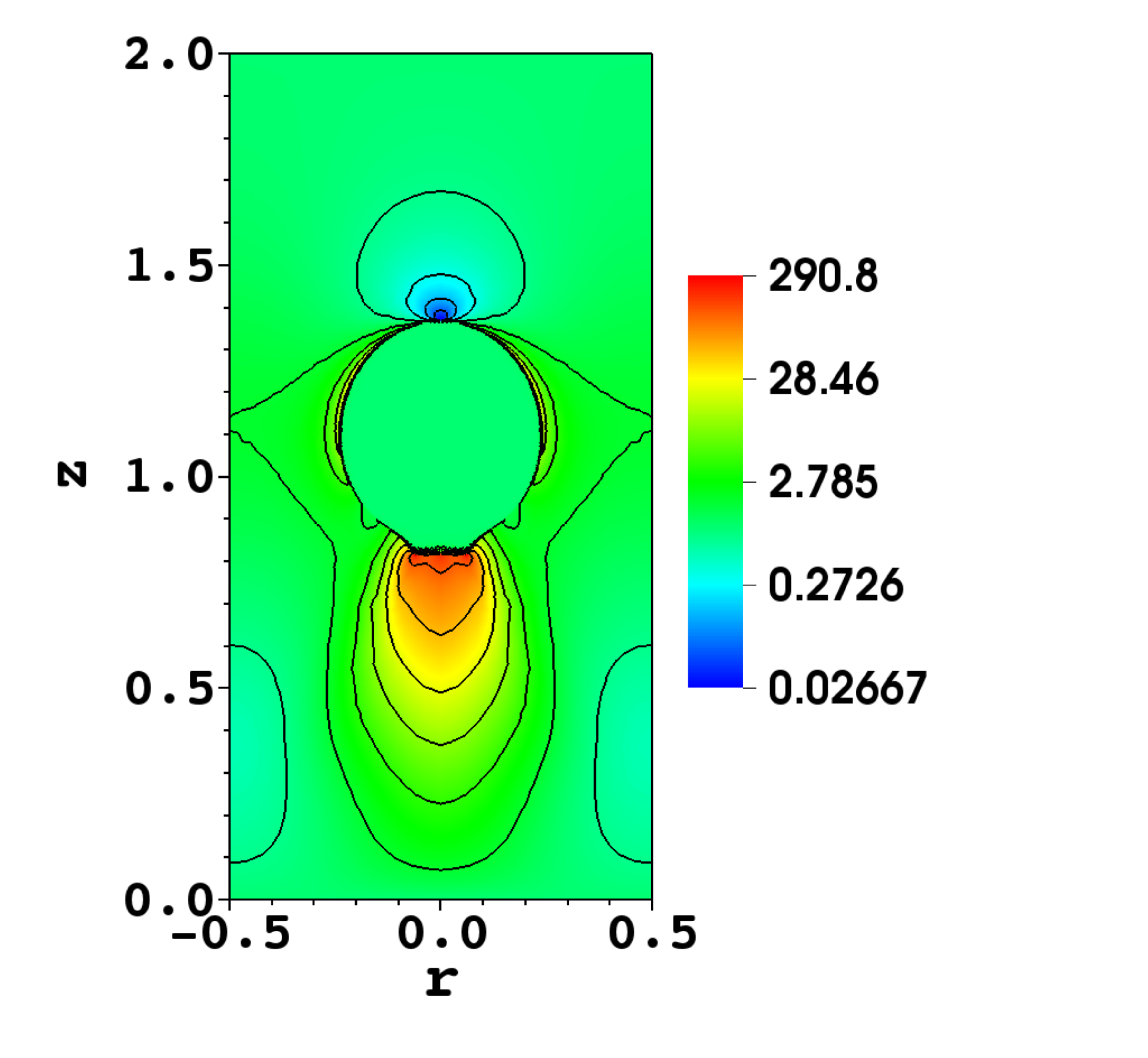}}}

\put(1.5,9.5){\makebox(3,6){\includegraphics[trim=1.6cm 1.5cm 7.6cm 1.0cm, clip=true,width=4.3cm]{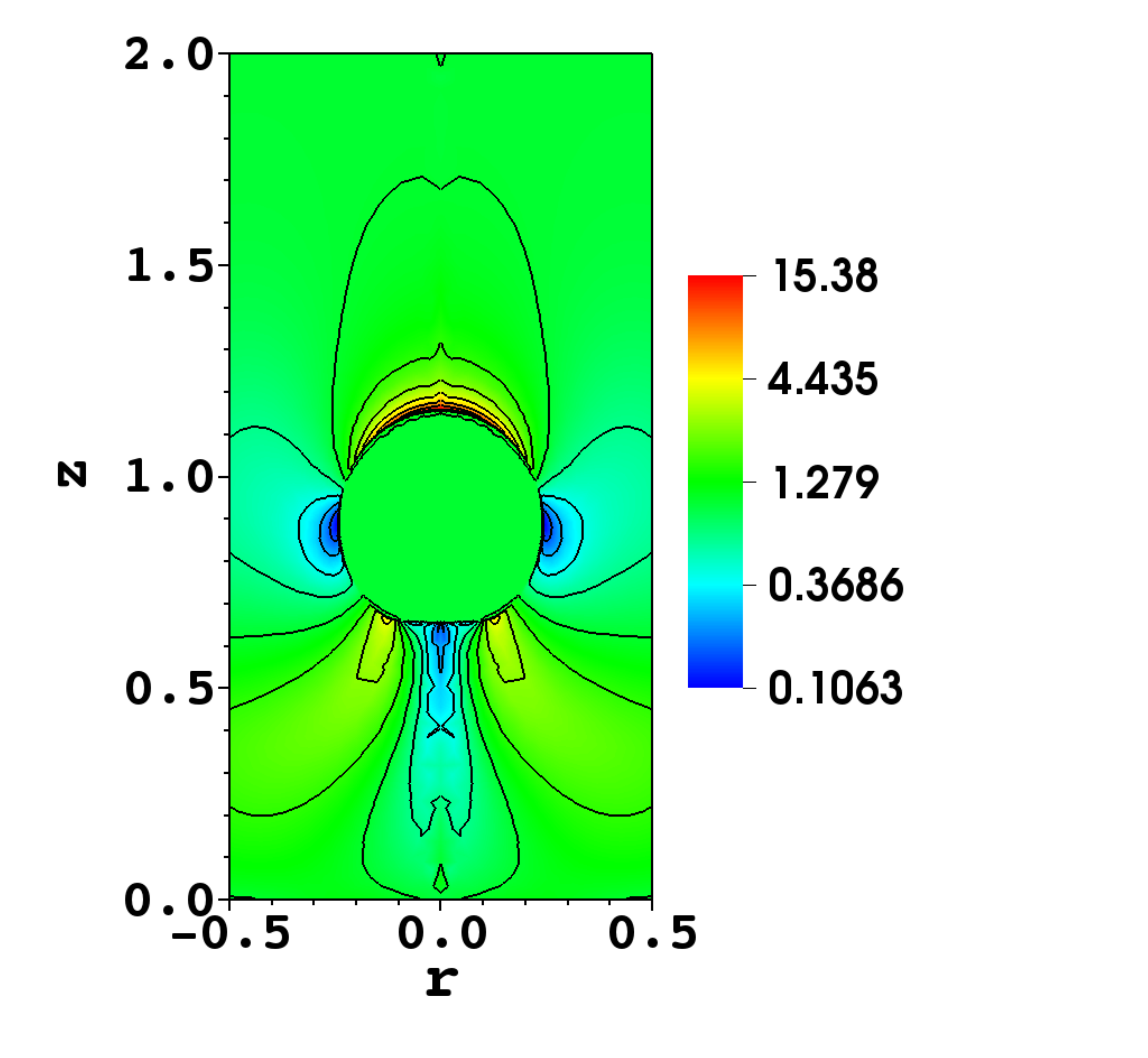}}}
\put(6.5,9.5){\makebox(3,6){\includegraphics[trim=1.6cm 1.5cm 7.6cm 1.0cm, clip=true,width=4.3cm]{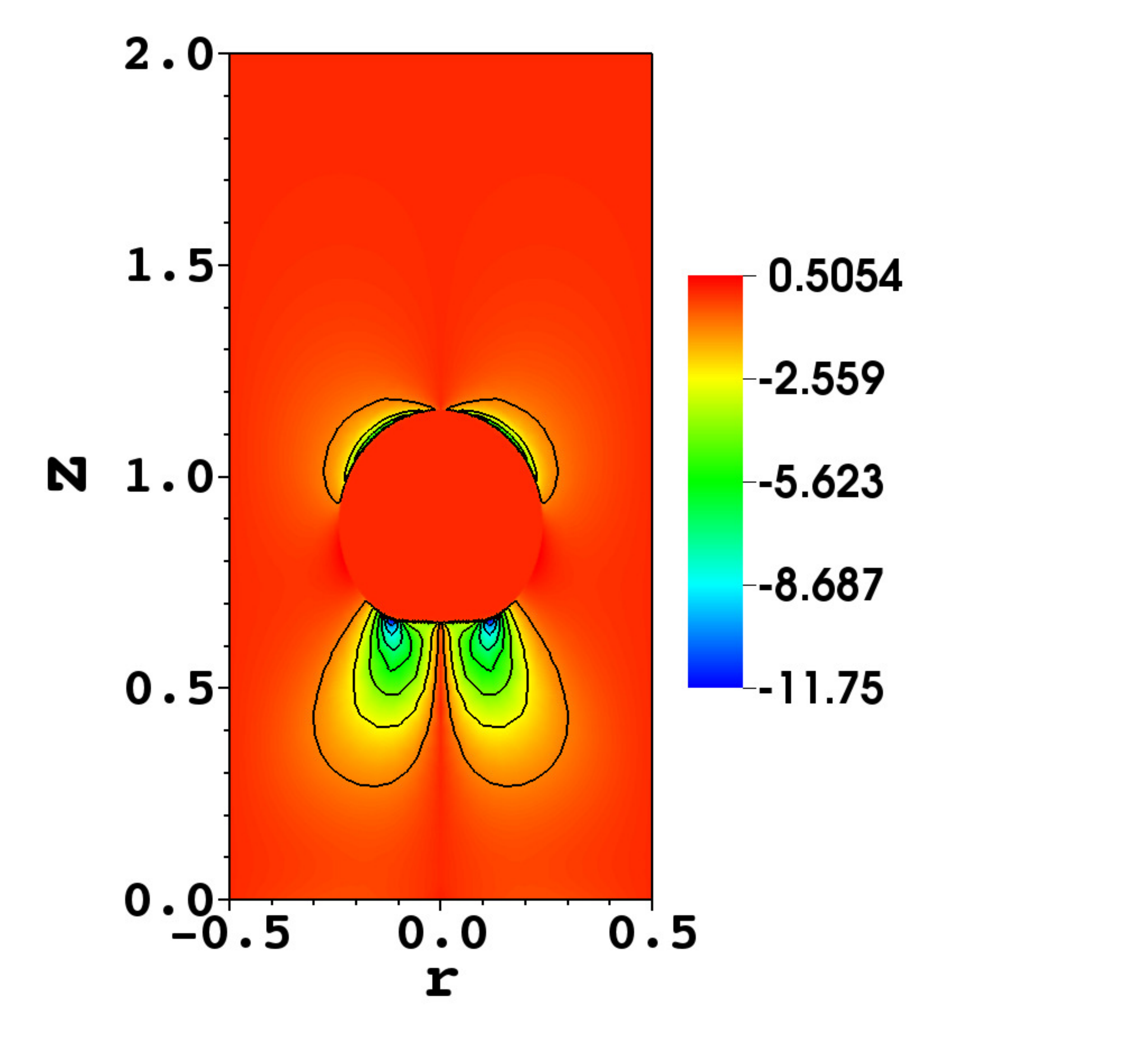}}}
\put(11.5,9.5){\makebox(3,6){\includegraphics[trim=1.6cm 1.5cm 7.6cm 1.0cm, clip=true,width=4.3cm]{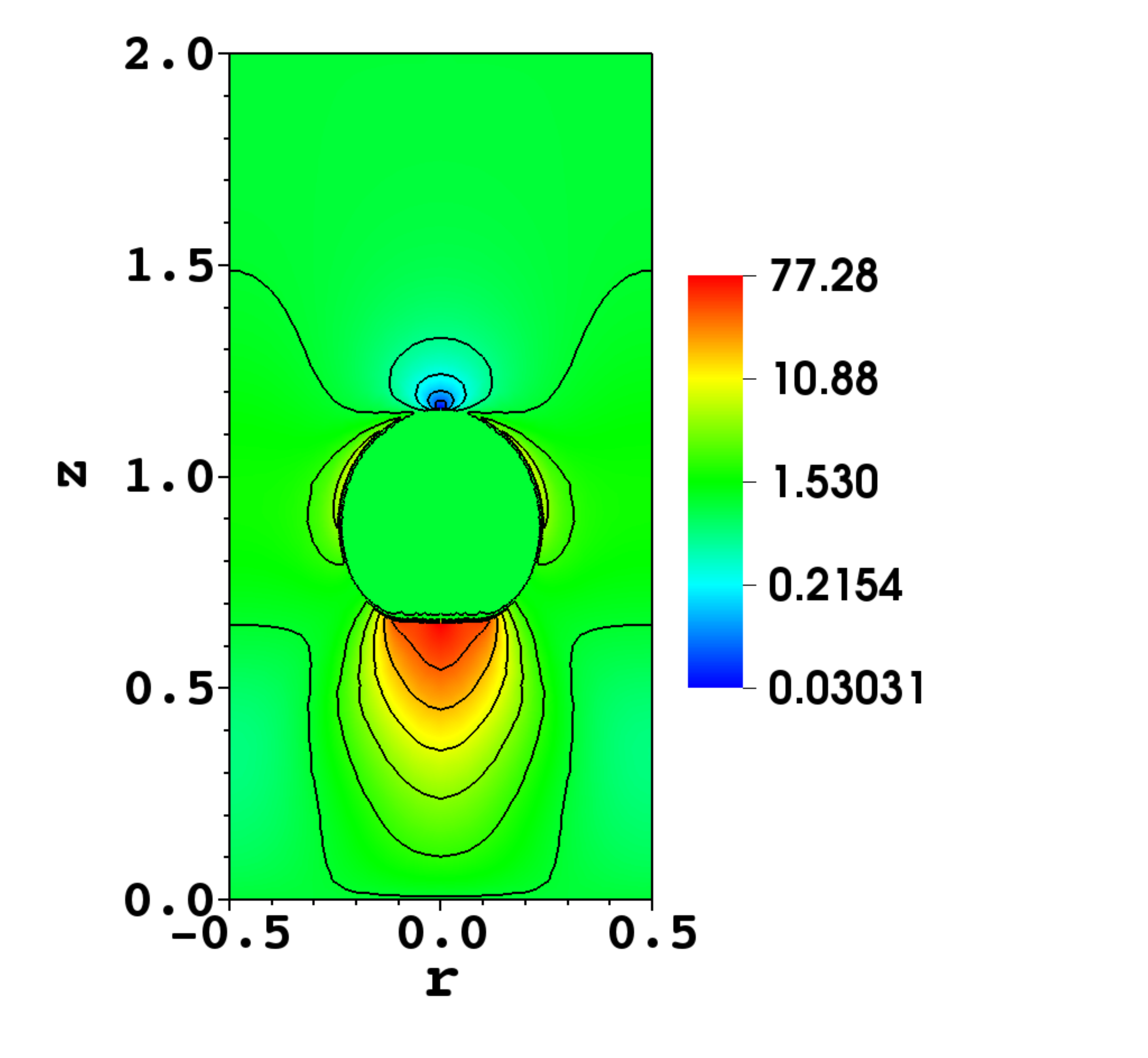}}}

\put(1.5,14.8){\makebox(3,6){\includegraphics[trim=1.6cm 1.5cm 7.6cm 1.0cm, clip=true,width=4.3cm]{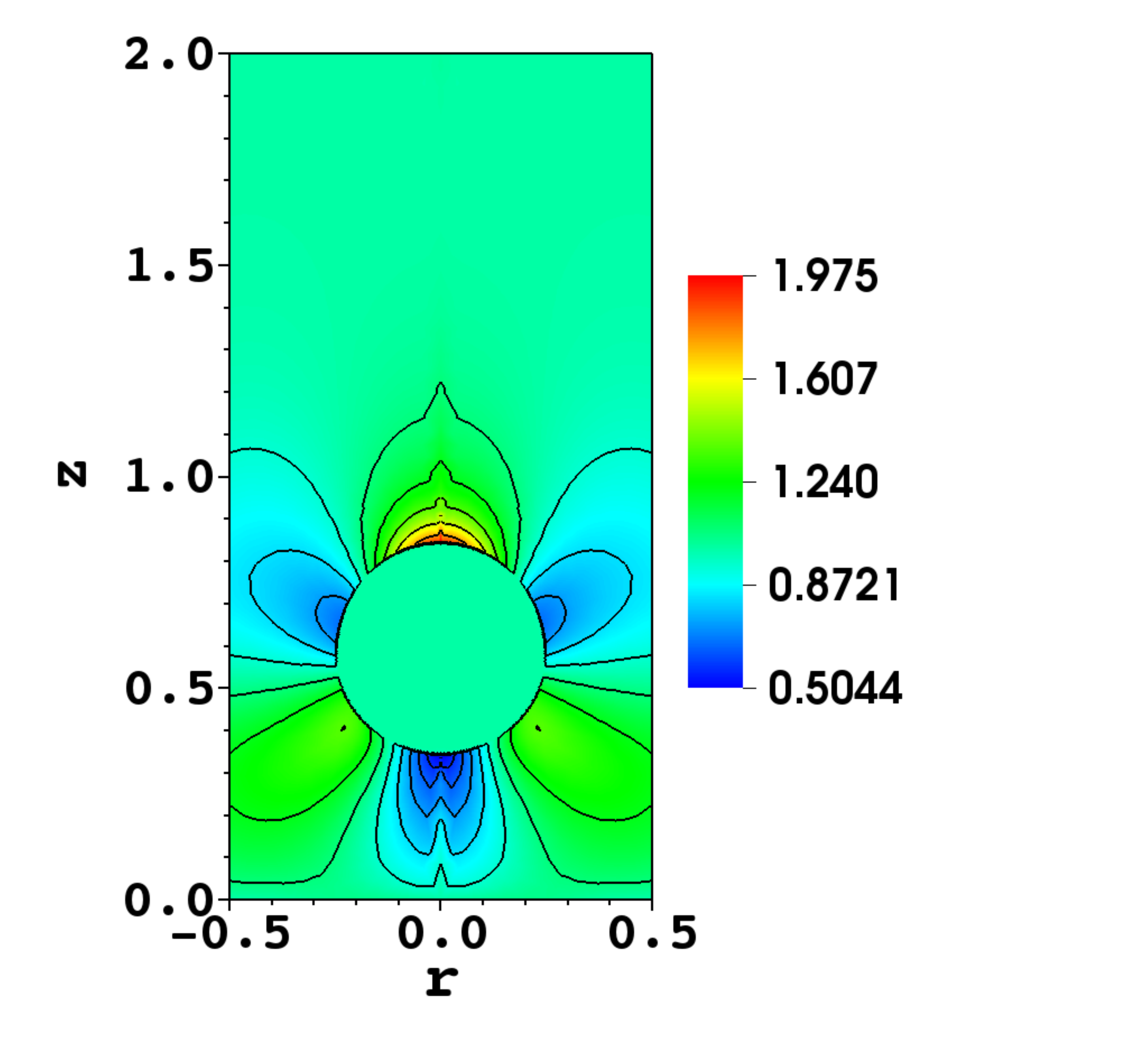}}}
\put(6.5,14.8){\makebox(3,6){\includegraphics[trim=1.6cm 1.5cm 7.6cm 1.0cm, clip=true,width=4.3cm]{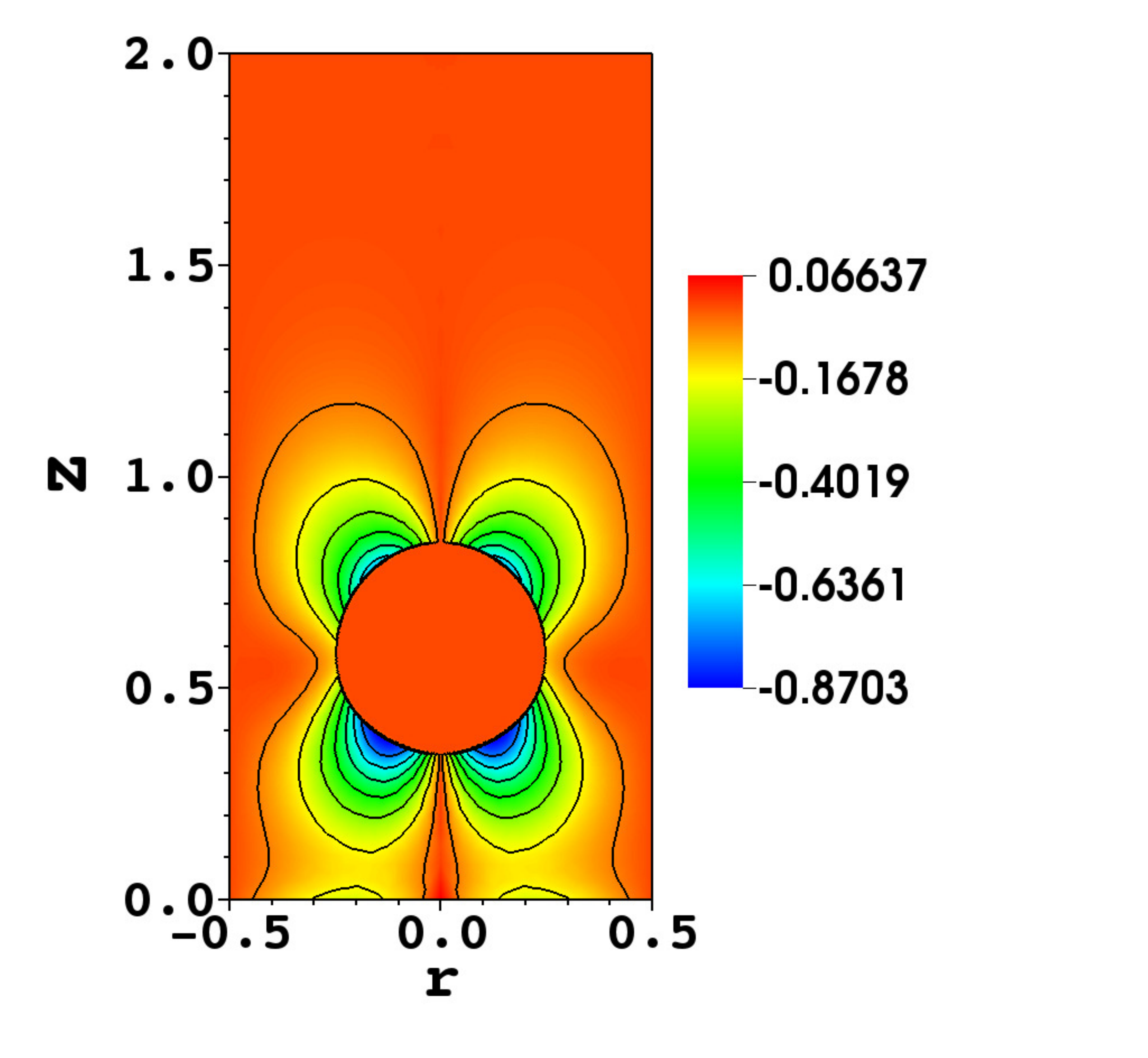}}}
\put(11.5,14.8){\makebox(3,6){\includegraphics[trim=1.6cm 1.5cm 7.6cm 1.0cm, clip=true,width=4.3cm]{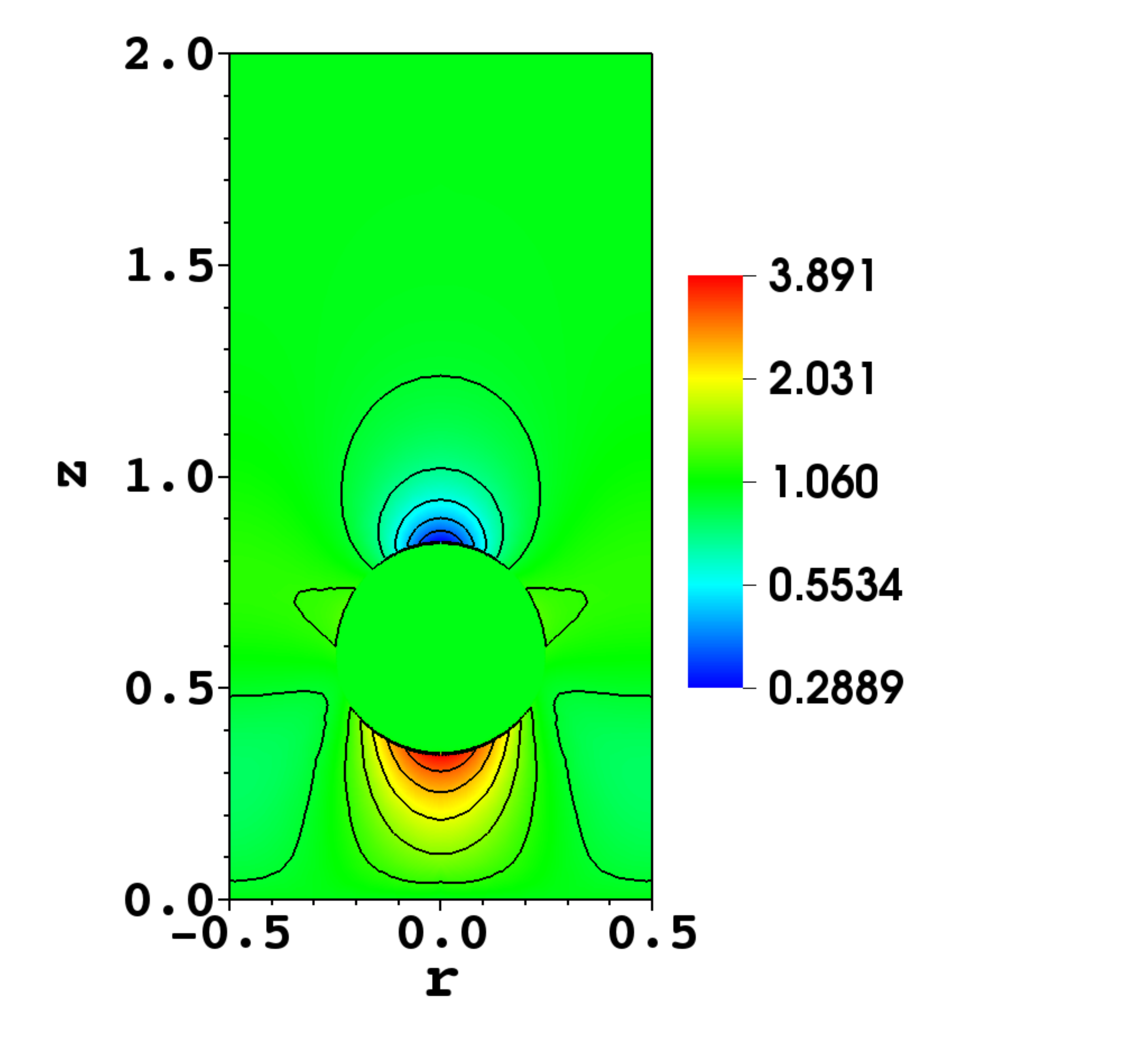}}}

\put(7.5,4.3){$t$~=~$9.0$}
\put(7.5,9.6){$t$~=~$6.0$}
\put(7.5,14.9){$t$~=~$4.0$}
\put(7.5,20.2){$t$~=~$1.0$}

\put(14.05,24.5){$\tau_{zz}$}
\put(14.05,19.2){$\tau_{zz}$}
\put(14.05,13.9){$\tau_{zz}$}
\put(14.05,8.6){$\tau_{zz}$}
\put(14.05,3.3){$\tau_{zz}$}

\put(9.05,24.5){$\tau_{rz}$}
\put(9.05,19.2){$\tau_{rz}$}
\put(9.05,13.9){$\tau_{rz}$}
\put(9.05,8.6){$\tau_{rz}$}
\put(9.05,3.3){$\tau_{rz}$}

\put(4.05,24.5){$\tau_{rr}$}
\put(4.05,19.2){$\tau_{rr}$}
\put(4.05,13.9){$\tau_{rr}$}
\put(4.05,8.6){$\tau_{rr}$}
\put(4.05,3.3){$\tau_{rr}$}

\end{picture}
\end{center}
\caption{Viscoelastic conformation stress profiles for a Newtonian bubble rising in a viscoelastic fluid with flow parameters $\Rey_2$~=~$10$, Eo~=~$400$, $\text{Wi}_2$~=~25, $\rho_1/\rho_2$~=~0.1, $\varepsilon$~=~10, $\beta_1$~=~1.0, $\beta_2$~=~0.75, $\alpha_2$~=~0.1, $D$~=~0.5 and $h_c$~=~2.0  at dimensionless times $t$~=~1.0, 4.0, 6.0 and 9.0.}
\label{Tau_VTKPlots_NV}
\end{figure*}
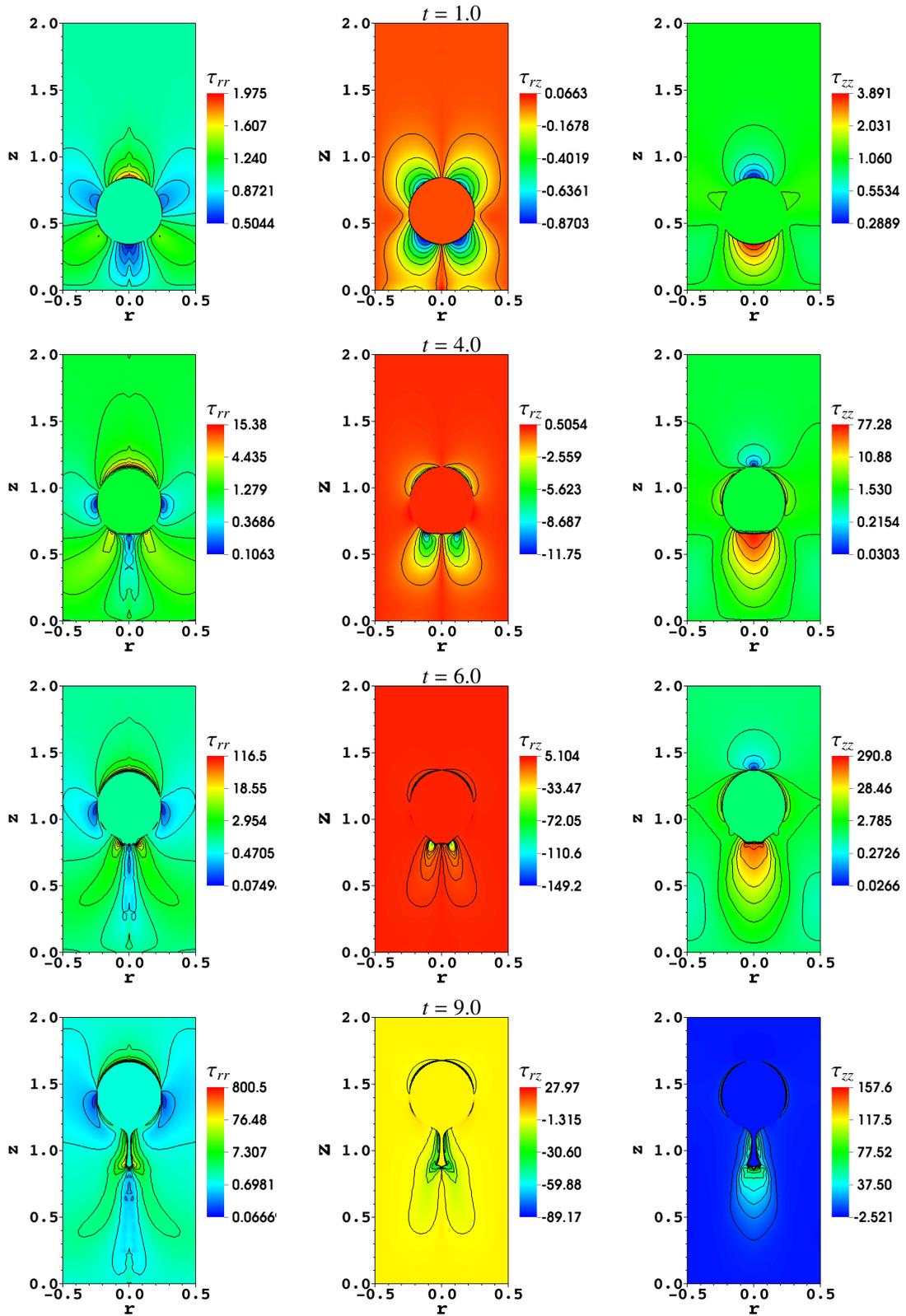

Fig.~\ref{Tau_VTKPlots_NV} presents the viscoelastic stress profiles for the base case flow parameters at dimensionless time instances $t$~=~1.0, 4.0, 6.0 and 9.0. 
At time $t$~=~0, the bubble is of a spherical shape with  initial velocities of the bubble and the bulk fluid assumed to be zero and the viscoelastic conformation stress tensor is set as $\btau_{p,0}$~=~$\mathbb{I}$. 
Initially, the buoyancy force generated by the density difference between two fluids accelerates the bubble in the opposite direction of the gravity, i.e. the bubble rises up in the bulk fluid column.
The transient behaviour of a buoyant bubble accelerating from rest in a viscoelastic fluid depends on its volume and the magnitudes of the viscous and viscoelastic stresses, which themselves depend on the fluid properties such as the viscosity and the relaxation time. 
The bubble is driven by the force of buoyancy, while the viscous and viscoelastic stresses resist its motion. 
If the deforming stresses at the interface are sufficiently smaller than the interfacial tension force, the bubble shape remains approximately spherical.
However, when these deforming stresses are significant the interface deforms and the bubble shape changes depending on the properties of the bulk fluid~: it deforms to an oblate shape in inertia-dominated flows and to a prolate shape with or without a cusp-like trailing end in flows in which viscoelasticity is important.

At $t$~=~1.0, we can observe that the maximum values of viscoelastic stress component $\tau_{rr}$ starts to accumulate at the front stagnation point, while $\tau_{rz}$ gets built up along the entire circumference of the bubble. 
However, the maximum values of $\tau_{zz}$ are concentrated at the rear stagnation point.  
The initial motion of the bubble is dominated by viscous stresses as the viscoelastic stresses take some time to build up.
Further, along the interface, the interfacial tension force dominates compared to the viscous and viscoelastic stresses.
Hence, the shape of the bubble is more spherical at $t$~=~1.0, similar to a Newtonian bubble rising in a Newtonian fluid column.
At $t$~=~4.0, we can observe that the peak magnitude of viscoelastic stresses have increased, but still the viscous stresses continue to dominate the flow dynamics and hence, the bubble shape remains more spherical.  

At time $t$~=~6.0, the bubble starts to become prolate and this is an indication that the viscoelastic stresses are starting to dominate the flow dynamics. 
In particular, the viscous and viscoelastic stresses overcome the interfacial tension. 
Further, the maximum values of $\tau_{zz}$ and minimum values of $\tau_{rr}$ are concentrated at the rear stagnation point.
Hence, the polymers near the trailing end of the bubble get stretched along the $z$ direction.
The extensional viscoelastic stresses in general being large in a thin section at the trailing end of the bubble can surmount the interfacial tension, hence forming a cusp-like trailing end.
The cusp-like trailing end becomes more and more obvious as the time progresses. 
Since, the maximum values of $\tau_{rr}$ and minimum values of $\tau_{zz}$ occur at the front stagnation point, the upstream axial flow experiences a strong turn tangential to the bubble surface so that the polymers are greatly extended in the radial directions.
Thus, the bubble doesn't experience noticeable deformation in the vicinity of its front end.
With further advancement in time, the viscoelastic stresses completely dominate the rising bubble dynamics. 
At $t$~=~9.0, $\tau_{zz}$ gets concentrated only in the rear stagnation point resulting in the trailing end of the bubble being extremely pulled out. 
Next, we perform a parametric study to examine the effects of viscosity ratio, Newtonian solvent ratio, Giesekus mobility factor and E\"{o}tv\"{o}s number on the rising bubble dynamics in a viscoelastic fluid column.

\subsubsection{Influence of viscosity ratio on the bubble dynamics}
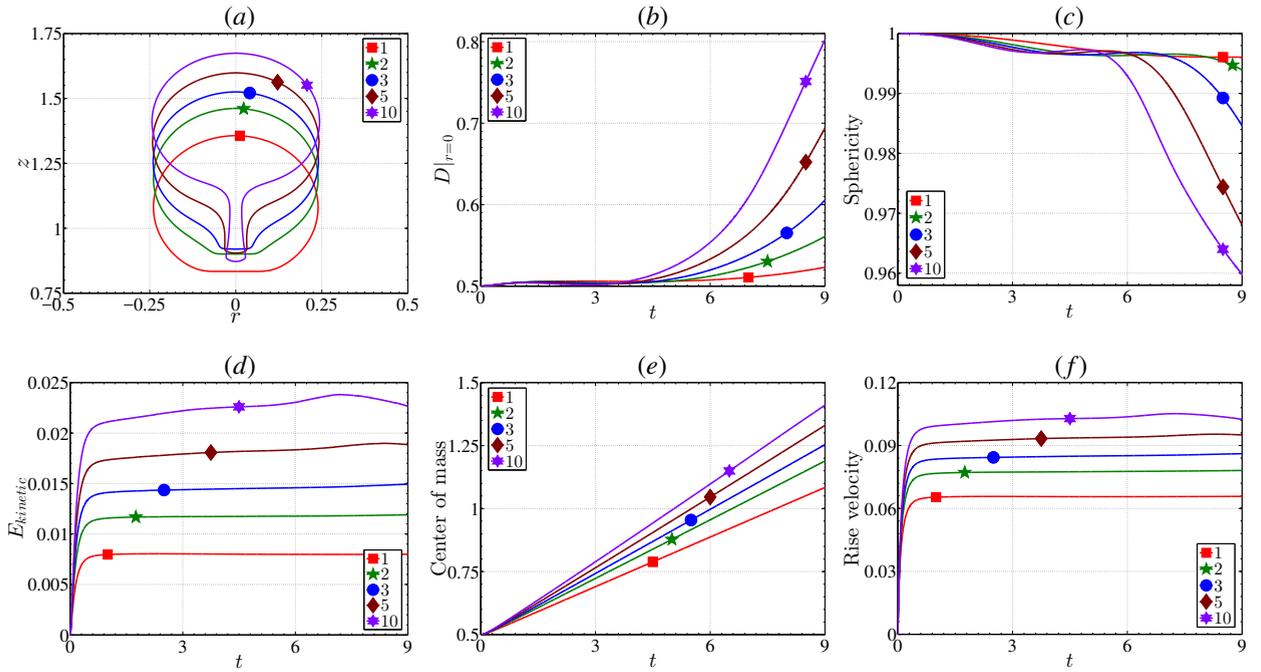
\begin{figure*}[ht!]
\begin{center}
\unitlength1cm
\begin{picture}(14.5,8.3)
\put(1.9,6.2){\makebox(0,0){\includegraphics[width=8.5cm,height=4.2cm,keepaspectratio]{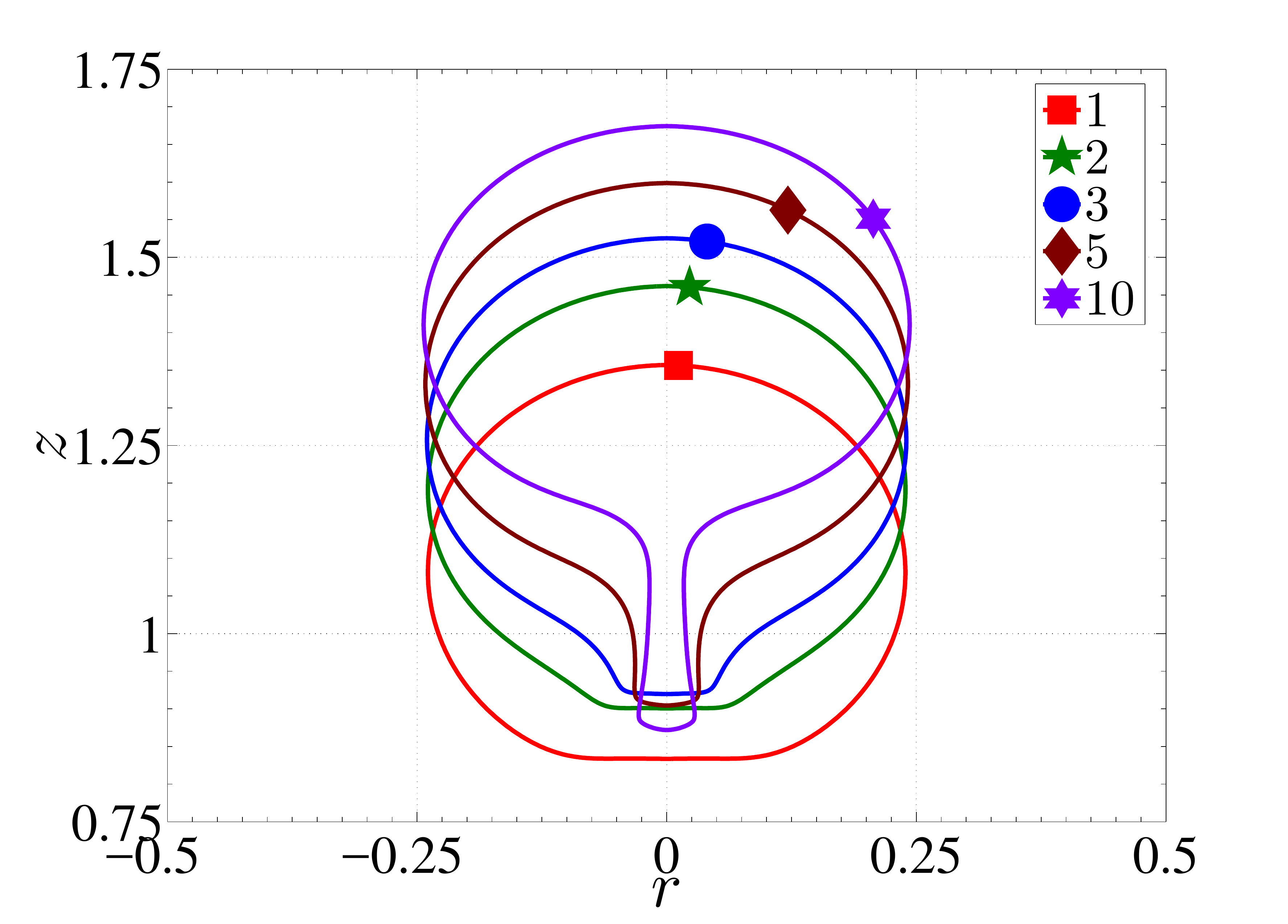}}}
\put(7.4,6.2){\makebox(0,0){\includegraphics[width=8.5cm,height=4.2cm,keepaspectratio]{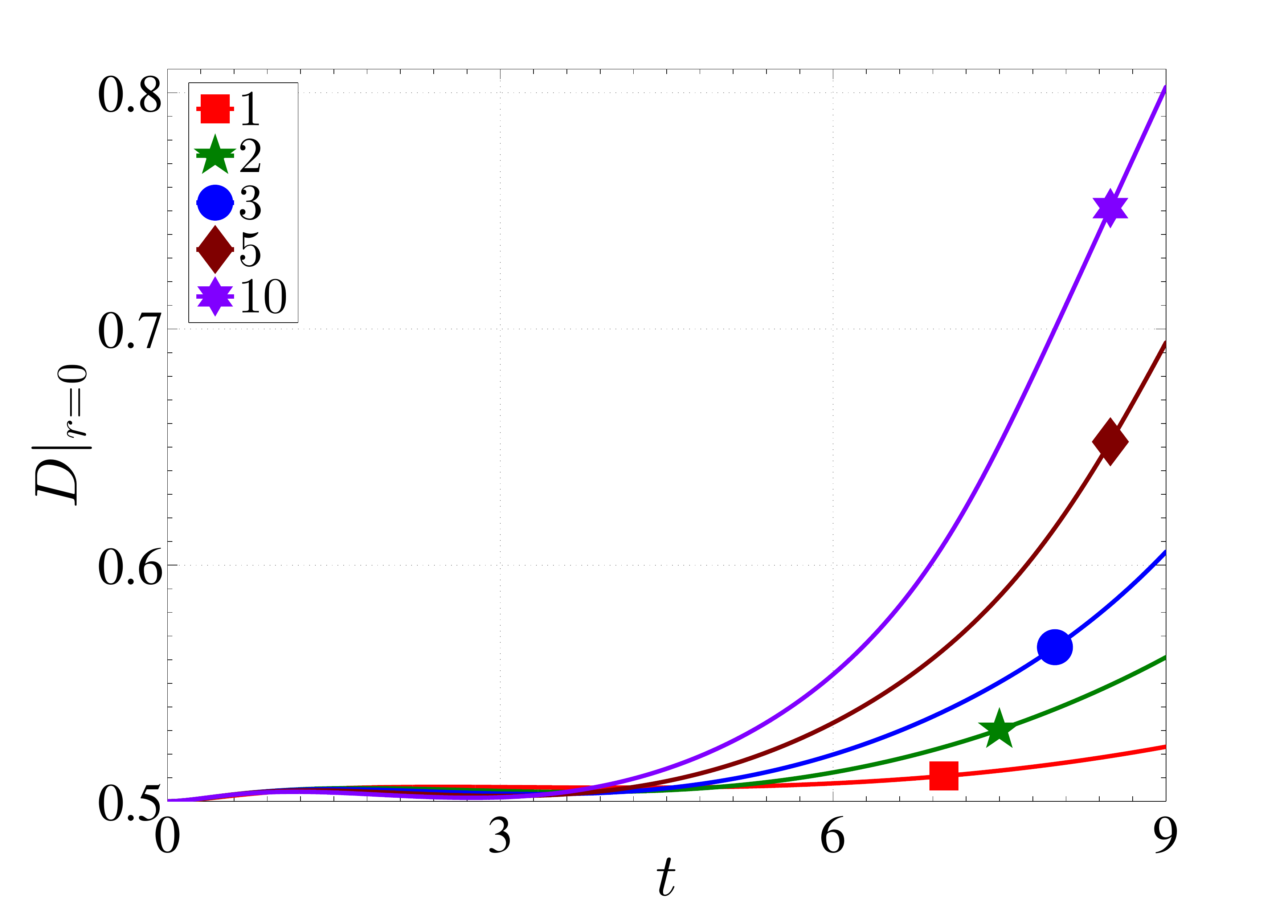}}}
\put(12.9,6.2){\makebox(0,0){\includegraphics[width=8.5cm,height=4.2cm,keepaspectratio]{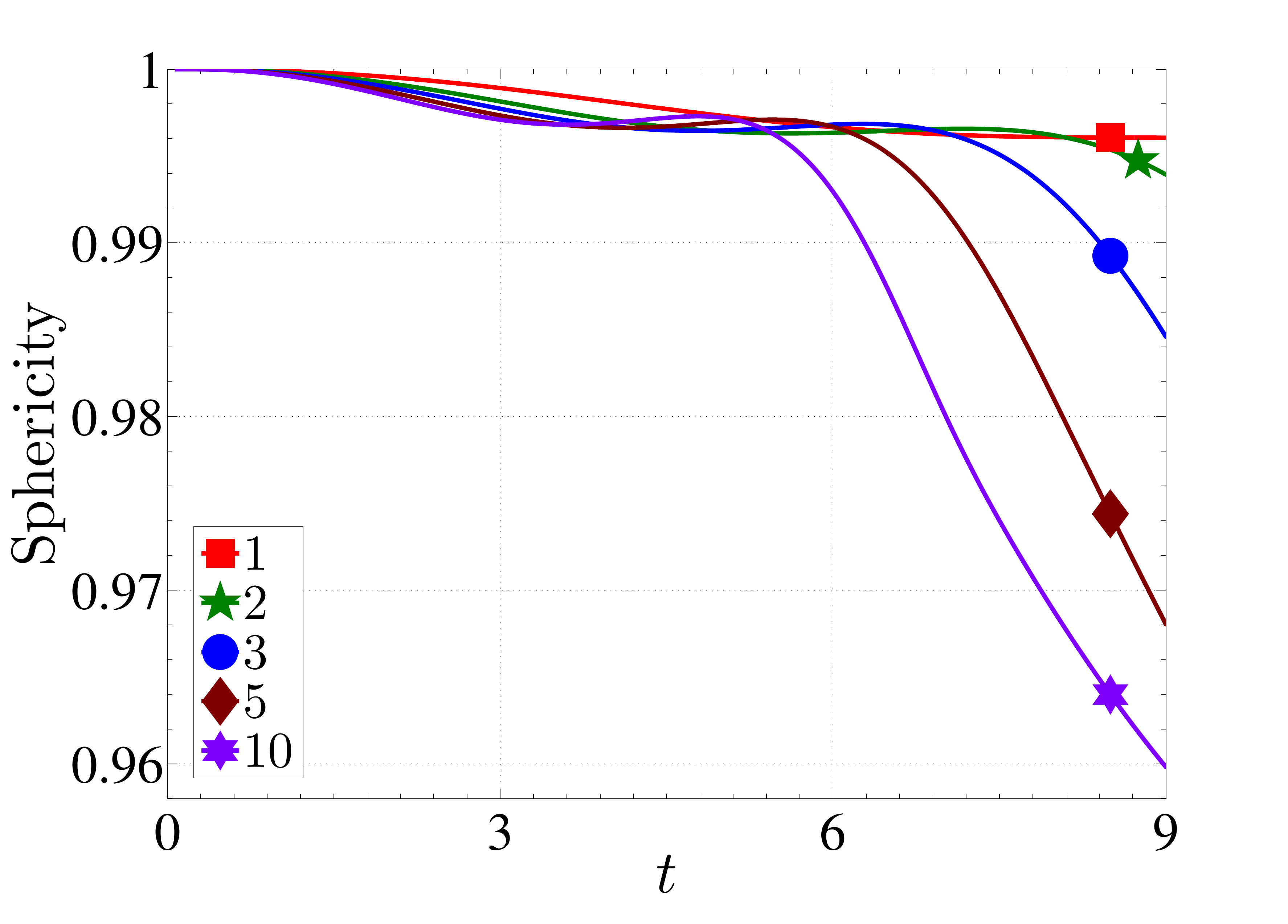}}}
\put(1.9,1.6){\makebox(0,0){\includegraphics[width=8.5cm,height=4.2cm,keepaspectratio]{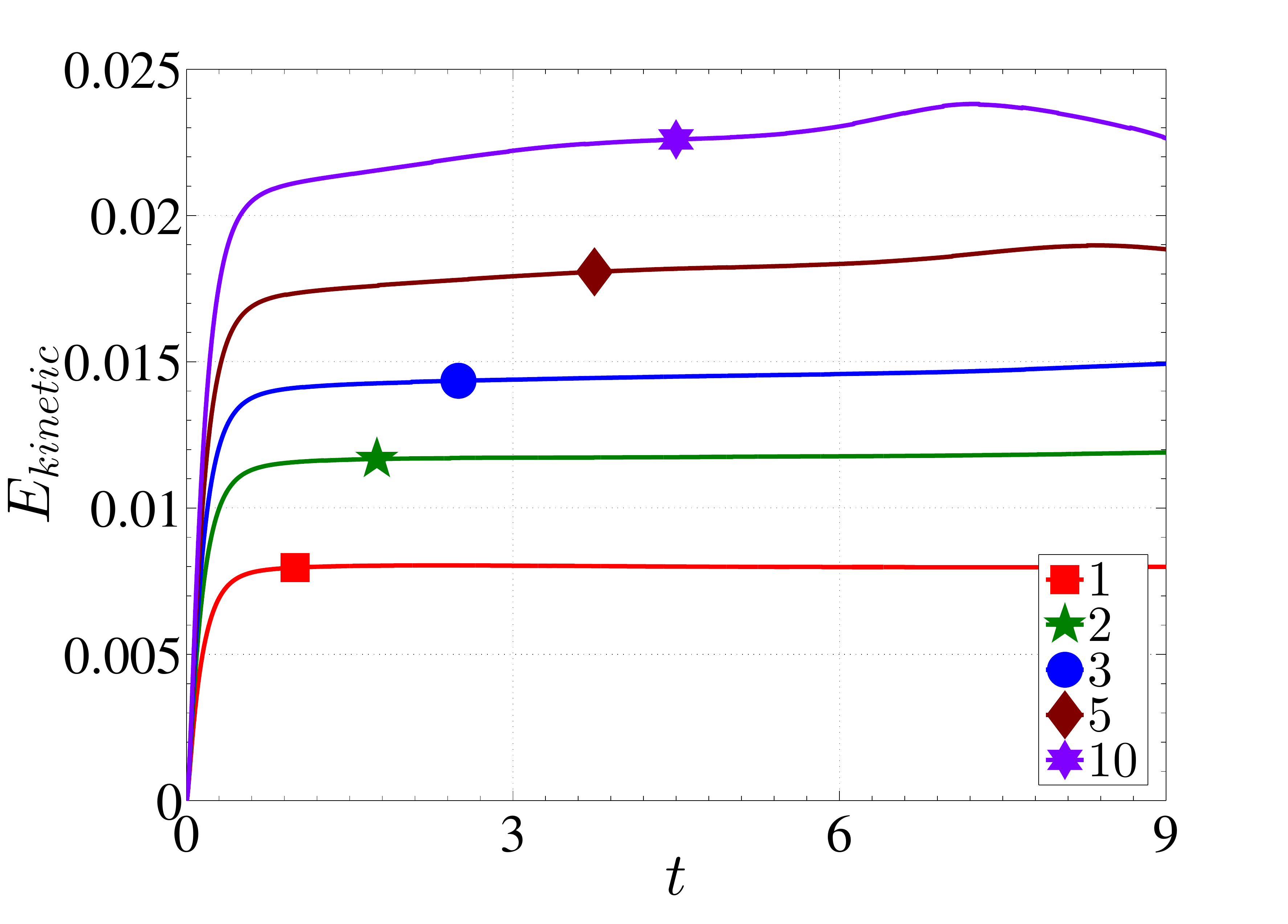}}}
\put(7.4,1.6){\makebox(0,0){\includegraphics[width=8.5cm,height=4.2cm,keepaspectratio]{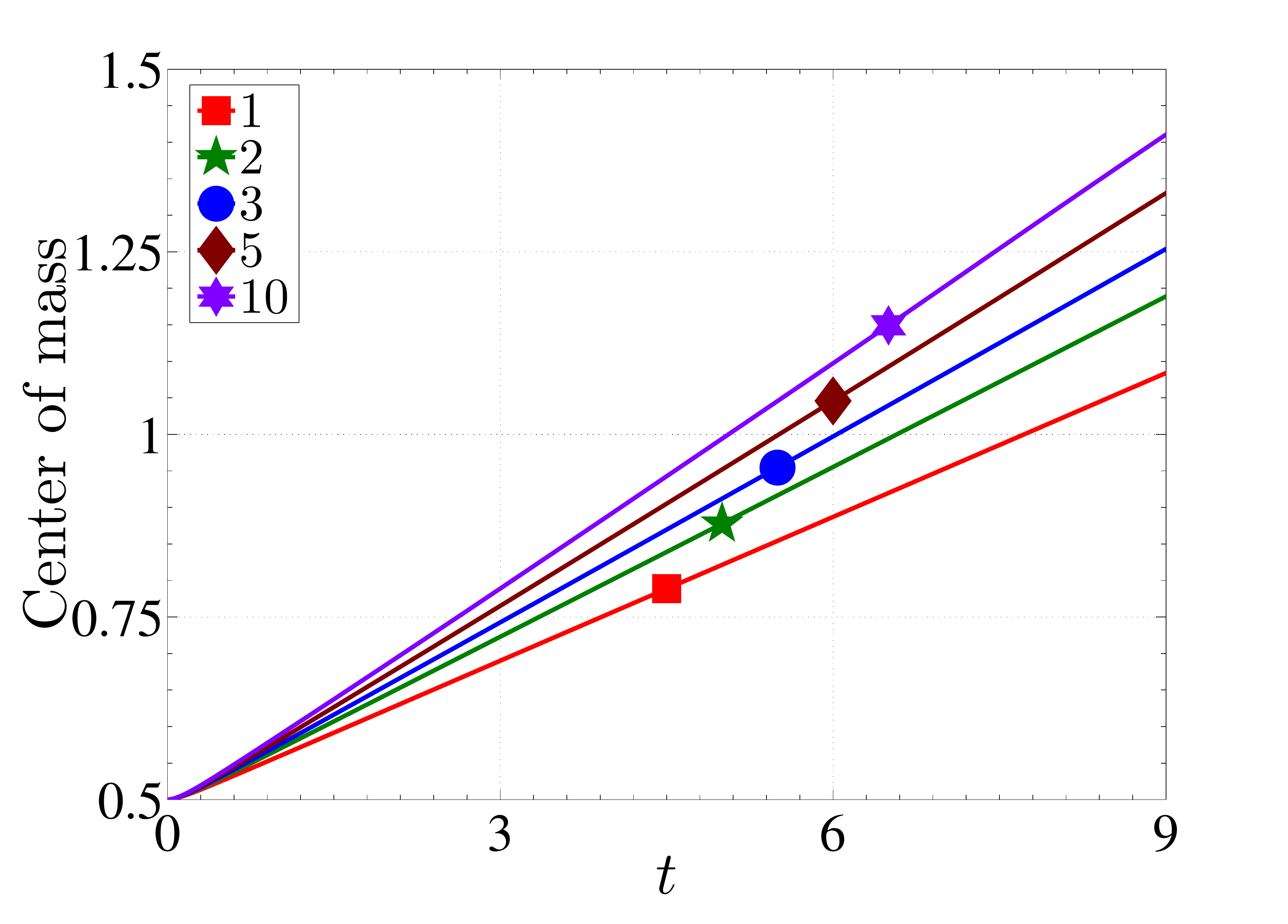}}}
\put(12.9,1.6){\makebox(0,0){\includegraphics[width=8.5cm,height=4.2cm,keepaspectratio]{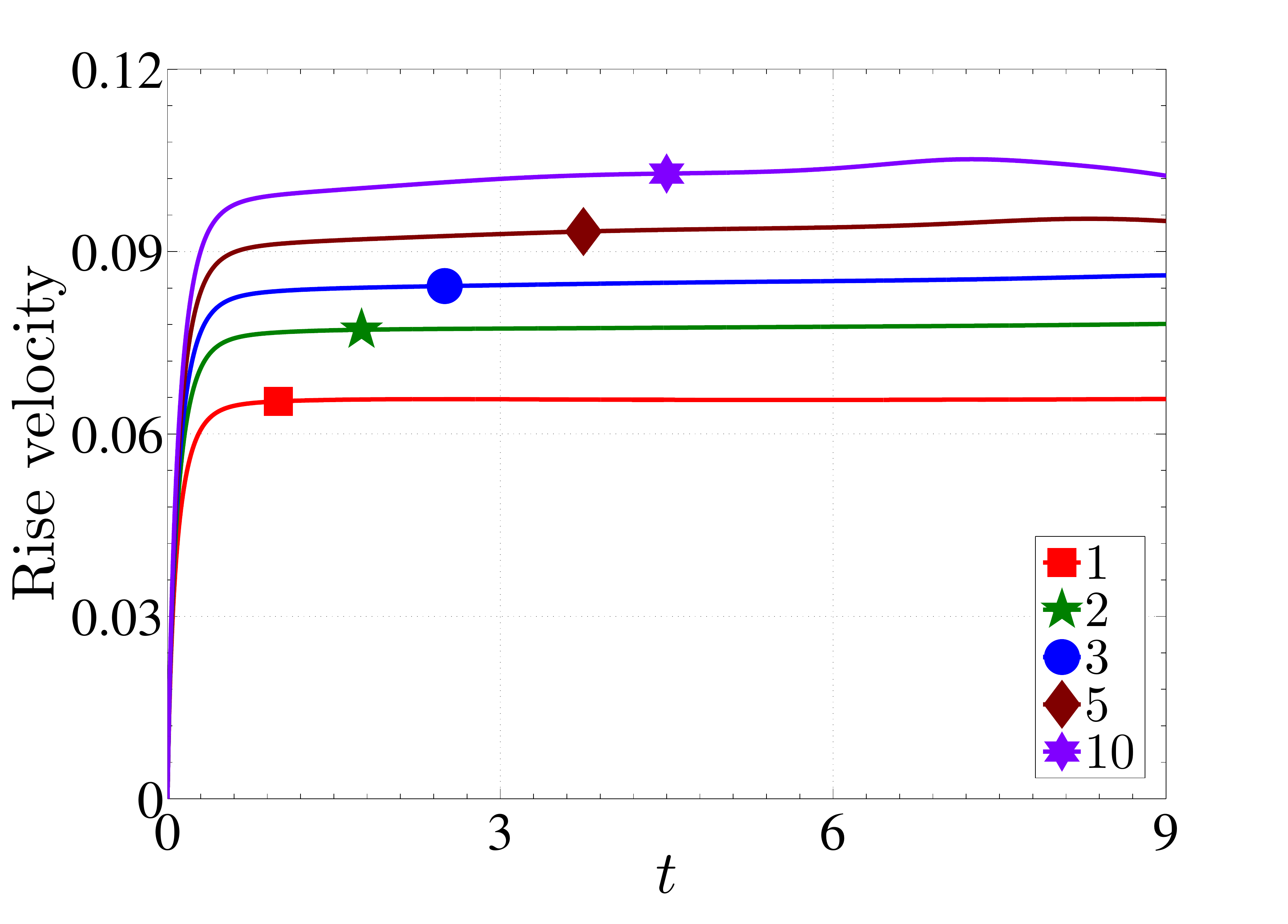}}}
\put(1.85,8.1){$(a)$}
\put(7.3, 8.1){$(b)$}
\put(12.8, 8.1){$(c)$}
\put(1.85,3.5){$(d)$}
\put(7.3, 3.5){$(e)$}
\put(12.8, 3.5){$(f)$}
\end{picture}
\end{center}
\caption{Influence of viscosity ratio for a Newtonian bubble rising in a viscoelastic fluid column~: (a)~bubble shape at $t$~=~9, (b)~diameter of the bubble at $r$~=~0, (c)~sphericity, (d)~kinetic energy, (e)~center of mass and (f)~rise velocity of the bubble for different viscosity ratios (i)~$\varepsilon$~=~1, (ii)~$\varepsilon$~=~2, (iii)~$\varepsilon$~=~3, (iv)~$\varepsilon$~=~5 and (v)~$\varepsilon$~=~10 with $\Rey_2$~=~$10$, Eo~=~$400$, $\text{Wi}_2$~=~25, $\rho_1/\rho_2$~=~0.1, $\beta_1$~=~1.0, $\beta_2$~=~0.75, $\alpha_2$~=~0.1, $D$~=~0.5 and $h_c$~=~2.0.} 
\label{Plots_ViscosityEffect_NV}
\end{figure*}

To study the influence of viscosity ratio on the rising bubble dynamics, we consider the base case flow parameters and vary only the viscosity ratio.
In particular, we vary only the total viscosity of the inner phase and keep all other parameters the same.
The following five different viscosity ratios are used in this study~: (i)~$\varepsilon$~=~1, (ii)~$\varepsilon$~=~2, (iii)~$\varepsilon$~=~3, (iv)~$\varepsilon$~=~5 and (v)~$\varepsilon$~=~10.
Fig.~\ref{Plots_ViscosityEffect_NV} presents the computational results for all the five variants of viscosity ratios.
By increasing the viscosity ratio, in principle we only increase the Reynolds number of the bubble while other parameters remain the same.
Hence, with an increase in the Reynolds number of the bubble, it forces the bubble to rise with a higher velocity and the same can be observed in Fig.~\ref{Plots_ViscosityEffect_NV}(f). 
Initially, the motion is inertia dominated due to buoyancy and hence, the rise velocity increases tremendously till about $t$~=~0.3.
After that, the viscous and viscoelastic stresses resist the buoyant force and we can observe an upward movement of the bubble with a steady rise velocity.
The kinetic energy of the bubble increases with an increase in the viscosity ratio, since it is accompanied by an increase in the rise velocity.
We can observe from Fig.~\ref{Plots_ViscosityEffect_NV}(d), that after the initial acceleration the temporal evolution of the kinetic energy of the bubble  curves seem to be parallel with an increase in the viscosity ratio.
Further, the bubble also rises higher with increased rise velocity and kinetic energy in the bubble and thus, the center of mass of the bubble is higher with an increase in the viscosity ratio, see Fig.~~\ref{Plots_ViscosityEffect_NV}(e).

Fig.~\ref{Plots_ViscosityEffect_NV}(a) depicts the bubble shapes at $t$~=~9. 
For high viscosity ratios, the bubble surface close to the trailing end becomes concave and a very long and narrow tail develops. 
This is due to the fact that, with an increase in the Reynolds number of the bubble, there is increased generation and accumulation of extensional viscoelastic stresses at the rear stagnation point.
Hence, at a given time the bubble with higher viscosity ratio will show greater extended trailing edge characteristics in the bubble and the same in observed in Fig.~\ref{Plots_ViscosityEffect_NV}(a).
However, for low viscosity ratios, the bubble does have an extended trailing edge but occurs at a later time as the viscoelastic stresses are accumulated slowly.  
Further, Fig.~\ref{Plots_ViscosityEffect_NV}(b) presents the temporal evolution of the diameter of the bubble at the axis of symmetry.
We can observe that till around $t$~=~4, the bubble rises with almost the same diameter, which indicates that the interfacial tension dominated over the viscous and viscoelastic stresses till $t$~=~4.
However, after $t$~=~4, the diameter of the bubble increases with an increase in the viscosity ratio, as viscoelastic stresses start to dominate the bubble shapes.
Further, Fig.~\ref{Plots_ViscosityEffect_NV}(c) depicts the temporal evolution of the sphericity of the bubble.
It is a good indicative of the bubble deformation.
As expected, we can observe that the sphericity of the bubble at $t$~=~9 decreases with an increase in the viscosity ratio.

\begin{figure*}
\begin{center}
\unitlength1cm
\begin{picture}(20,6.25)
\put(3.5,-0.3){\makebox(3,6){\includegraphics[trim=0.5cm 1.5cm 7.6cm 1.0cm, clip=true,width=5.5cm]{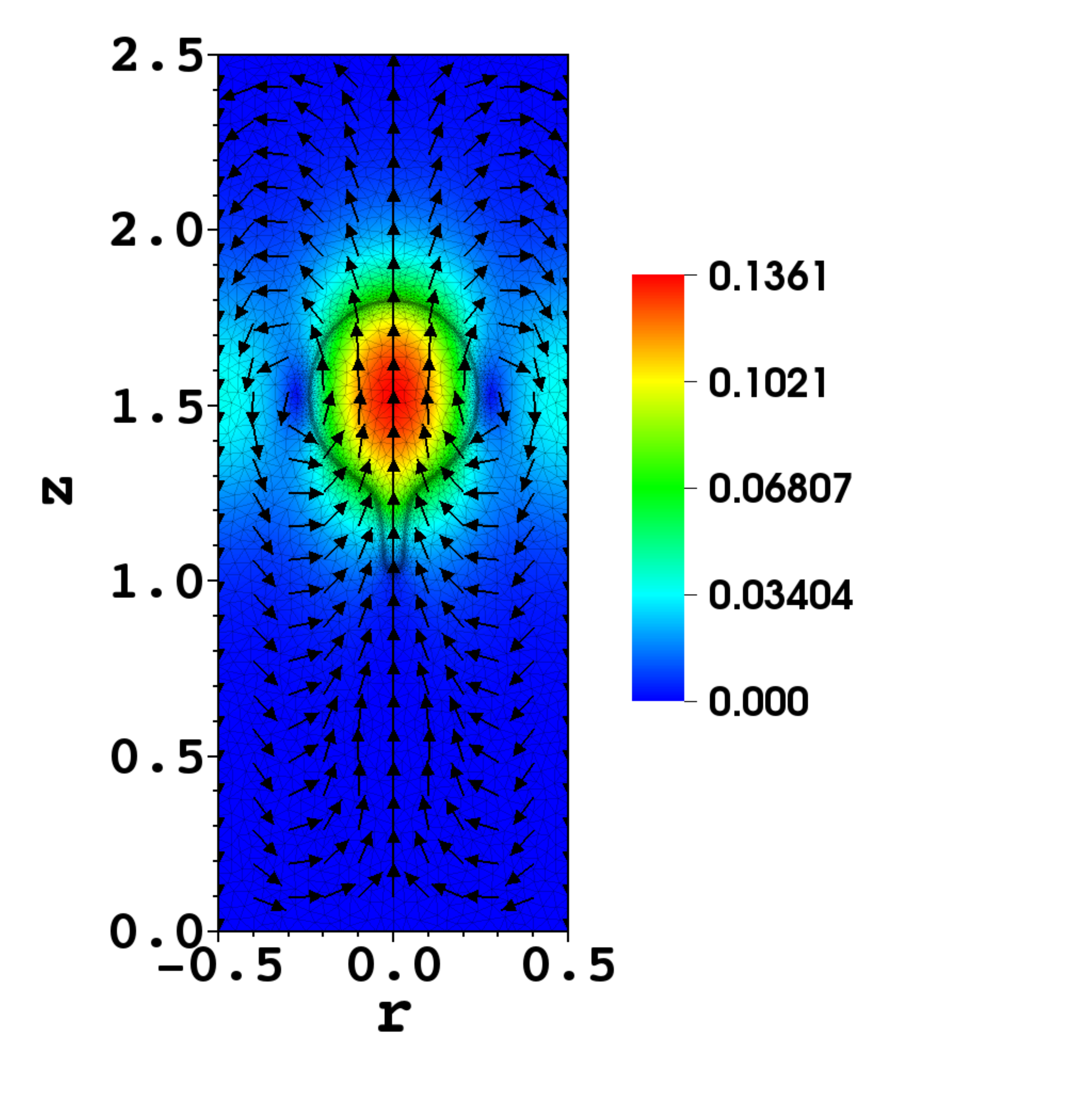}}}
\put(9.5,-0.3){\makebox(3,6){\includegraphics[trim=0.5cm 1.5cm 7.6cm 1.0cm, clip=true,width=5.5cm]{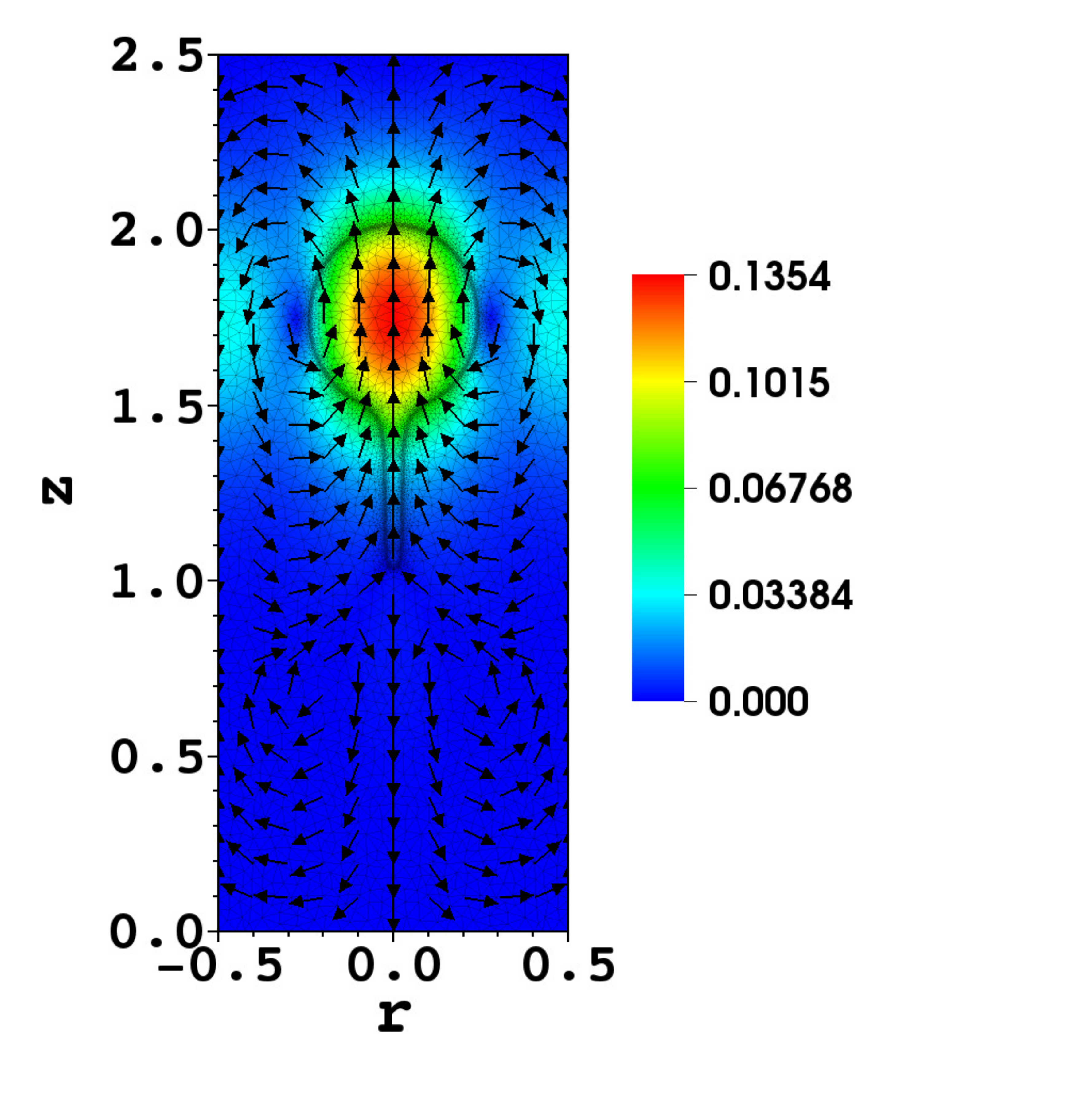}}}
\put(4.2,6.0){$t = 13.25$}
\put(10.25,6.0){$t = 16.0$}
\put(6.25,4.55){$|\bu|$}
\put(12.25,4.55){$|\bu|$}
\end{picture}
\end{center}
\caption{Magnitude of velocity profiles and velocity vectors at dimensionless times $t$~=~13.25 and 16.0 for a Newtonian bubble rising in a viscoelastic fluid column with flow parameters~: $\Rey_2$~=~$10$, Eo~=~$400$, $\text{Wi}_2$~=~25, $\varepsilon$~=~2, $\rho_1/\rho_2$~=~0.1, $\beta_1$~=~1.0, $\beta_2$~=~0.75, $\alpha_2$~=~0.1, $D$~=~0.5 and $h_c$~=~2.5.}
\label{NegativeWake_VTKPlots_NV}
\end{figure*}

The bubble rising in a viscoelastic fluid reveals an interesting flow phenomenon such that in the wake of the rising bubble, the velocity field very close to the trailing end is in the direction of the motion of the bubble whereas it reverses its direction at a small distance away from the trailing end, which is commonly referred to as negative wake. 
In the case of Newtonian fluids, the fluid velocity behind the bubble is always in the same direction as the bubble's motion.   
Fig.~\ref{NegativeWake_VTKPlots_NV} depicts the negative wake phenomenon. 
At $t$~=~13.25, the fluid velocity behind the bubble is in the same direction as the bubble's motion. 
However, immediately after $t$~=~13.25 the flow direction starts to reverse in the wake region and at $t$~=~16.0, we can observe that the flow direction has completely reversed at a small distance away from the trailing end.

\subsubsection{Influence of Newtonian solvent ratio on the bubble dynamics}

\begin{figure*}[ht!]
\begin{center}
\unitlength1cm
\begin{picture}(14.5,8.3)
\put(1.9,6.2){\makebox(0,0){\includegraphics[width=8.5cm,height=4.2cm,keepaspectratio]{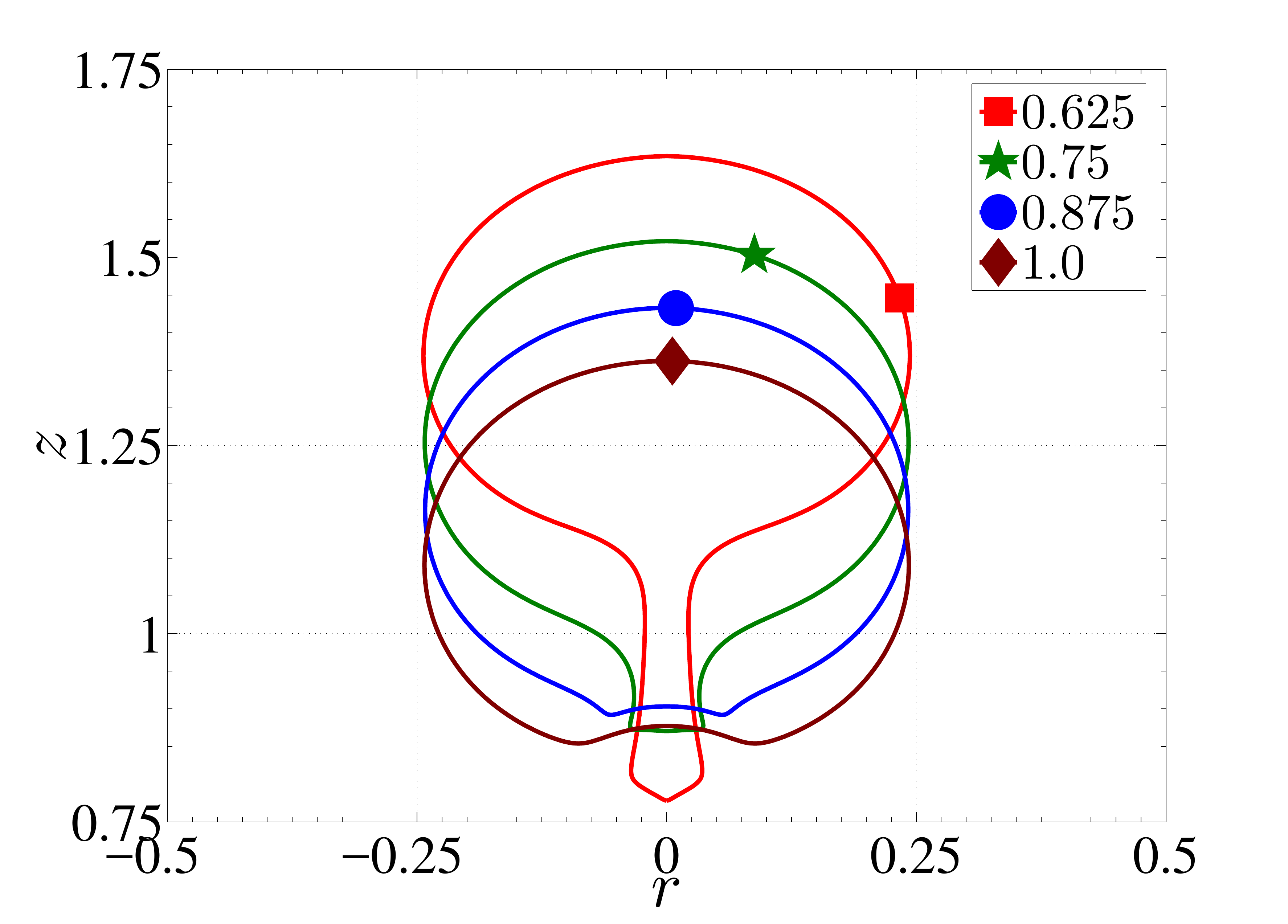}}}
\put(7.4,6.2){\makebox(0,0){\includegraphics[width=8.5cm,height=4.2cm,keepaspectratio]{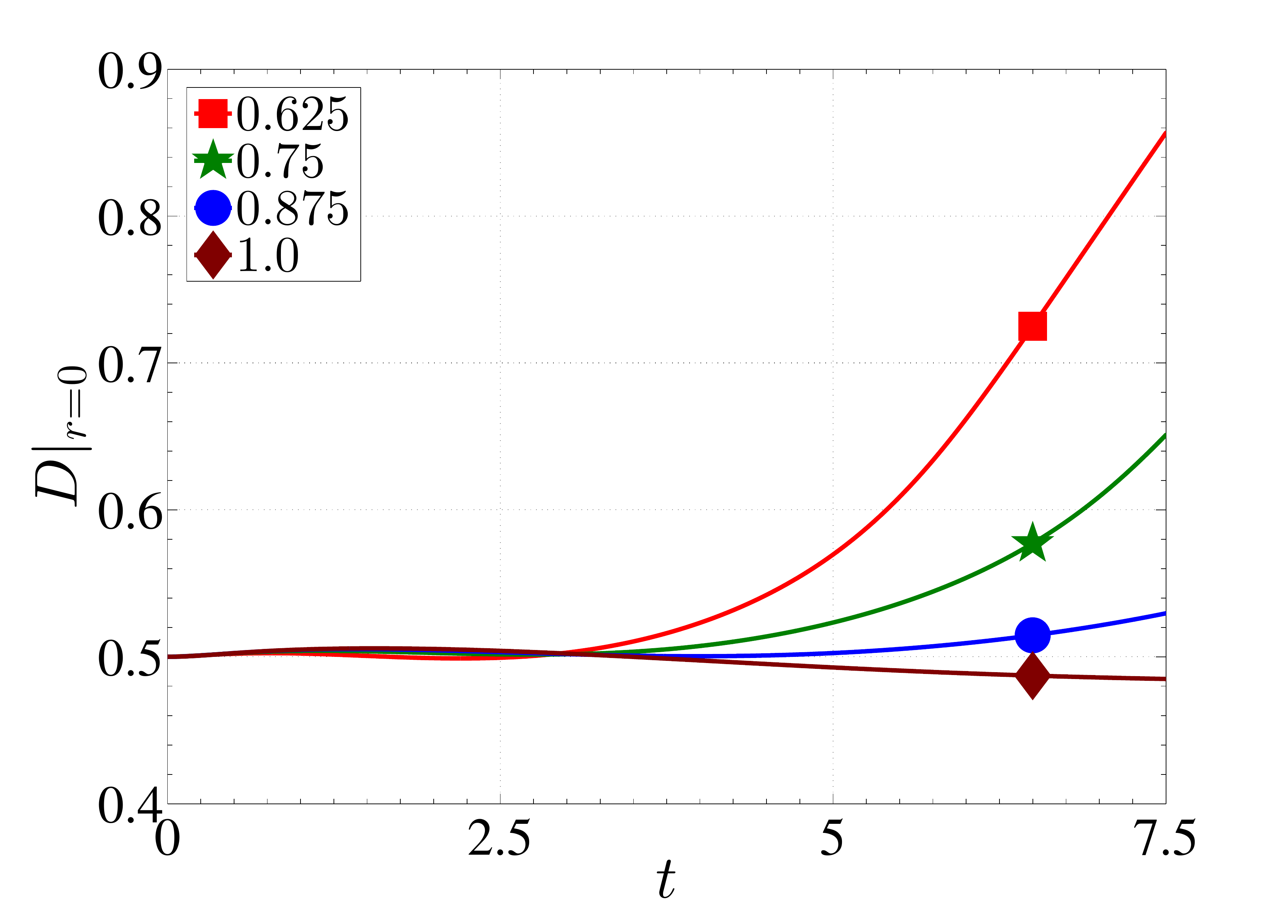}}}
\put(12.9,6.2){\makebox(0,0){\includegraphics[width=8.5cm,height=4.2cm,keepaspectratio]{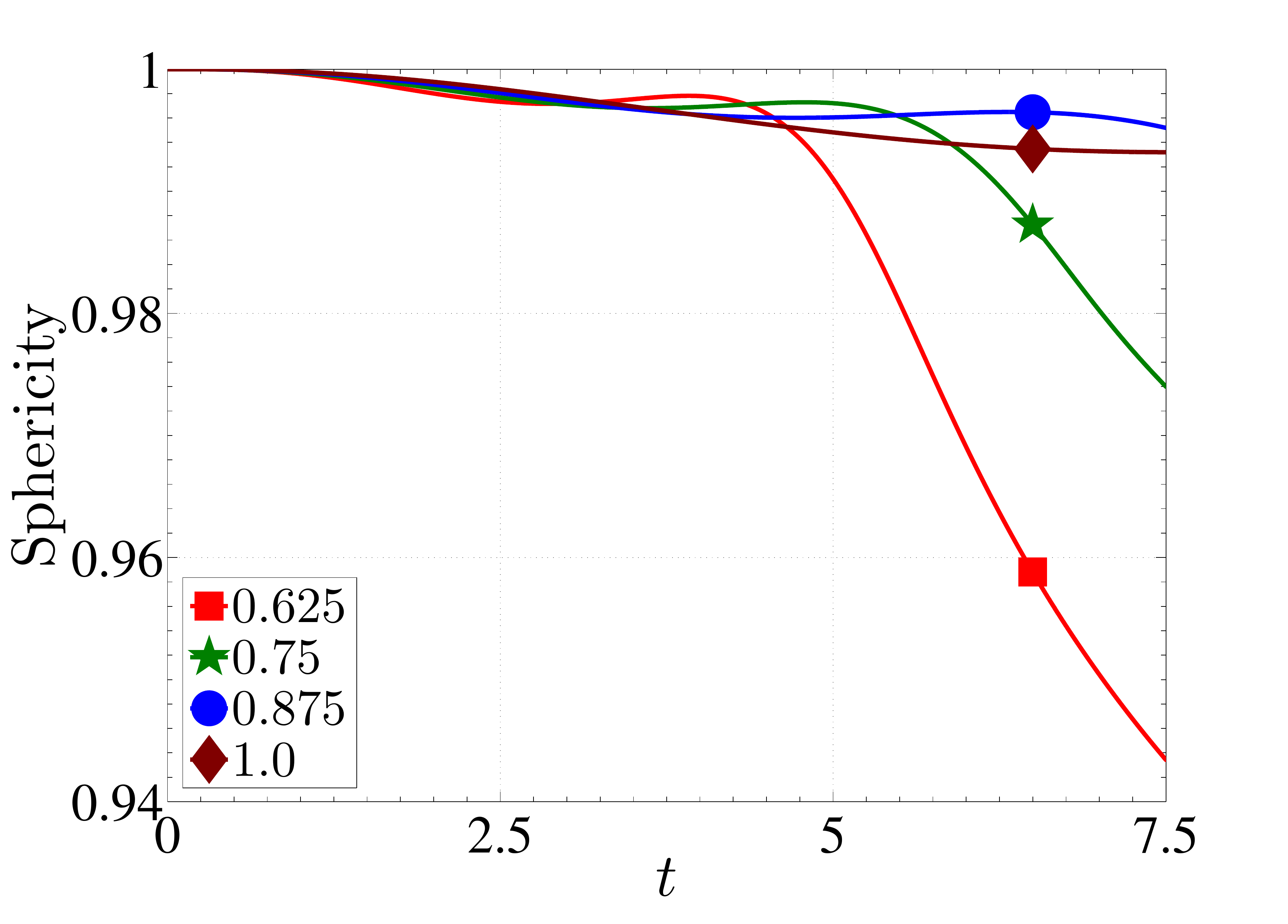}}}
\put(1.9,1.6){\makebox(0,0){\includegraphics[width=8.5cm,height=4.2cm,keepaspectratio]{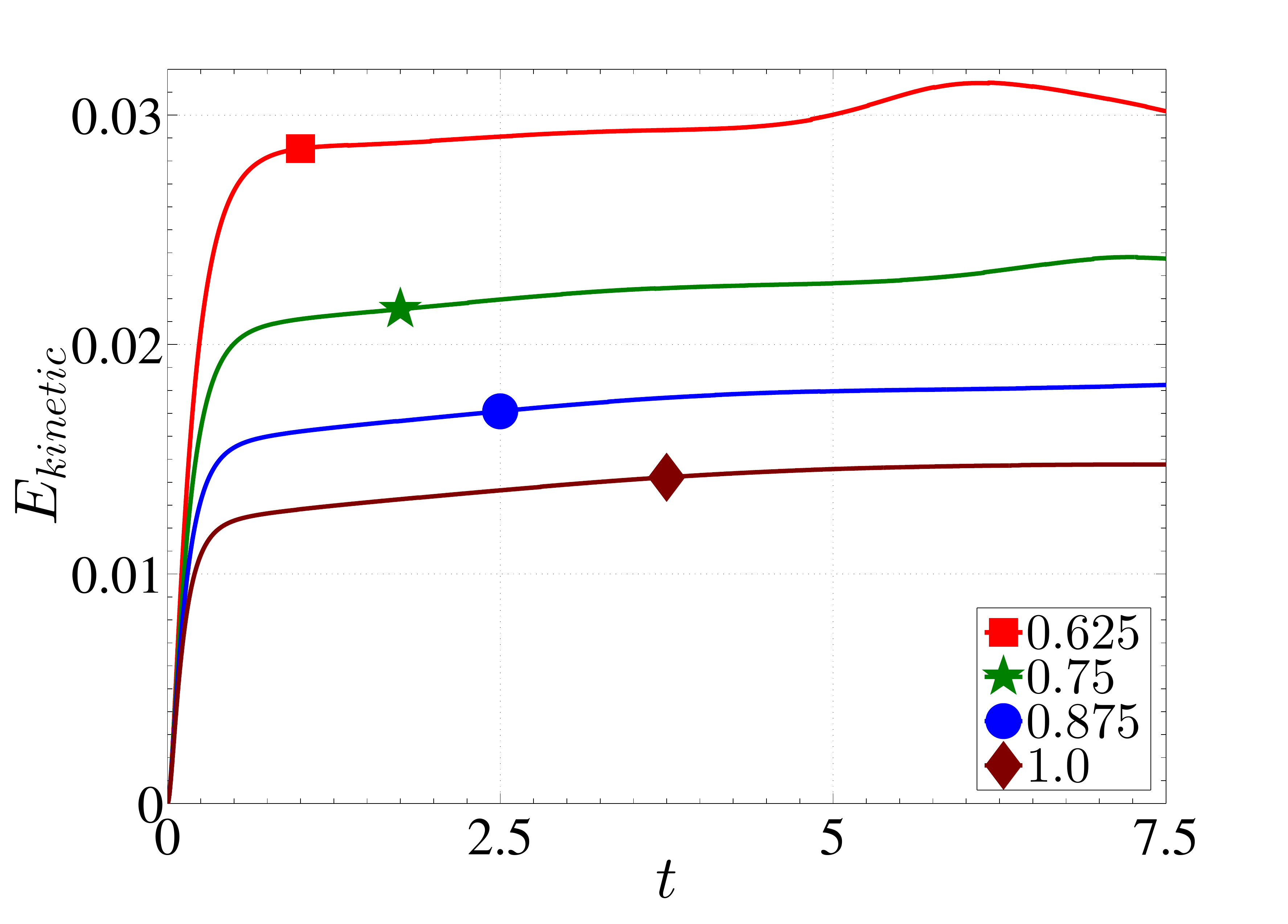}}}
\put(7.4,1.6){\makebox(0,0){\includegraphics[width=8.5cm,height=4.2cm,keepaspectratio]{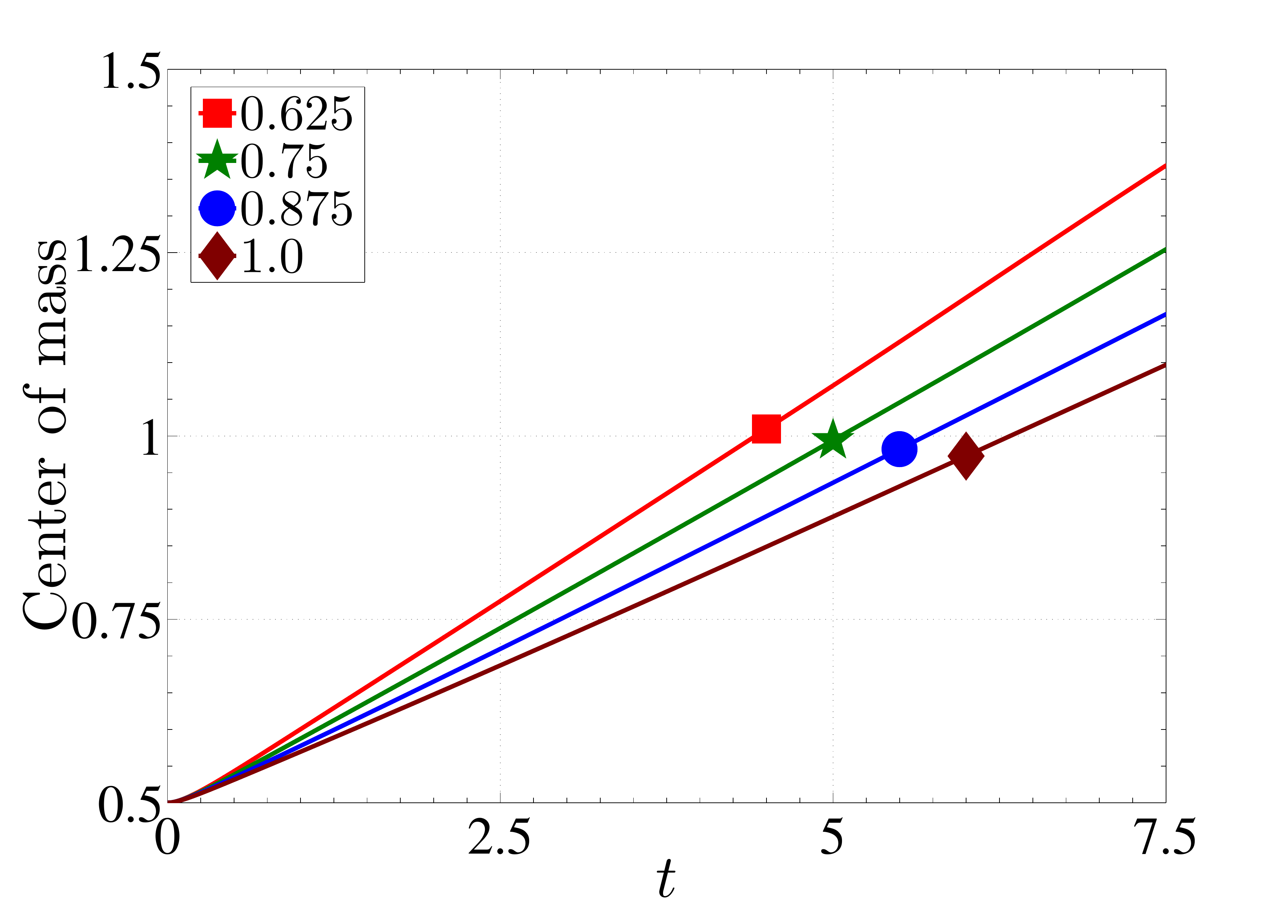}}}
\put(12.9,1.6){\makebox(0,0){\includegraphics[width=8.5cm,height=4.2cm,keepaspectratio]{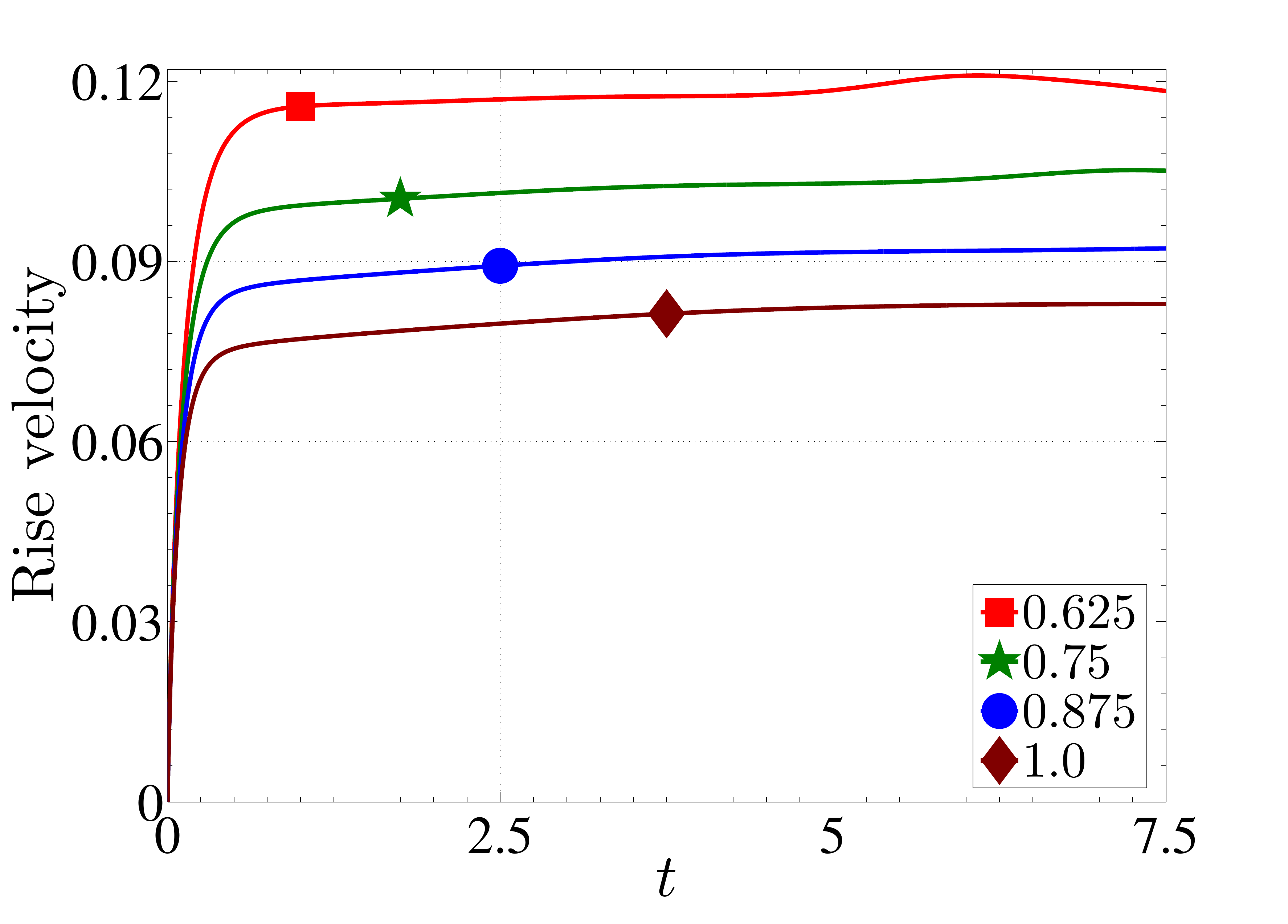}}}
\put(1.85,8.1){$(a)$}
\put(7.3, 8.1){$(b)$}
\put(12.8, 8.1){$(c)$}
\put(1.85,3.5){$(d)$}
\put(7.3, 3.5){$(e)$}
\put(12.8, 3.5){$(f)$}
\end{picture}
\end{center}
\caption{Influence of Newtonian solvent ratio for a Newtonian bubble rising in a viscoelastic fluid column~: (a)~bubble shape at $t$~=~7.5, (b)~diameter of the bubble at $r$~=~0, (c)~sphericity, (d)~kinetic energy, (e)~center of mass and (f)~rise velocity of the bubble for different Newtonian solvent ratios (i)~$\beta_2$~=~0.625, (ii)~$\beta_2$~=~0.75, (iii)~$\beta_2$~=~0.875 and (iv)~$\beta_2$~=~1.0 with $\Rey_2$~=~$10$, Eo~=~$400$, $\text{Wi}_2$~=~25, $\varepsilon$~=~10, $\rho_1/\rho_2$~=~0.1, $\beta_1$~=~1.0, $\alpha_2$~=~0.1, $D$~=~0.5 and $h_c$~=~2.0.} 
\label{Plots_BetaEffect_NV}
\end{figure*}
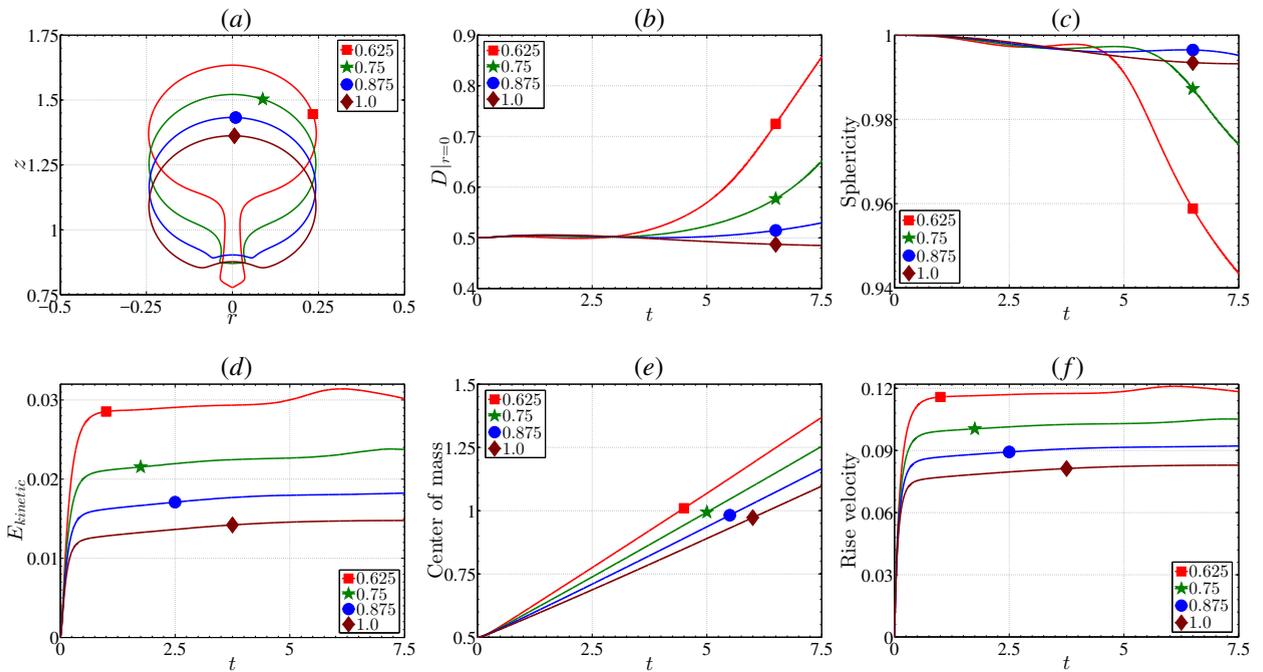

In this section, we study the influence of Newtonian solvent ratio on the rising Newtonian bubble dynamics in a viscoelastic fluid column. 
We consider the base case flow parameters and vary only the Newtonian solvent ratio of the bulk fluid column.
In particular, we vary the Newtonian solvent viscosity and polymeric viscosity of the bulk fluid but keep the total viscosity constant.
Four different values are used for the Newtonian solvent ratio in this study, which are as follows~: (i)~$\beta_2$~=~0.625, (ii)~$\beta_2$~=~0.75, (iii)~$\beta_2$~=~0.875 and (iv)~$\beta_2$~=~1.0.
Lower the Newtonian solvent ratio, greater is the polymeric viscosity and lesser is the Newtonian viscosity, thereby increasing the viscoelastic character of the fluid column.
Fig.~\ref{Plots_BetaEffect_NV} presents the numerical results for different Newtonian solvent ratios.
Note that the case $\beta_2$~=~1.0 represents a Newtonian bubble rising in a Newtonian fluid column. 
From Fig.~\ref{Plots_BetaEffect_NV}(a), we can observe that the bubble shape at the trailing end develops a longer and narrower tail  and also rises higher with decrease in the Newtonian solvent ratio.
With increased viscoelasticity in the bulk fluid, the extensional stresses at the rear stagnation point increases leading to a longer and narrower tail.
The greater rise in the bubble is accompanied by a higher center of mass, see Fig.~\ref{Plots_BetaEffect_NV}(e). 
Further, the kinetic energy and the rise velocity of the bubble increases with a decrease in the Newtonian solvent ratio, refer Fig.~\ref{Plots_BetaEffect_NV}(e) and (f). 
The curves become parallel after the viscous and viscoelastic stresses start to overcome the interfacial tension.
One interesting observation is that, the increase in the magnitude of the  kinetic energy and rise velocity of the bubble seems to be higher with decreasing Newtonian solvent ratio.
In Fig.~\ref{Plots_BetaEffect_NV}(b), we can observe that the diameter of the bubble at the axis of symmetry increases with a decrease in the Newtonian solvent ratio.
This occurs since with an increase in the viscoelastic character of the outer fluid column, the bubble develops a longer trailing edge due to greater extensional viscoelastic stresses near the rear stagnation point.
Further, the sphericity of the bubble decreases with a decrease in the Newtonian solvent ratio due to increased deformation at the rear end, see Fig.~\ref{Plots_BetaEffect_NV}(c).
 
\subsubsection{Influence of Giesekus mobility factor on the bubble dynamics}

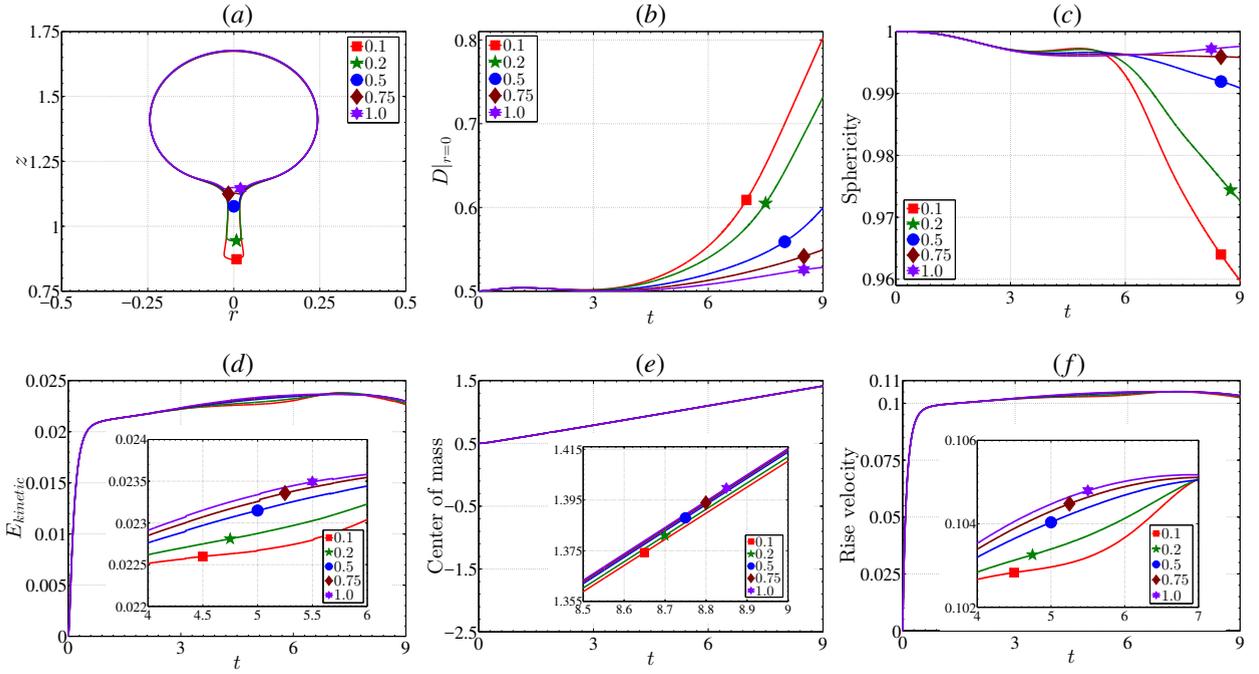
\begin{figure*}[ht!]
\begin{center}
\unitlength1cm
\begin{picture}(14.5,8.3)
\put(1.9,6.2){\makebox(0,0){\includegraphics[width=8.5cm,height=4.2cm,keepaspectratio]{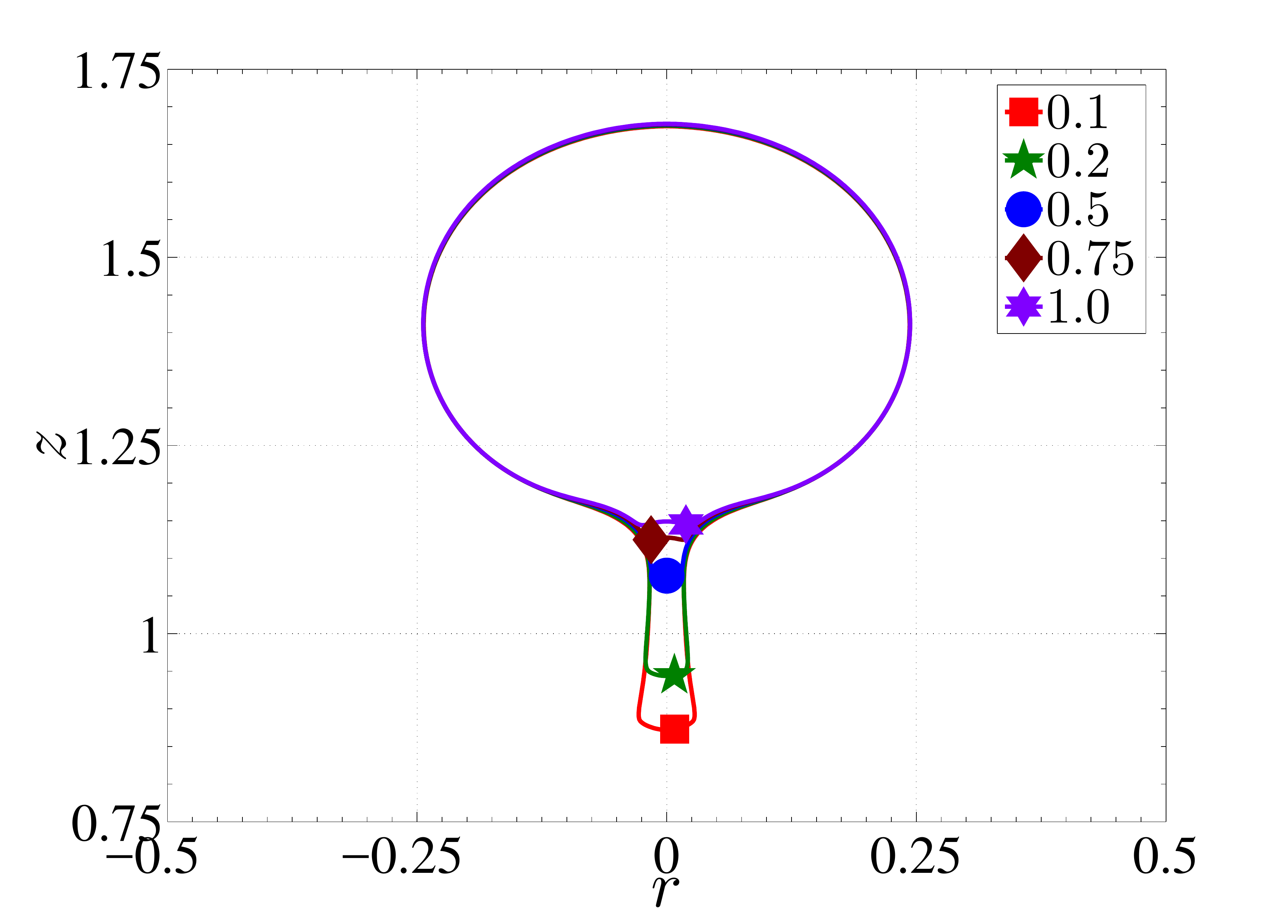}}}
\put(7.4,6.2){\makebox(0,0){\includegraphics[width=8.5cm,height=4.2cm,keepaspectratio]{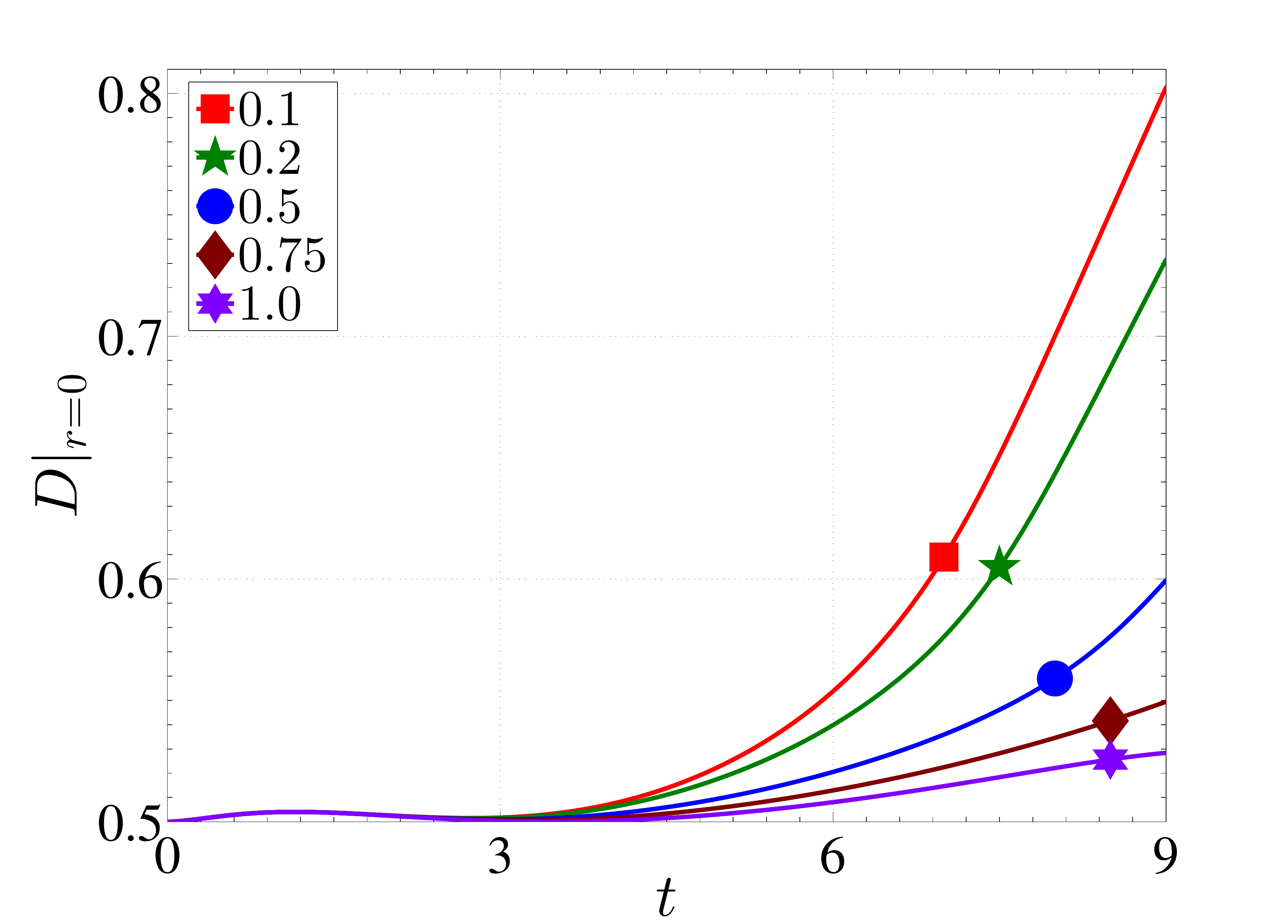}}}
\put(12.9,6.2){\makebox(0,0){\includegraphics[width=8.5cm,height=4.2cm,keepaspectratio]{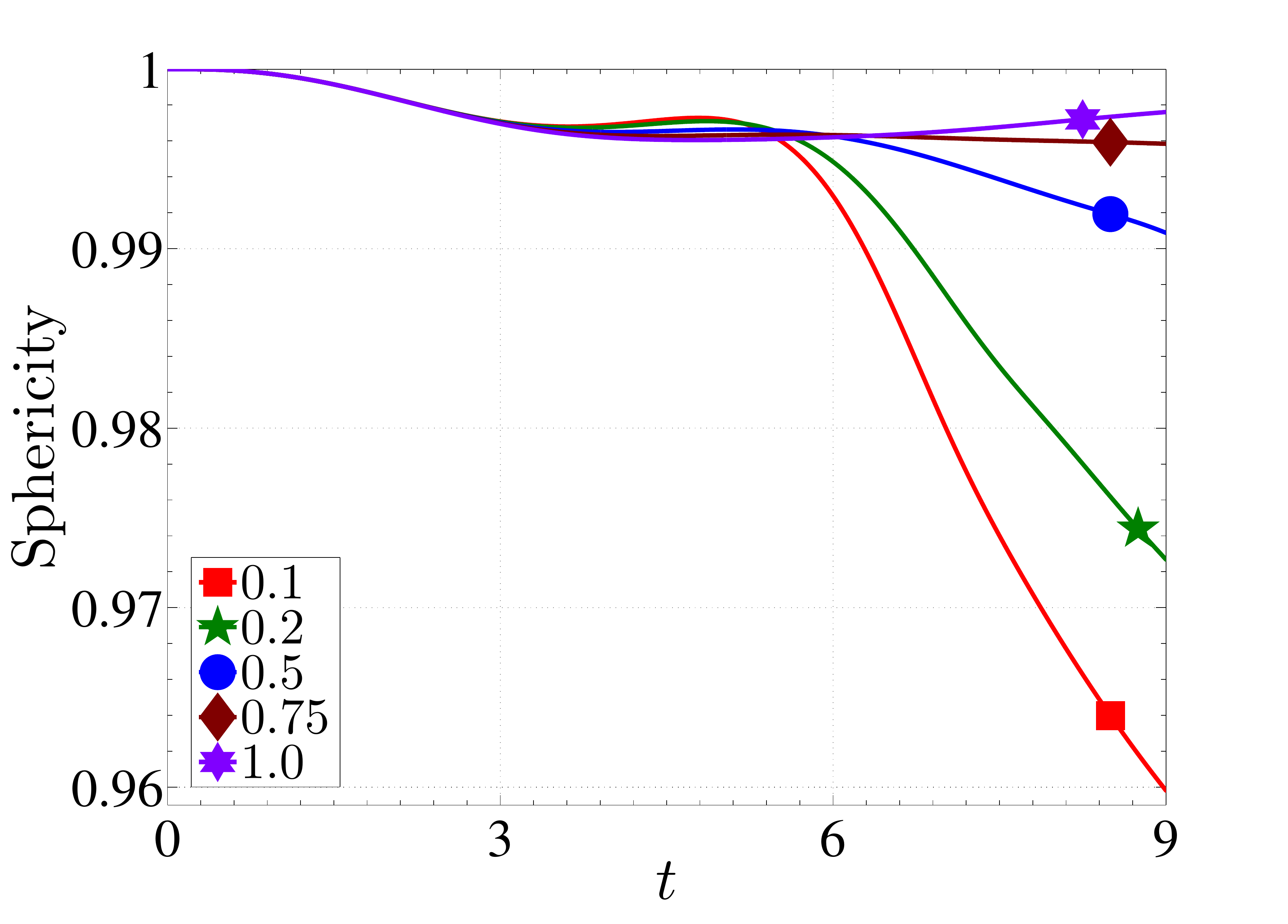}}}
\put(1.9,1.6){\makebox(0,0){\includegraphics[width=8.5cm,height=4.2cm,keepaspectratio]{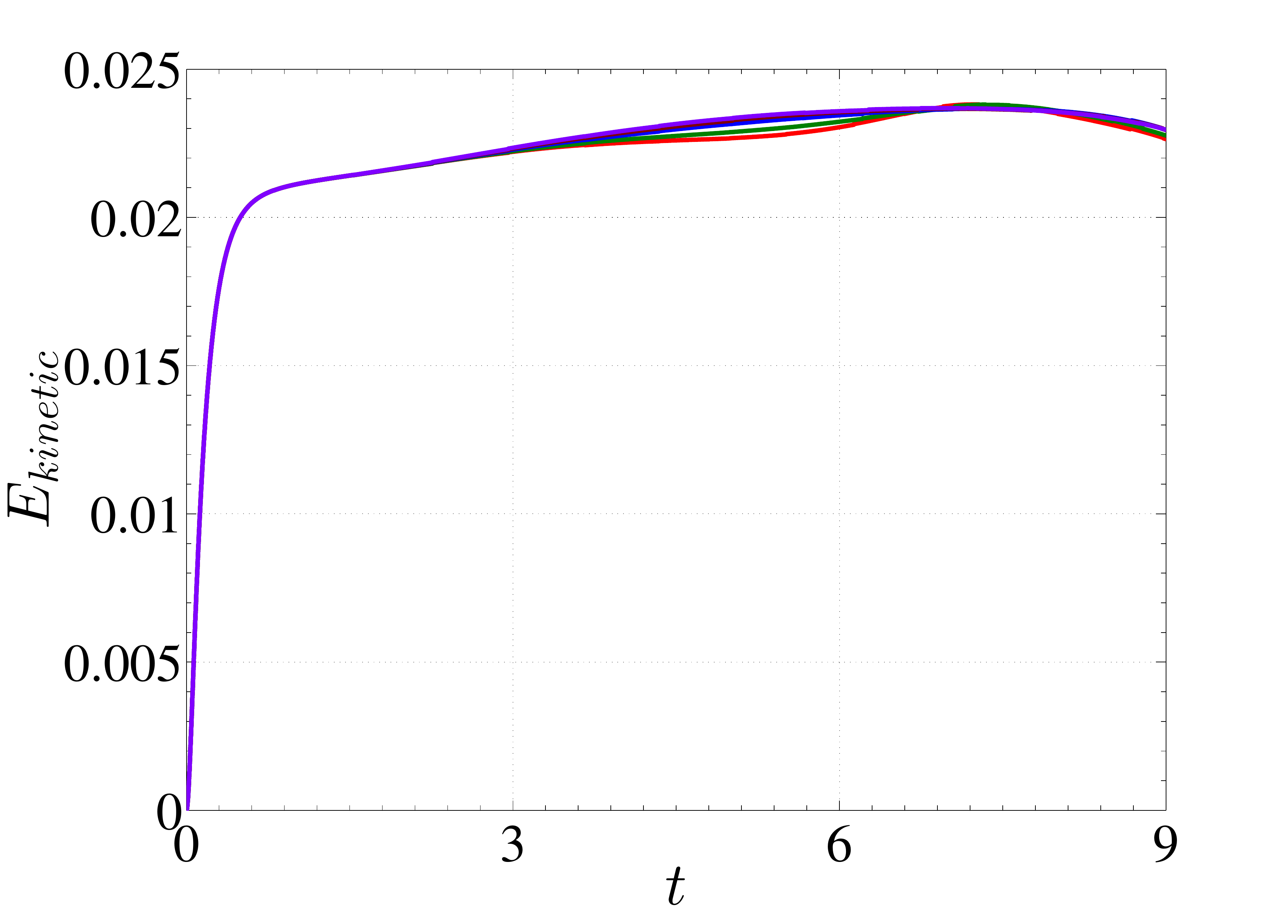}}}
\put(2.2,1.45){\makebox(0,0){\includegraphics[width=8.5cm,height=2.7cm,keepaspectratio]{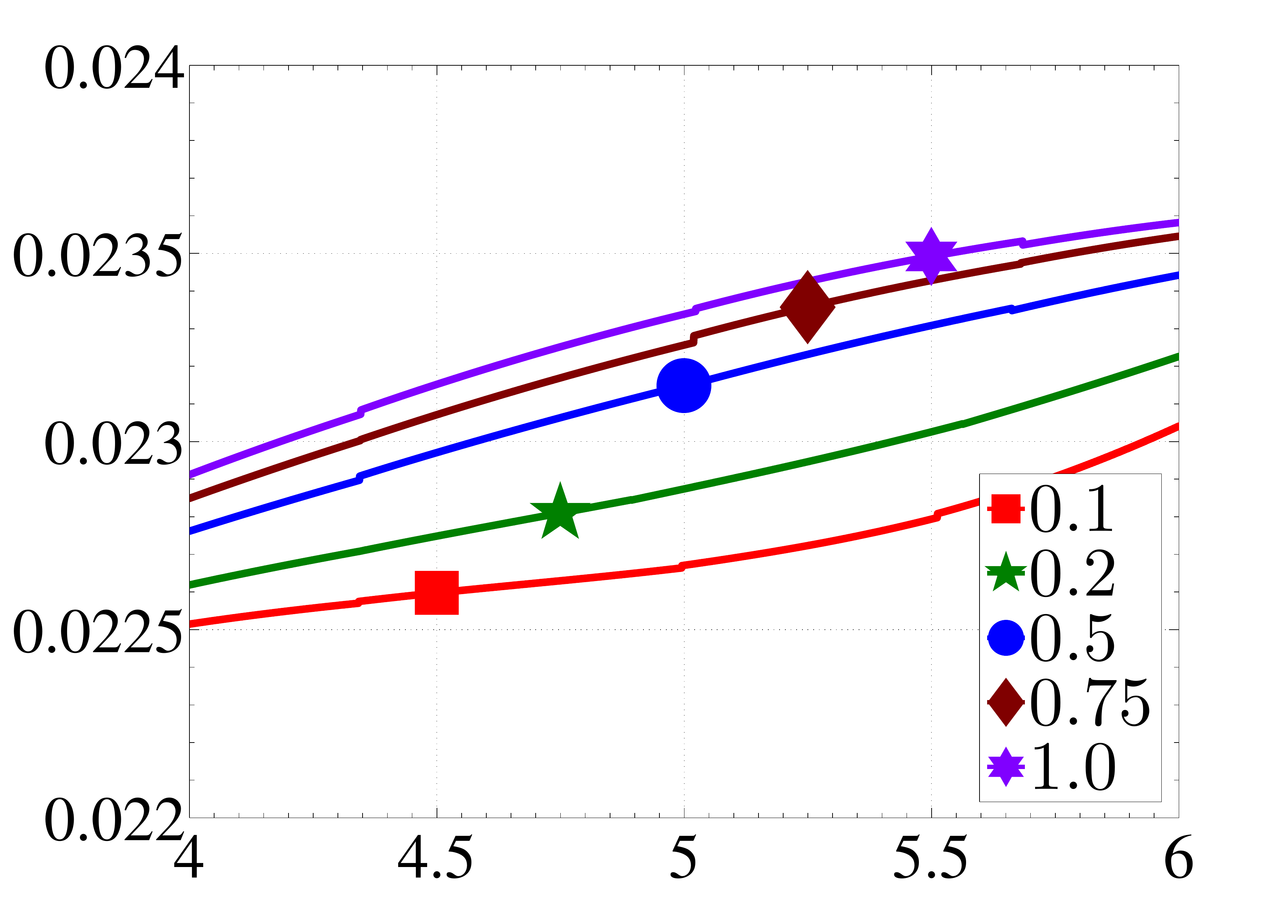}}}
\put(7.4,1.6){\makebox(0,0){\includegraphics[width=8.5cm,height=4.2cm,keepaspectratio]{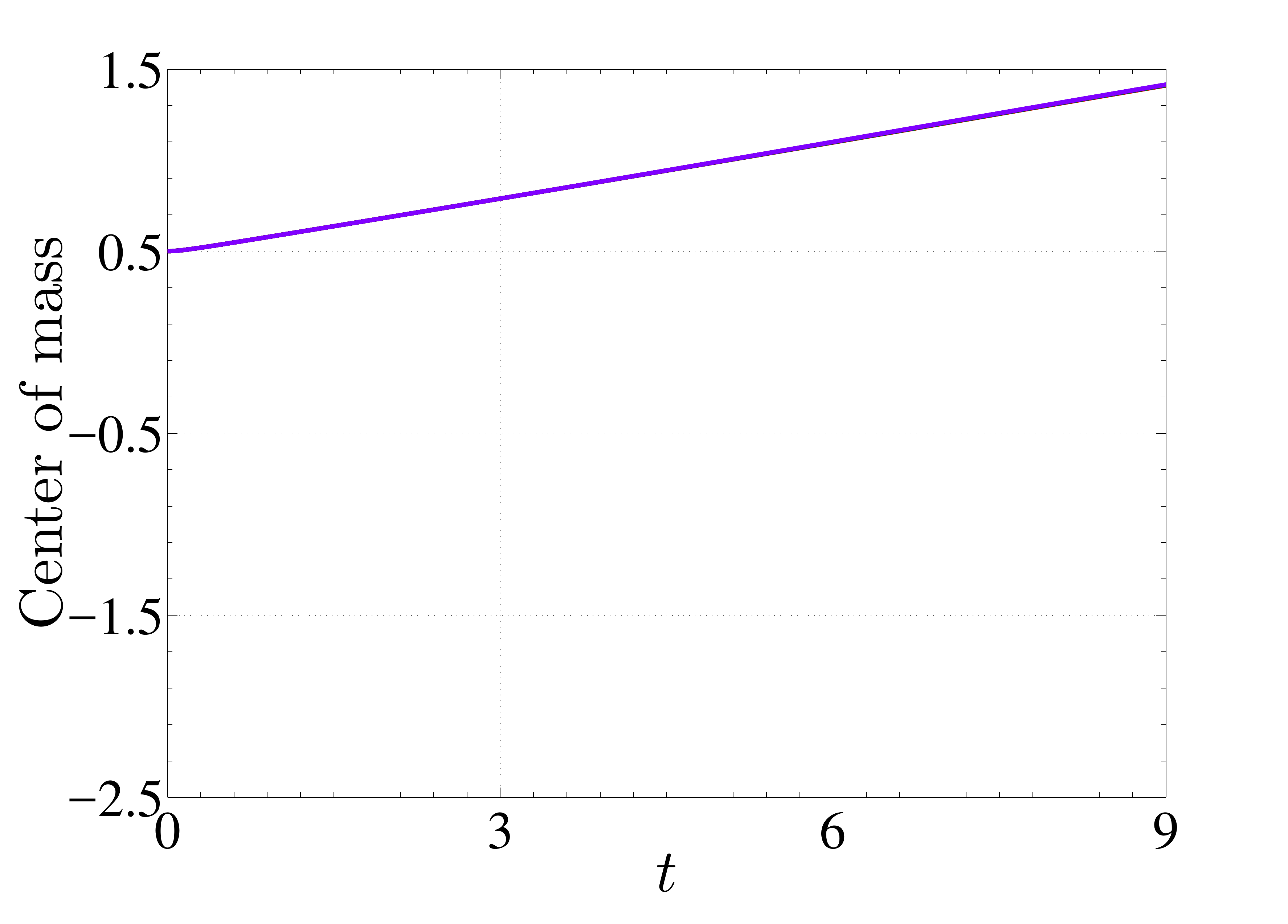}}}
\put(7.9,1.45){\makebox(0,0){\includegraphics[width=8.5cm,height=2.5cm,keepaspectratio]{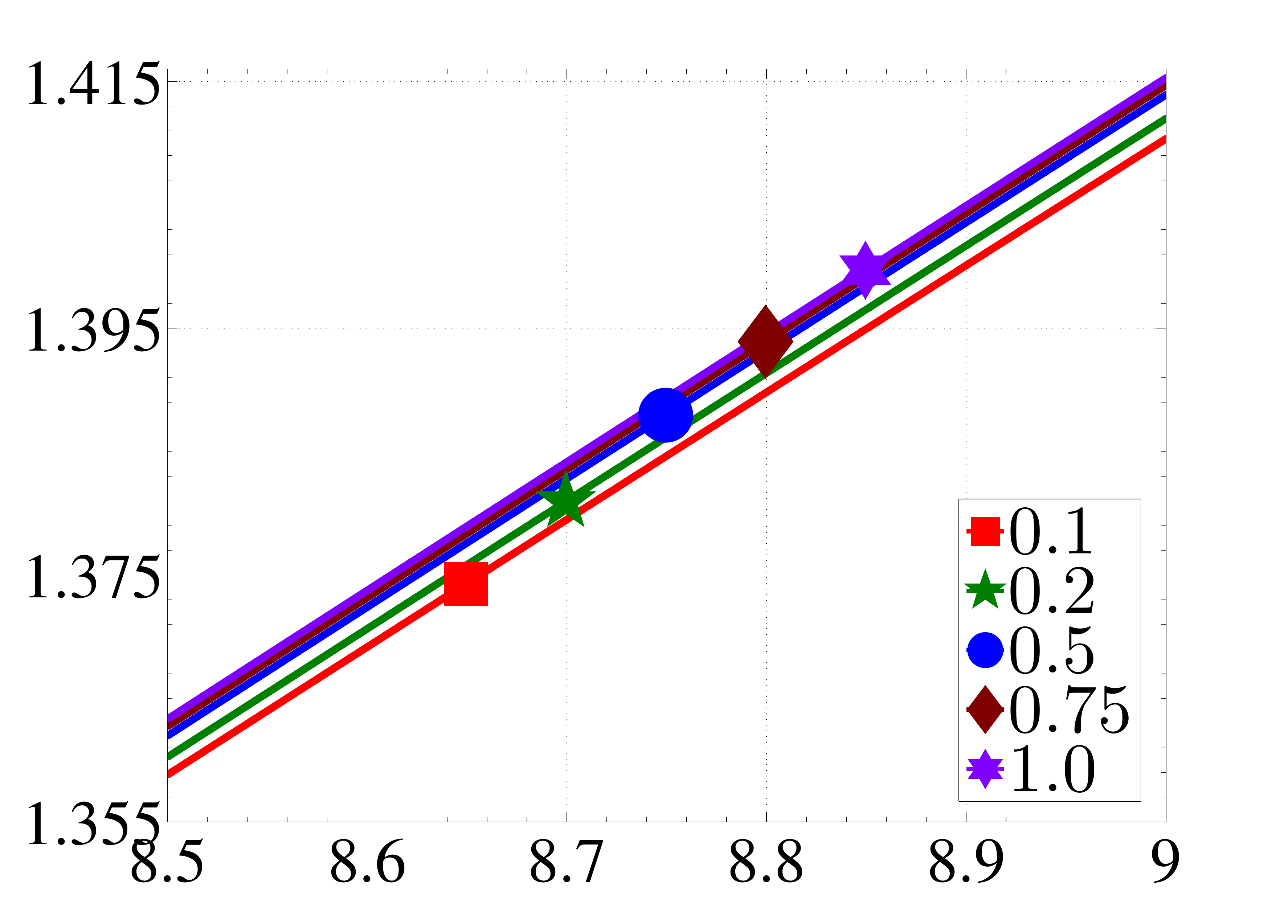}}}
\put(12.9,1.6){\makebox(0,0){\includegraphics[width=8.5cm,height=4.2cm,keepaspectratio]{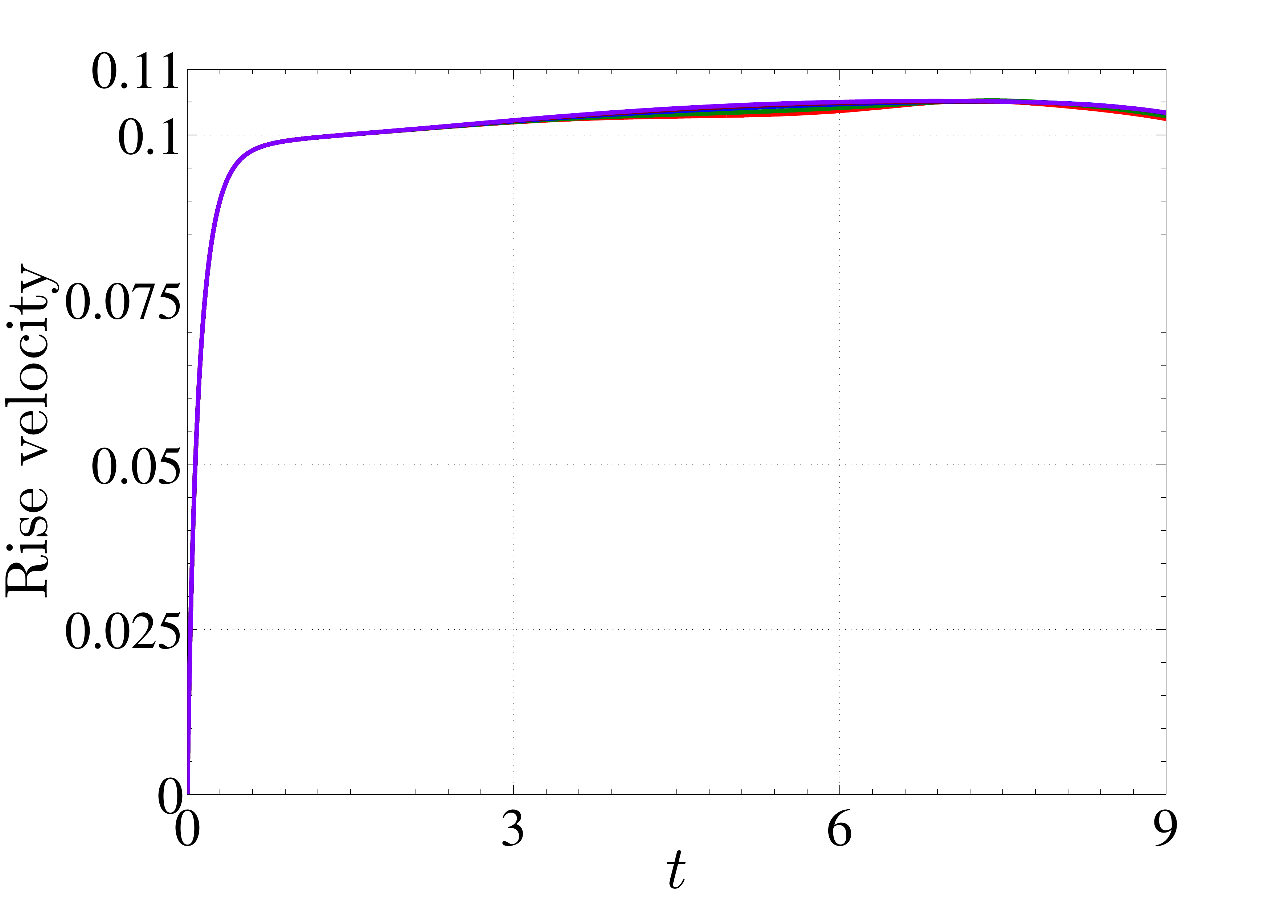}}}
\put(13.2,1.45){\makebox(0,0){\includegraphics[width=8.5cm,height=2.7cm,keepaspectratio]{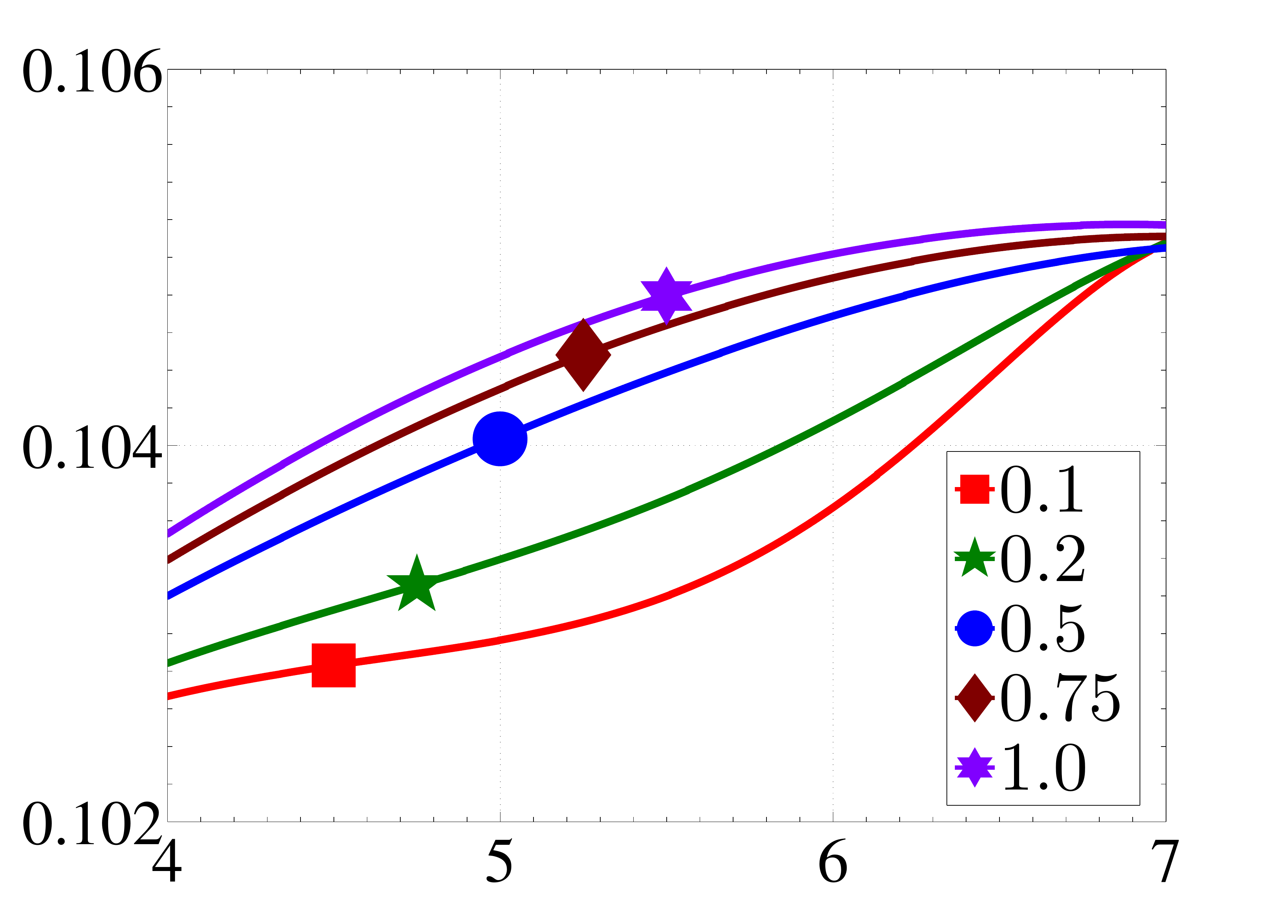}}}
\put(1.85,8.1){$(a)$}
\put(7.3, 8.1){$(b)$}
\put(12.8, 8.1){$(c)$}
\put(1.85,3.5){$(d)$}
\put(7.3, 3.5){$(e)$}
\put(12.8, 3.5){$(f)$}
\end{picture}
\end{center}
\caption{Influence of Giesekus mobility factor for a Newtonian bubble rising in a viscoelastic fluid column~: (a)~bubble shape at $t$~=~9, (b)~diameter of the bubble at $r$~=~0, (c)~sphericity, (d)~kinetic energy, (e)~center of mass and (f)~rise velocity of the bubble for different Giesekus mobility factors (i)~$\alpha_2$~=~0.1, (ii)~$\alpha_2$~=~0.2, (iii)~$\alpha_2$~=~0.5, (iv)~$\alpha_2$~=~0.75 and (v)~$\alpha_2$~=~1.0 with $\Rey_2$~=~$10$, Eo~=~$400$, $\text{Wi}_2$~=~25, $\varepsilon$~=~10, $\rho_1/\rho_2$~=~0.1, $\beta_1$~=~1.0, $\beta_2$~=~0.75, $D$~=~0.5 and $h_c$~=~2.0.} 
\label{Plots_AlphaEffect_NV}
\end{figure*}

To examine the influence of Giesekus mobility factor on the Newtonian bubble rising in a viscoelastic fluid column, we consider the following five different Giesekus factors~: (i)~$\alpha_2$~=~0.1, (ii)~$\alpha_2$~=~0.2, (iii)~$\alpha_2$~=~0.5, (iv)~$\alpha_2$~=~0.75 and (v)~$\alpha_2$~=~1.0.
The other flow parameters are the same as the base case.
Fig.~\ref{Plots_AlphaEffect_NV} presents the computational results for different Giesekus factors.
With an increase in the Giesekus factor, the shear thinning effects increases.
Hence, with increased shear thinning, the bubble is expected to have higher rise velocity and eventually greater kinetic energy.
From Fig.~\ref{Plots_AlphaEffect_NV}(d) and (f), we can observe that there is not much visible effect of Giesekus factor. 
However, from the zoomed plots, we can observe the shear thinning effect very clearly.
Increasing the Giesekus factor leads to a decrease in the magnitude of the viscoelastic stresses generated in the bulk fluid column.
Hence, from Fig.~\ref{Plots_AlphaEffect_NV}(a), we can observe that the trailing end of the bubble becomes flatter and the tail becomes shorter with an increase in the Giesekus factor.
Since the tail becomes shorter, the magnitude of the increase of the diameter of the bubble at the axis of symmetry decreases with an increase in the Giesekus factor, see Fig.~\ref{Plots_AlphaEffect_NV}(b). 
The sphericity of the bubble decreases with a decrease in the Giesekus factor due to large deformation at the tail end of the bubble.
Further from Fig.~\ref{Plots_AlphaEffect_NV}(e), we can observe that the center of the mass of the bubble is higher for larger values of Giesekus factor as the tail end of the bubble becomes shorter and less extended out.

\subsubsection{Influence of E\"{o}tv\"{o}s number on the bubble dynamics}

\begin{figure*}[ht!]
\begin{center}
\unitlength1cm
\begin{picture}(14.5,8.3)
\put(1.9,6.2){\makebox(0,0){\includegraphics[width=8.5cm,height=4.2cm,keepaspectratio]{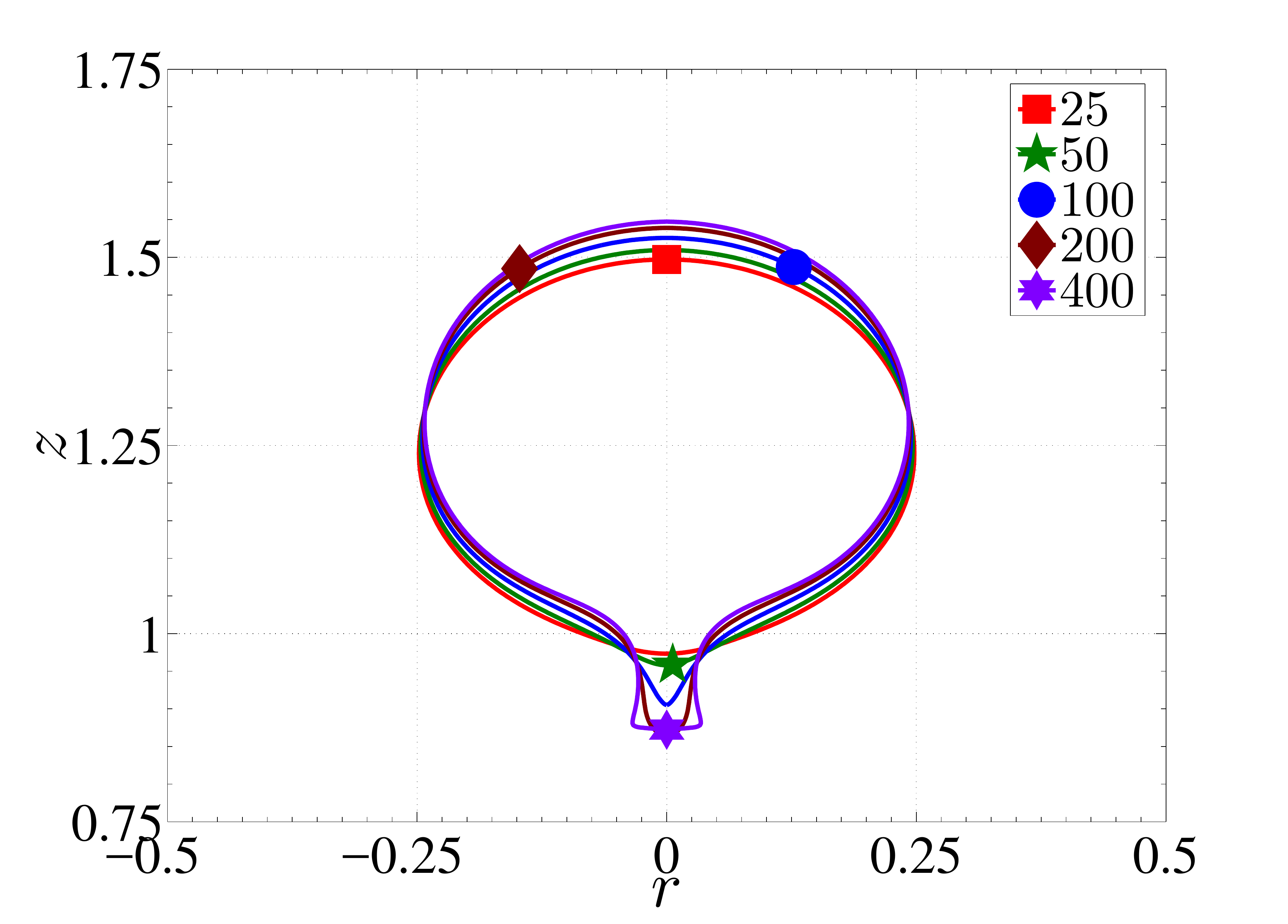}}}
\put(7.4,6.2){\makebox(0,0){\includegraphics[width=8.5cm,height=4.2cm,keepaspectratio]{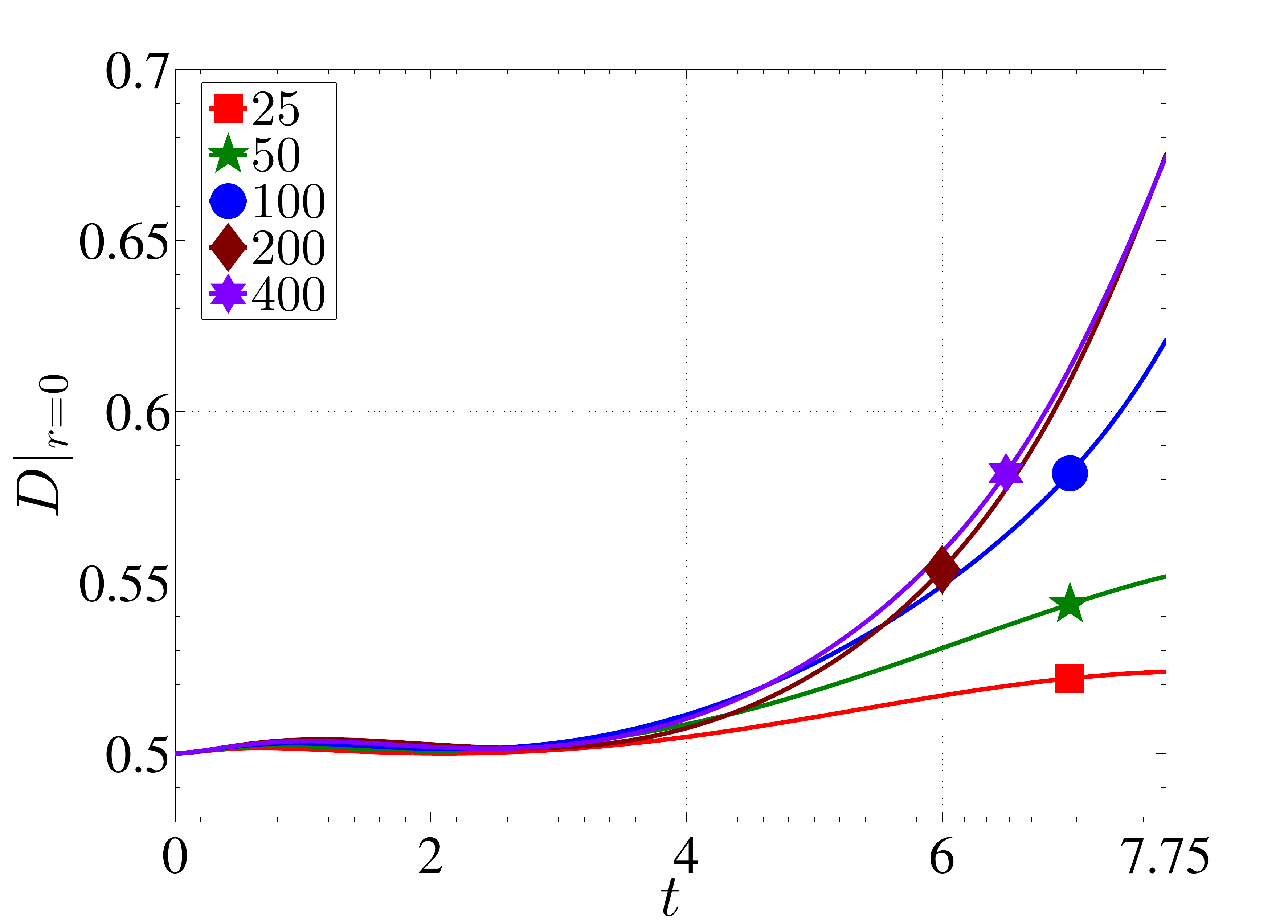}}}
\put(12.9,6.2){\makebox(0,0){\includegraphics[width=8.5cm,height=4.2cm,keepaspectratio]{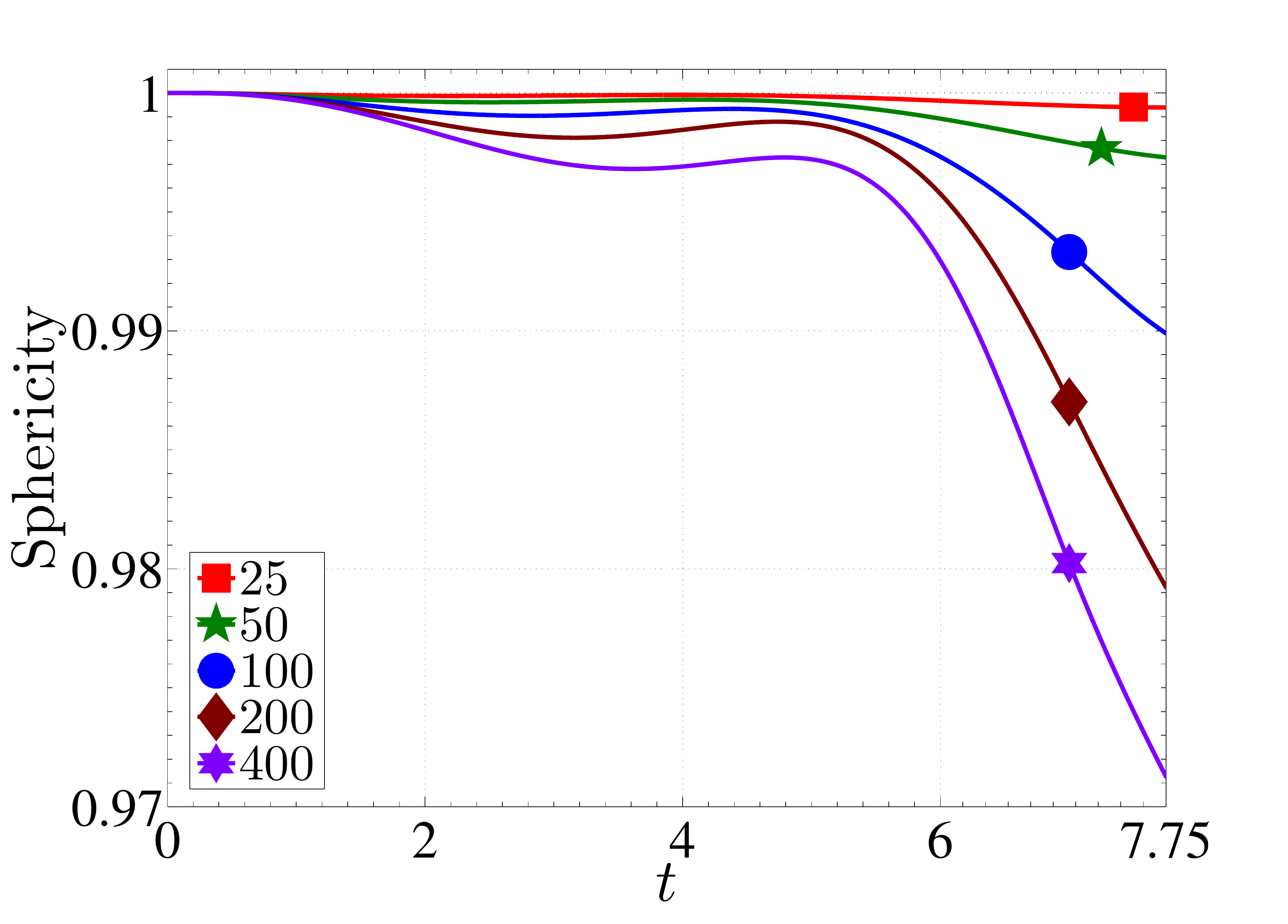}}}
\put(1.9,1.6){\makebox(0,0){\includegraphics[width=8.5cm,height=4.2cm,keepaspectratio]{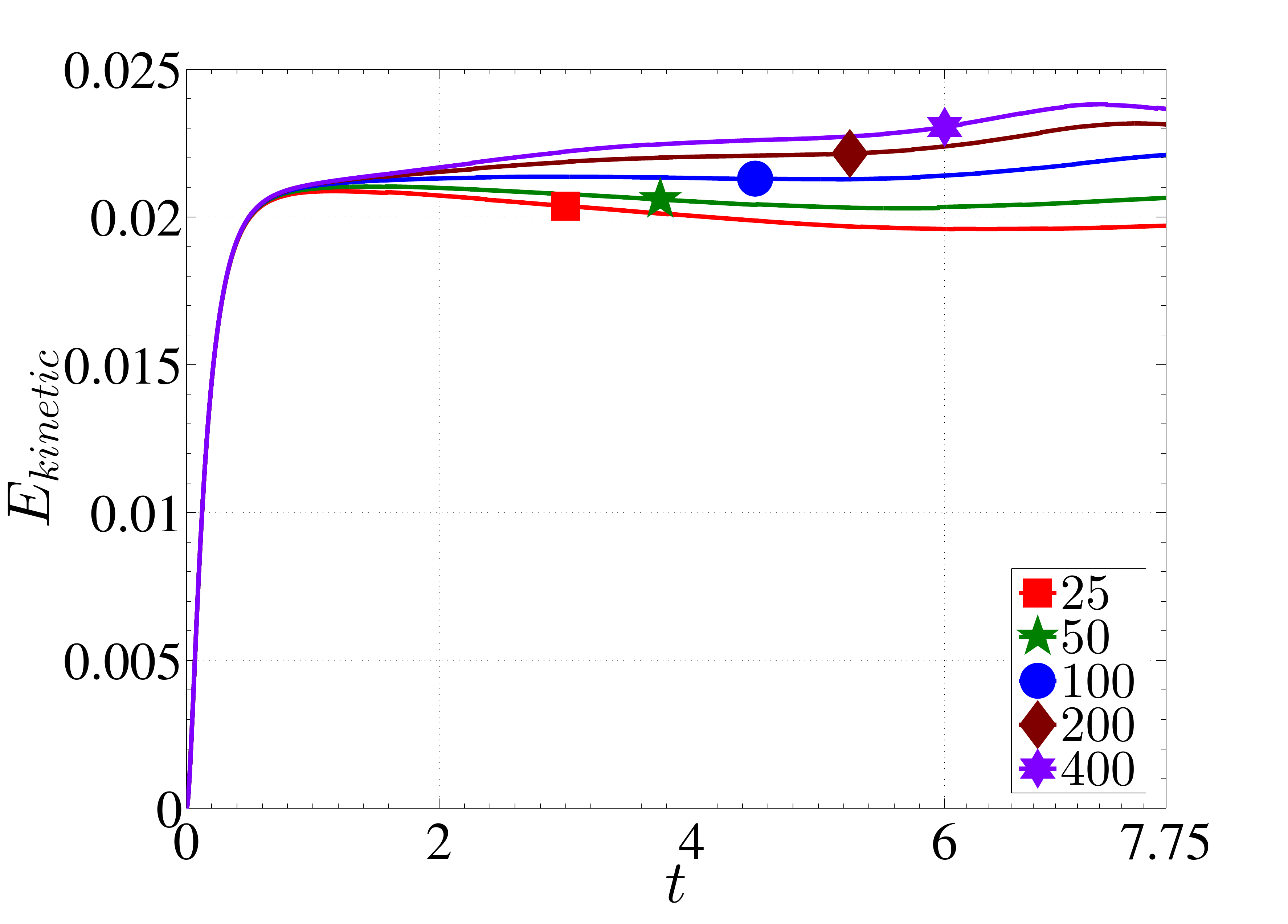}}}
\put(7.4,1.6){\makebox(0,0){\includegraphics[width=8.5cm,height=4.2cm,keepaspectratio]{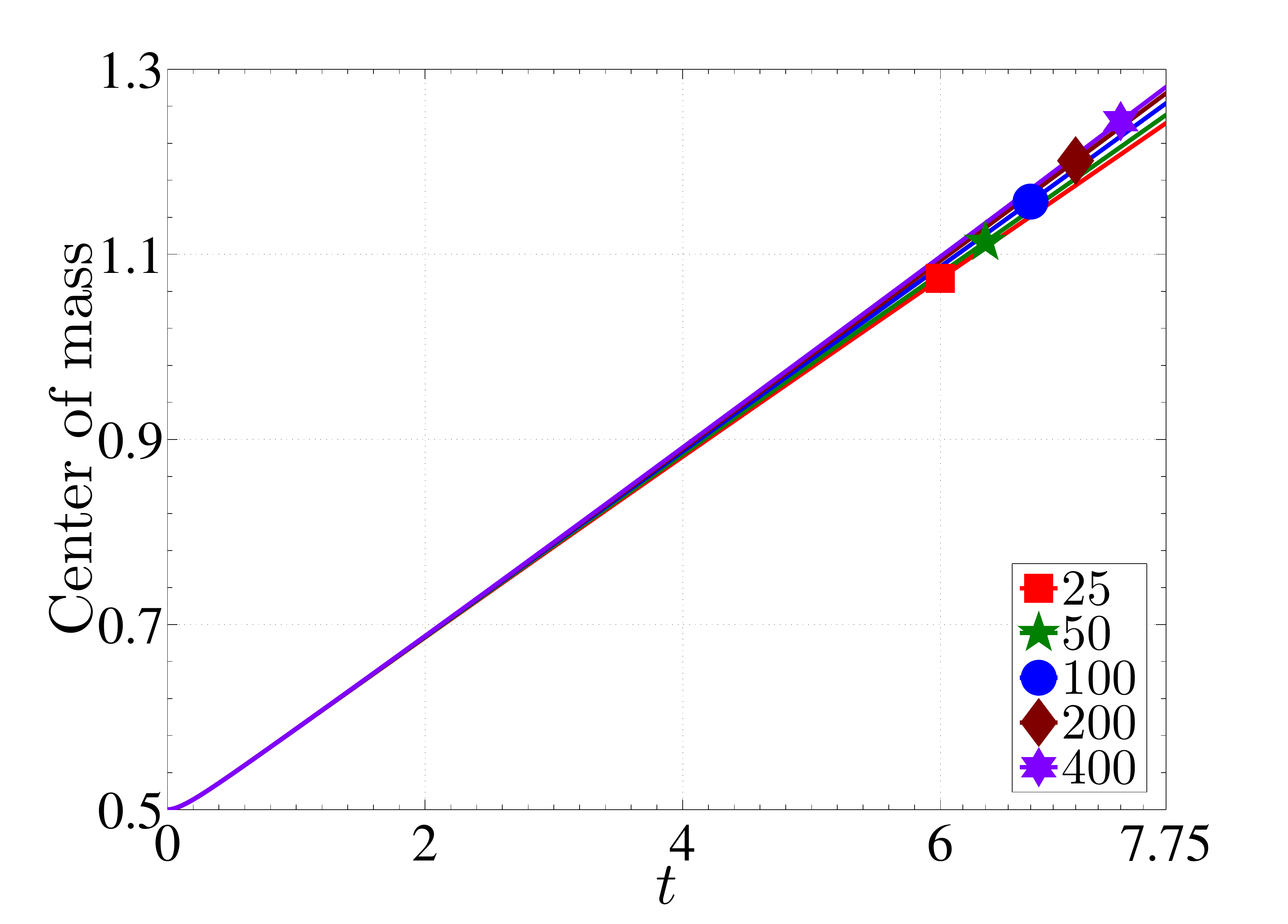}}}
\put(12.9,1.6){\makebox(0,0){\includegraphics[width=8.5cm,height=4.2cm,keepaspectratio]{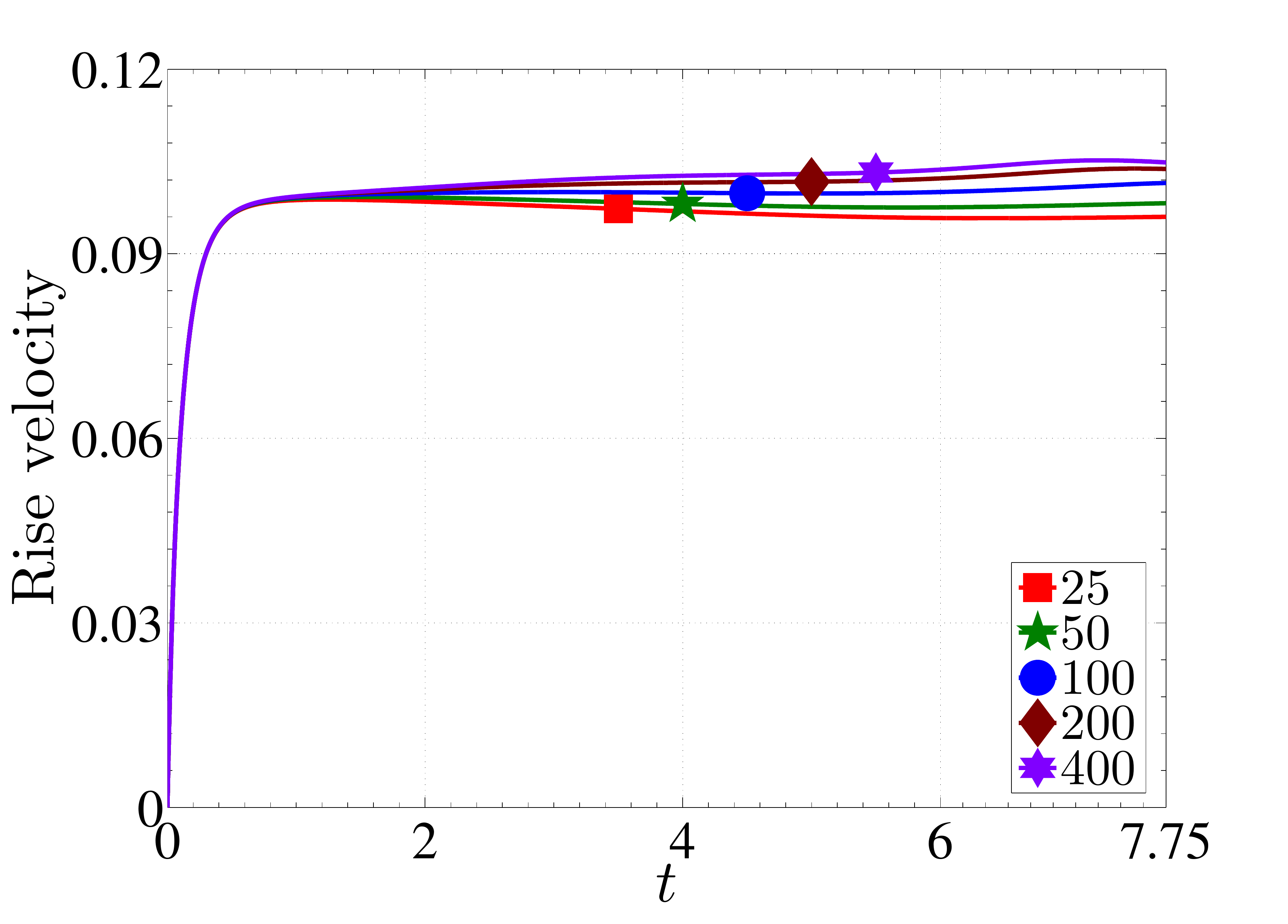}}}
\put(1.85,8.1){$(a)$}
\put(7.3, 8.1){$(b)$}
\put(12.8, 8.1){$(c)$}
\put(1.85,3.5){$(d)$}
\put(7.3, 3.5){$(e)$}
\put(12.8, 3.5){$(f)$}
\end{picture}
\end{center}
\caption{Influence of E\"{o}tv\"{o}s number for a Newtonian bubble rising in a viscoelastic fluid column~: (a)~bubble shape at $t$~=~7.75, (b)~diameter of the bubble at $r$~=~0, (c)~sphericity, (d)~kinetic energy, (e)~center of mass and (f)~rise velocity of the bubble for different E\"{o}tv\"{o}s numbers (i)~Eo~=~25, (ii)~Eo~=~50, (iii)~Eo~=~100, (iv)~Eo~=~200 and (v)~Eo~=~400 with $\Rey_2$~=~$10$, $\text{Wi}_2$~=~25, $\varepsilon$~=~10, $\rho_1/\rho_2$~=~0.1, $\beta_1$~=~1.0, $\beta_2$~=~0.75, $\alpha_2$~=~0.1, $D$~=~0.5 and $h_c$~=~2.0.} 
\label{Plots_WeberEffect_NV}
\end{figure*}
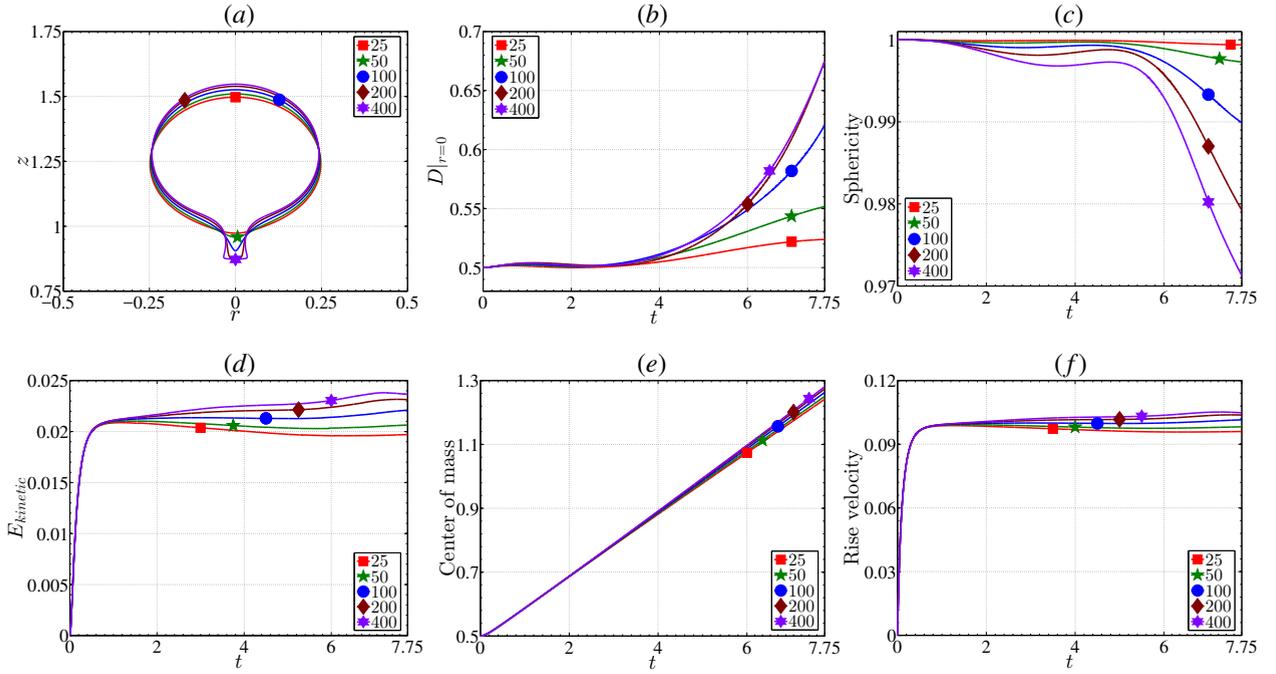

In this section, we study the influence of E\"{o}tv\"{o}s number on the rising Newtonian bubble dynamics in a viscoelastic fluid column.
We consider the base case flow parameters and vary only the E\"{o}tv\"{o}s number, i.e. vary the interfacial tension.
Five different values are used for the E\"{o}tv\"{o}s number in this study, which are as follows~: (i)~Eo~=~25, (ii)~Eo~=~50, (iii)~Eo~=~100, (iv)~Eo~=~200 and (v)~Eo~=~400.
Increasing the E\"{o}tv\"{o}s number, decreases the interfacial tension, thereby making the interface more easily deformable and thus increases the degree of interface stretching by the polymer stress.
In Fig.~\ref{Plots_WeberEffect_NV}(a), we can observe that at low E\"{o}tv\"{o}s numbers, the bubble shapes are more similar to a Newtonian bubble rising in a Newtonian fluid column.
In fact, with further advancement in time, they still do not deform as observed with high E\"{o}tv\"{o}s numbers.
This phenomenon can be explained by the fact that there exists a critical capillary number, beyond which the bubble experiences unsteady deformations in the form of an extended trialing edge.
For interface flows, capillary number is the ratio of E\"{o}tv\"{o}s number to the Reynolds number. 
Hence, by increasing the E\"{o}tv\"{o}s number, we actually increase the capillary number. 
From Fig.~\ref{Plots_WeberEffect_NV}(a), we can comment that the critical E\"{o}tv\"{o}s number for unsteady drop shapes for the considered flow parameters is between 50 and 100 as bubbles  beyond Eo~=~100 become cusp-like shaped.

Since, the extended trailing edge behaviour increases with an increase in the E\"{o}tv\"{o}s number, the diameter of the bubble at the axis of symmetry increases when the viscoelastic stresses start to overcome the interfacial tension, refer Fig.~\ref{Plots_WeberEffect_NV}(b).
However, till the motion is inertia dominated, there is not much effect of E\"{o}tv\"{o}s number on the diameter of the bubble.
Further, Fig.~\ref{Plots_WeberEffect_NV}(c) presents the temporal evolution of the sphericity of the bubble. 
It quite natural that, with increase in the E\"{o}tv\"{o}s number, the interface becomes more deformable and hence, the sphericity decreases.
Next, Fig.~\ref{Plots_WeberEffect_NV}(d) and (f) depicts the kinetic energy and rise velocity of the bubble. 
We can observe that they increase with an increase in the E\"{o}tv\"{o}s number.
Further, the center of mass of the bubble is higher for larger E\"{o}tv\"{o}s numbers, see Fig.~\ref{Plots_WeberEffect_NV}(e), as the bubble rises higher with greater rise velocity.

\subsection{Viscoelastic bubble rising in a Newtonian fluid column}
In this section, we consider a buoyancy driven 3D-axisymmetric viscoelastic bubble rising in a Newtonian fluid column. 
The base case parameters for studying the effects of various flow variables are defined as follows~: $\Rey_2$~=~10, Eo~=~400, $\text{Wi}_1$~=~10, $\rho_1/\rho_2$~=~0.1, $\varepsilon$~=~2, $\beta_1$~=~0.5, $\beta_2$~=~1.0, $\alpha_1$~=~0.1, $D$~=~0.5 and $h_c$~=~2.5.
During the triangulation, we limit the maximum area of each cell in the mesh to 0.001, which leads to 1198 and 2954 cells in the initial inner and outer domains respectively. 
The finite element spaces used in computations for the velocity / pressure / viscoelastic stress are  $P_2^{bubble}$ / $P_1^{disc}$ / $P_2^{bubble}$.
This choice of initial mesh and finite element spaces results in 41854 velocity, 12456 pressure and 62781 viscoelastic degrees of freedom. 
Further, we use a constant time step $\delta t = 0.0005$ and 400 degrees of freedom on the interface with initial mesh size $h_0 = 0.00392695$.
The stabilization constants used in computations are $c_1$~=~0.05, $c_2$~=~0.05 and $c_3$~=~0.05.

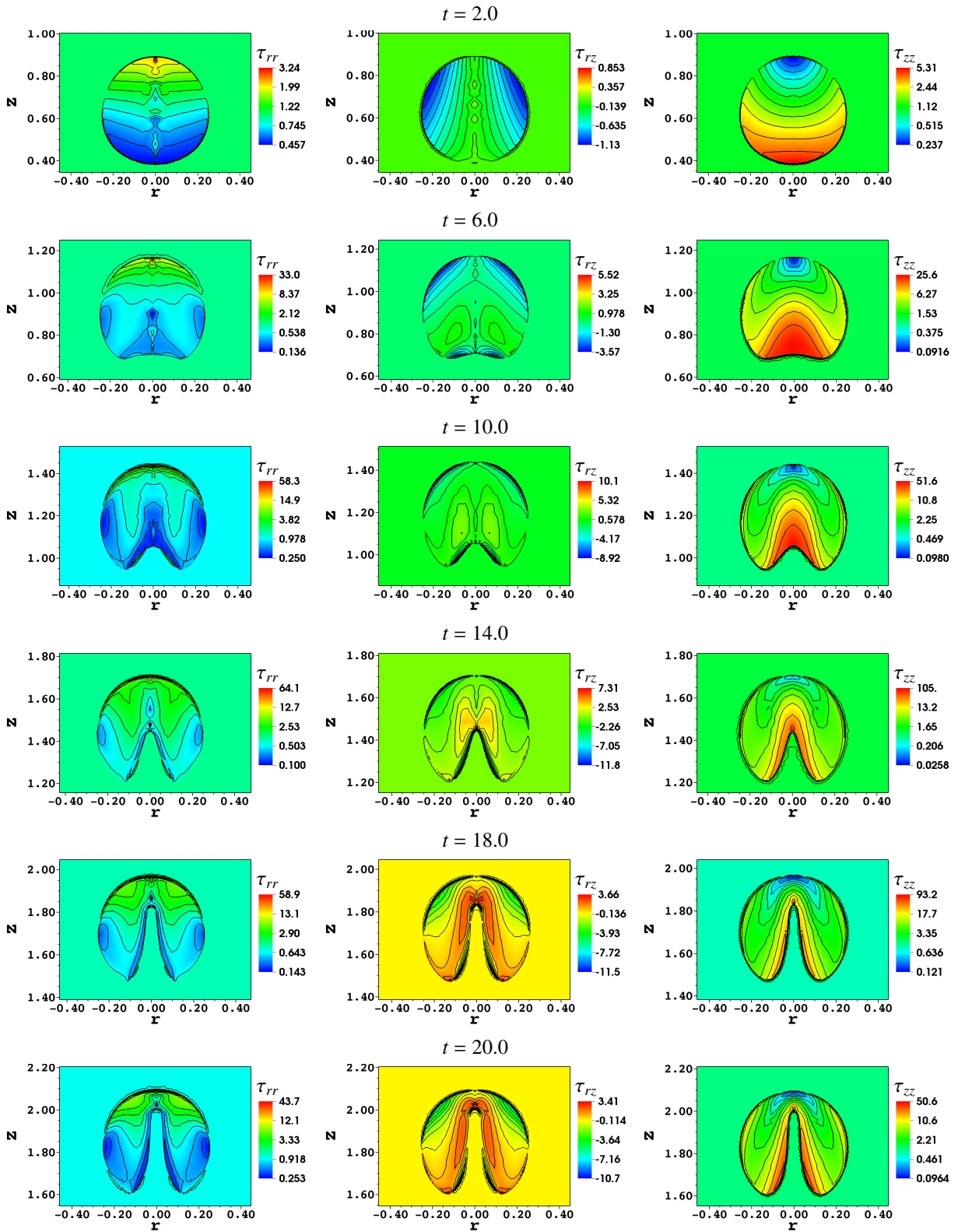
\begin{figure*}
\begin{center}
\unitlength1cm
\begin{picture}(20,20.5)

\put(1.2,-2.0){\makebox(3,6){\includegraphics[trim=1.0cm 0.0cm 0.0cm 0.7cm, clip=true,width=5.2cm]{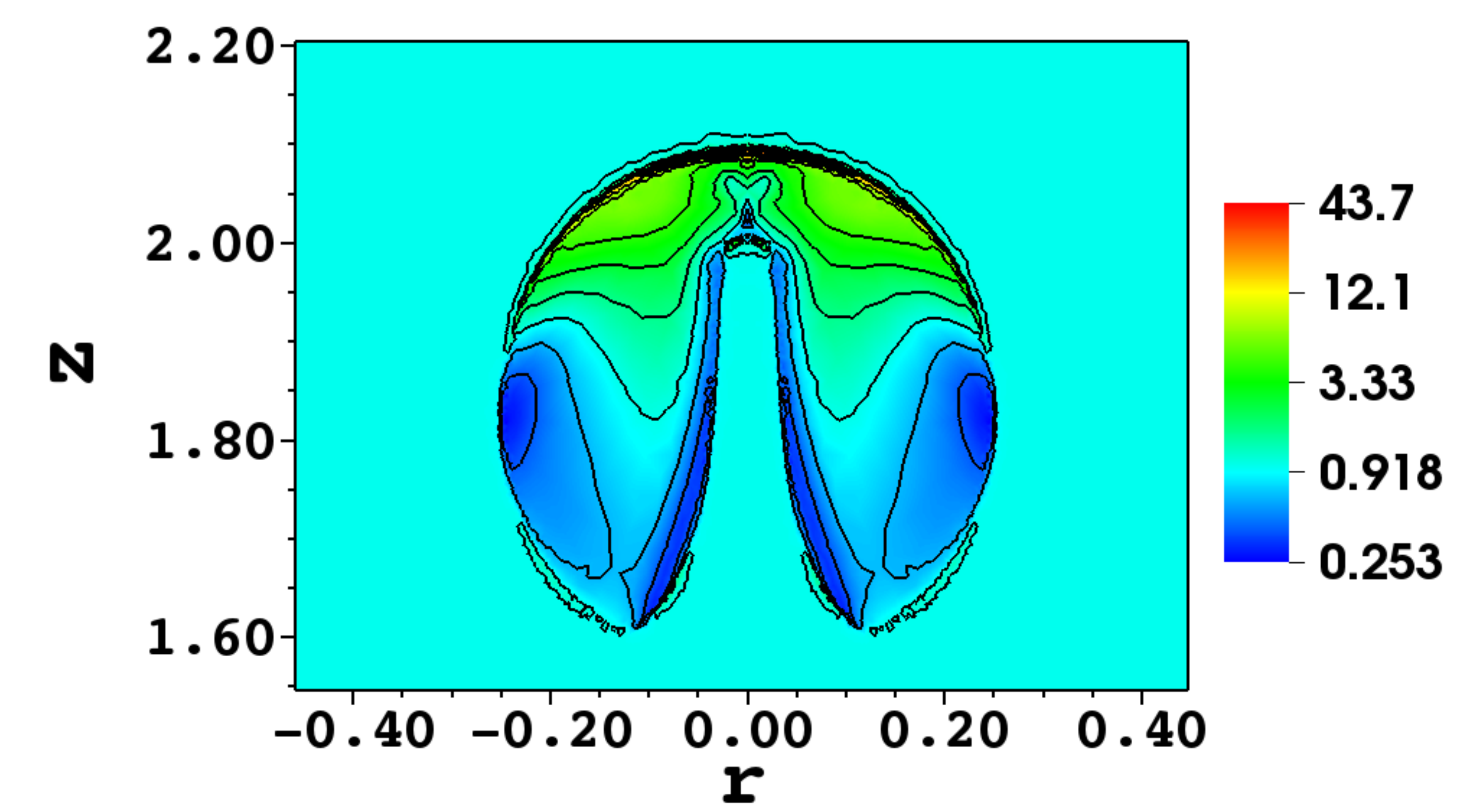}}}
\put(6.6,-2.0){\makebox(3,6){\includegraphics[trim=1.0cm 0.0cm 0.0cm 0.7cm, clip=true,width=5.2cm]{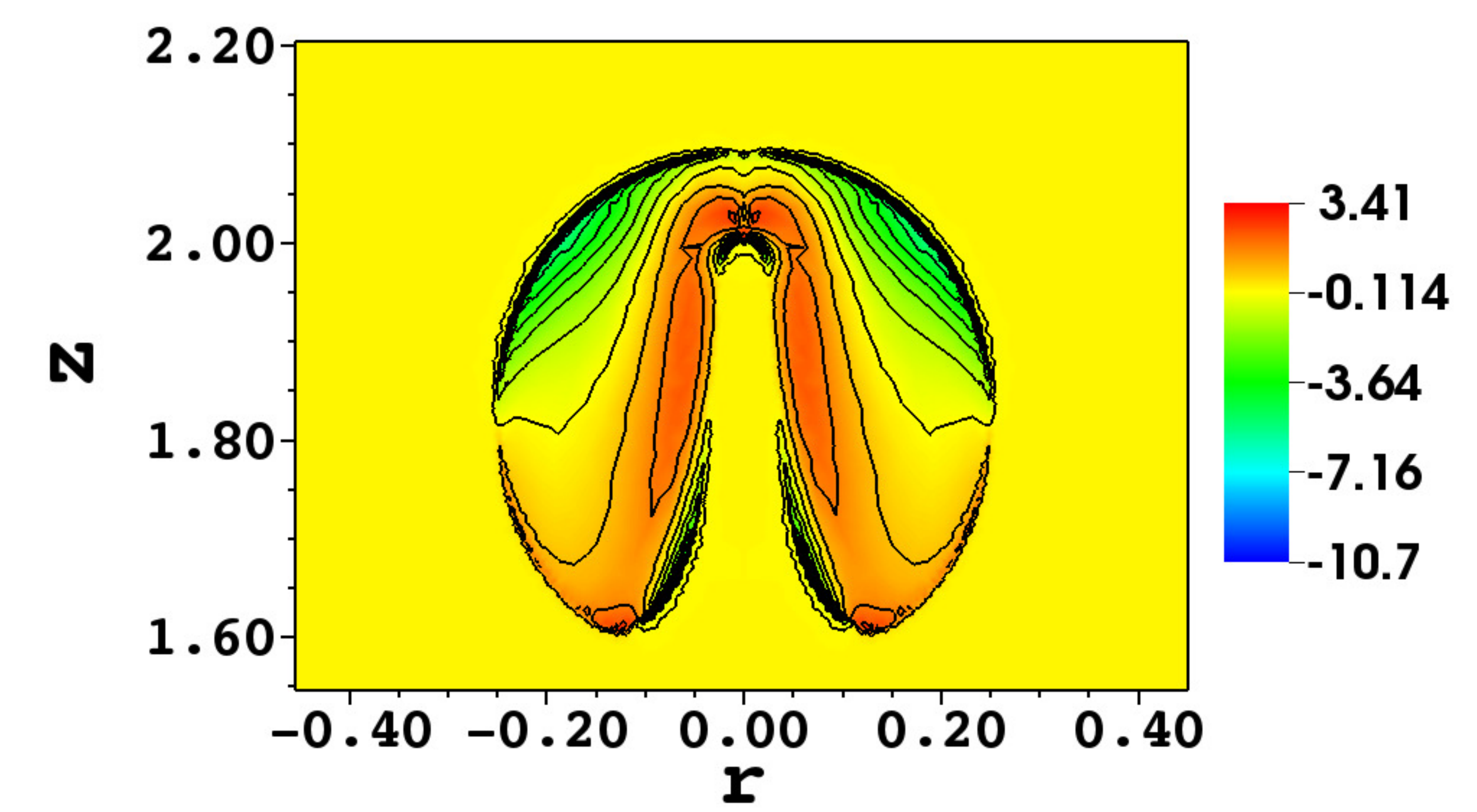}}}
\put(12.0,-2.0){\makebox(3,6){\includegraphics[trim=1.0cm 0.0cm 0.0cm 0.7cm, clip=true,width=5.2cm]{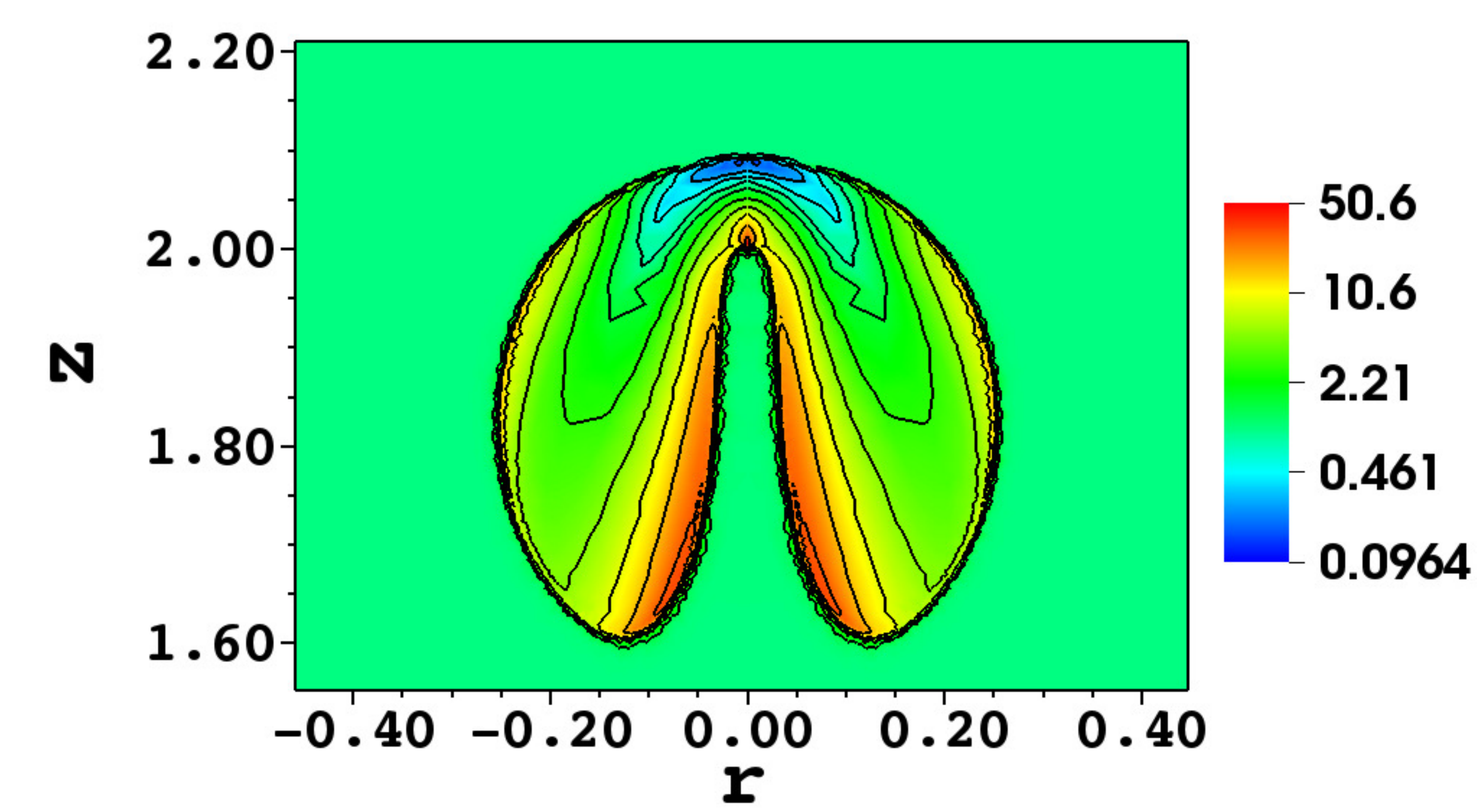}}}

\put(1.2,1.5){\makebox(3,6){\includegraphics[trim=1.0cm 0.0cm 0.0cm 0.7cm, clip=true,width=5.2cm]{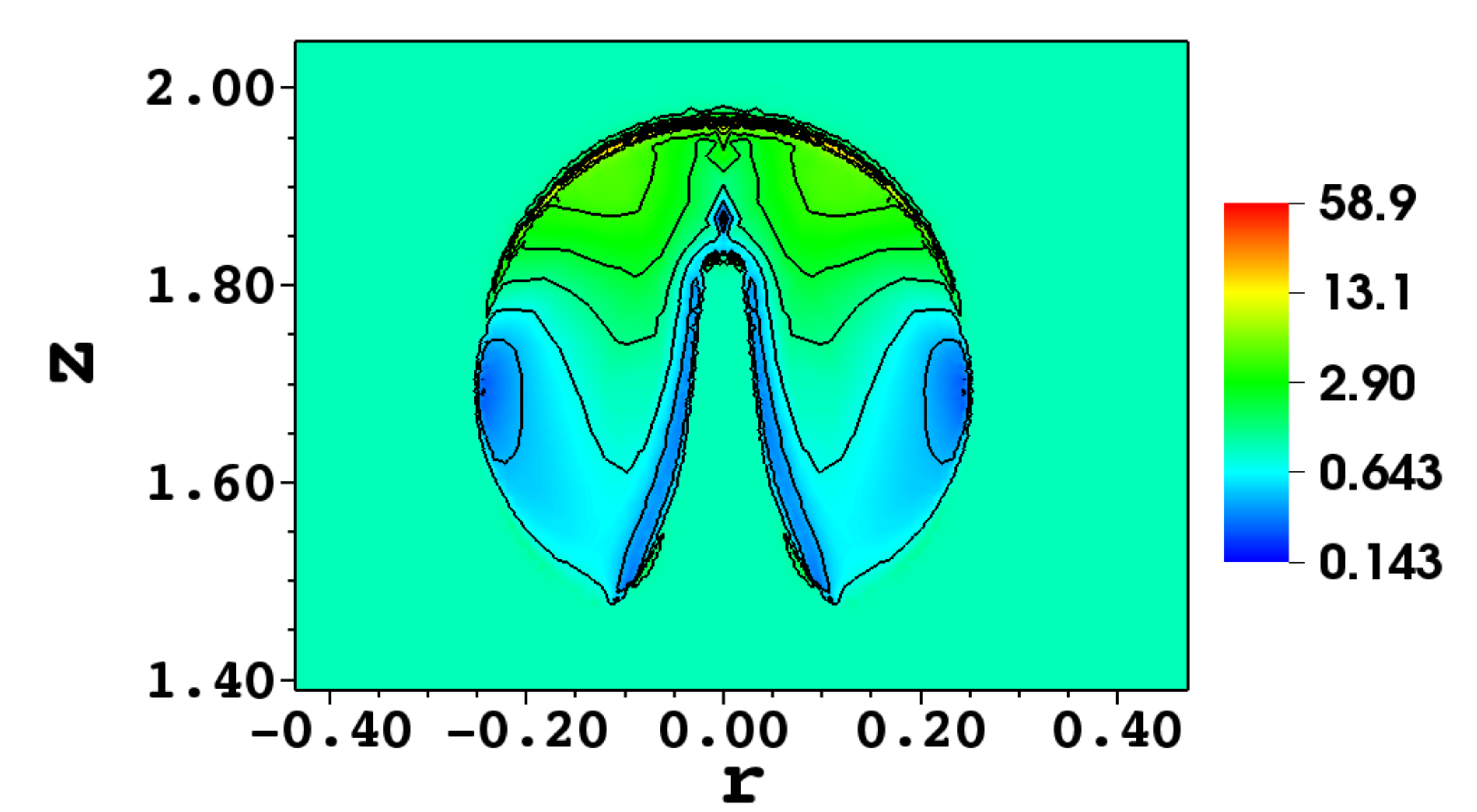}}}
\put(6.6,1.5){\makebox(3,6){\includegraphics[trim=1.0cm 0.0cm 0.0cm 0.7cm, clip=true,width=5.2cm]{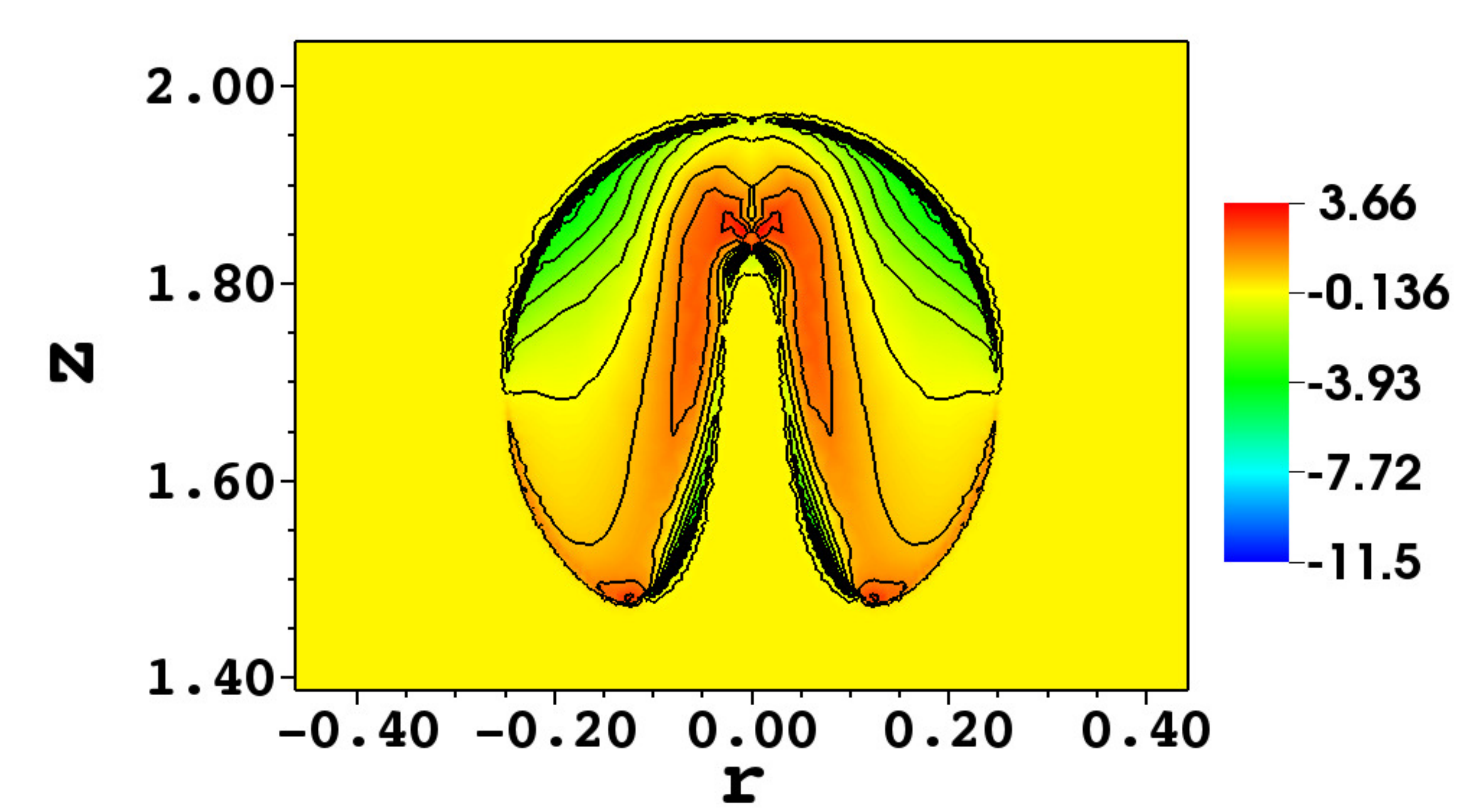}}}
\put(12.0,1.5){\makebox(3,6){\includegraphics[trim=1.0cm 0.0cm 0.0cm 0.7cm, clip=true,width=5.2cm]{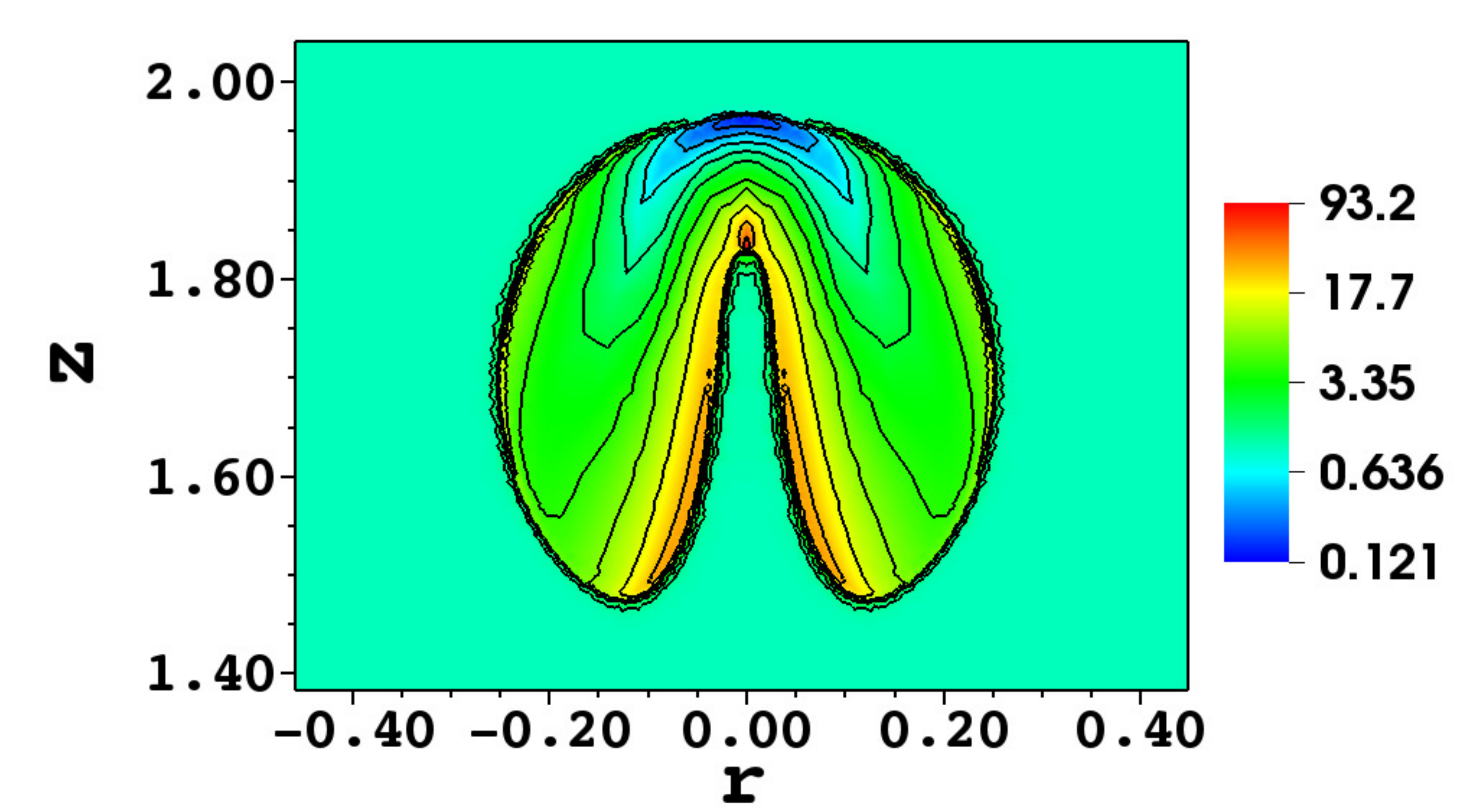}}}

\put(1.2,5.0){\makebox(3,6){\includegraphics[trim=1.0cm 0.0cm 0.0cm 0.7cm, clip=true,width=5.2cm]{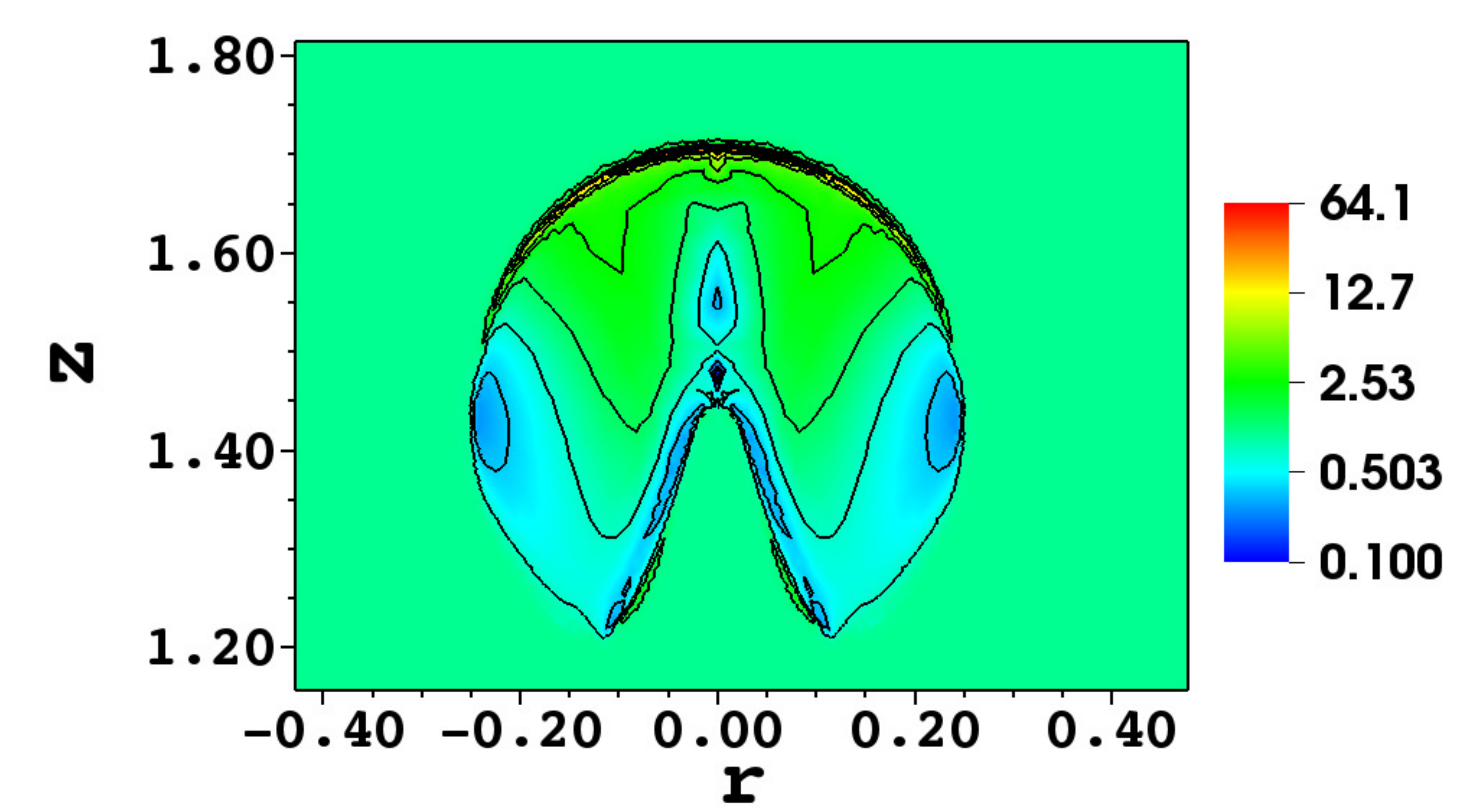}}}
\put(6.6,5.0){\makebox(3,6){\includegraphics[trim=1.0cm 0.0cm 0.0cm 0.7cm, clip=true,width=5.2cm]{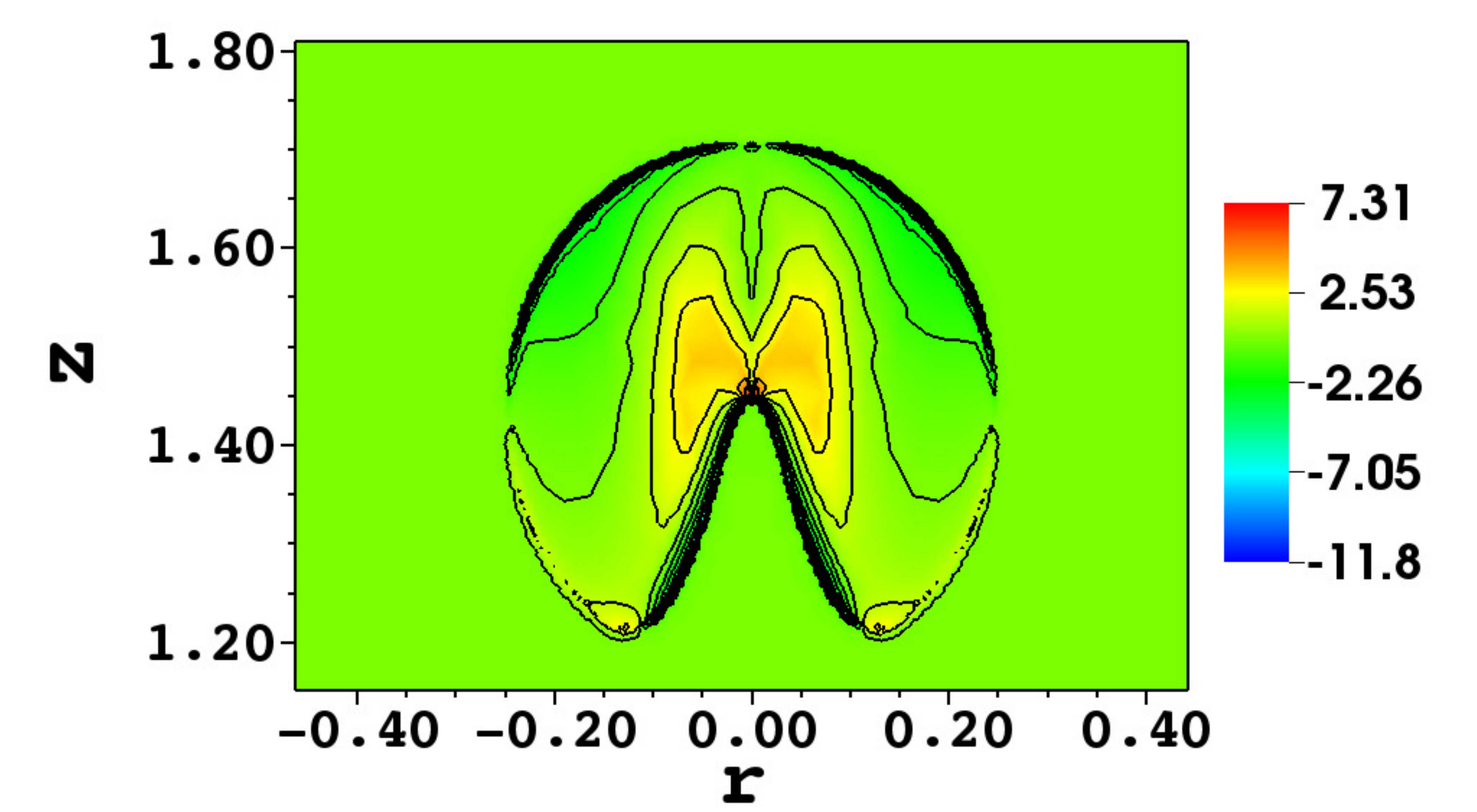}}}
\put(12.0,5.0){\makebox(3,6){\includegraphics[trim=1.0cm 0.0cm 0.0cm 0.7cm, clip=true,width=5.2cm]{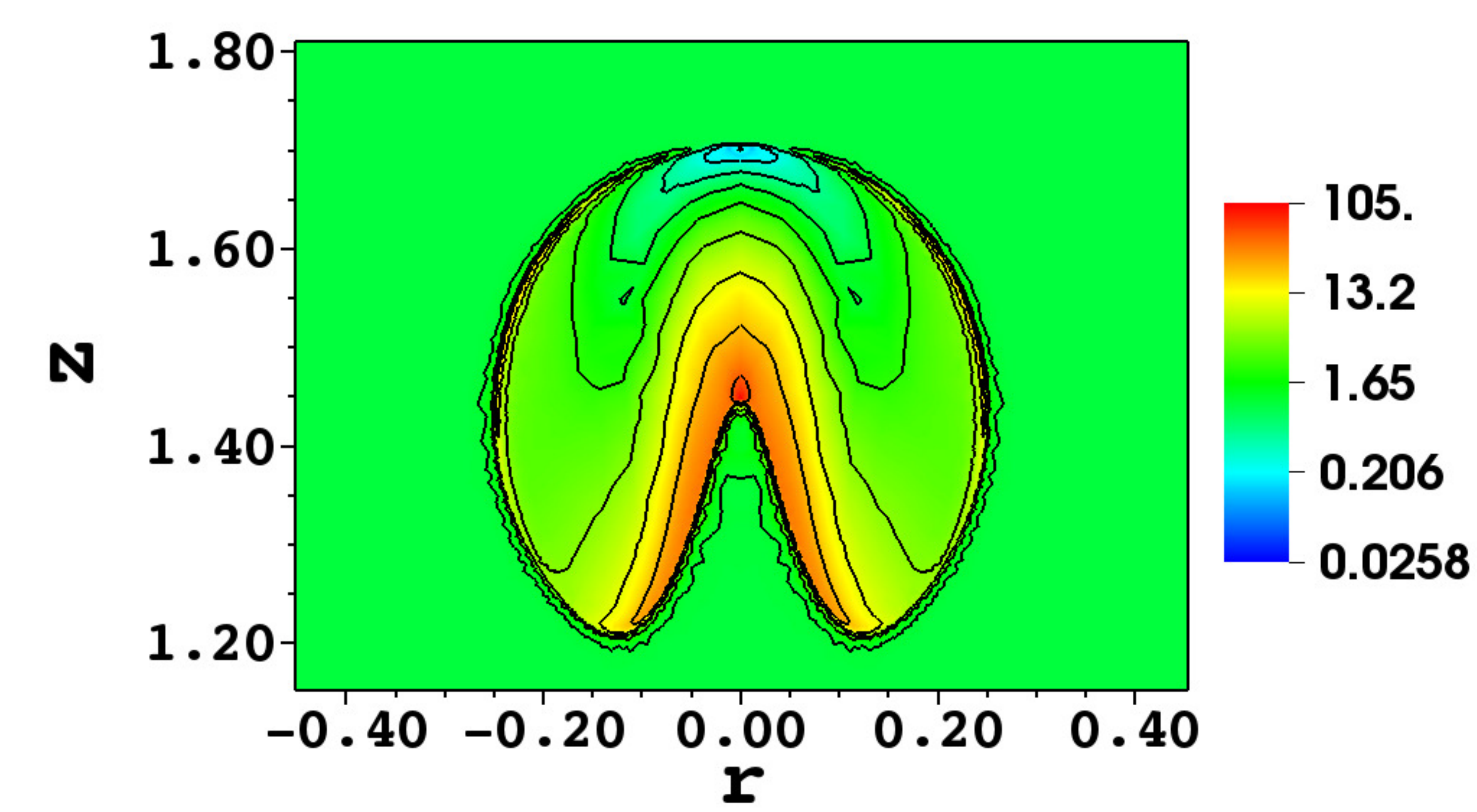}}}

\put(1.2,8.5){\makebox(3,6){\includegraphics[trim=1.0cm 0.0cm 0.0cm 0.7cm, clip=true,width=5.2cm]{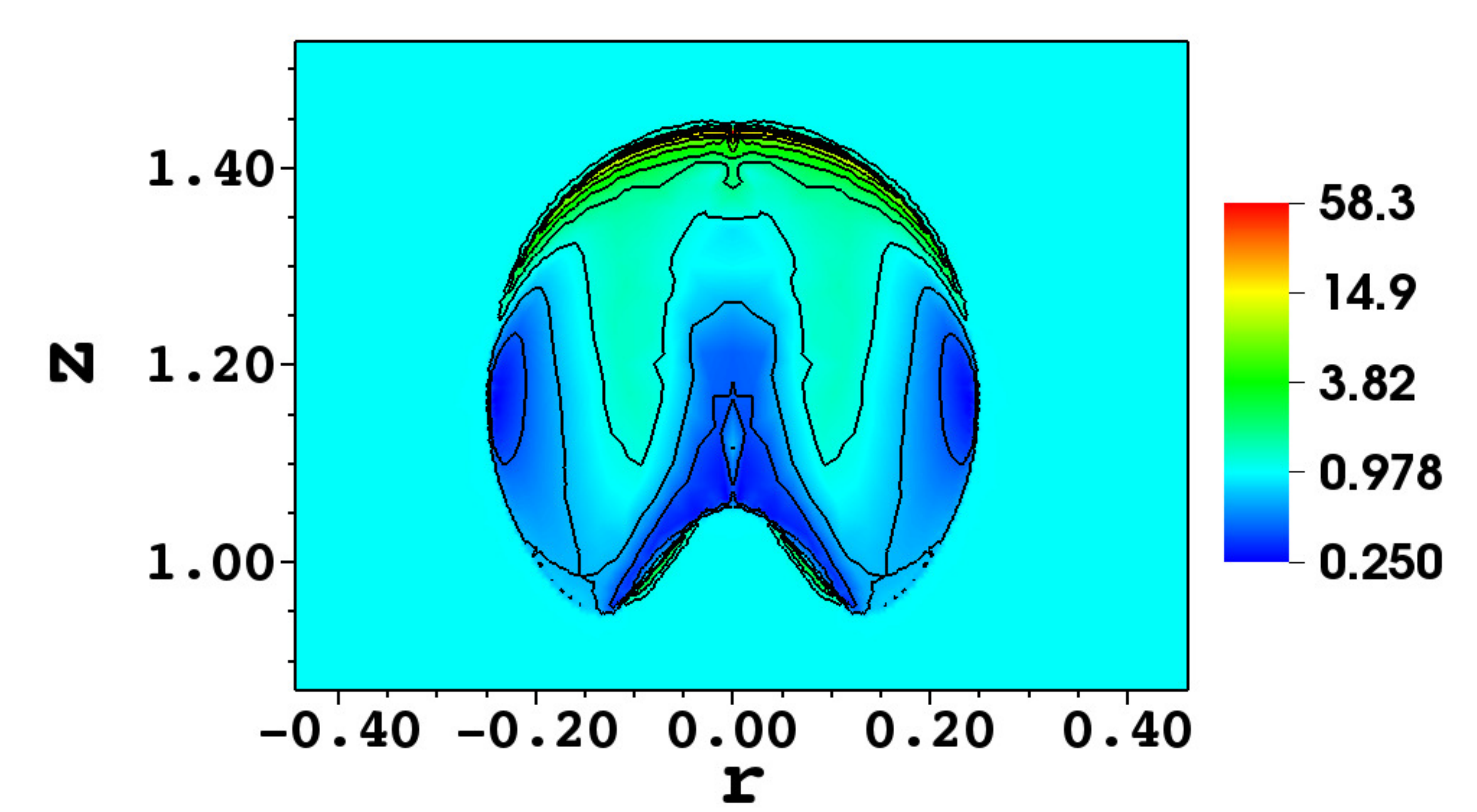}}}
\put(6.6,8.5){\makebox(3,6){\includegraphics[trim=1.0cm 0.0cm 0.0cm 0.7cm, clip=true,width=5.2cm]{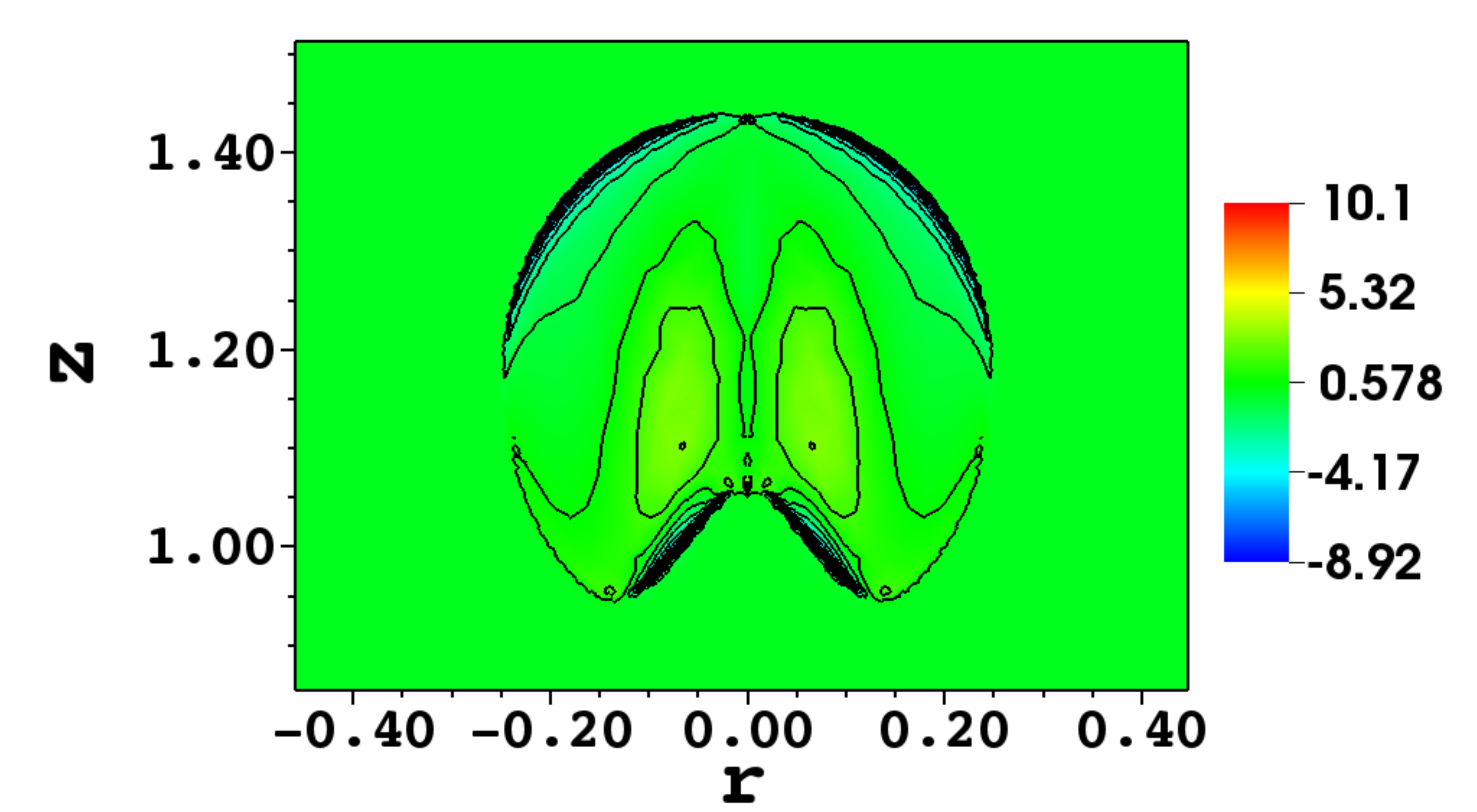}}}
\put(12.0,8.5){\makebox(3,6){\includegraphics[trim=1.0cm 0.0cm 0.0cm 0.7cm, clip=true,width=5.2cm]{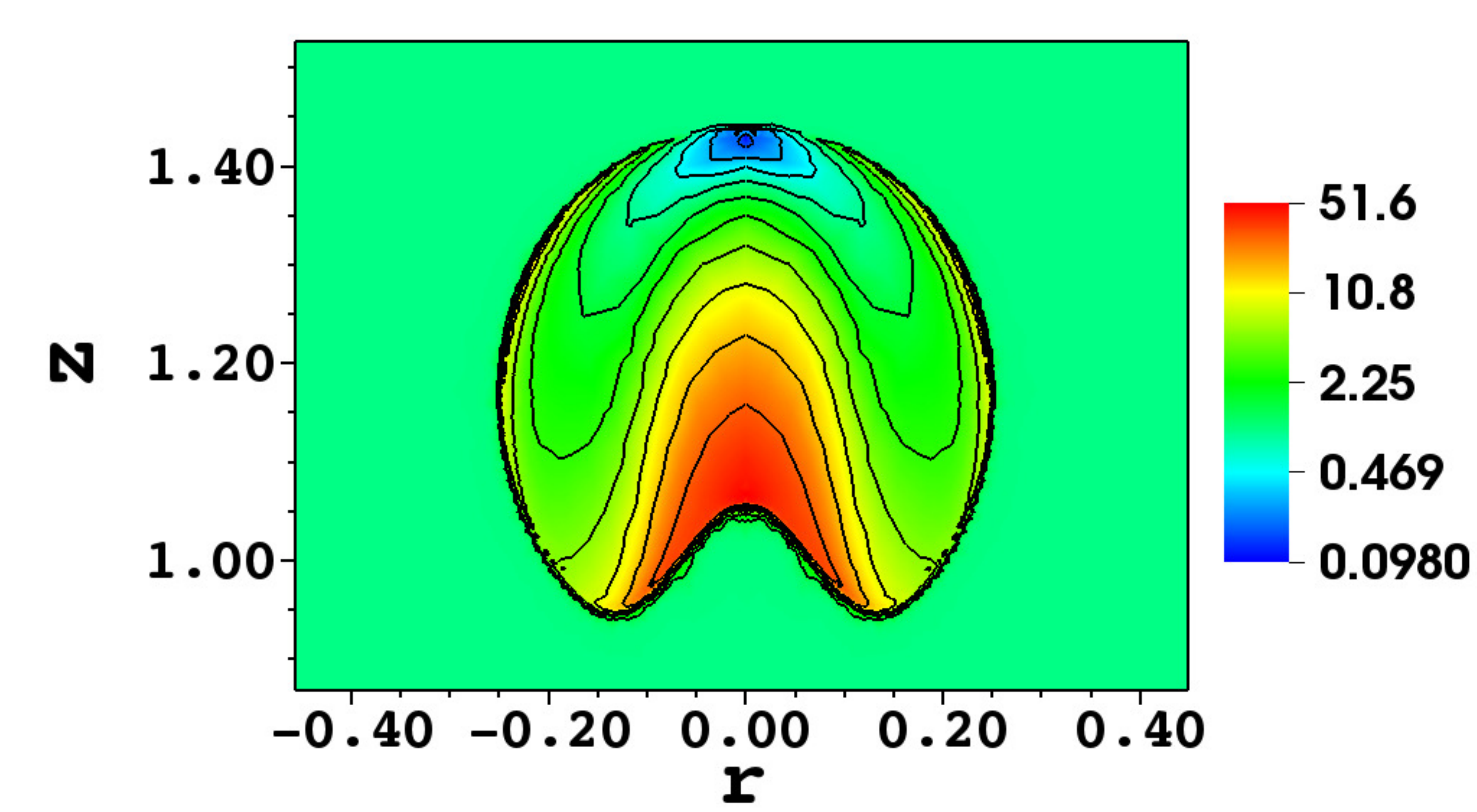}}}

\put(1.2,12.0){\makebox(3,6){\includegraphics[trim=1.0cm 0.0cm 0.0cm 0.7cm, clip=true,width=5.2cm]{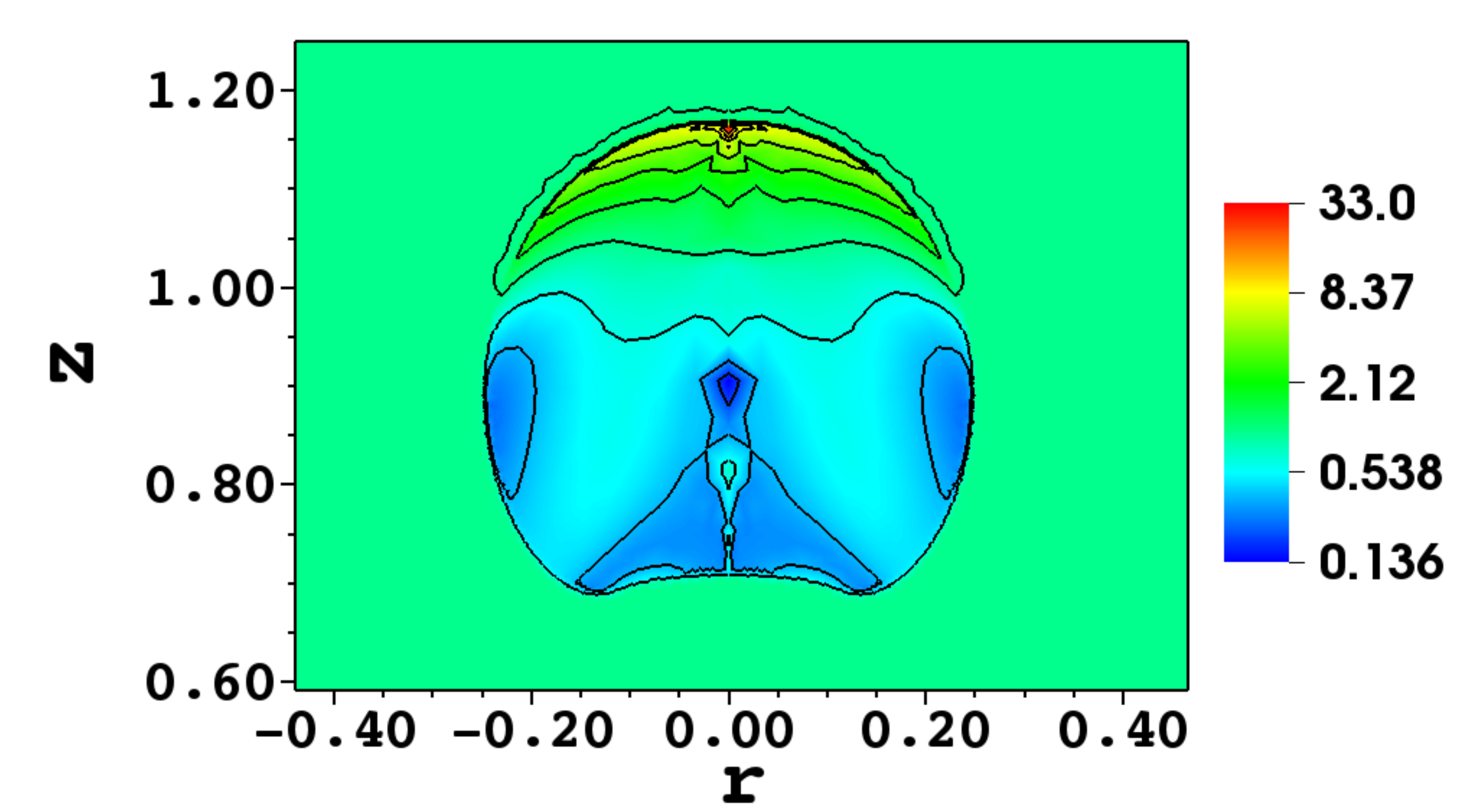}}}
\put(6.6,12.0){\makebox(3,6){\includegraphics[trim=1.0cm 0.0cm 0.0cm 0.7cm, clip=true,width=5.2cm]{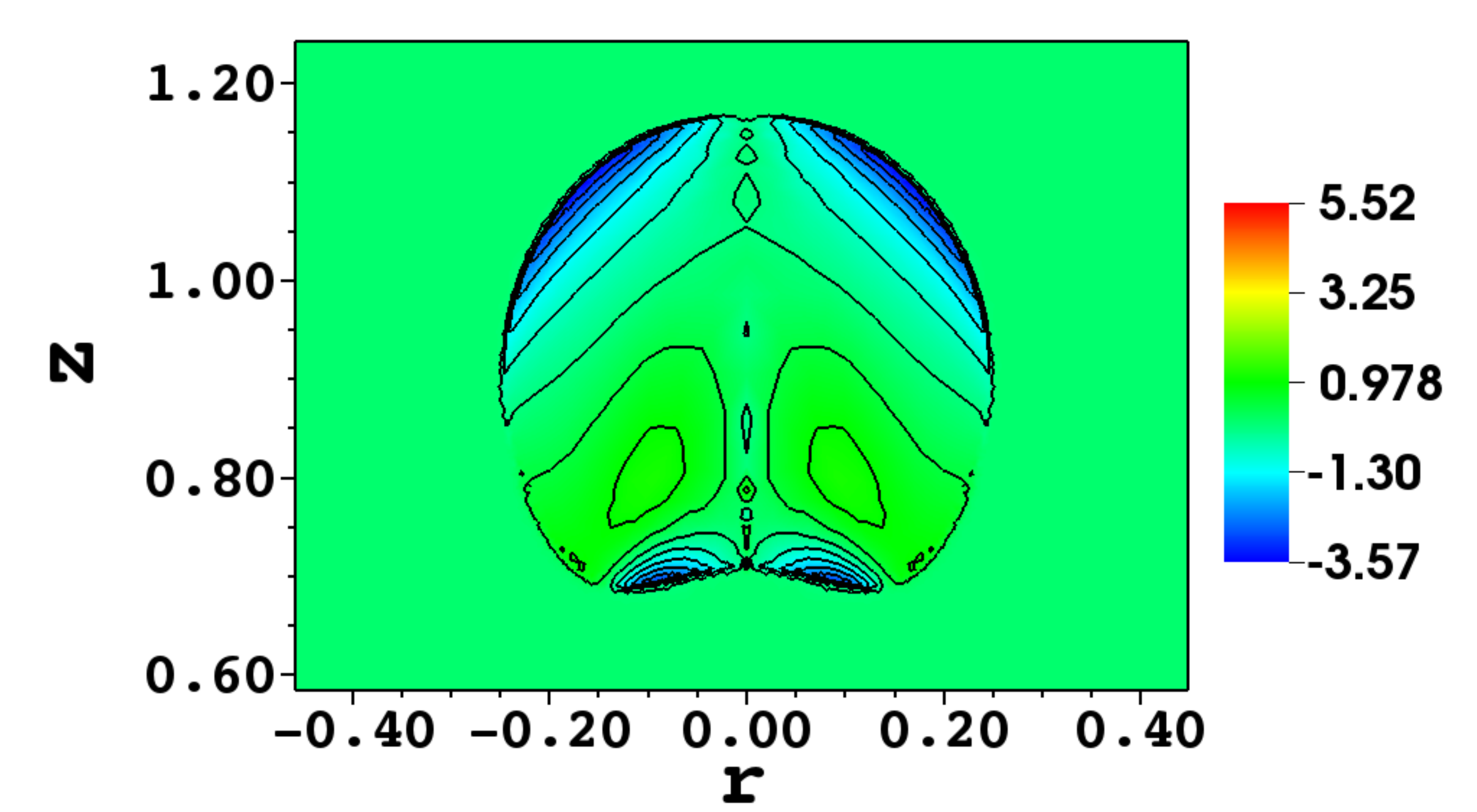}}}
\put(12.0,12.0){\makebox(3,6){\includegraphics[trim=1.0cm 0.0cm 0.0cm 0.7cm, clip=true,width=5.2cm]{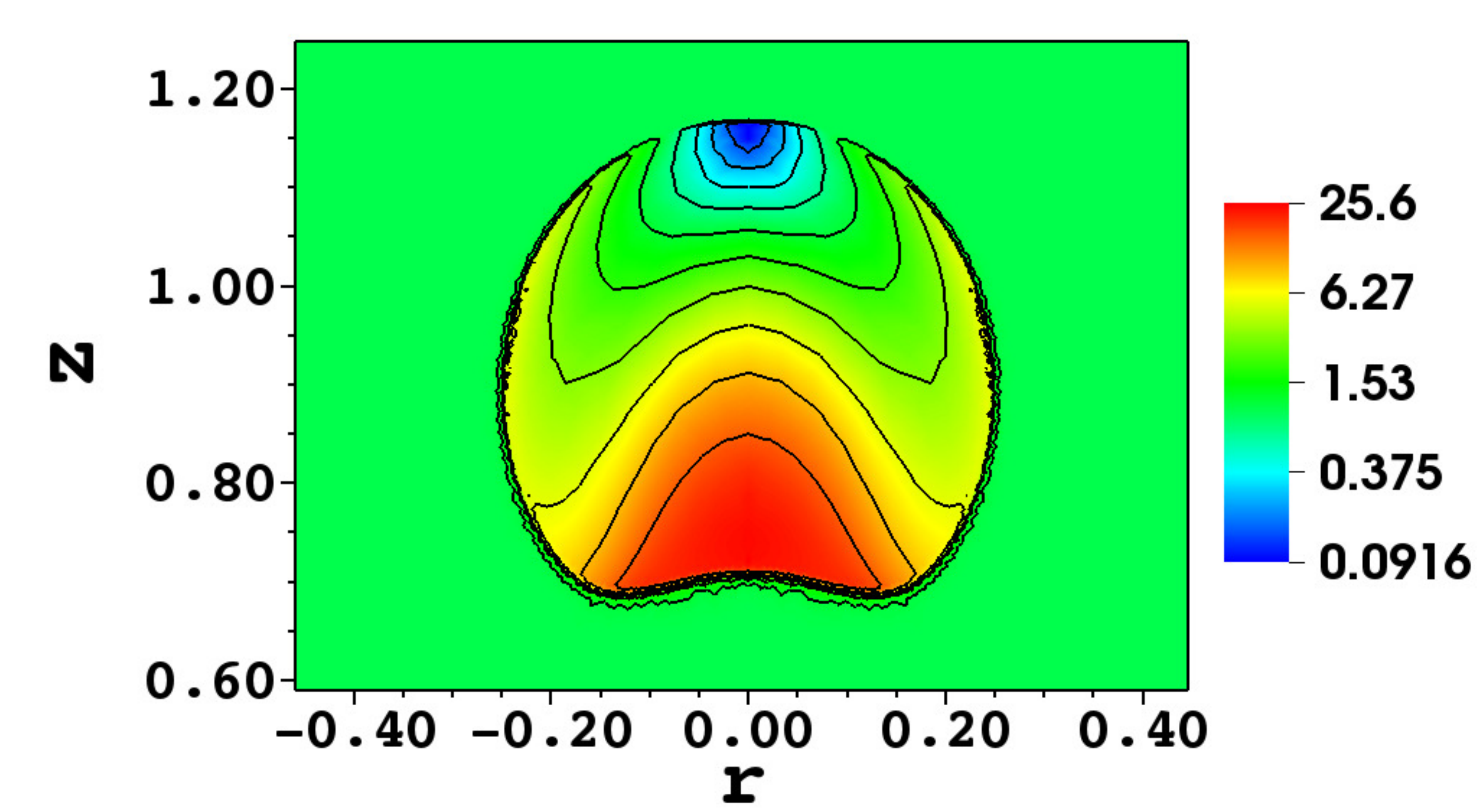}}}

\put(1.2,15.5){\makebox(3,6){\includegraphics[trim=1.0cm 0.0cm 0.0cm 0.7cm, clip=true,width=5.2cm]{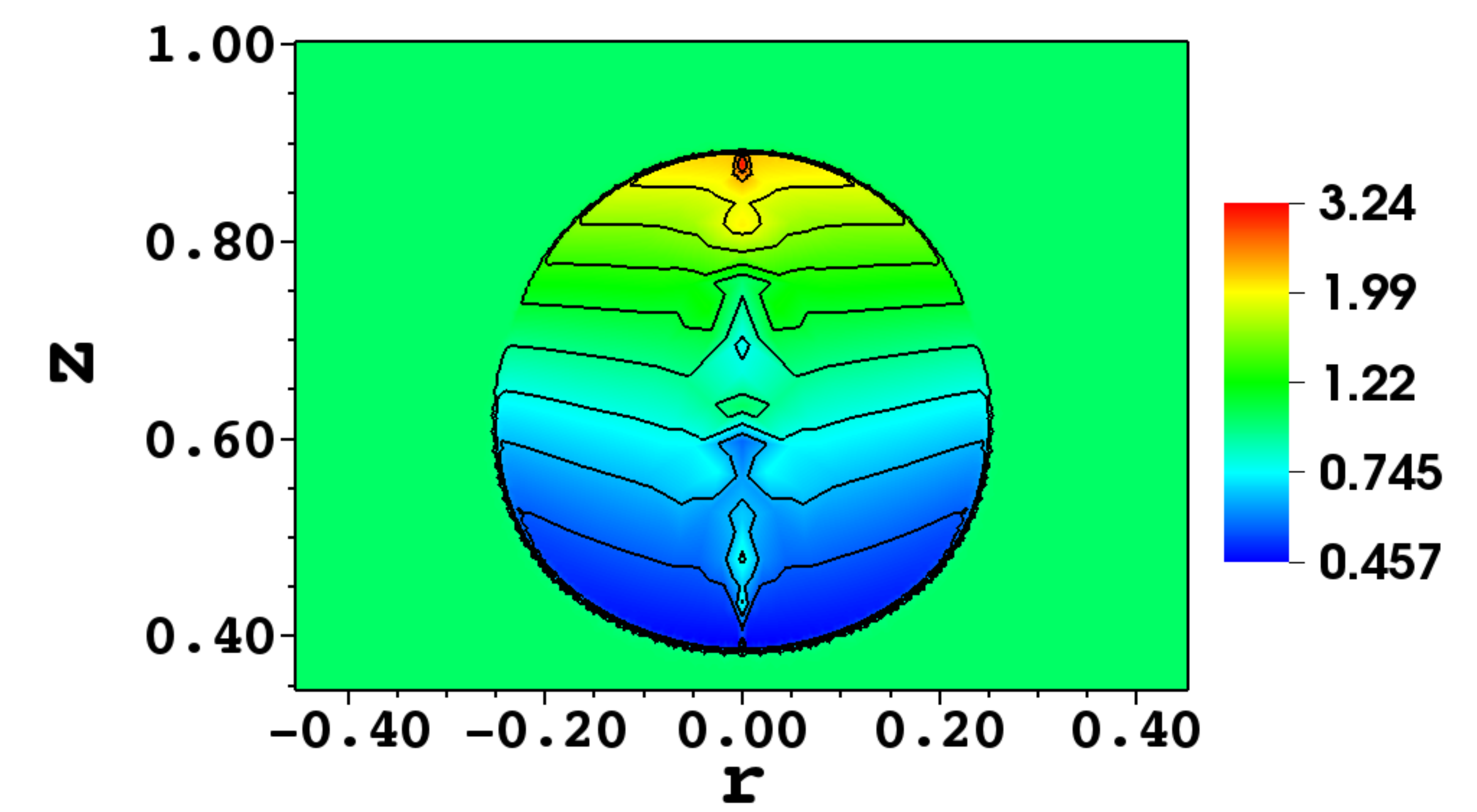}}}
\put(6.6,15.5){\makebox(3,6){\includegraphics[trim=1.0cm 0.0cm 0.0cm 0.7cm, clip=true,width=5.2cm]{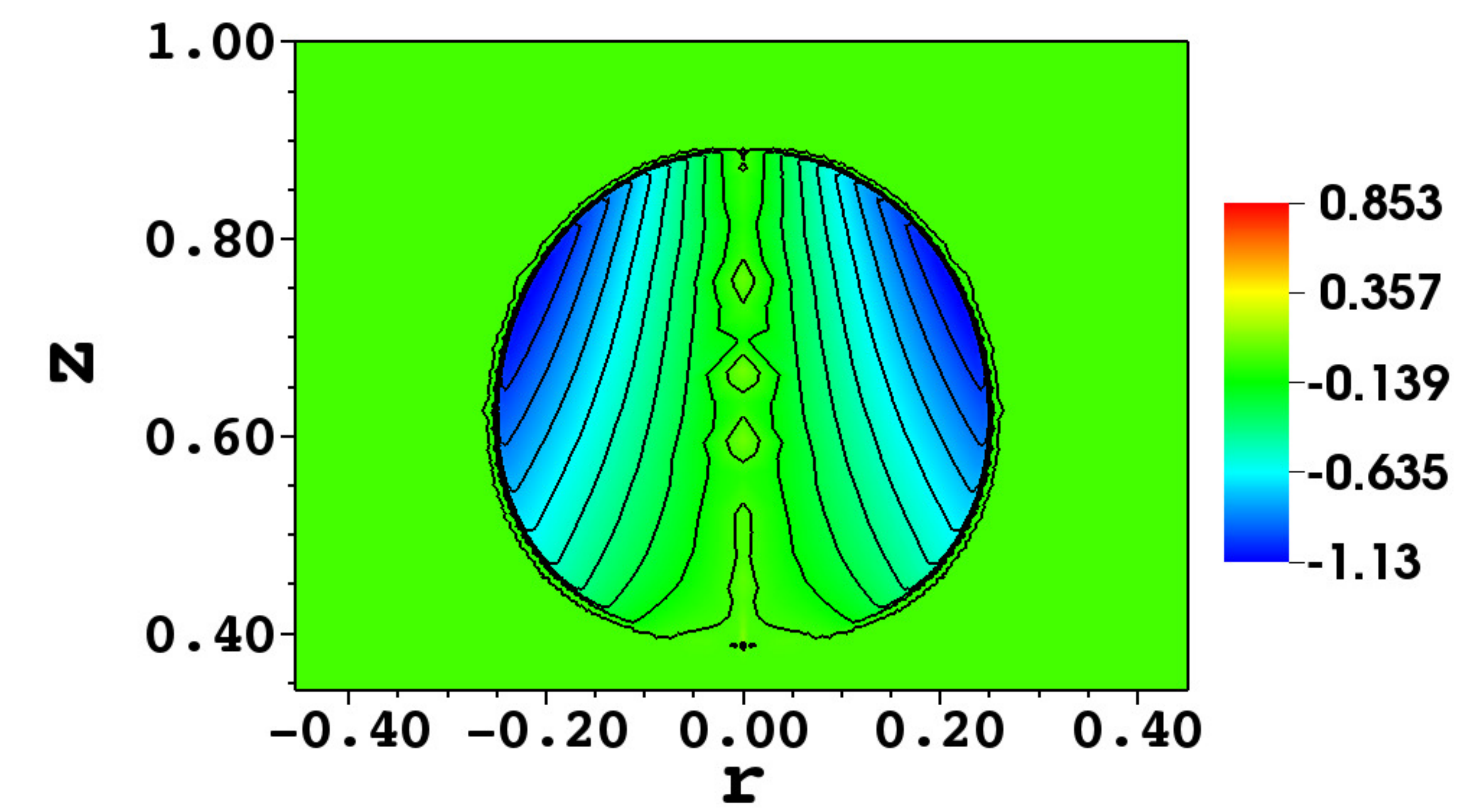}}}
\put(12.0,15.5){\makebox(3,6){\includegraphics[trim=1.0cm 0.0cm 0.0cm 0.7cm, clip=true,width=5.2cm]{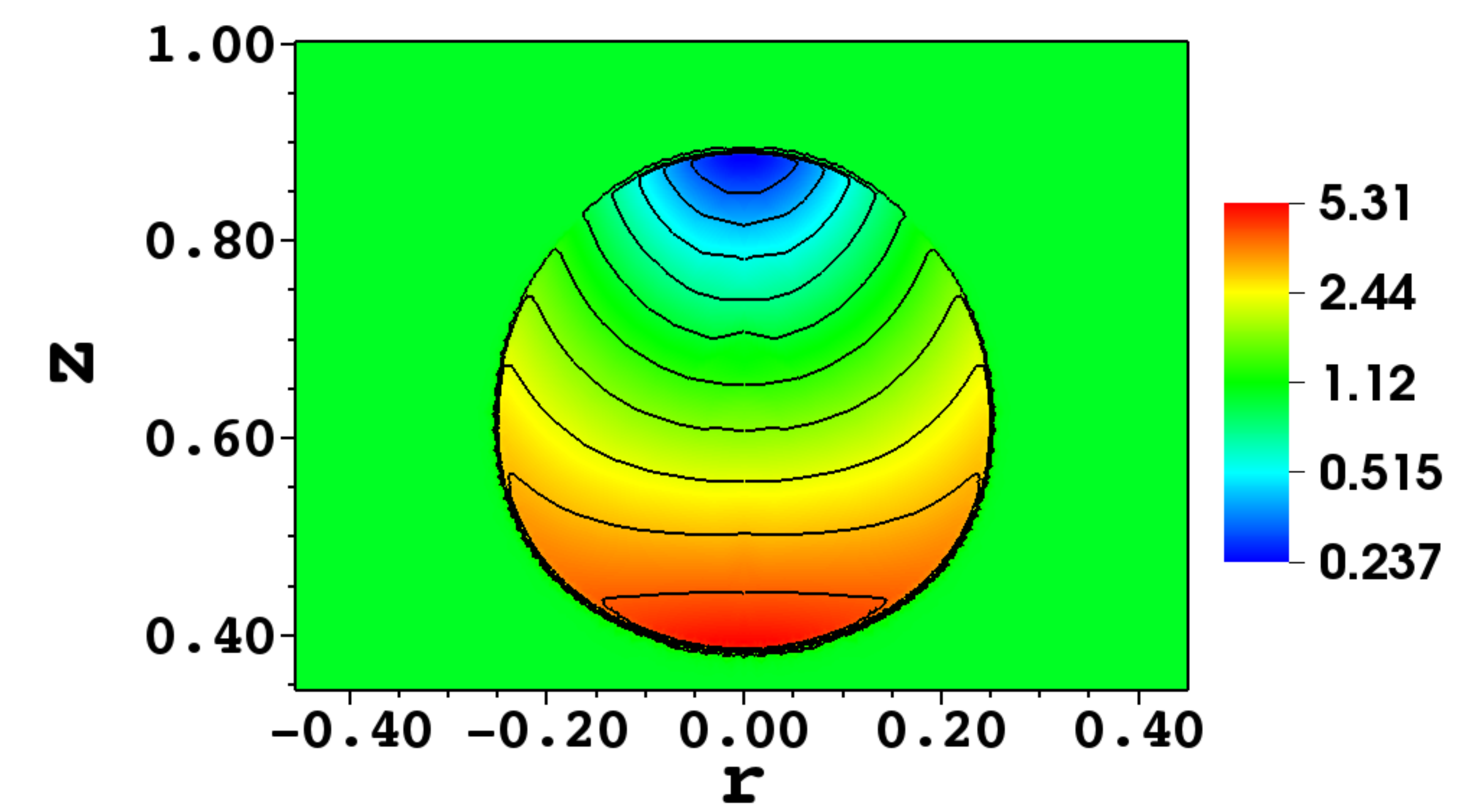}}}

\put(7.5,2.6){$t$~=~$20.0$}
\put(7.5,6.1){$t$~=~$18.0$}
\put(7.5,9.6){$t$~=~$14.0$}
\put(7.5,13.1){$t$~=~$10.0$}
\put(7.5,16.6){$t$~=~$6.0$}
\put(7.5,20.1){$t$~=~$2.0$}

\put(15.15,19.45){$\tau_{zz}$}
\put(15.15,15.95){$\tau_{zz}$}
\put(15.15,12.45){$\tau_{zz}$}
\put(15.15,8.95){$\tau_{zz}$}
\put(15.15,5.45){$\tau_{zz}$}
\put(15.15,1.95){$\tau_{zz}$}

\put(9.75,19.45){$\tau_{rz}$}
\put(9.75,15.95){$\tau_{rz}$}
\put(9.75,12.45){$\tau_{rz}$}
\put(9.75,8.95){$\tau_{rz}$}
\put(9.75,5.45){$\tau_{rz}$}
\put(9.75,1.95){$\tau_{rz}$}

\put(4.35,19.45){$\tau_{rr}$}
\put(4.35,15.95){$\tau_{rr}$}
\put(4.35,12.45){$\tau_{rr}$}
\put(4.35,8.95){$\tau_{rr}$}
\put(4.35,5.45){$\tau_{rr}$}
\put(4.35,1.95){$\tau_{rr}$}

\end{picture}
\end{center}
\caption{Viscoelastic conformation stress profiles for a viscoelastic bubble rising in a Newtonian fluid with flow parameters $\Rey_2$~=~10, Eo~=~400, $\text{Wi}_1$~=~10, $\rho_1/\rho_2$~=~0.1, $\varepsilon$~=~2, $\beta_1$~=~0.5, $\beta_2$~=~1.0, $\alpha_1$~=~0.1, $D$~=~0.5 and $h_c$~=~2.5  at dimensionless times $t$~=~2, 6, 10, 14, 18 and 20.}
\label{Tau_VTKPlots_VN}
\end{figure*}

Fig.~\ref{Tau_VTKPlots_VN} presents the viscoelastic stress profiles in the bubble for the base case flow parameters at dimensionless time instances $t$~=~2, 6, 10, 14, 18 and 20. 
Initially, the bubble is of a spherical shape with  $\bu_0$~=~0 and $\btau_{p,0}$~=~$\mathbb{I}$.
The viscoelastic bubble rises up in the bulk fluid column due to buoyancy force generated by the density difference between the two immiscible fluids.
As the bubble rises, the initial motion of the bubble is inertia dominated as viscoelastic stresses take some time to build up. 
Thus, at $t$~=~2, we can observe that the bubble shape is still more spherical.
However, at $t$~=~6, the bubble at the tail end starts to deform and it attains a cylindrical shape with a dimpled trailing end.
The viscous and viscoelastic stresses start to overcome the interfacial tension.
The maximum values of viscoelastic stress component $\tau_{rr}$ are concentrated in the top end of the bubble, while $\tau_{zz}$ is built up more near the tail end of the bubble.
The polymers inside the bubble is stretched along the flow direction.
Since the local flow direction is normal to the interface at the rear stagnation point, the polymer stress component $\tau_{zz}$ reaches its maximum value at the tail end of the bubble and pulls the interface inward. 
Since, the maximum values of $\tau_{rr}$ and minimum values of $\tau_{zz}$ occur at the top end of the bubble, the upstream axial flow experiences a strong turn tangential to the bubble surface so that the polymers are greatly extended in the radial directions.
Thus, the bubble doesn't experience noticeable deformation at its front end.
With further advancement in time, the viscoelastic stresses increases and this can be observed by looking at the maximum values of the stress components.
Hence, with time the bubble at the trailing end is more pulled up inward. 
However, beyond $t$~=~14, the magnitude of viscoelastic stresses start to decrease.
The simulations were stopped at $t$~=~20, as beyond that the bubble shall start to split and the assumption of no topological change in the computational domain shall fail when the bubble splits.

\subsubsection{Influence of viscosity ratio on the bubble dynamics}
\begin{figure*}[ht!]
\begin{center}
\unitlength1cm
\begin{picture}(14.5,8.3)
\put(1.9,6.2){\makebox(0,0){\includegraphics[width=8.5cm,height=4.2cm,keepaspectratio]{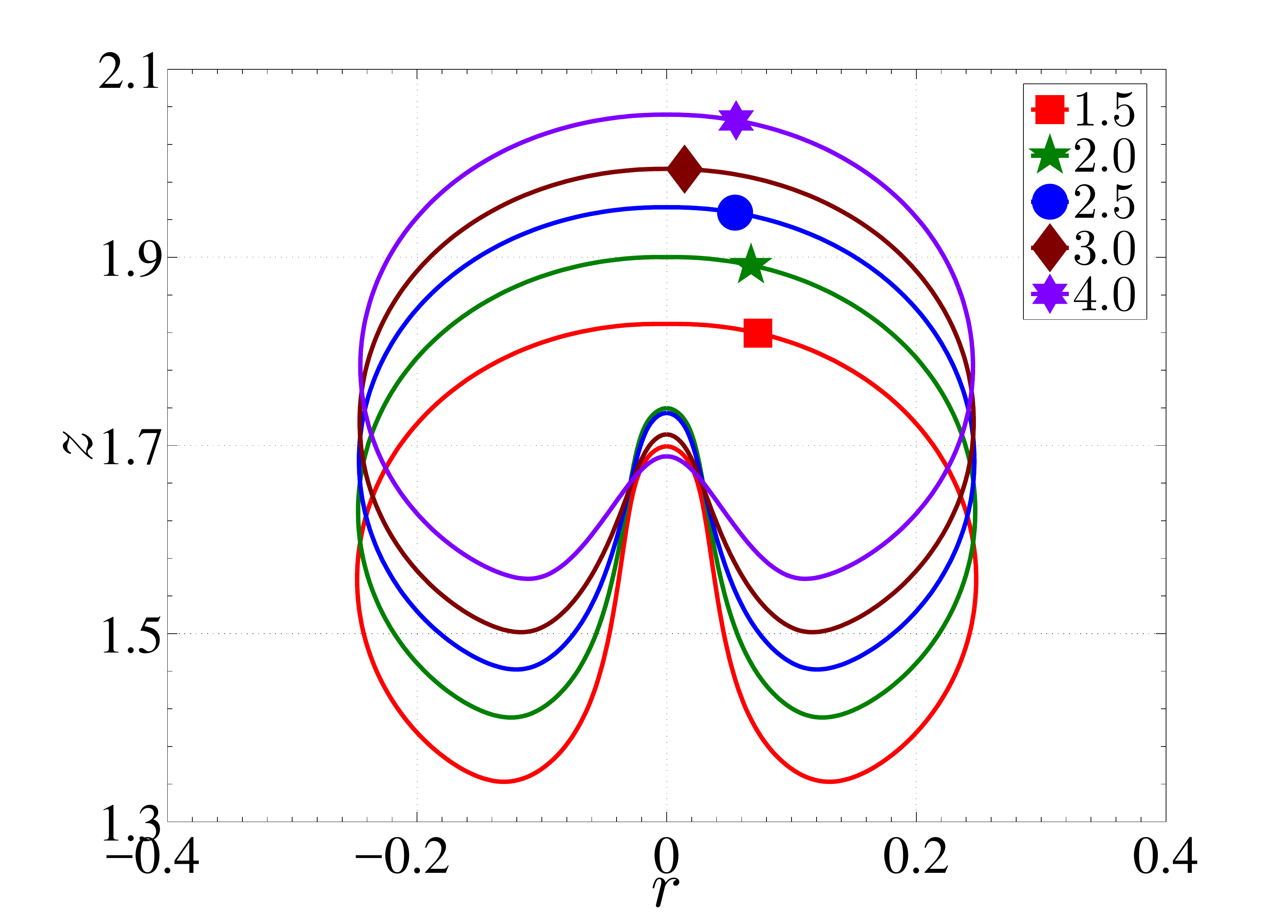}}}
\put(7.4,6.2){\makebox(0,0){\includegraphics[width=8.5cm,height=4.2cm,keepaspectratio]{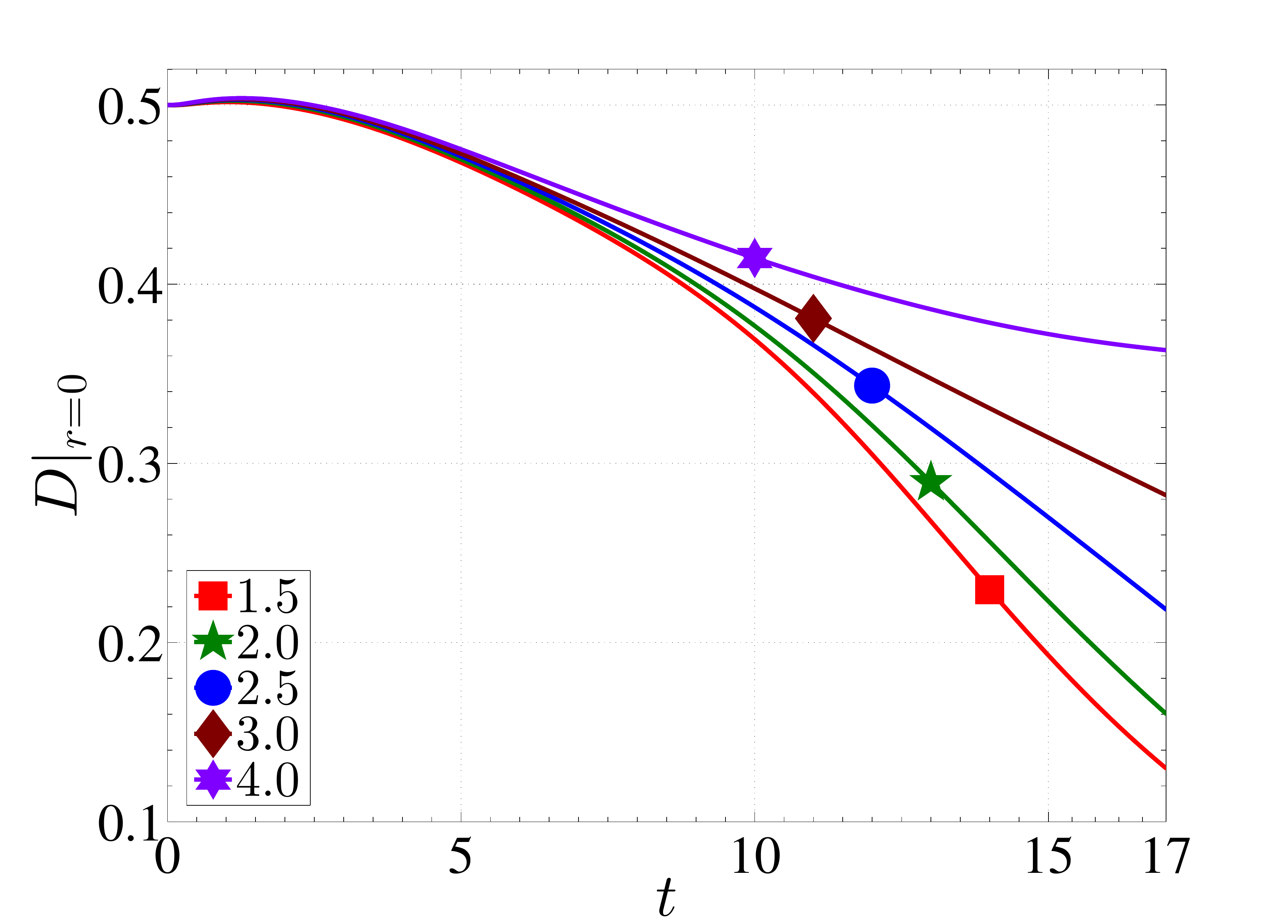}}}
\put(12.9,6.2){\makebox(0,0){\includegraphics[width=8.5cm,height=4.2cm,keepaspectratio]{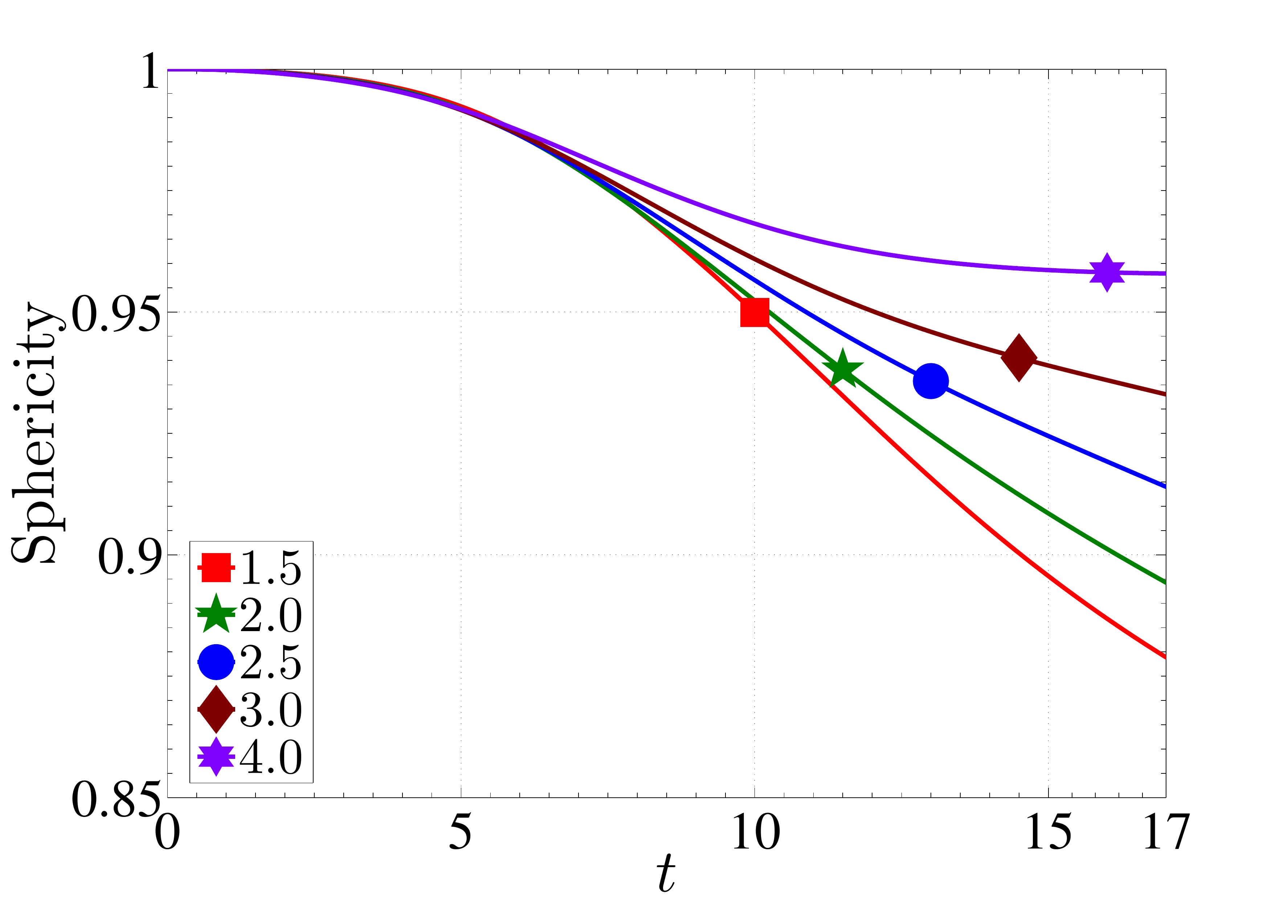}}}
\put(1.9,1.6){\makebox(0,0){\includegraphics[width=8.5cm,height=4.2cm,keepaspectratio]{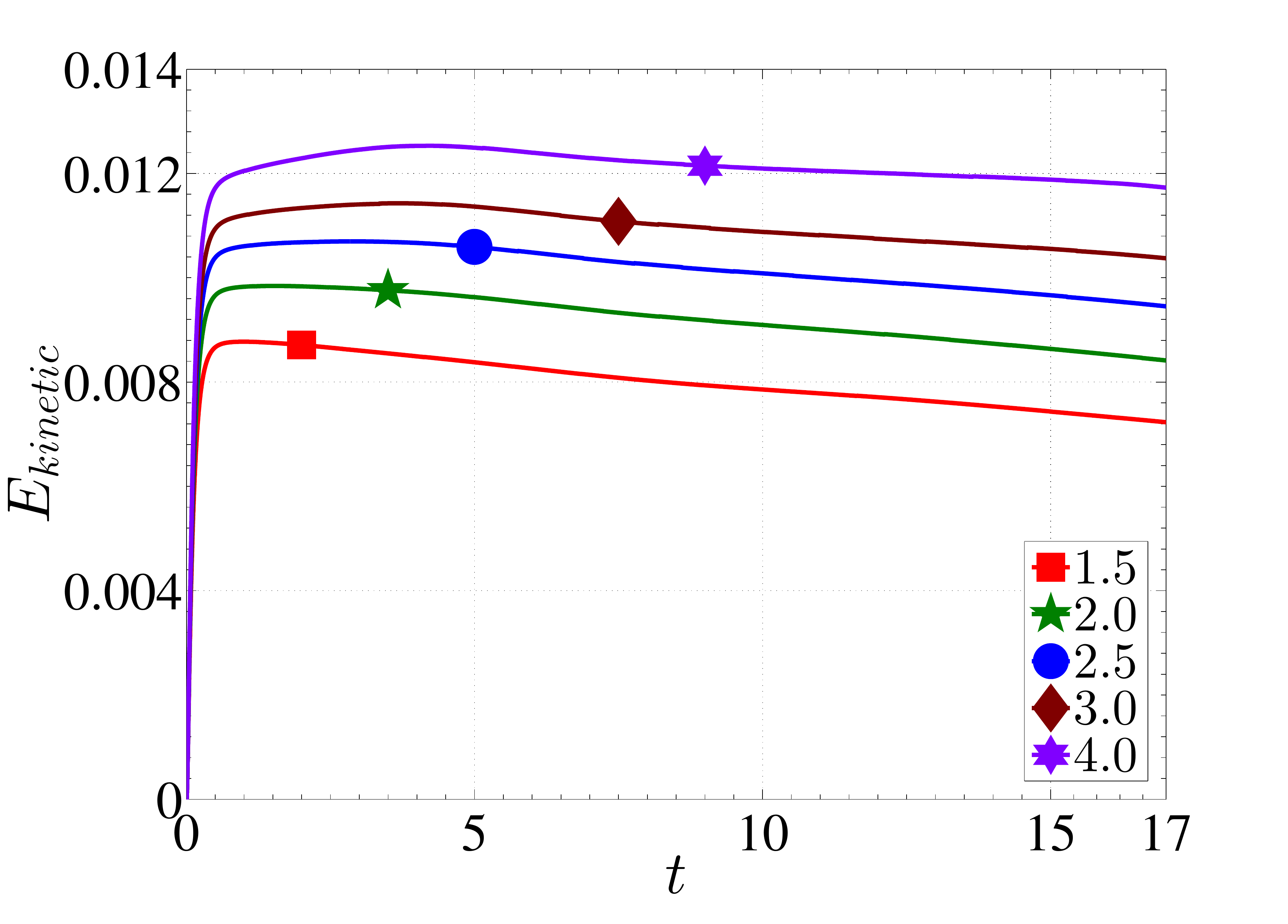}}}
\put(7.4,1.6){\makebox(0,0){\includegraphics[width=8.5cm,height=4.2cm,keepaspectratio]{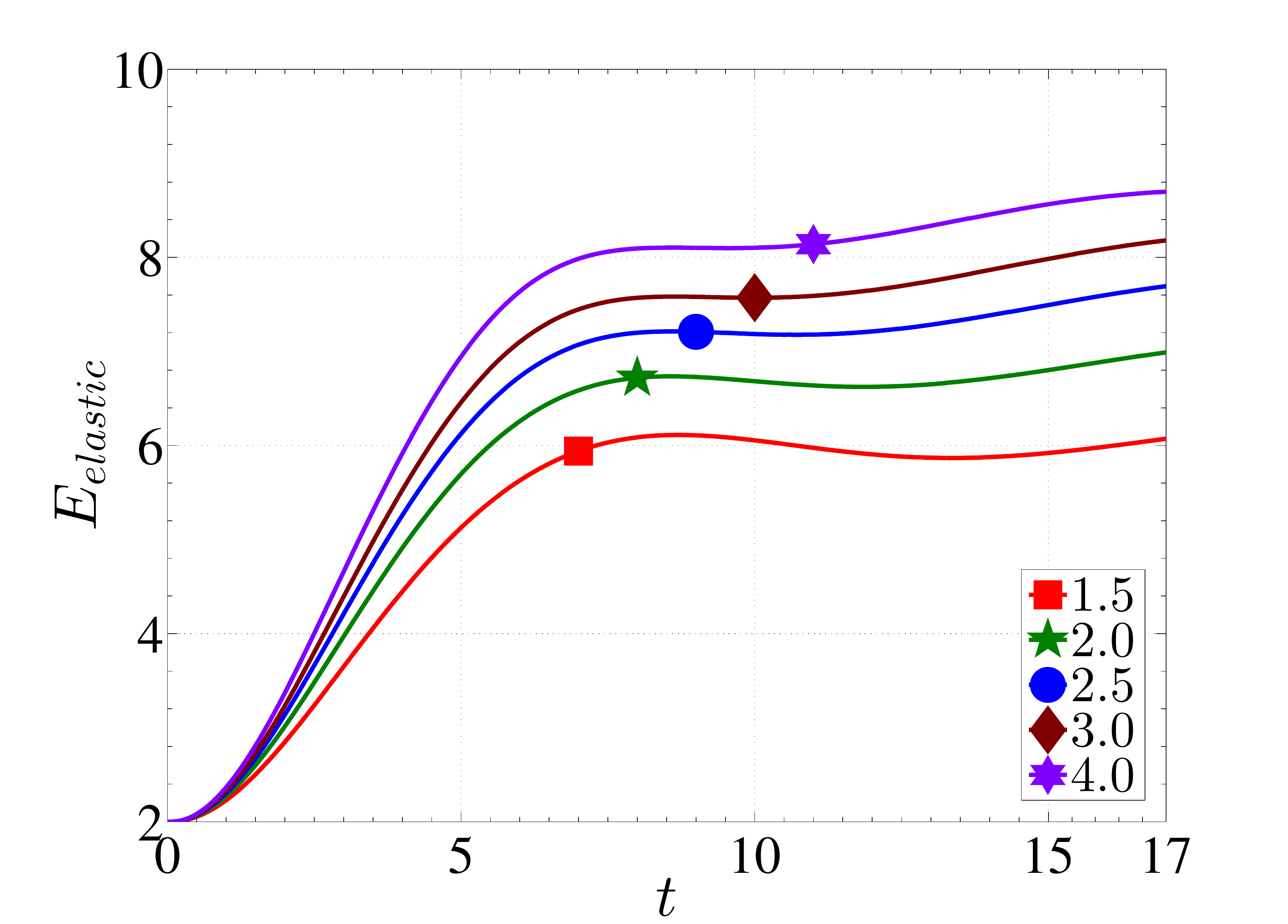}}}
\put(12.9,1.6){\makebox(0,0){\includegraphics[width=8.5cm,height=4.2cm,keepaspectratio]{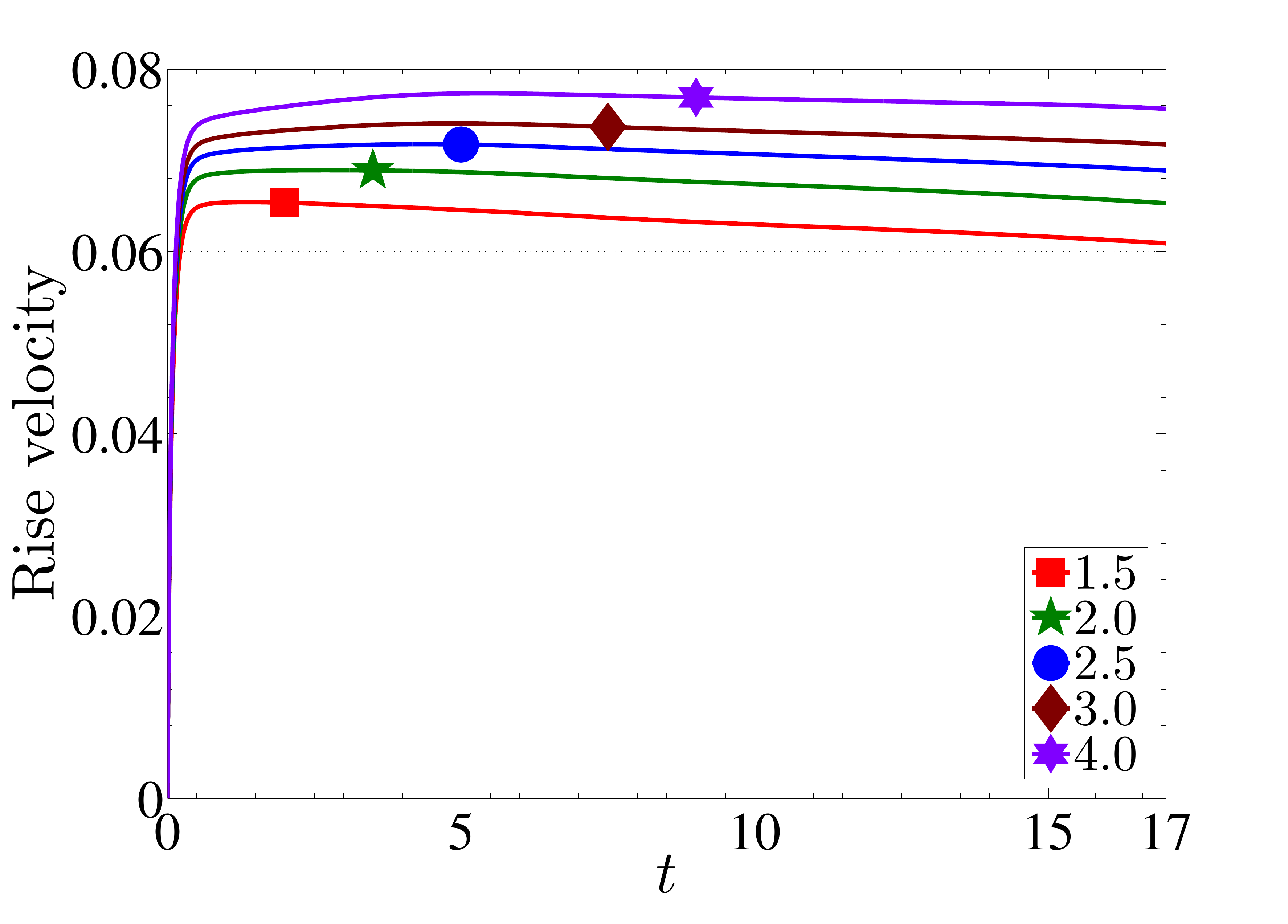}}}
\put(1.85,8.1){$(a)$}
\put(7.3, 8.1){$(b)$}
\put(12.8, 8.1){$(c)$}
\put(1.85,3.5){$(d)$}
\put(7.3, 3.5){$(e)$}
\put(12.8, 3.5){$(f)$}
\end{picture}
\end{center}
\caption{Influence of viscosity ratio for a viscoelastic bubble rising in a Newtonian fluid column~: (a)~bubble shape at $t$~=~17, (b)~diameter of the bubble at $r$~=~0, (c)~sphericity, (d)~kinetic energy, (e)~elastic energy and (f)~rise velocity of the bubble for different viscosity ratios (i)~$\varepsilon$~=~1.5, (ii)~$\varepsilon$~=~2.0, (iii)~$\varepsilon$~=~2.5, (iv)~$\varepsilon$~=~3.0 and (v)~$\varepsilon$~=~4.0 with $\Rey_2$~=~10, Eo~=~400, $\text{Wi}_1$~=~10, $\rho_1/\rho_2$~=~0.1, $\beta_1$~=~0.5, $\beta_2$~=~1.0, $\alpha_1$~=~0.1, $D$~=~0.5 and $h_c$~=~2.5.} 
\label{Plots_ViscosityEffect_VN}
\end{figure*}

In this section, we study the influence of viscosity ratio on the rising viscoelastic bubble dynamics.
We consider the base case flow parameters and vary only the viscosity ratio.
The following five different viscosity ratios are used in this study~: (i)~$\varepsilon$~=~1.5, (ii)~$\varepsilon$~=~2.0, (iii)~$\varepsilon$~=~2.5, (iv)~$\varepsilon$~=~3.0 and (v)~$\varepsilon$~=~4.0.
Fig.~\ref{Plots_ViscosityEffect_VN} presents the numerical results for different viscosity ratios.
With an increase in the viscosity ratio, the Reynolds number of the bubble increases and it forces the bubble to rise with a higher rise velocity and the same can be observed in  Fig.~\ref{Plots_ViscosityEffect_VN}(f).
Since, the bubble rises with a higher velocity, the kinetic energy will also be higher, refer Fig.~\ref{Plots_ViscosityEffect_VN}(d).
Fig.~\ref{Plots_ViscosityEffect_VN}(e) presents the temporal evolution of elastic energy in the bubble.
The elastic energy in the bubble depends on the viscoelastic stresses in the bubble. 
Since, the viscoelastic stresses are generated in regions of high gradients in the velocity field, more viscoelastic stresses would be generated for bubbles with higher Reynolds number.
Hence, with an increase in the viscosity ratio, we observe that the elastic energy in the bubble also increases.
Since, the bubble rises with a higher rise velocity, the position of the bubble shall also be higher and we observe the same in Fig.~\ref{Plots_ViscosityEffect_VN}(a).
Fig.~\ref{Plots_ViscosityEffect_VN}(b) presents the temporal evolution of the diameter of the bubble at the axis of symmetry.
We can observe that the effects of viscosity ratio is negligible till around $t$~=~5. 
After that, the diameter of the bubble decreases more at lower viscosity ratios and the same phenomenon is observed in the sphericity of the bubble in Fig.~\ref{Plots_ViscosityEffect_VN}(c).

\subsubsection{Influence of Newtonian solvent ratio on the bubble dynamics}
\begin{figure*}[ht!]
\begin{center}
\unitlength1cm
\begin{picture}(14.5,8.3)
\put(1.9,6.2){\makebox(0,0){\includegraphics[width=8.5cm,height=4.2cm,keepaspectratio]{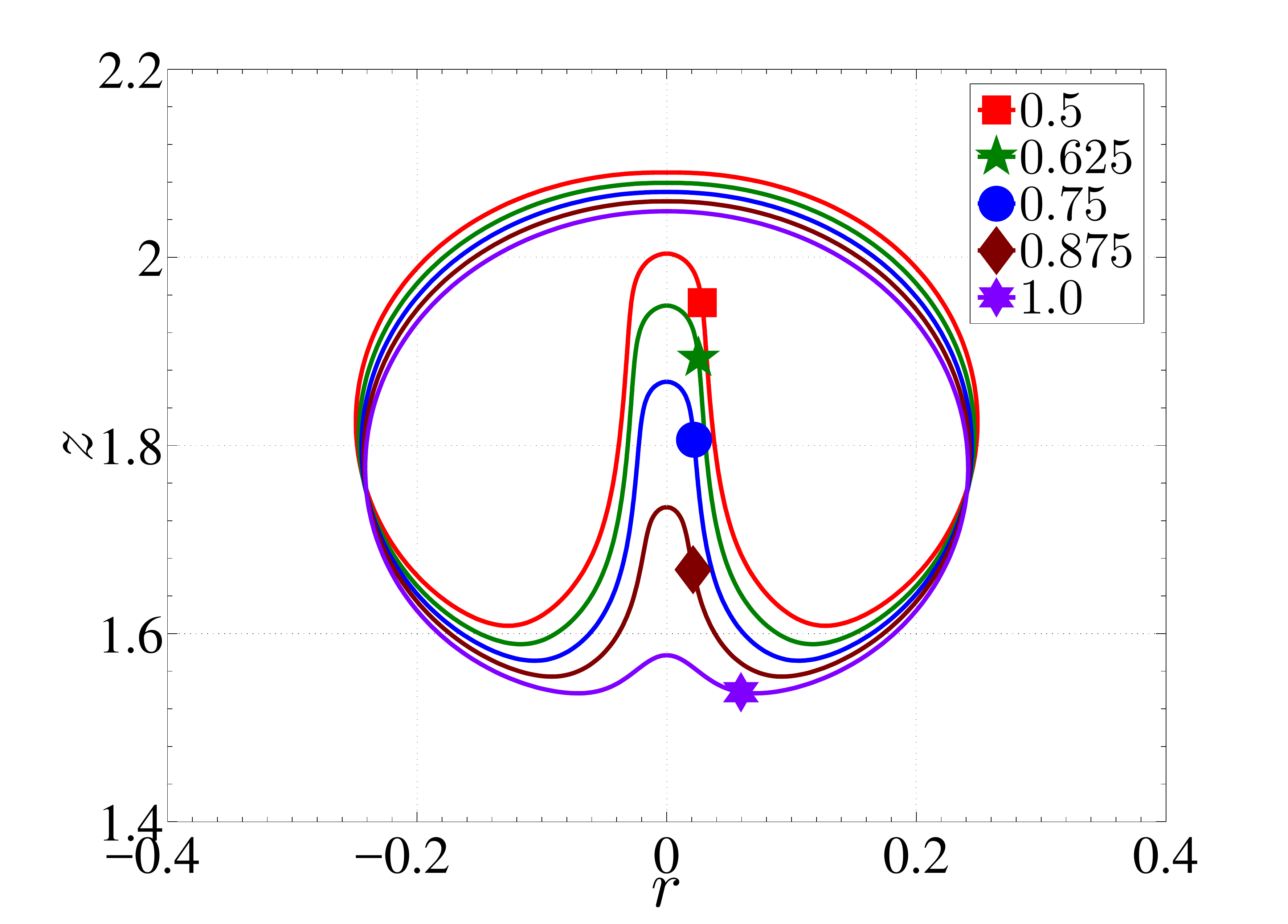}}}
\put(7.4,6.2){\makebox(0,0){\includegraphics[width=8.5cm,height=4.2cm,keepaspectratio]{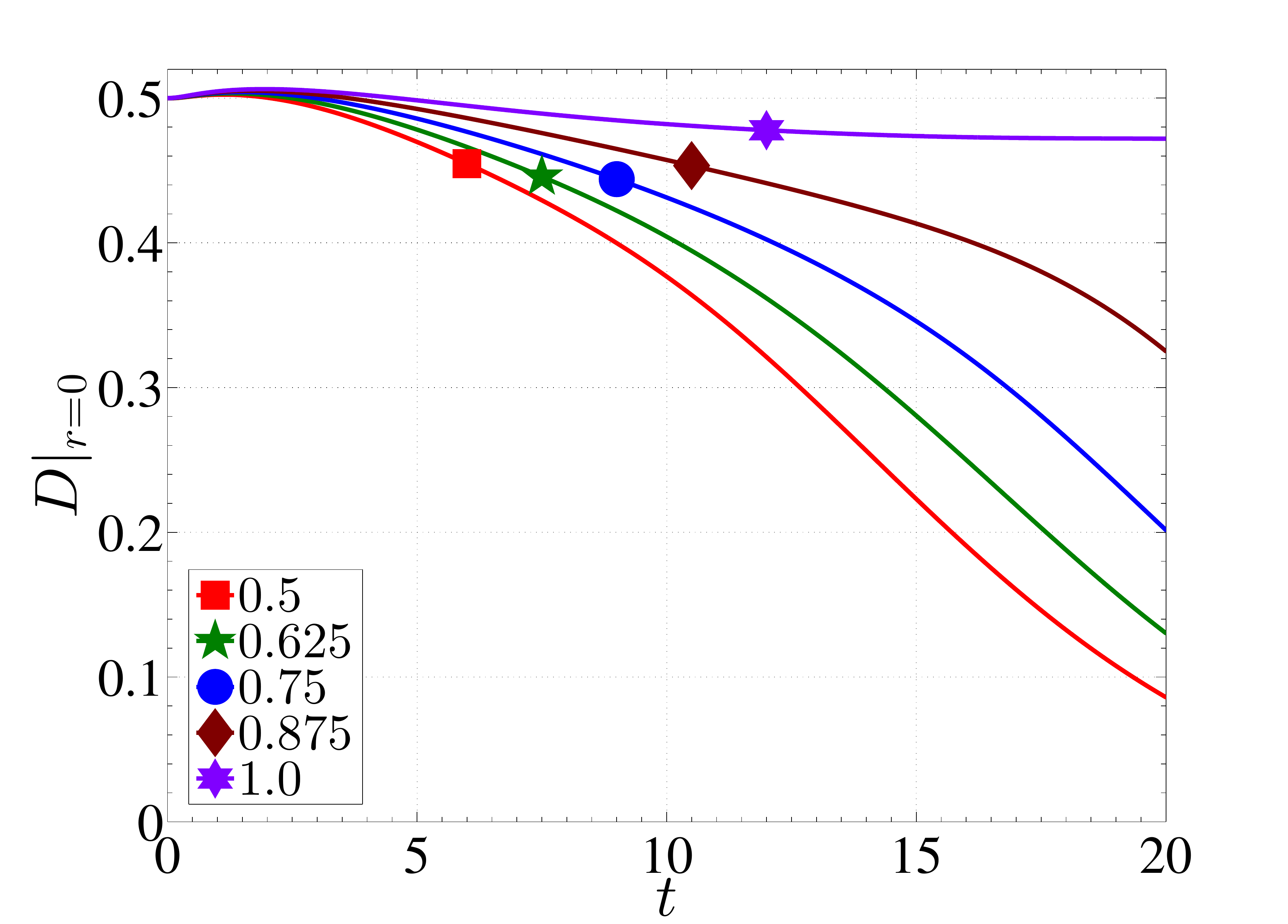}}}
\put(12.9,6.2){\makebox(0,0){\includegraphics[width=8.5cm,height=4.2cm,keepaspectratio]{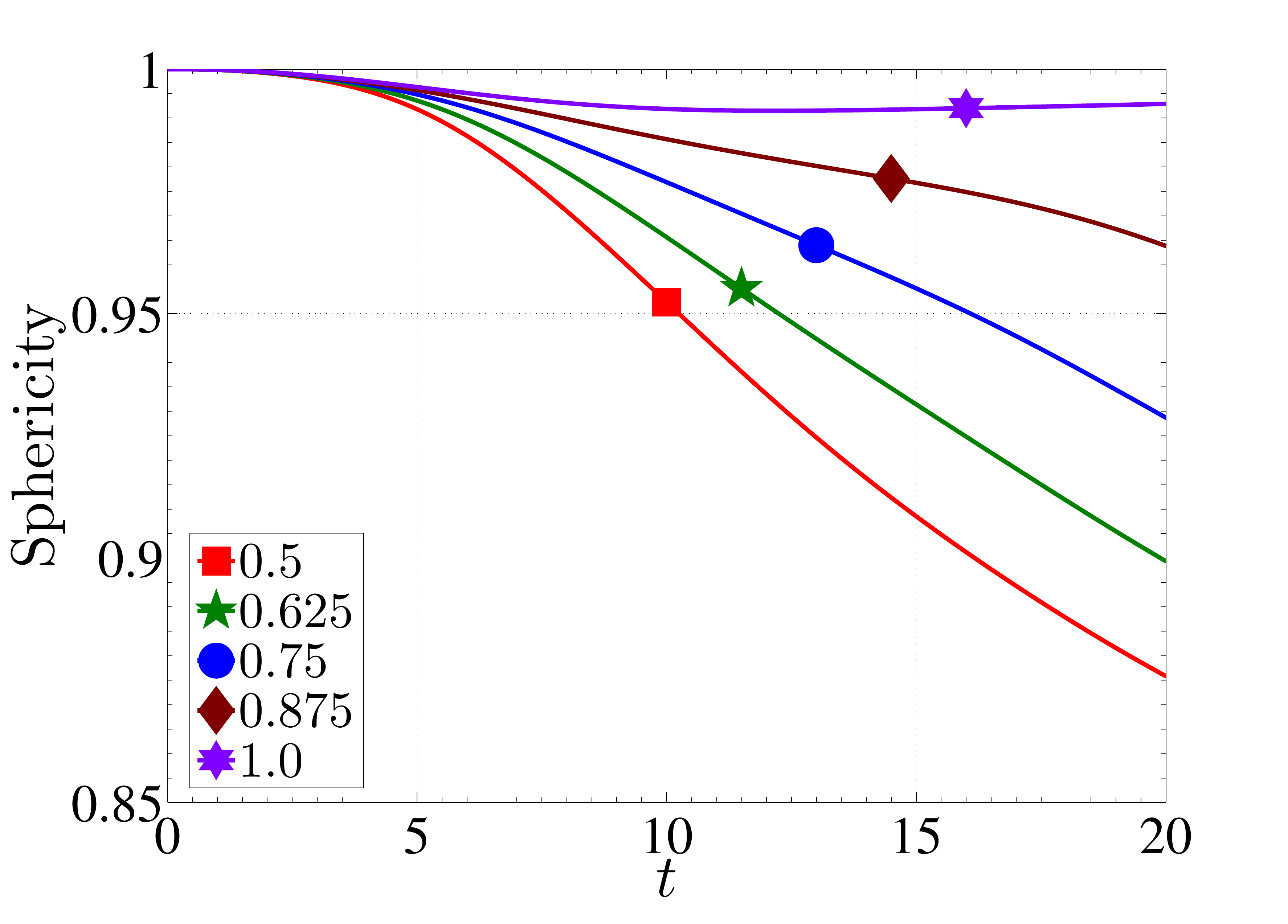}}}
\put(1.9,1.6){\makebox(0,0){\includegraphics[width=8.5cm,height=4.2cm,keepaspectratio]{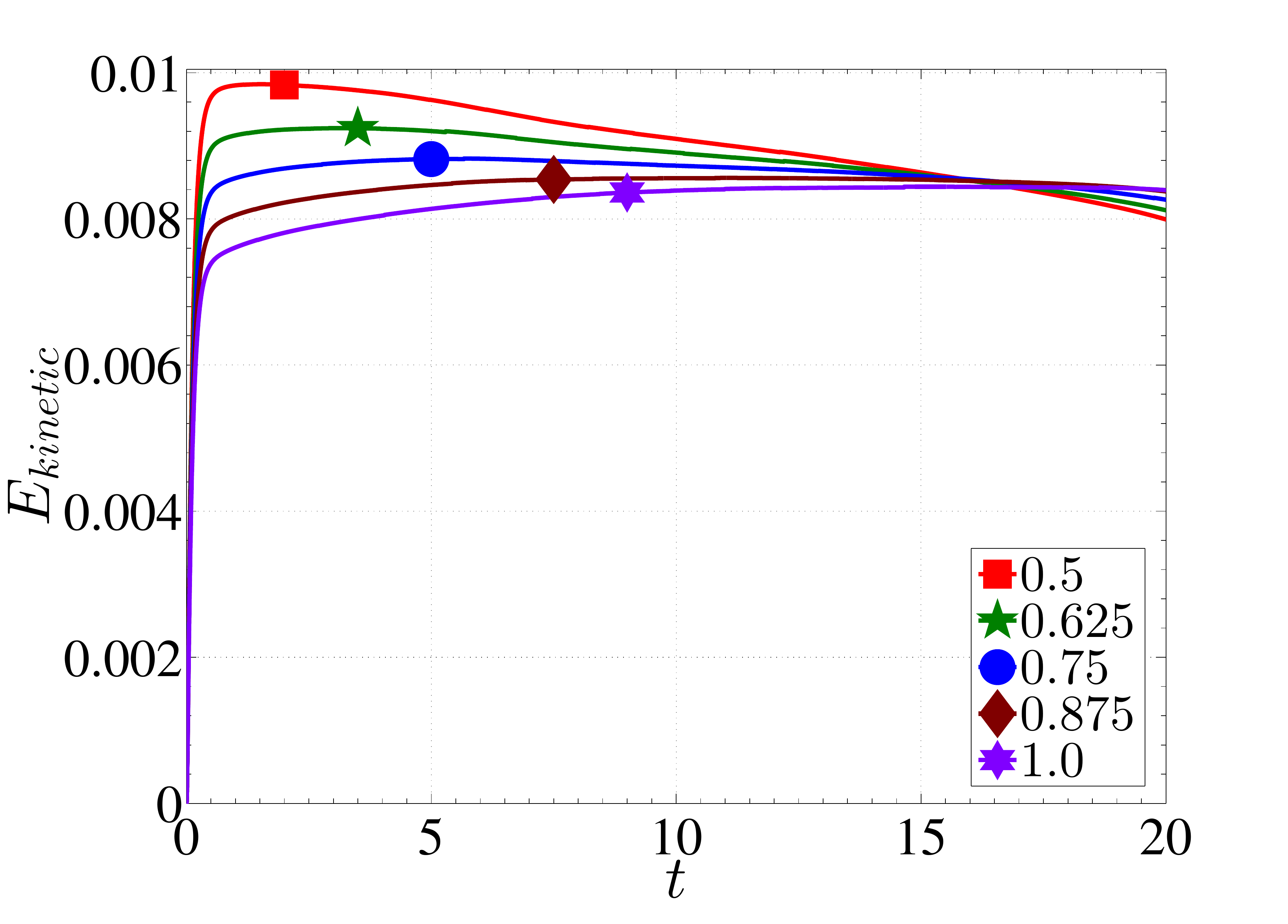}}}
\put(7.4,1.6){\makebox(0,0){\includegraphics[width=8.5cm,height=4.2cm,keepaspectratio]{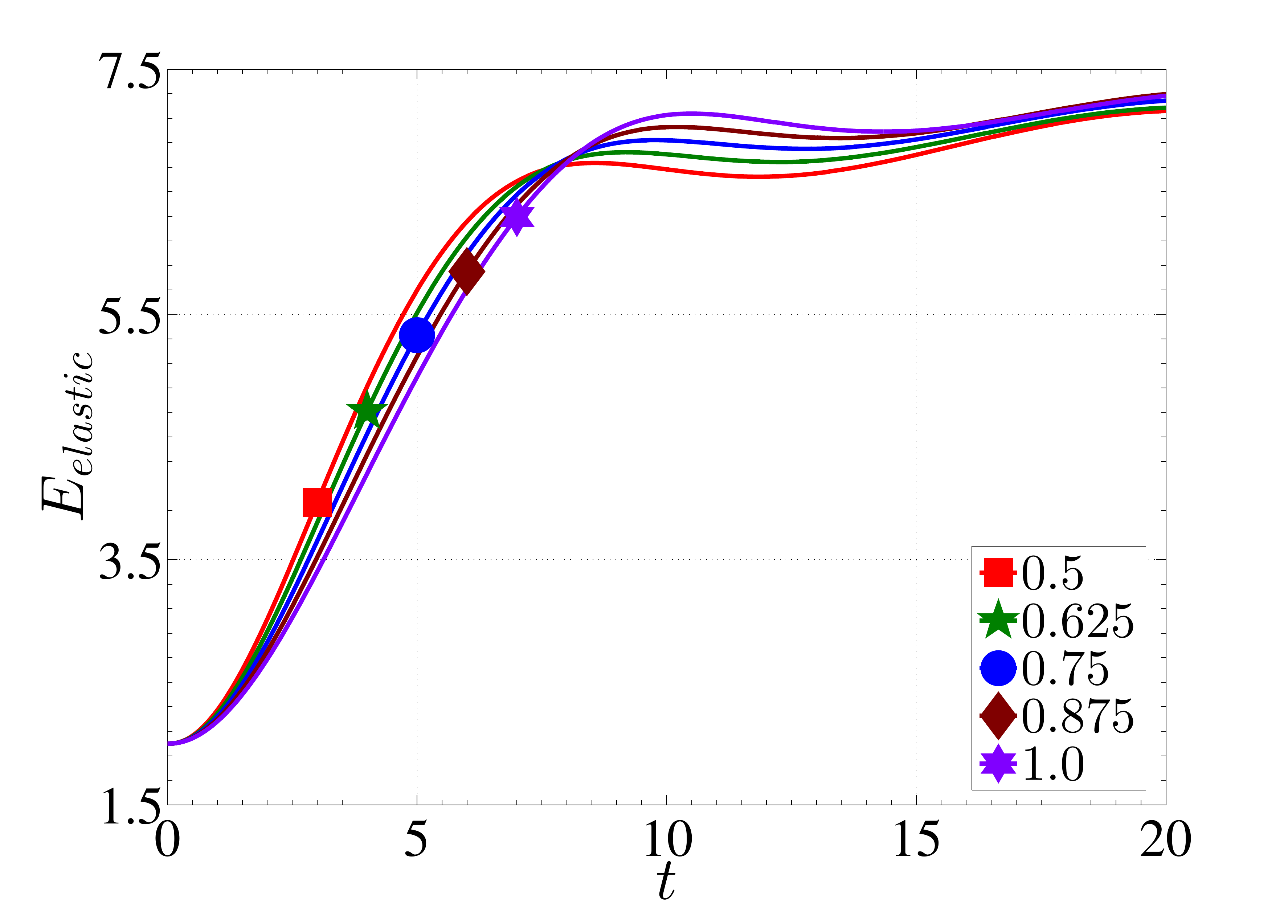}}}
\put(12.9,1.6){\makebox(0,0){\includegraphics[width=8.5cm,height=4.2cm,keepaspectratio]{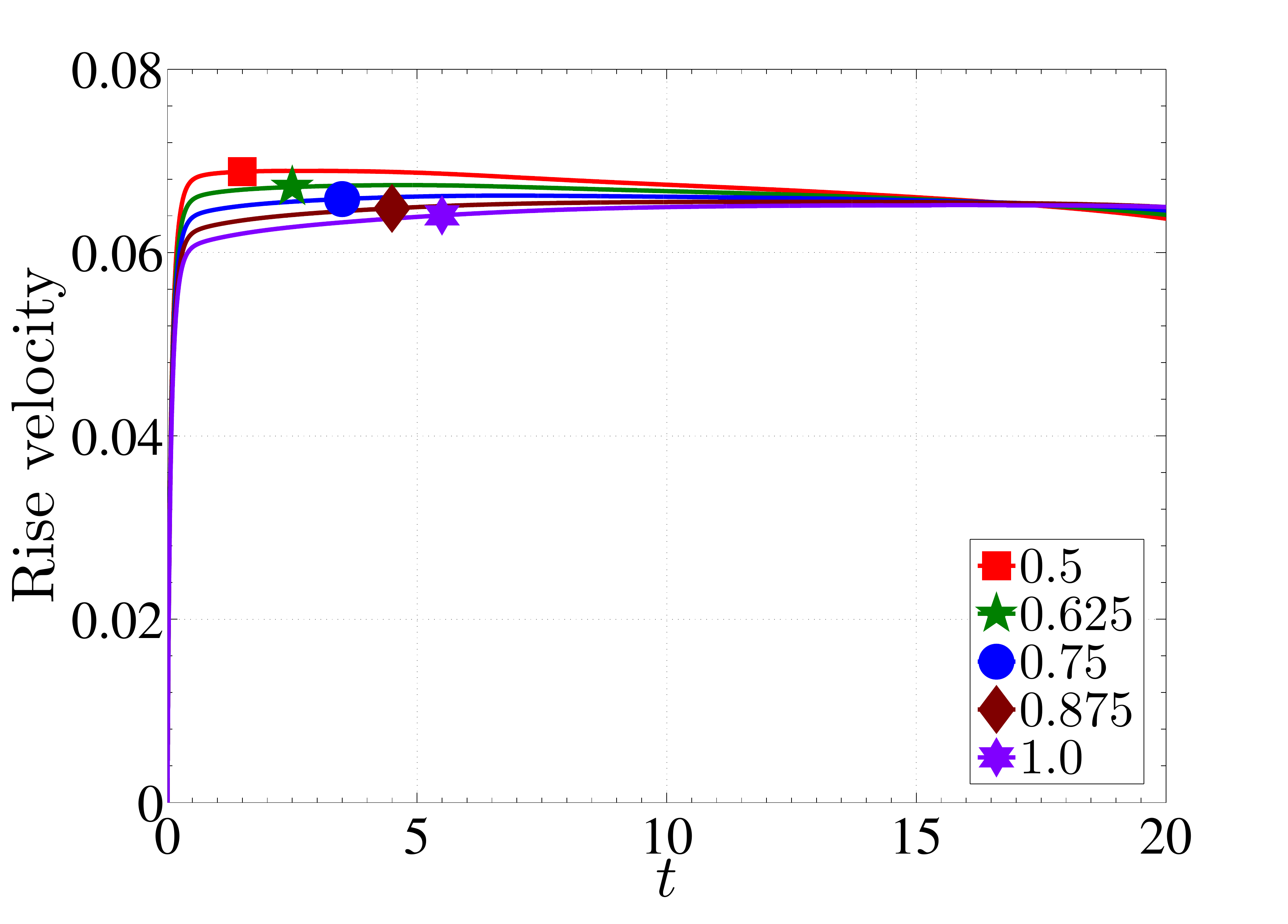}}}
\put(1.85,8.1){$(a)$}
\put(7.3, 8.1){$(b)$}
\put(12.8, 8.1){$(c)$}
\put(1.85,3.5){$(d)$}
\put(7.3, 3.5){$(e)$}
\put(12.8, 3.5){$(f)$}
\end{picture}
\end{center}
\caption{Influence of Newtonian solvent ratio for a viscoelastic bubble rising in a Newtonian fluid column~: (a)~bubble shape at $t$~=~20, (b)~diameter of the bubble at $r$~=~0, (c)~sphericity, (d)~kinetic energy, (e)~elastic energy and (f)~rise velocity of the bubble for different Newtonian solvent ratios (i)~$\beta_1$~=~0.5, (ii)~$\beta_1$~=~0.625, (iii)~$\beta_1$~=~0.75, (iv)~$\beta_1$~=~0.875 and (v)~$\beta_1$~=~1.0 with $\Rey_2$~=~10, Eo~=~400, $\text{Wi}_1$~=~10, $\rho_1/\rho_2$~=~0.1, $\varepsilon$~=~2, $\beta_2$~=~1.0, $\alpha_1$~=~0.1, $D$~=~0.5 and $h_c$~=~2.5.} 
\label{Plots_BetaEffect_VN}
\end{figure*}
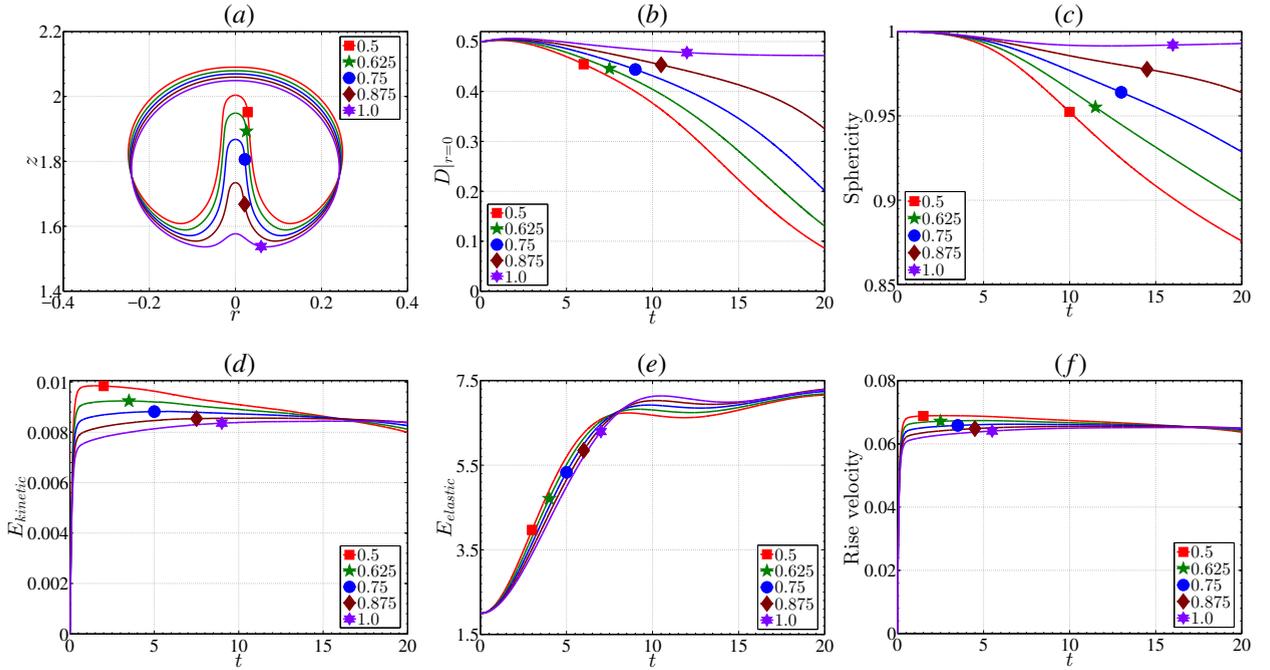

To study the influence of Newtonian solvent ratio on the rising bubble dynamics, we consider the base case flow parameters and vary only the Newtonian solvent ratio of the bubble.
We consider the following five different Newtonian solvent ratios in this study~: (i)~$\beta_1$~=~0.5, (ii)~$\beta_1$~=~0.625, (iii)~$\beta_1$~=~0.75, (iv)~$\beta_1$~=~0.875 and (v)~$\beta_1$~=~1.0.
The case $\beta_1$~=~1.0, represents a Newtonian bubble rising in a Newtonian fluid column. 
Fig.~\ref{Plots_BetaEffect_VN} presents the computational results for different Newtonian solvent ratios.
Lower the Newtonian solvent ratio, greater is the polymeric viscosity and lesser is the Newtonian viscosity, thereby increasing the viscoelastic character of the fluid column.
Hence with an increase in the viscoelastic character of the bubble, it deforms more at the trailing end.
In Fig.~\ref{Plots_BetaEffect_VN}(a), we can observe that the degree of dimpleness increases with decreasing Newtonian solvent ratio.
Thus, the diameter of the bubble at the axis of symmetry as well decreases with a decrease in the Newtonian solvent ratio, see Fig.~\ref{Plots_BetaEffect_VN}(b).
Similar behavior is also observed in the sphericity of the bubble.
Further, initially the kinetic energy and rise velocity of the bubble increases with a decrease in the Newtonian solvent ratio.
However, after around $t$~=~17, the trend reverses.
Fig.~\ref{Plots_BetaEffect_VN}(e) presents the temporal evolution of the elastic energy in the bubble.
Till $t$~=~8.0, the magnitude of increase in the elastic energy in the bubble increases with a decrease in the Newtonian solvent ratio. 
However, after $t$~=~8.0 the trend reverses.

\subsubsection{Influence of Giesekus mobility factor on the bubble dynamics}

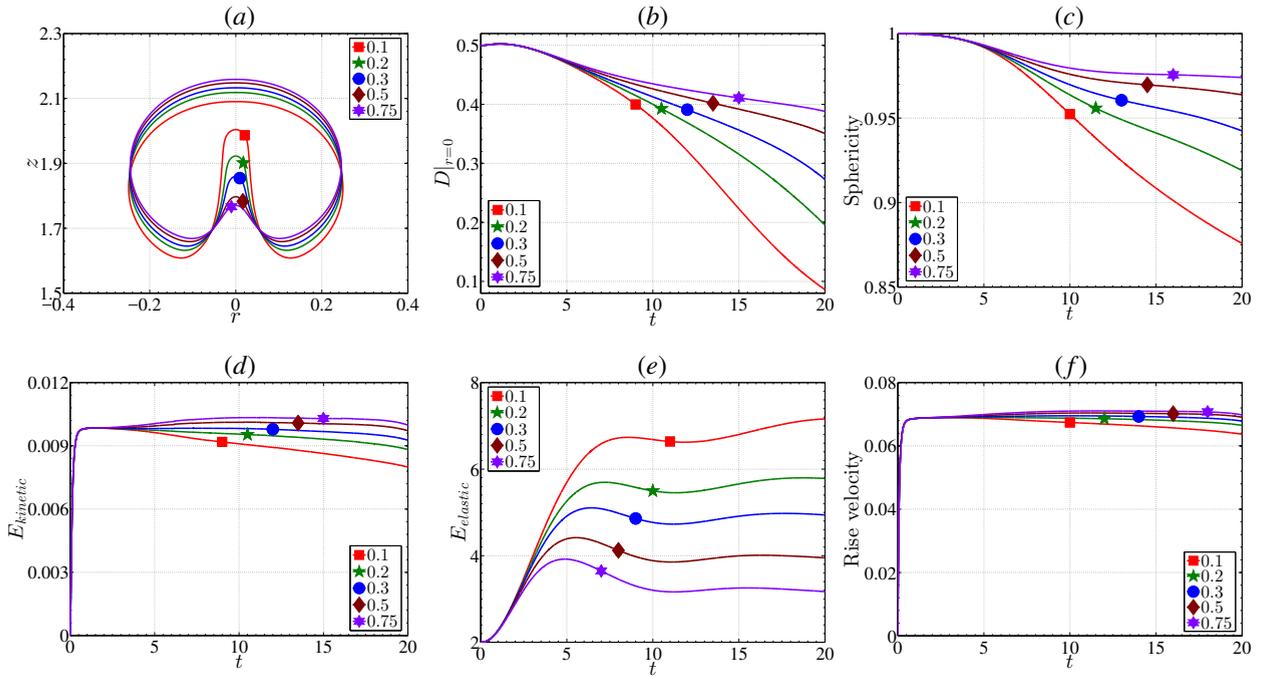
\begin{figure*}[ht!]
\begin{center}
\unitlength1cm
\begin{picture}(14.5,8.3)
\put(1.9,6.2){\makebox(0,0){\includegraphics[width=8.5cm,height=4.2cm,keepaspectratio]{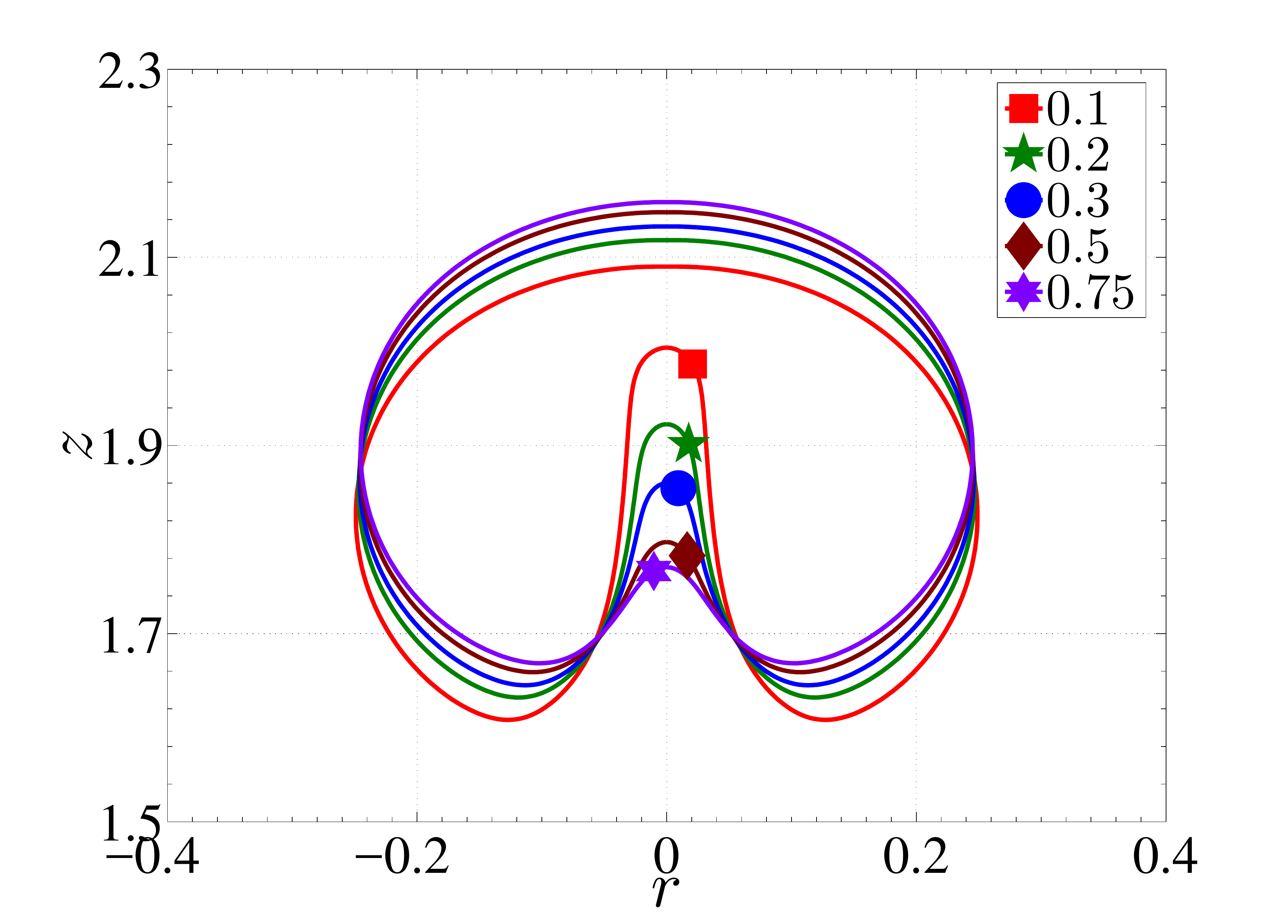}}}
\put(7.4,6.2){\makebox(0,0){\includegraphics[width=8.5cm,height=4.2cm,keepaspectratio]{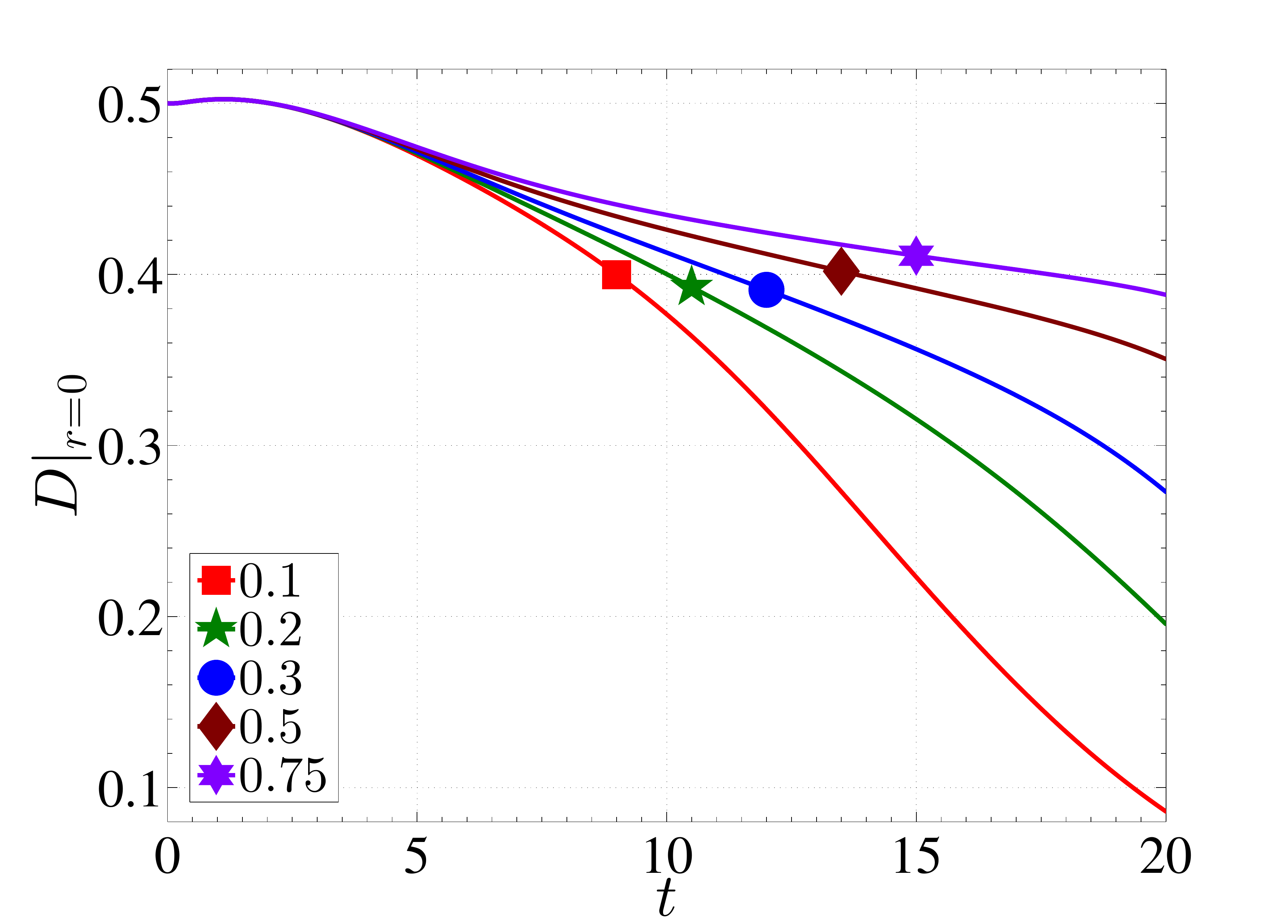}}}
\put(12.9,6.2){\makebox(0,0){\includegraphics[width=8.5cm,height=4.2cm,keepaspectratio]{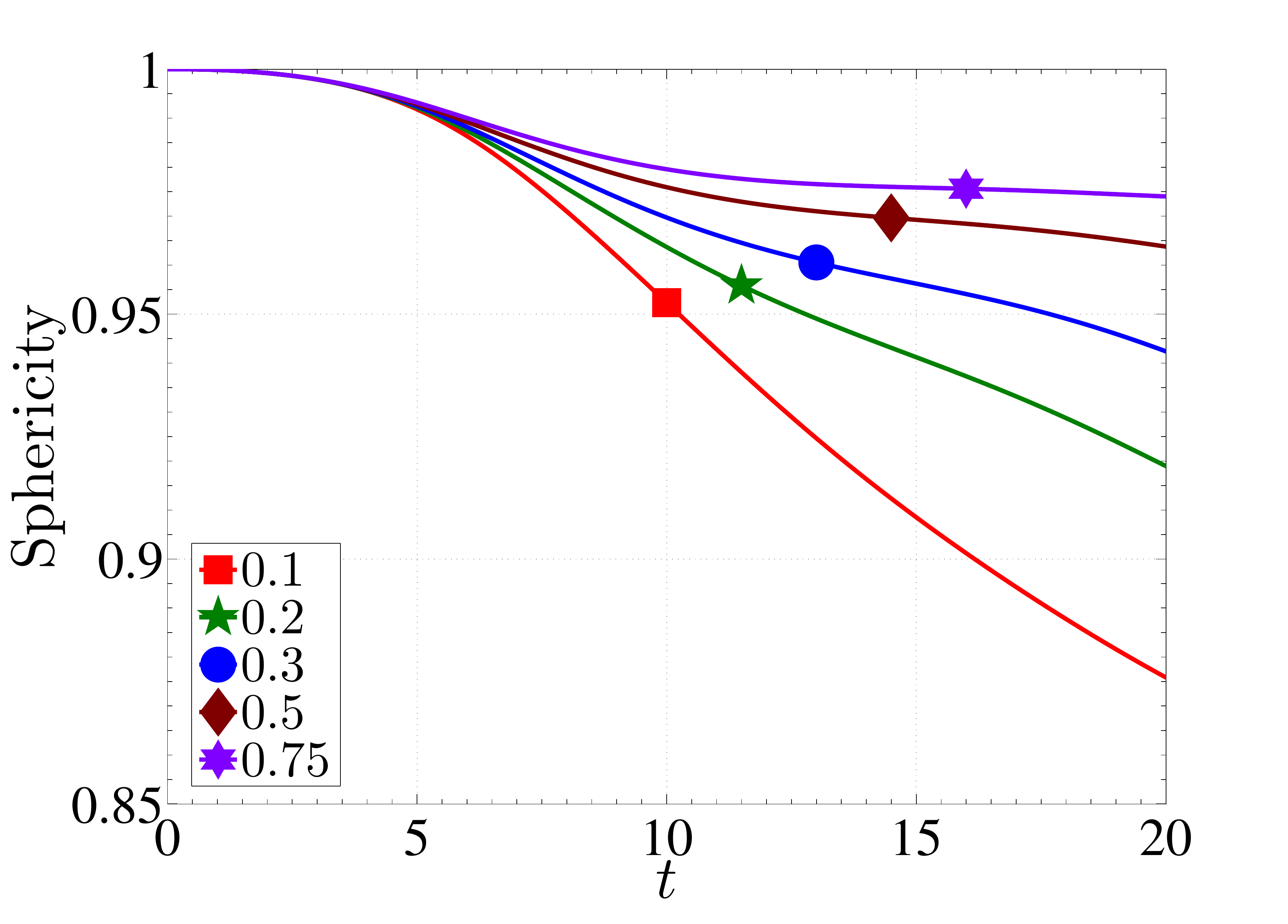}}}
\put(1.9,1.6){\makebox(0,0){\includegraphics[width=8.5cm,height=4.2cm,keepaspectratio]{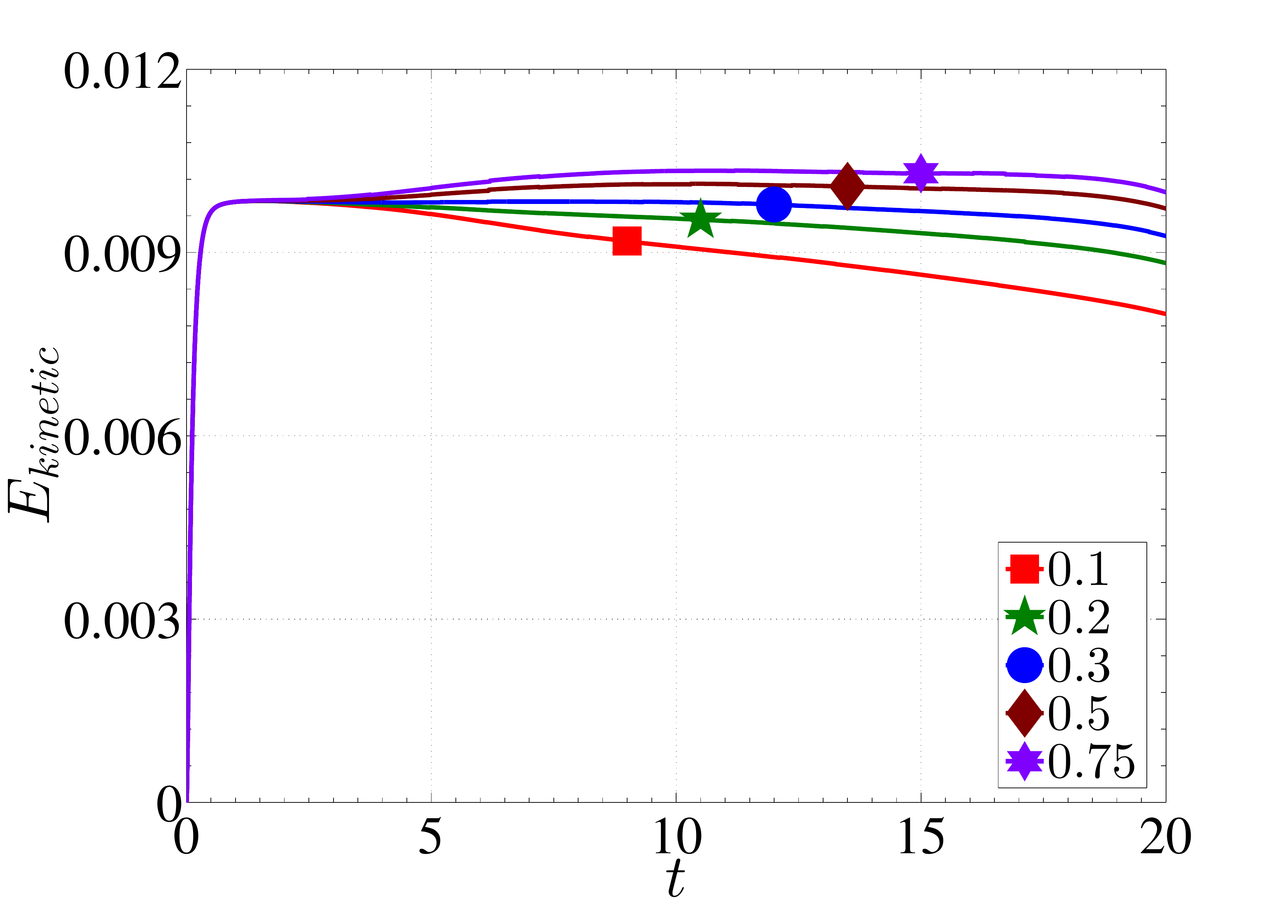}}}
\put(7.4,1.6){\makebox(0,0){\includegraphics[width=8.5cm,height=4.2cm,keepaspectratio]{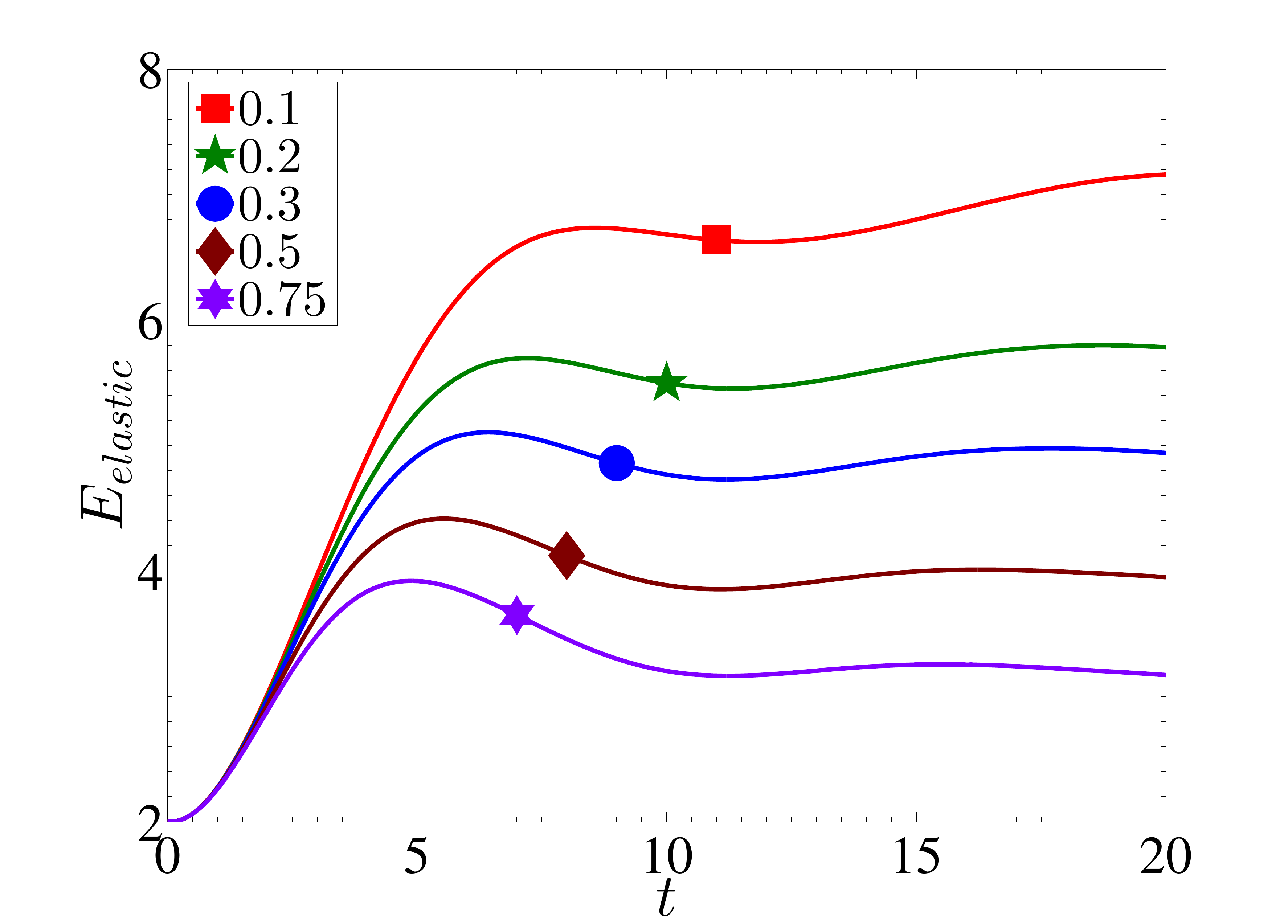}}}
\put(12.9,1.6){\makebox(0,0){\includegraphics[width=8.5cm,height=4.2cm,keepaspectratio]{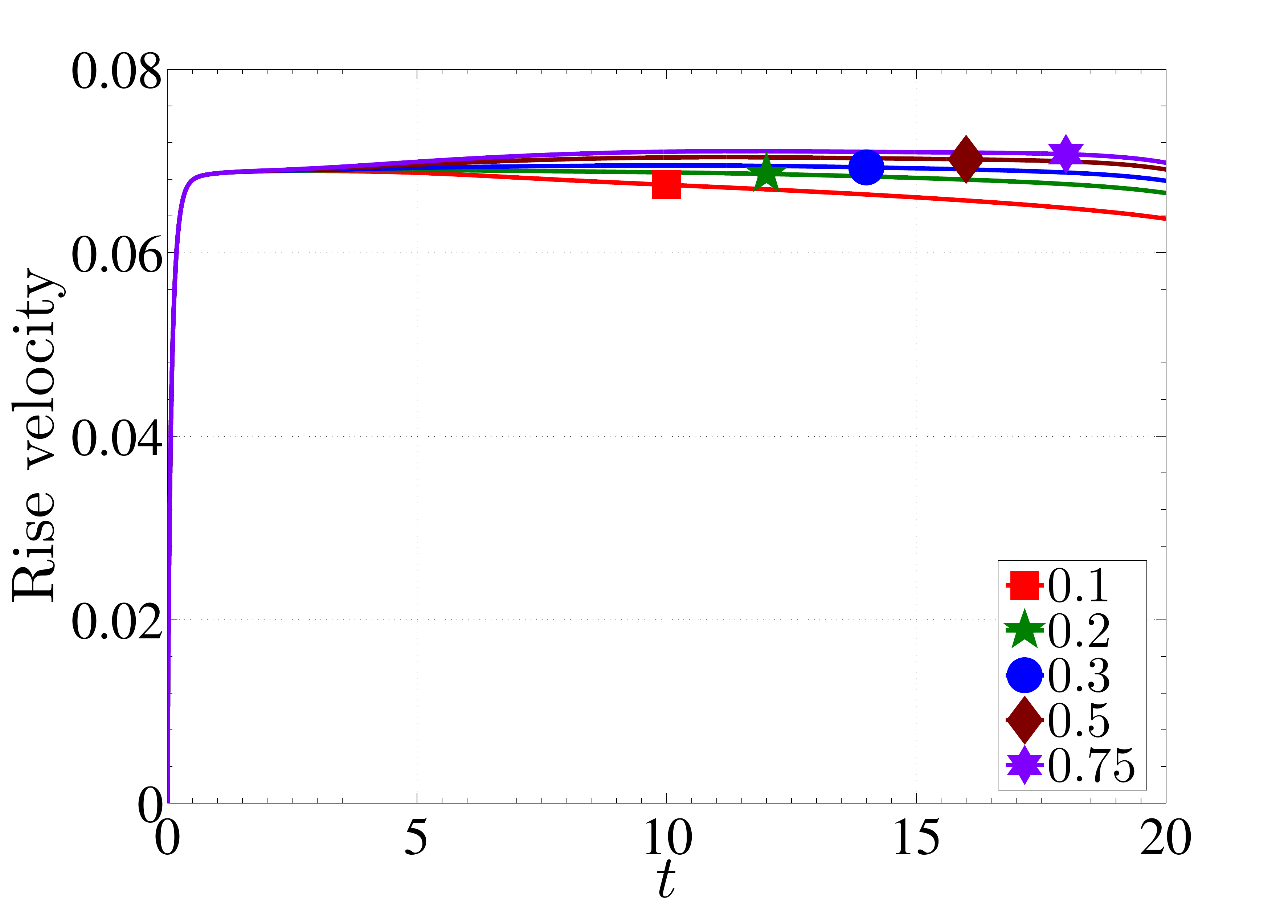}}}
\put(1.85,8.1){$(a)$}
\put(7.3, 8.1){$(b)$}
\put(12.8, 8.1){$(c)$}
\put(1.85,3.5){$(d)$}
\put(7.3, 3.5){$(e)$}
\put(12.8, 3.5){$(f)$}
\end{picture}
\end{center}
\caption{Influence of Giesekus mobility factor for a viscoelastic bubble rising in a Newtonian fluid column~: (a)~bubble shape at $t$~=~20, (b)~diameter of the bubble at $r$~=~0, (c)~sphericity, (d)~kinetic energy, (e)~elastic energy and (f)~rise velocity of the bubble for different Giesekus factors (i)~$\alpha_1$~=~0.1, (ii)~$\alpha_1$~=~0.2, (iii)~$\alpha_1$~=~0.3, (iv)~$\alpha_1$~=~0.5 and (v)~$\alpha_1$~=~0.75 with $\Rey_2$~=~10, Eo~=~400, $\text{Wi}_1$~=~10, $\rho_1/\rho_2$~=~0.1, $\varepsilon$~=~2, $\beta_1$~=~0.5, $\beta_2$~=~1.0, $D$~=~0.5 and $h_c$~=~2.5.} 
\label{Plots_AlphaEffect_VN}
\end{figure*}

In this section, we study the influence of Giesekus mobility factor on the viscoelastic bubble rising in a Newtonian fluid column.
We consider the base case flow parameters and use the following five different Giesekus factors~: (i)~$\alpha_1$~=~0.1, (ii)~$\alpha_1$~=~0.2, (iii)~$\alpha_1$~=~0.3, (iv)~$\alpha_1$~=~0.5 and (v)~$\alpha_1$~=~0.75.
Fig.~\ref{Plots_AlphaEffect_VN} presents the numerical results for different Giesekus factors.
Initially, the motion of the bubble is inertia dominated and the Giesekus factor comes into play only when the viscoelastic stresses dominate the flow.
Hence, there is no effect of Giesekus factor on the bubble dynamics till about $t$~=~3.0.
However, after that the rise velocity and kinetic energy in the bubble increases with an increase in the Giesekus factor, as shear thinning effects increases.
Increasing the Giesekus factor leads to a decrease in the magnitude of the viscoelastic stresses generated in the bubble.
Hence, from Fig.~\ref{Plots_AlphaEffect_VN}(e) we can observe that the magnitude of increase in the elastic energy decreases with an increase in the Giesekus factor.
Further, from Fig.~\ref{Plots_AlphaEffect_VN}(a) we can observe that at low Giesekus factor, the effect of dimpleness is higher.
Thus, the diameter and sphericity of the bubble decreases more with a decrease in the Giesekus factor, refer Fig.~\ref{Plots_AlphaEffect_VN}(b) and (c) respectively.

\subsubsection{Influence of E\"{o}tv\"{o}s number on the bubble dynamics}
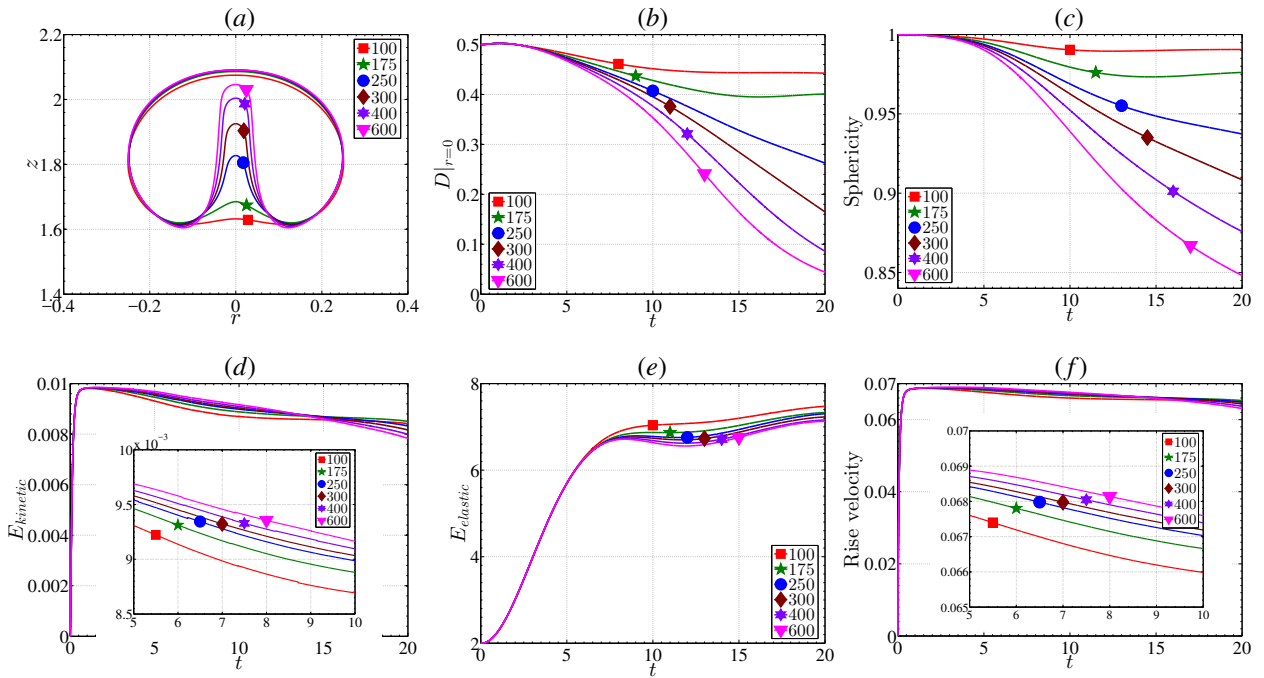
\begin{figure*}[ht!]
\begin{center}
\unitlength1cm
\begin{picture}(14.5,8.3)
\put(1.9,6.2){\makebox(0,0){\includegraphics[width=8.5cm,height=4.2cm,keepaspectratio]{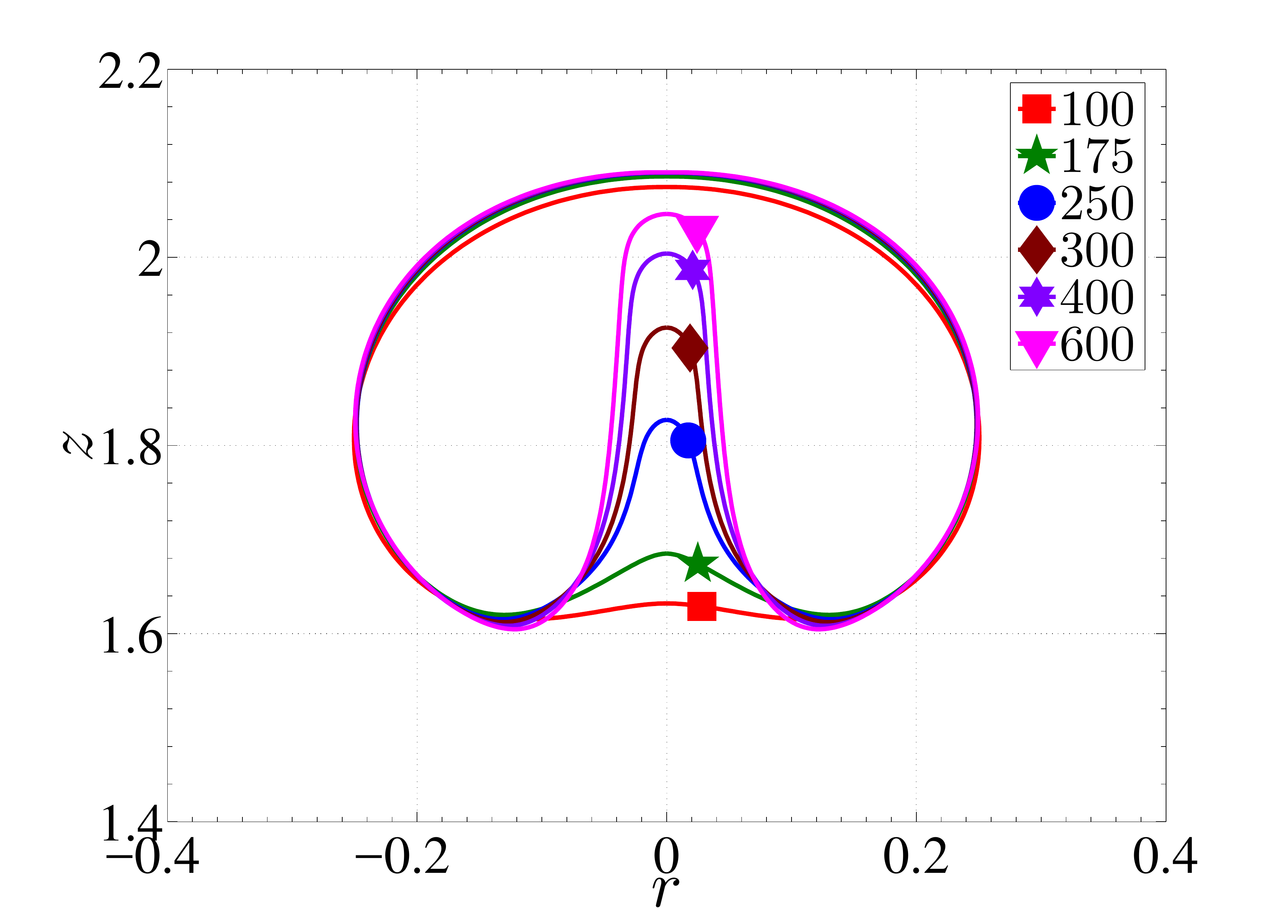}}}
\put(7.4,6.2){\makebox(0,0){\includegraphics[width=8.5cm,height=4.2cm,keepaspectratio]{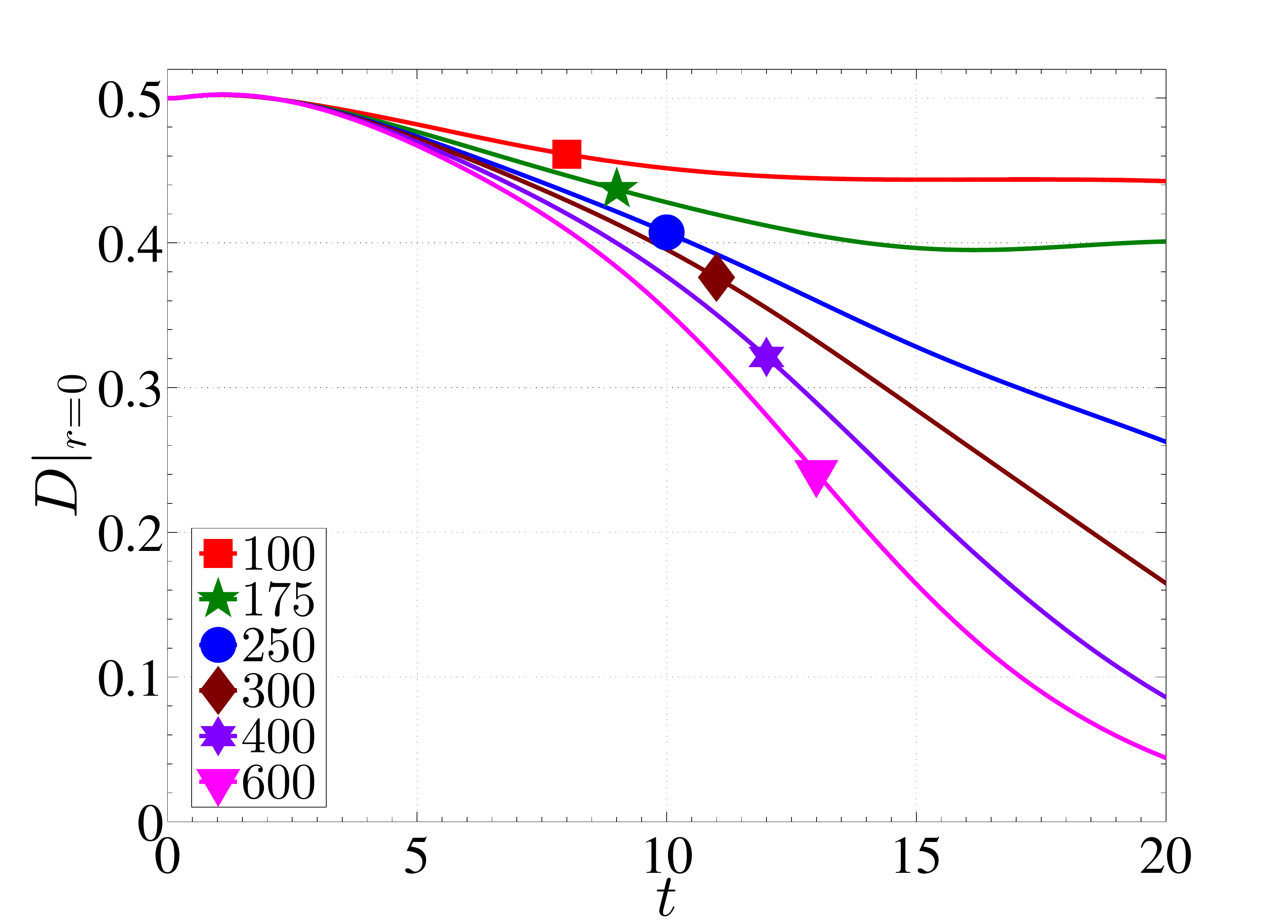}}}
\put(12.9,6.2){\makebox(0,0){\includegraphics[width=8.5cm,height=4.2cm,keepaspectratio]{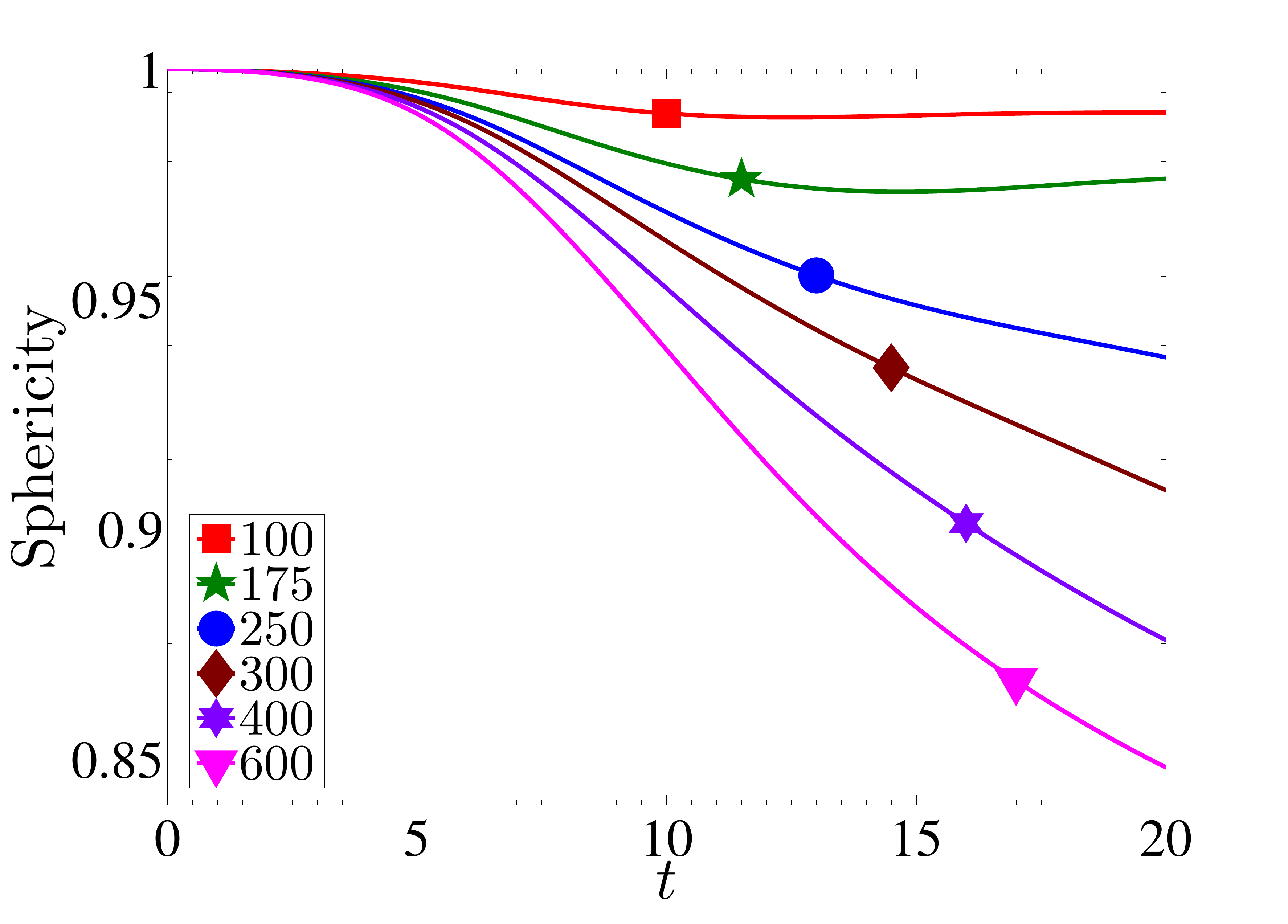}}}
\put(1.9,1.6){\makebox(0,0){\includegraphics[width=8.5cm,height=4.2cm,keepaspectratio]{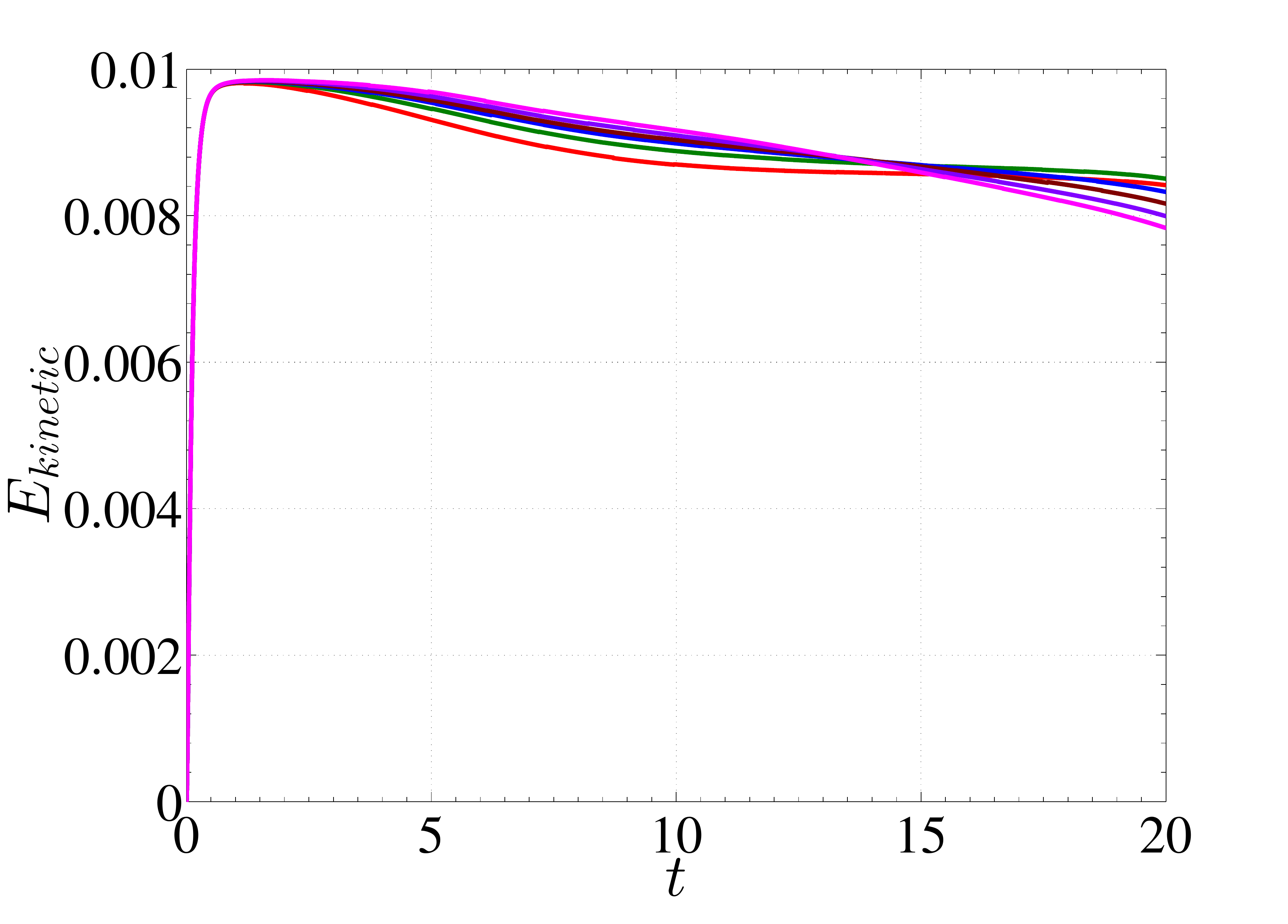}}}
\put(2.05,1.4){\makebox(0,0){\includegraphics[width=8.5cm,height=2.7cm,keepaspectratio]{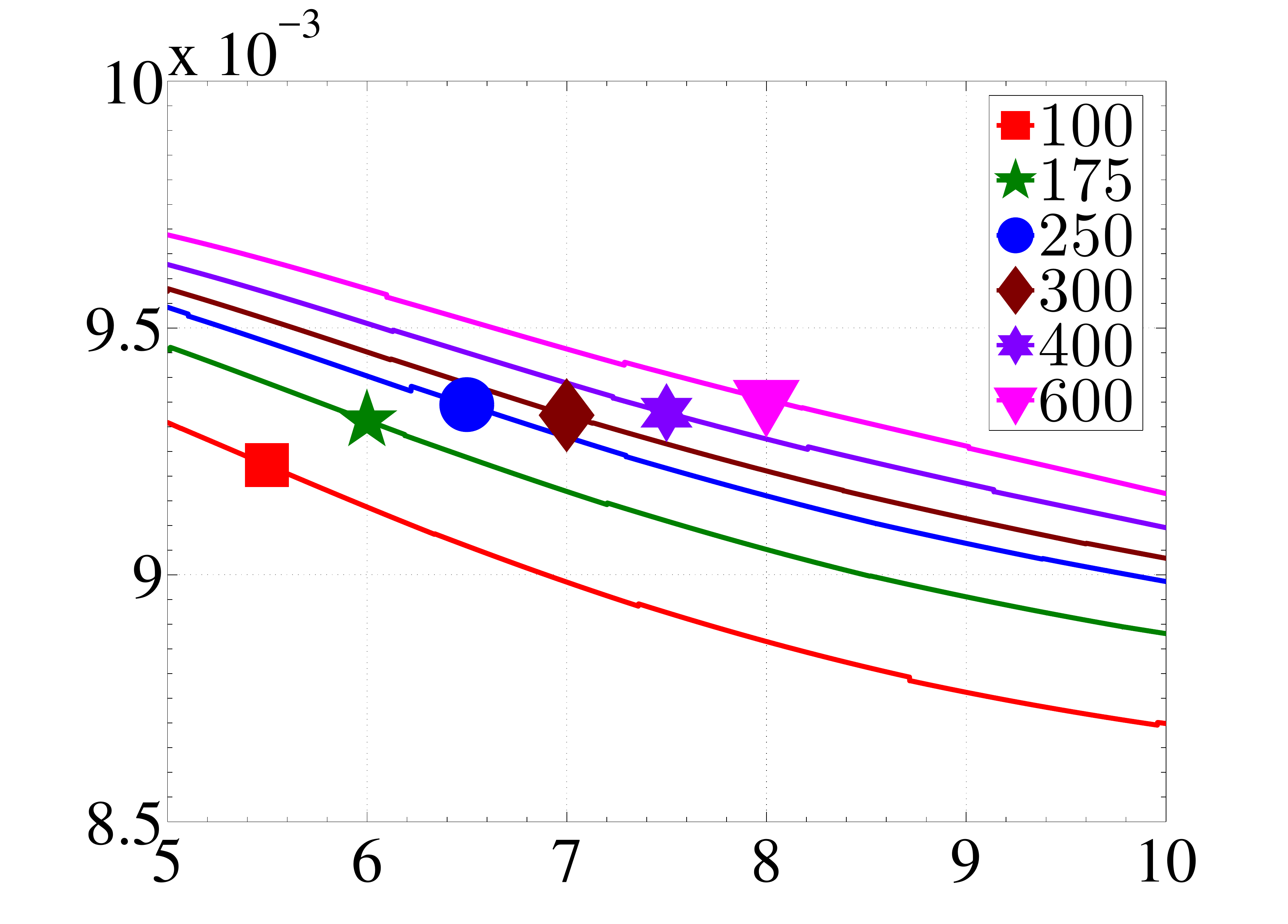}}}
\put(7.4,1.6){\makebox(0,0){\includegraphics[width=8.5cm,height=4.2cm,keepaspectratio]{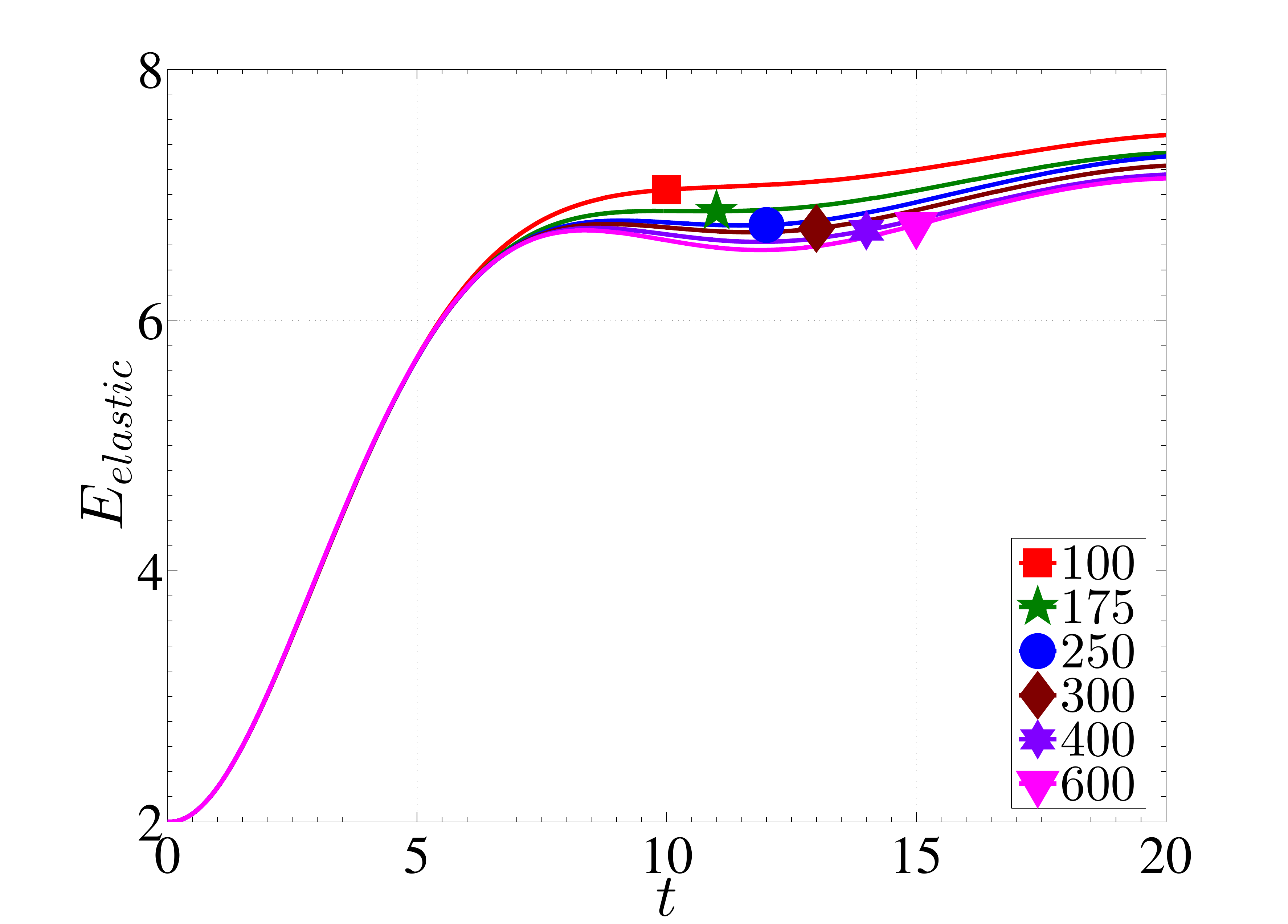}}}
\put(12.9,1.6){\makebox(0,0){\includegraphics[width=8.5cm,height=4.2cm,keepaspectratio]{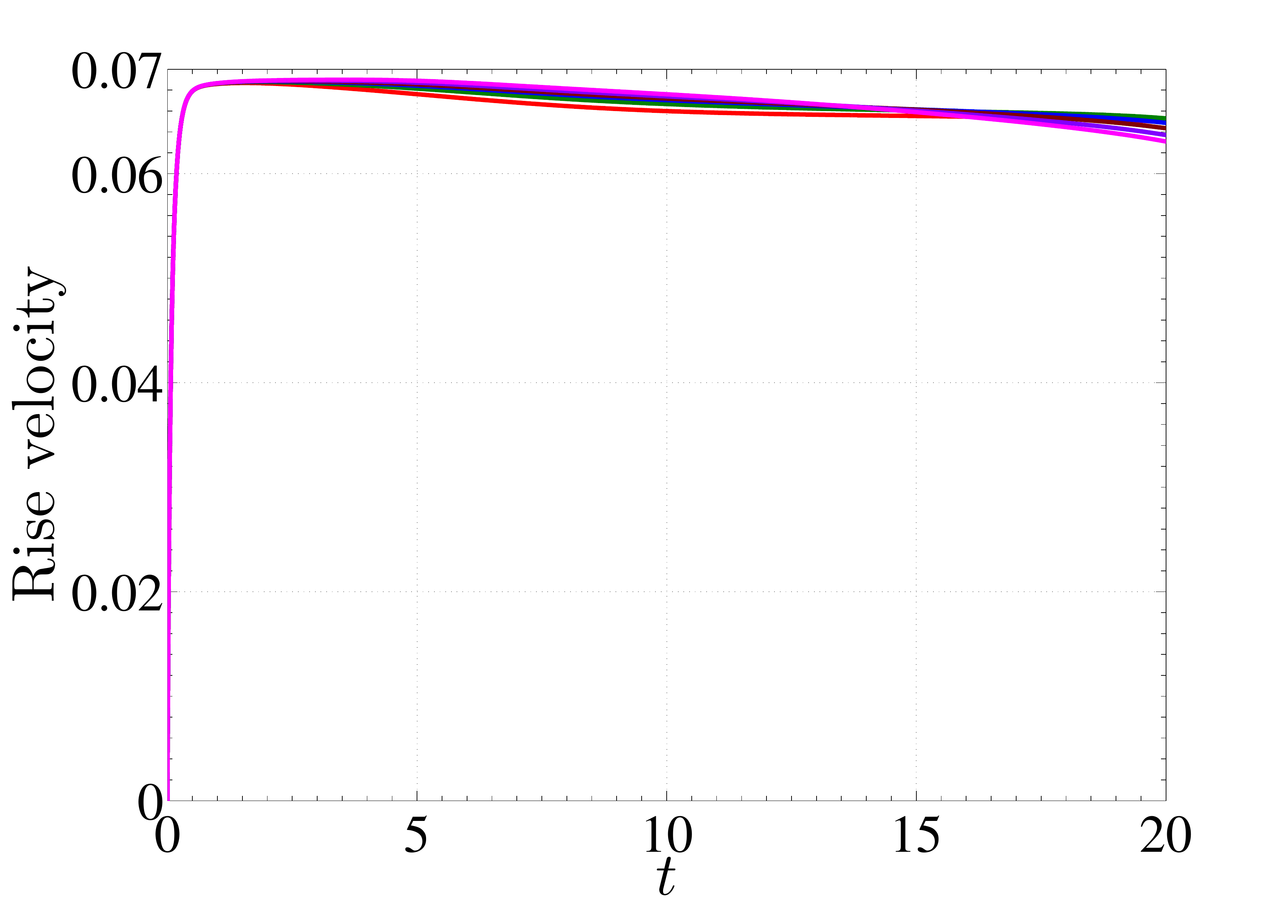}}}
\put(13.15,1.55){\makebox(0,0){\includegraphics[width=8.5cm,height=2.85cm,keepaspectratio]{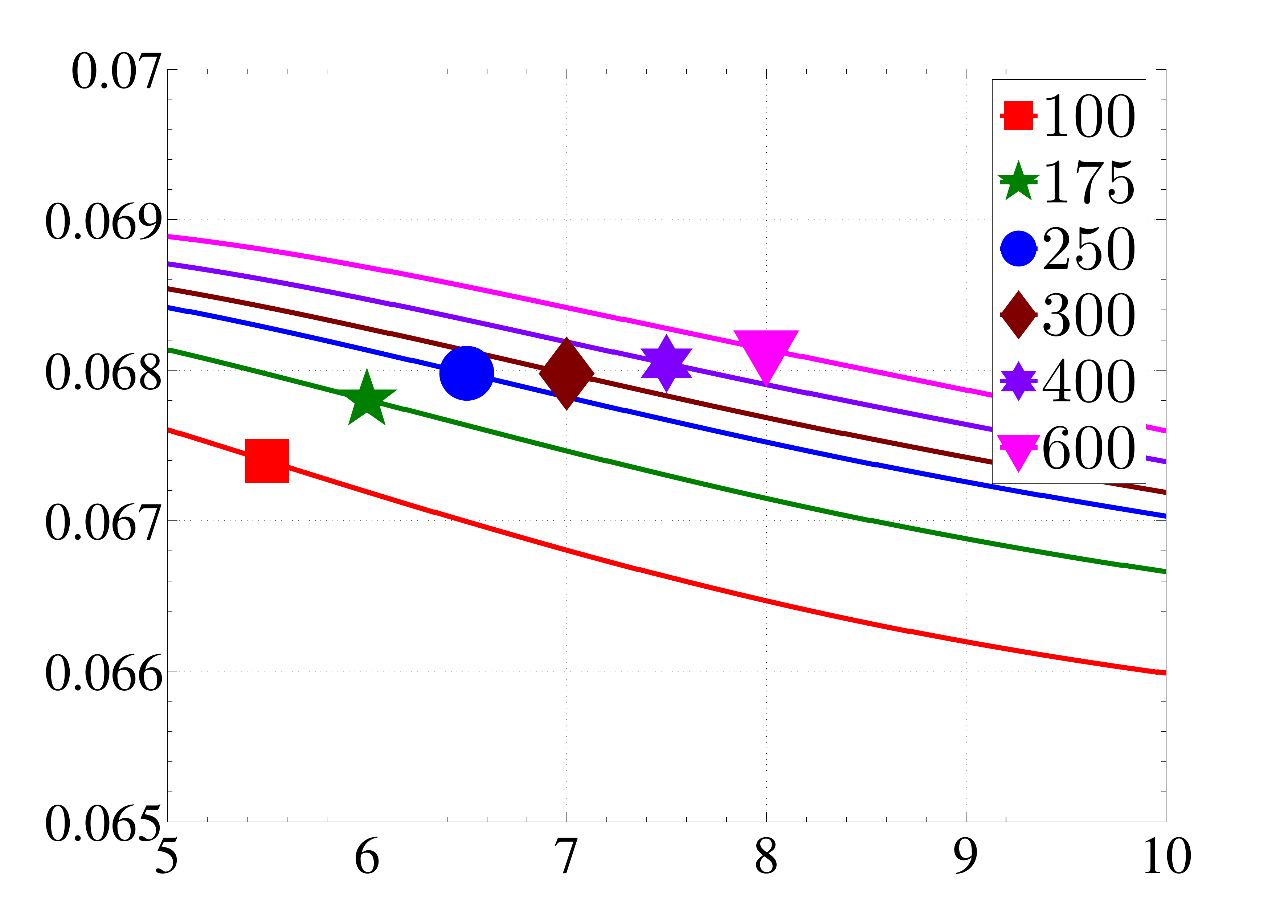}}}
\put(1.85,8.1){$(a)$}
\put(7.3, 8.1){$(b)$}
\put(12.8, 8.1){$(c)$}
\put(1.85,3.5){$(d)$}
\put(7.3, 3.5){$(e)$}
\put(12.8, 3.5){$(f)$}
\end{picture}
\end{center}
\caption{Influence of E\"{o}tv\"{o}s number for a viscoelastic bubble rising in a Newtonian fluid column~: (a)~bubble shape at $t$~=~20, (b)~diameter of the bubble at $r$~=~0, (c)~sphericity, (d)~kinetic energy, (e)~elastic energy and (f)~rise velocity of the bubble for different E\"{o}tv\"{o}s numbers (i)~Eo~=~100, (ii)~Eo~=~175, (iii)~Eo~=~250, (iv)~Eo~=~300, (v)~Eo~=~400 and (vi)~Eo~=~600  with $\Rey_2$~=~10, $\text{Wi}_1$~=~10, $\rho_1/\rho_2$~=~0.1, $\varepsilon$~=~2, $\beta_1$~=~0.5, $\beta_2$~=~1.0, $\alpha_1$~=~0.1, $D$~=~0.5 and $h_c$~=~2.5.}
\label{Plots_WeberEffect_VN}
\end{figure*}

In this section, we study the influence of E\"{o}tv\"{o}s number on the rising viscoelastic bubble dynamics in a Newtonian fluid column.
We consider the base case flow parameters and vary only the E\"{o}tv\"{o}s number, i.e. vary the interfacial tension.
Six different values are used for the E\"{o}tv\"{o}s number in this study, which are as follows~: (i)~Eo~=~100, (ii)~Eo~=~175, (iii)~Eo~=~250, (iv)~Eo~=~300, (v)~Eo~=~400 and (vi)~Eo~=~600.
Increasing the E\"{o}tv\"{o}s number, decreases the interfacial tension, thereby making the interface more easily deformable.
Thus, from Fig.~\ref{Plots_WeberEffect_VN}(a) we can observe that at high E\"{o}tv\"{o}s numbers, the bubble is more dimpled.
In fact at low E\"{o}tv\"{o}s numbers, the bubble shapes are more similar to a Newtonian bubble rising in a Newtonian fluid column.
With further advancement in time, bubbles with low  E\"{o}tv\"{o}s numbers  still do not deform as observed with high E\"{o}tv\"{o}s numbers.
This is due to the fact that there exists a critical capillary number, beyond which the bubble experiences unsteady deformations in the form of a dimpled shape.
From Fig.~\ref{Plots_WeberEffect_VN}(a), we can comment that the critical E\"{o}tv\"{o}s number  for unsteady drop shapes is between 175 and 250 for the considered flow parameters.
Since the trailing end of the bubble is pulled more with an increase in the E\"{o}tv\"{o}s number, the diameter of the bubble at the axis of symmetry and the sphericity of the bubble decreases, see Fig.~\ref{Plots_WeberEffect_VN}(b) and (c), respectively.
Further, Fig.~\ref{Plots_WeberEffect_VN}(e) presents the temporal evolution of the elastic energy in the bubble. 
Till around $t$~=~6, there is no effect of E\"{o}tv\"{o}s number on the elastic energy in the bubble.
However, after that the magnitude of increase in the elastic energy of the bubble decreases with an increase in the E\"{o}tv\"{o}s number.

\section{Summary and observations}\label{Summary}
A finite element scheme using the arbitrary Lagrangian-Eulerian~(ALE) approach was presented for computations of 3D-axisymmetric viscoelastic two-phase flows.
The coupled Navier--Stokes and the Giesekus constitutive equations which describe the viscoelastic flow dynamics were solved monolithically  using the proposed numerical scheme.
The highlights of the numerical scheme are the tangential gradient operator technique for the curvature approximation with semi-implicit treatment, the ALE approach with moving meshes to track the interface, 3D-axisymmetric variational form using cylindrical coordinates and the three-field local projection stabilized formulation. 
This stabilized scheme had allowed to use equal order interpolation for the velocity and the viscoelastic stress, whereas inf-stable finite elements were used for the velocity and the pressure.
First order implicit Euler method was used for the time discretization.
Further, the linear elastic mesh update technique was used to displace the inner mesh points of the computational domain and it avoids quick distortion of the mesh.

The numerical scheme was first validated for a 2D planar Newtonian bubble rising in a Newtonian fluid column using benchmark results in the literature.
Next, a grid independence test was performed for the benchmark configuration to obtain a suitable mesh for grid-independent numerical solutions. 
A comprehensive numerical investigation was performed for a Newtonian bubble rising in a viscoelastic fluid and a viscoelastic bubble rising in a Newtonian fluid.
The effects of the viscosity ratio, Newtonian solvent ratio, Giesekus mobility factor and E\"{o}tv\"{o}s number on the rising bubble dynamics were analyzed.
The observations are summarized as follows~:
The numerical study showed that beyond a critical E\"{o}tv\"{o}s number, a Newtonian bubble rising in a viscoelastic fluid experiences an extended trailing edge with a cusp-like shape.
For interface flows with high viscosity ratios / low Newtonian solvent ratio / low Giesekus mobility factor / high E\"{o}tv\"{o}s numbers, the effect of viscoelasticity increases leading to an even longer and sharper trailing edge.
Further, we had observed a negative wake phenomena where the velocity at the vicinity of the trailing end is in the direction of the bubble but slightly further away from the trailing end the velocity reverses its direction. 
Next, when a viscoelastic bubble rises in a Newtonian fluid we had observed an indentation around the rear stagnation point with a dimpled shape.
With low viscosity ratios / low Newtonian solvent ratio / low Giesekus mobility factor / high E\"{o}tv\"{o}s numbers, the effect of viscoelasticity increases leading to the rear end of the bubble being pulled up more.

\section*{Acknowledgements}
The work of Jagannath Venkatesan is supported by the Tata Consultancy Services~(TCS), India through the TCS Research Scholarship Program.

\bibliographystyle{model1-num-names}
\bibliography{masterlit}

\begin{thebibliography}{58}
\expandafter\ifx\csname natexlab\endcsname\relax\def\natexlab#1{#1}\fi
\providecommand{\url}[1]{\texttt{#1}}
\providecommand{\href}[2]{#2}
\providecommand{\path}[1]{#1}
\providecommand{\DOIprefix}{doi:}
\providecommand{\ArXivprefix}{arXiv:}
\providecommand{\URLprefix}{URL: }
\providecommand{\Pubmedprefix}{pmid:}
\providecommand{\doi}[1]{\href{http://dx.doi.org/#1}{\path{#1}}}
\providecommand{\Pubmed}[1]{\href{pmid:#1}{\path{#1}}}
\providecommand{\bibinfo}[2]{#2}
\ifx\xfnm\relax \def\xfnm[#1]{\unskip,\space#1}\fi
%Type = Article
\bibitem[{Liu et~al.(1995)Liu, Liao, and Joseph}]{Liu95}
\bibinfo{author}{Y.~J. Liu}, \bibinfo{author}{T.~Y. Liao},
  \bibinfo{author}{D.~D. Joseph},
\newblock \bibinfo{title}{A two-dimensional cusp at the trailing edge of an air
  bubble rising in a viscoelastic liquid},
\newblock \bibinfo{journal}{J. Fluid Mech.} \bibinfo{volume}{304}
  (\bibinfo{year}{1995}) \bibinfo{pages}{321--342}.
%Type = Article
\bibitem[{Sostarecz and Belmonte(2003)}]{Sostarecz03}
\bibinfo{author}{M.~C. Sostarecz}, \bibinfo{author}{A.~Belmonte},
\newblock \bibinfo{title}{Motion and shape of a viscoelastic drop falling
  through a viscous fluid},
\newblock \bibinfo{journal}{J. Fluid Mech.} \bibinfo{volume}{497}
  (\bibinfo{year}{2003}) \bibinfo{pages}{235--252}.
%Type = Article
\bibitem[{Pilz and Brenn(2007)}]{Pilz07}
\bibinfo{author}{C.~Pilz}, \bibinfo{author}{G.~Brenn},
\newblock \bibinfo{title}{On the critical bubble volume at the rise velocity
  jump discontinuity in viscoelastic liquids},
\newblock \bibinfo{journal}{J. Non-Newtonian Fluid Mech.} \bibinfo{volume}{145}
  (\bibinfo{year}{2007}) \bibinfo{pages}{124--138}.
%Type = Article
\bibitem[{Amirnia et~al.(2013)Amirnia, de~Bruyn, Bergougnou, and
  Margaritis}]{Amirnia13}
\bibinfo{author}{S.~Amirnia}, \bibinfo{author}{J.~R. de~Bruyn},
  \bibinfo{author}{M.~A. Bergougnou}, \bibinfo{author}{A.~Margaritis},
\newblock \bibinfo{title}{Continuous rise velocity of air bubbles in
  non-{N}ewtonian biopolymer solutions},
\newblock \bibinfo{journal}{Chemical Engineering Science} \bibinfo{volume}{94}
  (\bibinfo{year}{2013}) \bibinfo{pages}{60--68}.
%Type = Article
\bibitem[{Xu et~al.(2017)Xu, Zhang, Liu, Wang, Wei, and Liu}]{Xu17}
\bibinfo{author}{X.~Xu}, \bibinfo{author}{J.~Zhang}, \bibinfo{author}{F.~Liu},
  \bibinfo{author}{X.~Wang}, \bibinfo{author}{W.~Wei},
  \bibinfo{author}{Z.~Liu},
\newblock \bibinfo{title}{Rising behavior of single bubble in infinite stagnant
  non-{N}ewtonian liquids},
\newblock \bibinfo{journal}{International Journal of Muliphase Flow}
  \bibinfo{volume}{95} (\bibinfo{year}{2017}) \bibinfo{pages}{84--90}.
%Type = Article
\bibitem[{Izbassarov and Muradoglu(2015)}]{Daulet15}
\bibinfo{author}{D.~Izbassarov}, \bibinfo{author}{M.~Muradoglu},
\newblock \bibinfo{title}{A front-tracking method for computational modeling of
  viscoelastic two-phase flow systems},
\newblock \bibinfo{journal}{J. Non-Newtonian Fluid Mech.} \bibinfo{volume}{223}
  (\bibinfo{year}{2015}) \bibinfo{pages}{122--140}.
%Type = Article
\bibitem[{Figueiredo et~al.(2016)Figueiredo, Oishi, Afonso, Tasso, and
  Cuminato}]{Figueiredo16}
\bibinfo{author}{R.~A. Figueiredo}, \bibinfo{author}{C.~M. Oishi},
  \bibinfo{author}{A.~M. Afonso}, \bibinfo{author}{I.~V.~M. Tasso},
  \bibinfo{author}{J.~A. Cuminato},
\newblock \bibinfo{title}{A two-phase solver for complex fluids: {S}tudies of
  the {W}eissenberg effect},
\newblock \bibinfo{journal}{Int. J. Multiphase Flow} \bibinfo{volume}{84}
  (\bibinfo{year}{2016}) \bibinfo{pages}{98--115}.
%Type = Article
\bibitem[{Walters and Phillips(2016)}]{Walters16}
\bibinfo{author}{M.~J. Walters}, \bibinfo{author}{T.~N. Phillips},
\newblock \bibinfo{title}{A non-singular boundary element method for modelling
  bubble dynamics in viscoelastic fluids},
\newblock \bibinfo{journal}{J. Non-Newtonian Fluid Mech.} \bibinfo{volume}{235}
  (\bibinfo{year}{2016}) \bibinfo{pages}{109--124}.
%Type = Article
\bibitem[{Habla et~al.(2011)Habla, Marschall, Hinrichsen, Dietsche, Jasak, and
  Favero}]{Habla11}
\bibinfo{author}{F.~Habla}, \bibinfo{author}{H.~Marschall},
  \bibinfo{author}{O.~Hinrichsen}, \bibinfo{author}{L.~Dietsche},
  \bibinfo{author}{H.~Jasak}, \bibinfo{author}{J.~L. Favero},
\newblock \bibinfo{title}{Numerical simulation of viscoelastic two-phase flows
  using open{FOAM}},
\newblock \bibinfo{journal}{Chemical Engineering Science} \bibinfo{volume}{66}
  (\bibinfo{year}{2011}) \bibinfo{pages}{5487--5496}.
%Type = Article
\bibitem[{Zainali et~al.(2013)Zainali, Tofighi, Shadloo, and
  Yildiz}]{Zainali13}
\bibinfo{author}{A.~Zainali}, \bibinfo{author}{N.~Tofighi},
  \bibinfo{author}{M.~S. Shadloo}, \bibinfo{author}{M.~Yildiz},
\newblock \bibinfo{title}{Numerical investigation of {N}ewtonian and
  non-{N}ewtonian multiphase flows using {ISPH} method},
\newblock \bibinfo{journal}{Comput. Methods Appl. Mech. Engrg.}
  \bibinfo{volume}{254} (\bibinfo{year}{2013}) \bibinfo{pages}{99--113}.
%Type = Article
\bibitem[{Oldroyd(1950)}]{Oldroyd}
\bibinfo{author}{J.~G. Oldroyd},
\newblock \bibinfo{title}{On the formulation of rheological equations of
  state},
\newblock \bibinfo{journal}{Proc. R. Soc. Lond. A} \bibinfo{volume}{200}
  (\bibinfo{year}{1950}) \bibinfo{pages}{523--541}.
%Type = Article
\bibitem[{Giesekus(1982)}]{Giesekus}
\bibinfo{author}{H.~Giesekus},
\newblock \bibinfo{title}{A simple constitutive equation for polymeric fluids
  based on the concept of deformation-dependent tensorial mobility},
\newblock \bibinfo{journal}{J. Non-Newtonian Fluid Mech.} \bibinfo{volume}{11}
  (\bibinfo{year}{1982}) \bibinfo{pages}{69--109}.
%Type = Article
\bibitem[{Bird et~al.(1980)Bird, Dotson, and Johnson}]{FENEP}
\bibinfo{author}{R.~B. Bird}, \bibinfo{author}{P.~J. Dotson},
  \bibinfo{author}{N.~L. Johnson},
\newblock \bibinfo{title}{Polymer solution rheology based on a finitely
  extensible bead-spring chain model},
\newblock \bibinfo{journal}{J. Non-Newtonian Fluid Mech.} \bibinfo{volume}{7}
  (\bibinfo{year}{1980}) \bibinfo{pages}{213--235}.
%Type = Article
\bibitem[{Chilcott and Rallison(1988)}]{FENECR}
\bibinfo{author}{M.~D. Chilcott}, \bibinfo{author}{J.~M. Rallison},
\newblock \bibinfo{title}{Creeping flow of dilute polymer solutions past
  cylinders and spheres},
\newblock \bibinfo{journal}{J. Non-Newtonian Fluid Mech.} \bibinfo{volume}{29}
  (\bibinfo{year}{1988}) \bibinfo{pages}{381--432}.
%Type = Article
\bibitem[{Thien and Tanner(1977)}]{PTT}
\bibinfo{author}{N.~P. Thien}, \bibinfo{author}{R.~I. Tanner},
\newblock \bibinfo{title}{A new constitutive equation derived from network
  theory},
\newblock \bibinfo{journal}{J. Non-Newtonian Fluid Mech.} \bibinfo{volume}{2}
  (\bibinfo{year}{1977}) \bibinfo{pages}{353--365}.
%Type = Article
\bibitem[{Verbeeten et~al.(2001)Verbeeten, Peters, and Baaijens}]{XPP}
\bibinfo{author}{W.~M.~H. Verbeeten}, \bibinfo{author}{G.~W.~M. Peters},
  \bibinfo{author}{F.~P.~T. Baaijens},
\newblock \bibinfo{title}{Differential constitutive equations for polymer
  melts: The extended {P}om-{P}om model},
\newblock \bibinfo{journal}{J. Rheol.} \bibinfo{volume}{45}
  (\bibinfo{year}{2001}) \bibinfo{pages}{823--843}.
%Type = Article
\bibitem[{Pillapakkam and Singh(2001)}]{Pillapakkam01}
\bibinfo{author}{S.~B. Pillapakkam}, \bibinfo{author}{P.~Singh},
\newblock \bibinfo{title}{A {L}evel-{S}et {M}ethod for {Co}mputing {S}olutions
  to {V}iscoelastic {T}wo-{P}hase {F}low},
\newblock \bibinfo{journal}{J. Comp. Phys.} \bibinfo{volume}{174}
  (\bibinfo{year}{2001}) \bibinfo{pages}{552--578}.
%Type = Article
\bibitem[{Pillapakkam et~al.(2007)Pillapakkam, Singh, Blackmore, and
  Aubry}]{Pillapakkam07}
\bibinfo{author}{S.~B. Pillapakkam}, \bibinfo{author}{P.~Singh},
  \bibinfo{author}{D.~Blackmore}, \bibinfo{author}{N.~Aubry},
\newblock \bibinfo{title}{Transient and steady state of a rising bubble in a
  viscoelastic fluid},
\newblock \bibinfo{journal}{J. Fluid Mech.} \bibinfo{volume}{589}
  (\bibinfo{year}{2007}) \bibinfo{pages}{215--252}.
%Type = Article
\bibitem[{Chinyoka et~al.(2005)Chinyoka, Renardy, Renardy, and
  Khismatullin}]{Chinyoka05}
\bibinfo{author}{T.~Chinyoka}, \bibinfo{author}{Y.~Y. Renardy},
  \bibinfo{author}{M.~Renardy}, \bibinfo{author}{D.~B. Khismatullin},
\newblock \bibinfo{title}{Two-dimensional study of drop deformation under
  simple shear for {O}ldroyd-{B} liquids},
\newblock \bibinfo{journal}{J. Non-Newtonian Fluid Mech.} \bibinfo{volume}{130}
  (\bibinfo{year}{2005}) \bibinfo{pages}{45--56}.
%Type = Article
\bibitem[{Harvie et~al.(2008)Harvie, Cooper-White, and Davidson}]{Harvie08}
\bibinfo{author}{D.~J.~E. Harvie}, \bibinfo{author}{J.~J. Cooper-White},
  \bibinfo{author}{M.~R. Davidson},
\newblock \bibinfo{title}{Deformation of a viscoelastic droplet passing through
  a microfluidic contraction},
\newblock \bibinfo{journal}{J. Non-Newtonian Fluid Mech.} \bibinfo{volume}{155}
  (\bibinfo{year}{2008}) \bibinfo{pages}{67--79}.
%Type = Article
\bibitem[{Yue et~al.(2005)Yue, Feng, Liu, and Shen}]{Yue05}
\bibinfo{author}{P.~Yue}, \bibinfo{author}{J.~J. Feng},
  \bibinfo{author}{C.~Liu}, \bibinfo{author}{J.~Shen},
\newblock \bibinfo{title}{Diffuse-interface simulations of drop coalescence and
  retraction in viscoelastic fluids},
\newblock \bibinfo{journal}{J. Non-Newtonian Fluid Mech.} \bibinfo{volume}{129}
  (\bibinfo{year}{2005}) \bibinfo{pages}{163--176}.
%Type = Article
\bibitem[{Yue et~al.(2006)Yue, Zhou, Feng, Ollivier-Gooch, and Hu}]{Yue06}
\bibinfo{author}{P.~Yue}, \bibinfo{author}{C.~Zhou}, \bibinfo{author}{J.~J.
  Feng}, \bibinfo{author}{C.~F. Ollivier-Gooch}, \bibinfo{author}{H.~H. Hu},
\newblock \bibinfo{title}{Phase-field simulations of interfacial dynamics in
  viscoelastic fluids using finite elements with adaptive meshing},
\newblock \bibinfo{journal}{J. Comp. Phys.} \bibinfo{volume}{219}
  (\bibinfo{year}{2006}) \bibinfo{pages}{47--67}.
%Type = Article
\bibitem[{Zhang et~al.(2010)Zhang, Wang, and Tang}]{Zhang10}
\bibinfo{author}{Y.~Zhang}, \bibinfo{author}{H.~Wang},
  \bibinfo{author}{T.~Tang},
\newblock \bibinfo{title}{Simulating {T}wo-{P}hase {V}iscoelastic {F}lows
  {U}sing {M}oving {F}inite {E}lement {M}ethods},
\newblock \bibinfo{journal}{Commun. Comput. Phys.} \bibinfo{volume}{7}
  (\bibinfo{year}{2010}) \bibinfo{pages}{333--349}.
%Type = Article
\bibitem[{You et~al.(2008)You, Borhan, and Haj-Hariri}]{You08}
\bibinfo{author}{R.~You}, \bibinfo{author}{A.~Borhan},
  \bibinfo{author}{H.~Haj-Hariri},
\newblock \bibinfo{title}{A finite volume formulation for simulating drop
  motion in a viscoelastic two-phase system},
\newblock \bibinfo{journal}{J. Non-Newtonian Fluid Mech.} \bibinfo{volume}{153}
  (\bibinfo{year}{2008}) \bibinfo{pages}{109--129}.
%Type = Article
\bibitem[{You et~al.(2009)You, Haj-Hariri, and Borhan}]{You09}
\bibinfo{author}{R.~You}, \bibinfo{author}{H.~Haj-Hariri},
  \bibinfo{author}{A.~Borhan},
\newblock \bibinfo{title}{Confined drop motion in viscoelastic two-phase
  systems},
\newblock \bibinfo{journal}{Physics of Fluids} \bibinfo{volume}{21}
  (\bibinfo{year}{2009}) \bibinfo{pages}{013102}.
%Type = Article
\bibitem[{Chung et~al.(2008)Chung, Hulsen, Kim, Ahn, and Lee}]{Chung08}
\bibinfo{author}{C.~Chung}, \bibinfo{author}{M.~A. Hulsen},
  \bibinfo{author}{J.~M. Kim}, \bibinfo{author}{K.~H. Ahn},
  \bibinfo{author}{S.~J. Lee},
\newblock \bibinfo{title}{Numerical study on the effect of viscoelasticity on
  drop deformation in simple shear and 5:1:5 planar contraction/expansion
  microchannel},
\newblock \bibinfo{journal}{J. Non-Newtonian Fluid Mech.} \bibinfo{volume}{155}
  (\bibinfo{year}{2008}) \bibinfo{pages}{80--93}.
%Type = Article
\bibitem[{Chung et~al.(2009)Chung, Kim, Hulsen, Ahn, and Lee}]{Chung09}
\bibinfo{author}{C.~Chung}, \bibinfo{author}{J.~M. Kim}, \bibinfo{author}{M.~A.
  Hulsen}, \bibinfo{author}{K.~H. Ahn}, \bibinfo{author}{S.~J. Lee},
\newblock \bibinfo{title}{Effect of viscoelasticity on drop dynamics in 5:1:5
  contraction/expansion microchannel flow},
\newblock \bibinfo{journal}{Chemical Engineering Science} \bibinfo{volume}{64}
  (\bibinfo{year}{2009}) \bibinfo{pages}{4515--4524}.
%Type = Article
\bibitem[{Mukherjee and Sarkar(2010)}]{Mukherjee10}
\bibinfo{author}{S.~Mukherjee}, \bibinfo{author}{K.~Sarkar},
\newblock \bibinfo{title}{Effects of viscoelasticity on the retraction of a
  sheared drop},
\newblock \bibinfo{journal}{J. Non-Newtonian Fluid Mech.} \bibinfo{volume}{165}
  (\bibinfo{year}{2010}) \bibinfo{pages}{340--349}.
%Type = Article
\bibitem[{Mukherjee and Sarkar(2011)}]{Mukherjee11}
\bibinfo{author}{S.~Mukherjee}, \bibinfo{author}{K.~Sarkar},
\newblock \bibinfo{title}{Viscoelastic drop falling through a viscous medium},
\newblock \bibinfo{journal}{Physics of Fluids} \bibinfo{volume}{23}
  (\bibinfo{year}{2011}) \bibinfo{pages}{013101}.
%Type = Article
\bibitem[{.Vahabi and Sadeghy(2014)}]{Vahabi14}
\bibinfo{author}{M.~.Vahabi}, \bibinfo{author}{K.~Sadeghy},
\newblock \bibinfo{title}{On the {U}se of {SPH} {M}ethod for {S}imulating {G}as
  {B}ubbles {R}ising in {V}iscoelastic {L}iquids},
\newblock \bibinfo{journal}{Nihon Reoroji Gakkaishi} \bibinfo{volume}{42}
  (\bibinfo{year}{2014}) \bibinfo{pages}{309--319}.
%Type = Article
\bibitem[{Lind and Phillips(2010)}]{Lind10}
\bibinfo{author}{S.~J. Lind}, \bibinfo{author}{T.~N. Phillips},
\newblock \bibinfo{title}{The effect of viscoelasticity on a rising gas
  bubble},
\newblock \bibinfo{journal}{J. Non-Newtonian Fluid Mech.} \bibinfo{volume}{165}
  (\bibinfo{year}{2010}) \bibinfo{pages}{852--865}.
%Type = Article
\bibitem[{Izbassarov and Muradoglu(2016)}]{Daulet16}
\bibinfo{author}{D.~Izbassarov}, \bibinfo{author}{M.~Muradoglu},
\newblock \bibinfo{title}{A computational study of two-phase viscoelastic
  systems in a capillary tube with a sudden contraction/expansion},
\newblock \bibinfo{journal}{Physics of Fluids} \bibinfo{volume}{28}
  (\bibinfo{year}{2016}) \bibinfo{pages}{012110}.
%Type = Article
\bibitem[{Ganesan et~al.(2007)Ganesan, Matthies, and Tobiska}]{GMT}
\bibinfo{author}{S.~Ganesan}, \bibinfo{author}{G.~Matthies},
  \bibinfo{author}{L.~Tobiska},
\newblock \bibinfo{title}{On spurious velocities in incompressible flow
  problems with interfaces},
\newblock \bibinfo{journal}{Comput. Methods Appl. Mech. Engrg.}
  \bibinfo{volume}{196} (\bibinfo{year}{2007}) \bibinfo{pages}{1193--1202}.
%Type = Article
\bibitem[{Ganesan(2015)}]{GANJCP15}
\bibinfo{author}{S.~Ganesan},
\newblock \bibinfo{title}{Simulations of impinging droplets with
  surfactant-dependent dynamic contact angle},
\newblock \bibinfo{journal}{J. Comput. Phys.} \bibinfo{volume}{301}
  (\bibinfo{year}{2015}) \bibinfo{pages}{178--200}.
%Type = Article
\bibitem[{Ganesan and Tobiska(2008)}]{GAN07}
\bibinfo{author}{S.~Ganesan}, \bibinfo{author}{L.~Tobiska},
\newblock \bibinfo{title}{An accurate finite element scheme with moving meshes
  for computing 3{D}-axisymmetric interface flows},
\newblock \bibinfo{journal}{Int. J. Numer. Methods Fluids} \bibinfo{volume}{57}
  (\bibinfo{year}{2008}) \bibinfo{pages}{119--138}.
%Type = Article
\bibitem[{Venkatesan and Ganesan(2018)}]{ViscoelasticDropLPS18}
\bibinfo{author}{J.~Venkatesan}, \bibinfo{author}{S.~Ganesan},
\newblock \bibinfo{title}{Computational modeling of impinging viscoelastic
  droplets},
\newblock \bibinfo{journal}{J. Comp. Phys.}  (\bibinfo{year}{2018})
  \bibinfo{pages}{submitted}.
%Type = Article
\bibitem[{Brooks and Hughes(1982)}]{BH82}
\bibinfo{author}{A.~N. Brooks}, \bibinfo{author}{T.~J.~R. Hughes},
\newblock \bibinfo{title}{Streamline upwind/{P}etrov-{G}alerkin formulations
  for convection dominated flows with particular emphasis on the incompressible
  {N}avier-{S}tokes equations},
\newblock \bibinfo{journal}{Comput. Methods Appl. Mech. Eng.}
  \bibinfo{volume}{32} (\bibinfo{year}{1982}) \bibinfo{pages}{199--259}.
%Type = Article
\bibitem[{Guenette and Fortin(1995)}]{Guenette95}
\bibinfo{author}{R.~Guenette}, \bibinfo{author}{M.~Fortin},
\newblock \bibinfo{title}{A new mixed finite element method for computing
  viscoelastic flows},
\newblock \bibinfo{journal}{J. Non-Newtonian Fluid Mech.} \bibinfo{volume}{60}
  (\bibinfo{year}{1995}) \bibinfo{pages}{27--52}.
%Type = Article
\bibitem[{Fortin et~al.(2000)Fortin, Guenette, and Pierre}]{Fortin00}
\bibinfo{author}{A.~Fortin}, \bibinfo{author}{R.~Guenette},
  \bibinfo{author}{R.~Pierre},
\newblock \bibinfo{title}{On the discrete {EVSS} method},
\newblock \bibinfo{journal}{Comput. Methods Appl. Mech. Engrg.}
  \bibinfo{volume}{189} (\bibinfo{year}{2000}) \bibinfo{pages}{121--139}.
%Type = Article
\bibitem[{Forin and Fortin(1989)}]{Fortin89b}
\bibinfo{author}{M.~Forin}, \bibinfo{author}{A.~Fortin},
\newblock \bibinfo{title}{A new approach for the {FEM} simulation of
  viscoelastic flows},
\newblock \bibinfo{journal}{J. Non-Newtonian Fluid Mech.} \bibinfo{volume}{32}
  (\bibinfo{year}{1989}) \bibinfo{pages}{295--310}.
%Type = Article
\bibitem[{Coronado et~al.(2006)Coronado, Arora, Behr, and Pasquali}]{BEHR06}
\bibinfo{author}{O.~M. Coronado}, \bibinfo{author}{D.~Arora},
  \bibinfo{author}{M.~Behr}, \bibinfo{author}{M.~Pasquali},
\newblock \bibinfo{title}{Four-field {G}alerkin/least-squares formulation for
  viscoelastic fluids},
\newblock \bibinfo{journal}{J. Non-Newtonian Fluid Mech.} \bibinfo{volume}{140}
  (\bibinfo{year}{2006}) \bibinfo{pages}{132–--144}.
%Type = Article
\bibitem[{Kwack and Masud(2010)}]{MASUD10}
\bibinfo{author}{J.~Kwack}, \bibinfo{author}{A.~Masud},
\newblock \bibinfo{title}{A three-field formulation for incompressible
  viscoelastic fluids},
\newblock \bibinfo{journal}{Int. J. of Eng. Sci.} \bibinfo{volume}{48}
  (\bibinfo{year}{2010}) \bibinfo{pages}{1413–--1432}.
%Type = Article
\bibitem[{Castillo and Codina(2014)}]{Codina14}
\bibinfo{author}{E.~Castillo}, \bibinfo{author}{R.~Codina},
\newblock \bibinfo{title}{Variational multi-scale stabilized formulations for
  the stationary three-field incompressible viscoelastic flow problem},
\newblock \bibinfo{journal}{Comput. Methods Appl. Mech. Eng.}
  \bibinfo{volume}{279} (\bibinfo{year}{2014}) \bibinfo{pages}{579--605}.
%Type = Article
\bibitem[{Fattal and Kupferman(2005)}]{Fattal05}
\bibinfo{author}{R.~Fattal}, \bibinfo{author}{R.~Kupferman},
\newblock \bibinfo{title}{Time-dependent simulation of viscoelastic flows at
  high {W}eissenberg number using the log-conformation representation},
\newblock \bibinfo{journal}{J. Non-Newtonian Fluid Mech.} \bibinfo{volume}{126}
  (\bibinfo{year}{2005}) \bibinfo{pages}{23--37}.
%Type = Article
\bibitem[{Venkatesan and Ganesan(2017)}]{VJSGLPS17}
\bibinfo{author}{J.~Venkatesan}, \bibinfo{author}{S.~Ganesan},
\newblock \bibinfo{title}{A three-field local projection stabilized formulation
  for computations of {O}ldroyd-{B} viscoelastic fluid flows},
\newblock \bibinfo{journal}{J. Non-Newtonian Fluid Mech.} \bibinfo{volume}{247}
  (\bibinfo{year}{2017}) \bibinfo{pages}{90--106}.
%Type = Incollection
\bibitem[{Shewchuk(1996)}]{TRI96}
\bibinfo{author}{J.~R. Shewchuk},
\newblock \bibinfo{title}{Triangle: {E}ngineering a {2D} {Q}uality {M}esh
  {G}enerator and {D}elaunay {T}riangulator},
\newblock in: \bibinfo{editor}{M.~C. Lin}, \bibinfo{editor}{D.~Manocha} (Eds.),
  \bibinfo{booktitle}{Applied Computational Geometry: Towards Geometric
  Engineering}, volume \bibinfo{volume}{1148} of
  \textit{\bibinfo{series}{Lecture Notes in Computer Science}},
  \bibinfo{publisher}{Springer-Verlag}, \bibinfo{year}{1996}, pp.
  \bibinfo{pages}{203--222}. \bibinfo{note}{From the First ACM Workshop on
  Applied Computational Geometry}.
%Type = Article
\bibitem[{Shewchuk(2002)}]{TRI02}
\bibinfo{author}{J.~R. Shewchuk},
\newblock \bibinfo{title}{Delaunay refinement algorithms for triangular mesh
  generation},
\newblock \bibinfo{journal}{Computational Geometry} \bibinfo{volume}{22}
  (\bibinfo{year}{2002}) \bibinfo{pages}{21–--74}.
%Type = Article
\bibitem[{Becker and Braack(2001)}]{BB01}
\bibinfo{author}{R.~Becker}, \bibinfo{author}{M.~Braack},
\newblock \bibinfo{title}{A finite element pressure gradient stabilization for
  the {S}tokes equations based on local projections},
\newblock \bibinfo{journal}{Calcolo} \bibinfo{volume}{38}
  (\bibinfo{year}{2001}) \bibinfo{pages}{173--199}.
%Type = Incollection
\bibitem[{Becker and Braack(2004)}]{BB04}
\bibinfo{author}{R.~Becker}, \bibinfo{author}{M.~Braack},
\newblock \bibinfo{title}{A two-level stabilization scheme for the
  {N}avier-{S}tokes equations},
\newblock in: \bibinfo{editor}{M.~Feistauer},
  \bibinfo{editor}{V.~Dolej\v{s}\'{\i}}, \bibinfo{editor}{P.~Knobloch},
  \bibinfo{editor}{K.~Najzar} (Eds.), \bibinfo{booktitle}{Numerical mathematics
  and advanced applications}, \bibinfo{publisher}{Springer-Verlag (Berlin)},
  \bibinfo{year}{2004}, pp. \bibinfo{pages}{123--130}.
%Type = Article
\bibitem[{Braack and Burman(2006)}]{BB06}
\bibinfo{author}{M.~Braack}, \bibinfo{author}{E.~Burman},
\newblock \bibinfo{title}{Local projection stabilization for the {O}seen
  problem and its interpretation as a variational multiscale method},
\newblock \bibinfo{journal}{SIAM J. Numer. Anal.} \bibinfo{volume}{43}
  (\bibinfo{year}{2006}) \bibinfo{pages}{2544--2566}.
%Type = Article
\bibitem[{Matthies et~al.(2007)Matthies, Skrzypacz, and Tobiska}]{MAT07}
\bibinfo{author}{G.~Matthies}, \bibinfo{author}{P.~Skrzypacz},
  \bibinfo{author}{L.~Tobiska},
\newblock \bibinfo{title}{A unified convergence analysis for local projection
  stabilisations applied to the {O}seen problem},
\newblock \bibinfo{journal}{Math. Model. Numer. Anal.} \bibinfo{volume}{41}
  (\bibinfo{year}{2007}) \bibinfo{pages}{713--742}.
%Type = Article
\bibitem[{Ganesan et~al.(2008)Ganesan, Matthies, and Tobiska}]{GAN08}
\bibinfo{author}{S.~Ganesan}, \bibinfo{author}{G.~Matthies},
  \bibinfo{author}{L.~Tobiska},
\newblock \bibinfo{title}{Local projection stabilization of equal order
  interpolation applied to the {S}tokes problem},
\newblock \bibinfo{journal}{Math. of Comput.} \bibinfo{volume}{77}
  (\bibinfo{year}{2008}) \bibinfo{pages}{2039--2060}.
%Type = Article
\bibitem[{Ganesan and Tobiska(2010)}]{GAN10}
\bibinfo{author}{S.~Ganesan}, \bibinfo{author}{L.~Tobiska},
\newblock \bibinfo{title}{Stabilization by {L}ocal {P}rojection for
  {C}onvection-{D}iffusion and {I}ncompressible {F}low {P}roblems},
\newblock \bibinfo{journal}{J. Sci. Comput.} \bibinfo{volume}{43}
  (\bibinfo{year}{2010}) \bibinfo{pages}{326--342}.
%Type = Article
\bibitem[{B\"{a}nsch(2001)}]{EB1}
\bibinfo{author}{E.~B\"{a}nsch},
\newblock \bibinfo{title}{Finite element discretization of the
  {N}avier-{S}tokes equations with a free capillary surface},
\newblock \bibinfo{journal}{Numer. Math.} \bibinfo{volume}{88}
  (\bibinfo{year}{2001}) \bibinfo{pages}{203--235}.
%Type = Article
\bibitem[{Amestoy et~al.(2001)Amestoy, Duff, Koster, and L'Excellent}]{MUMPS1}
\bibinfo{author}{P.~R. Amestoy}, \bibinfo{author}{I.~S. Duff},
  \bibinfo{author}{J.~Koster}, \bibinfo{author}{J.-Y. L'Excellent},
\newblock \bibinfo{title}{A fully asynchronous multifrontal solver using
  distributed dynamic scheduling},
\newblock \bibinfo{journal}{SIAM Journal on Matrix Analysis and Applications}
  \bibinfo{volume}{23} (\bibinfo{year}{2001}) \bibinfo{pages}{15--41}.
%Type = Article
\bibitem[{Amestoy et~al.(2006)Amestoy, Guermouche, L'Excellent, and
  Pralet}]{MUMPS2}
\bibinfo{author}{P.~R. Amestoy}, \bibinfo{author}{A.~Guermouche},
  \bibinfo{author}{J.-Y. L'Excellent}, \bibinfo{author}{S.~Pralet},
\newblock \bibinfo{title}{Hybrid scheduling for the parallel solution of linear
  systems},
\newblock \bibinfo{journal}{Parallel Computing} \bibinfo{volume}{32}
  (\bibinfo{year}{2006}) \bibinfo{pages}{136--156}.
%Type = Article
\bibitem[{Wilbrandt et~al.(2017)Wilbrandt, Bartsch, Ahmed, Alia, Anker, Blank,
  Caiazzo, Ganesan, Giere, Matthies, Meesala, Shamim, Venkatesan, and
  John}]{ParMooN1}
\bibinfo{author}{U.~Wilbrandt}, \bibinfo{author}{C.~Bartsch},
  \bibinfo{author}{N.~Ahmed}, \bibinfo{author}{N.~Alia},
  \bibinfo{author}{F.~Anker}, \bibinfo{author}{L.~Blank},
  \bibinfo{author}{A.~Caiazzo}, \bibinfo{author}{S.~Ganesan},
  \bibinfo{author}{S.~Giere}, \bibinfo{author}{G.~Matthies},
  \bibinfo{author}{R.~Meesala}, \bibinfo{author}{A.~Shamim},
  \bibinfo{author}{J.~Venkatesan}, \bibinfo{author}{V.~John},
\newblock \bibinfo{title}{Par{M}oo{N} - {A} modernized program package based on
  mapped finite elements},
\newblock \bibinfo{journal}{Comput. and Maths. with Appl.} \bibinfo{volume}{74}
  (\bibinfo{year}{2017}) \bibinfo{pages}{74--88}.
%Type = Article
\bibitem[{Hysing et~al.(2009)Hysing, Turek, Kuzmin, Parolini, Burman, Ganesan,
  and Tobiska}]{HYS09}
\bibinfo{author}{S.~Hysing}, \bibinfo{author}{S.~Turek},
  \bibinfo{author}{D.~Kuzmin}, \bibinfo{author}{N.~Parolini},
  \bibinfo{author}{E.~Burman}, \bibinfo{author}{S.~Ganesan},
  \bibinfo{author}{L.~Tobiska},
\newblock \bibinfo{title}{Quantitative benchmark computations of
  two-dimensional bubble dynamics},
\newblock \bibinfo{journal}{Int. J. Numer. Meth. Fluids} \bibinfo{volume}{60}
  (\bibinfo{year}{2009}) \bibinfo{pages}{1259--1288}.

\end{thebibliography}

\end{document}